# Phenomenology and Cosmology of Supersymmetric Grand Unified Theories

**Constantinos Pallis**

A Dissertation in Physics
Supervisor: Prof. George Lazarides

*Physics Division*
*School of Technology*
*Aristotle University of Thessaloniki*
*Thessaloniki, GR 540 06, Greece*

**June 2000**

# Summary

The Minimal Supersymmetric Standard Model is the most promising extension of the Standard Model. On the other hand, the Standard Big Bag Cosmology combined with inflation provides a consistent picture for the universe evolution. The combination of these two theories can successfully address the Cold Dark Matter problem. Indeed, the Lightest Supersymmetric Particle is the most plausible Cold Dark Matter candidate.

In this thesis, we calculate the cosmological relic density of a Bino-like Lightest Supersymmetric Particle in the framework of the Minimal Supersymmetric Standard Model. We include annihilation and coannihilation effects of the Bino with the lightest stau and other sleptons, which happen to have comparable masses with it. Coannihilation turns out to be of crucial importance for reducing the Bino relic density to an acceptable level. Requiring the Bino relic density to be in the cosmologically allowed region, derived from the mixed or the pure (in the presence of a nonzero cosmological constant) Cold Dark Matter scenarios for large scale structure formation in the universe, one can restrict the relative mass splitting between the Lightest and the Next-to-Lightest Supersymmetric Particle. Phenomenological constraints, also, result from the inclusion of the supersymmetric corrections to the CP-even Higgs boson and $b$-quark masses and the branching ratio of $b \to s\gamma$. We impose these constraints on the parameter space of two versions of the Minimal Supersymmetric Standard Model, employing radiative electroweak breaking with universal boundary conditions and gauge coupling unification.

In the first version of the model we assume Yukawa coupling unification and boundary conditions from gravity-mediated supersymmetry breaking. For $\mu < 0$, the branching ratio of $b \to s\gamma$ is compatible with data but the $b$-quark mass after including supersymmetric corrections exceeds the experimental limits. The Bino mass can range between 215 GeV and 770 GeV with the lightest stau mass being 8-0% larger. For $\mu > 0$, the predicted $b$-quark mass is experimentally acceptable and there is a sizable fraction of the parameter space allowed by $b \to s\gamma$, where Bino relic density is below the upper bound from Cold Dark Matter considerations.

In the second version of the model we assume boundary conditions from the Hořava-Witten Theory and restrict the parameter space by simultaneously imposing the phenomenological and cosmological constraints. Complete and $t-b$ Yukawa unification can be excluded. Also, $b-\tau$ Yukawa unification is not so favored since it, generally, requires almost degenerate lightest and next-to-lightest sparticle masses. The no Yukawa unification case is the most natural one since it can avoid this degeneracy. The lightest sparticle mass can range between 70 GeV and 670 GeV with the lightest stau mass being 93-0% larger.

# Publications

# Contents

**ΔΙΔΑΚΤΟΡΙΚΗ ΔΙΑΤΡΙΒΗ**

Κωνσταντίνου Πάλλη

# Φαινομενολογία και Κοσμολογία των Υπερσυμμετρικών Ενοποιημένων Θεωριών Βαθμίδας

*Αριστοτέλειο Πανεπιστήμιο Θεσσαλονίκης*
*Πολυτεχνική Σχολή*
*Γενικό Τμήμα*
*Τομέας Φυσικής*

**Ιούνιος 2000**

*Αφιερώνεται στους γονείς μου*

## ΕΥΧΑΡΙΣΤΙΕΣ

Για τους μη μυημένους, και όχι μόνο για αυτούς, το τμήμα των ευχαριστιών είναι από τα πιο ενδιαφέροντα σε μια διδακτορική διατριβή. Ίσως, γιατί είναι το πιο ανθρώπινο σημείο ενός επιστημονικού εγχειρήματος. Δε θα αρνηθώ τη συμμετοχή στη διαδικασία αυτή, αν και πιστεύω ότι τα πραγματικά αισθήματα ξεπερνούν τυπικές διαδικασίες και εκφράζονται, ή μάλλον αποδεικνύονται, αυθόρμητα, πηγαία και καθημερινά με πράξεις ζωής.

Ένα μεγάλο ευχαριστώ θα ήθελα να απευθύνω στον επιβλέποντα καθηγητή μου, κ. Γ. Λαζαρίδη. Η συμβολή του στην κατεύθυνση που πήρε η ερευνητική εργασία, η καθοδήγησή του κατά τη διάρκειά της και η συμπαράστασή του για την έγκαιρη ολοκλήρωση και την όσο το δυνατόν αρτιότερη παρουσίαση της διατριβής ήταν ανεκτίμητες.

Εξίσου θερμές ευχαριστίες θα ήθελα να απευθύνω στον κ. Ν. Βλάχο. Η καθοδήγησή του στη μελέτη της Κβαντικής Θεωρίας Πεδίου και οι εποικοδομητικές παρατηρήσεις του, κατά τη διάρκεια πολύωρων παρουσιάσεών μου στον πίνακα, υπήρξαν πηγές πολύτιμων γνώσεων.

Το ερευνητικό μέρος της διατριβής δε θα ολοκληρωνόταν, όμως, χωρίς τη συνεργασία του M. Gómez. Αισθάνομαι την ανάγκη, και από τη θέση αυτή, να τον ευχαριστήσω για τη βοήθεια και την εμψύχωσή του σε δύσκολες στιγμές της προσπάθειάς μου.

Επίσης, θα ήταν παράλειψη να μην ευχαριστήσω τον κ. Ν. Γκανούλη, για την τεχνική, κυρίως, υποστήριξή του σε θέματα υπολογιστών, τον κ. Π. Πορφυριάδη για την εγκατάσταση εύχρηστων προγραμμάτων, τον κ. Γ. Τσέγα για χρήσιμες συμβουλές προγραμματισμού και τον κ. B. Ananthanarayan για την παροχή διαφωτιστικών εξηγήσεων στην εκκίνηση των αριθμητικών μου αναζητήσεων.

Τέλος, θα ήθελα να ευχαριστήσω τα μέλη της εξεταστικής επιτροπής, Ιων. Βέργαδο και Αθ. Λαχανά, για την υπόδειξη ωφέλιμων διορθώσεων στο κείμενο της διατριβής και τους κυρίους Γ. Ζουπάνο, Κ. Μπαχά, Α. Μπολοβίνο, Γ. Ρίζο, Θ. Τομαρά και Μ. Χατζή για το ενδιαφέρον τους σε διάφορα στάδια των μεταπτυχιακών μου σπουδών.

Οικονομική αρωγή κατά τη διάρκεια της διατριβής μού παρείχε το Ίδρυμα Κρατικών Υποτροφιών (Ι. Κ. Υ) και το πρόγραμμα ΠΕΝΕΔ/95 Κ.Α.1795 του Υπουργείου Παιδείας.

# ΠΕΡΙΛΗΨΗ

Το Ελάχιστα Υπερσυμμετρικό Καθιερωμένο Πρότυπο αποτελεί την πιο δημοφιλή θεωρία για την εξέλιξη της Φυσικής των Στοιχειωδών Σωματιδίων επέκεινα του Καθιερωμένου Πρότυπου. Από την πλευρά της Κοσμολογίας, η Θεωρία της Μεγάλης Έκρηξης με τη συνδρομή του Πληθωρισμού αποτελεί, κατα γενική ομολογία, μια επιτυχημένη προσπάθεια ερμηνείας του Θαύματος της Δημιουργίας. Η αξιοποίηση του θεωρούμενου σωματιδιακού προτύπου με σκοπό τη διευθέτηση του κοσμολογικού προβλήματος της Σκοτεινής Υλής επιτυγχάνεται με την υποψηφιότητα για Ψυχρή Σκοτεινή Υλή του Ελαφρότατου Υπερσυμμετρικού Σωματίου. Στη διατριβή αυτή, υπολογίζεται η παράμετρος κοσμολογικής πυκνότητας ενός Ελαφρότατου Υπερσυμμετρικού Σωματίου μορφής Bino, συμπεριλαμβάνοντας διαδικασίες όχι μόνο αλληλοκαταστροφής αλλά και συγγενικής καταστροφής με τα παρόμοιας μάζας sleptons. Ο συνυπολογισμός των διαδικασιών αυτών είναι σημαντικός για τη μείωση της παραμέτρου κοσμολογικής πυκνότητας στα επιβαλλόμενα όρια από τα σενάρια δομής της Σκοτεινής Ύλης, για ορισμένη περιοχή παραμέτρων του μελετούμενου σωματιδιακού προτύπου. Επιπλέον, φαινομενολογικοί περιορισμοί επιβάλλονται στον παραμετρικό χώρο, οι οποίοι προέρχονται από τις υπερσυμμετρικές διορθώσεις των μαζών του $b$-quark και των μποζονίων Higgs καθώς και από τον ανηγμένο λόγο της διαδικασίας $b \to s\gamma$. Το πλαίσιο μελέτης εφαρμόζεται σε δύο εκδοχές του Ελάχιστα Υπερσυμμετρικού Καθιερωμένου Προτύπου με παγκόσμιες αρχικές συνθήκες, ενοποίηση των ζεύξεων Βαθμίδας και παραβίαση της ηλεκτρασθενούς συμμετρίας μέσω κβαντικών διορθώσεων. Στην πρώτη εκδοχή του προτύπου οι αρχικές συνθήκες προέρχονται από την κατάρρευση της Υπερβαρύτητας και υποτίθεται ενοποίηση των ζεύξεων Yukawa. Η μάζα του Bino μπορεί να κυμαίνεται από 215 GeV ως 770 GeV με τη μάζα του αμέσως βαρύτερου stau να μπορεί να είναι από 8-0% μεγαλύτερη. Η συμβατότητα των φαινομενολογικών περιορισμών, που προέρχονται από τη μάζα του $b$-quark και του ανηγμένου λόγου της διαδικασίας $b \to s\gamma$, επίσης, ελέγχεται. Στη δεύτερη εκδοχή του προτύπου οι αρχικές συνθήκες προέρχονται από τη Θεωρία Hořava-Witten και εξετάζεται η ταυτόχρονη ικανοποίηση των φαινομενολογικών και κοσμολογικών περιορισμών με πλήρη, μερική ή μη ενοποίηση των ζεύξεων Yukawa. Η πλήρης και η $b-t$ ενοποίηση μπορούν να απορριφθούν ενώ η $b-\tau$ απαιτεί σχεδόν απόλυτο εκφυλισμό των μαζών του Bino με το ελαφρύτερο stau. Η πιο φυσική περίπτωση είναι η μη ενοποίηση των ζεύξεων Yukawa, κατά τη οποία η μάζα του ελαφρότατου sparticle μπορεί να κυμαίνεται από 70 GeV ως 670 GeV με τη μάζα του αμέσως βαρύτερου sparticle να μπορεί να είναι από 93-0% μεγαλύτερη.

# Περιεχόμενα

# Κεφάλαιο 1

# Εισαγωγή

Ο τίτλος της διδακτορικής διατριβής που παρουσιάζεται, είναι Φαινομενολογία και Κοσμολογία των Υπερσυμμετρικών Ενοποιημένων Θεωριών Βαθμίδας. Ο τίτλος θα ήταν πιο ακριβής αν αναφερόταν η ταυτότητα των θεωριών βαθμίδας που ελέγχονται και τα φαινομενολογικά και κοσμολογικά κριτήρια που χρησιμοποιούνται. Η παροχή αυτών των επεξηγηματικών διευκρινήσεων καθώς και η παρουσίαση της δομής του κειμένου είναι ο σκοπός αυτού του εισαγωγικού σημειώματος.

Σημείο εκκίνησης κάθε νεωτερισμού στη Σύγχρονη Φυσική Στοιχειωδών Σωματιδίων είναι το Καθιερωμένο Πρότυπο (: SM). Καθιερωμένο ως κοινά αποδεκτό. Βασίζεται στις πειραματικά ελεγμένες θεωρίες των ηλεκτρομαγνητικών (QED), των ασθενών (θεωρία Fermi) και των ισχυρών αλληλεπιδράσεων (QCD). Το πρώτο είδος αλληλεπίδρασης ευθύνεται για διαδικασίες μεταξύ ηλεκτρικά φορτισμένων σωματίων, το δεύτερο για τη μετατροπή νετρονίων σε προτόνια και το τρίτο για διαδικασίες μεταξύ αδρονίων, σωματίων με φορτίο χρώματος. Η επιτυχία του SM συνίσταται σε δύο κυρίως σημεία: Επίτυγχάνεται μια κοινή περιγραφή και των τριών αλληλεπιδράσεων με χρήση μίας λαγκραζιανής αναλοίωτης κάτω από την τοπική συμμετρία $SU(3)_c \times SU(2)_L \times U(1)$ και επιλύεται το πρόβλημα της επανακανονικοποιησιμότητας της θεωρίας των ασθενών αλληλεπιδράσεων με το μηχανισμό Higgs, κατά τον οποίο μερικά από τα μποζόνια βαθμίδας εφοδιάζονται με μάζα με ταυτόχρονη κατάρρευση της αρχικής συμμέτριας στην $SU(3)_c \times U(1)_{EM}$. Ο μηχανισμός Higgs δεν έχει επιβεβαιωθεί πλήρως, γιατί προβλέπει την ύπαρξη βαθμωτού σωματίου το οποίο δεν έχει παρατηρηθεί ακόμα. Εκτός από το σωμάτιο αυτό, μετά την προσφάτως επιτευχθείσα ανίχνευση και του top quark, όλο το υπόλοιπο σωματιδιακό φάσμα του SM έχει επαληθευτεί πειραματικά. Αυτό δομείται από φερμιονικά πεδία ύλης (λεπτόνια και quarks διατεταγμένα σε τρεις γενεές) και τους μποζονικούς διαδότες των αλληλεπιδράσεων.

Παρά την αδιαμφισβήτητα σημαντική επιτυχία του SM, είναι κοινά αποδεκτό ότι δεν μπορεί να αποτελέσει την τελική απάντηση στα προβλήματα της Φυσικής Στοιχειωδών Σωματιδίων. Η βαρύτητα δεν περιέχεται καθόλου στο φορμαλισμό και η ενοποιημένη εικόνα που προσφέρεται για την περιγραφή των υπολοίπων αλληλεπιδράσεων είναι ψευδεπίγραφη, αφού οι σταθερές ζεύξης που σχετίζονται με τις αλληλεπιδράσεις αυτές είναι διαφορέτικές μεταξύ τους. Με σκοπό την άρση του δεύτερου μειονέκτηματος διατυπώθηκαν οι Μεγαλο-Ενοποιημένες Θεωρίες Βαθμίδας (: GUT). Οι θεωρίες αυτές βασίζονται σε ομάδες συμμετρίας οι οποίες περιέχουν αυτή του SM. Με τη υιοθέτηση κάποιας από αυτές τις ομάδες συμμετρίας είναι δυνατόν σε υψηλή κλίμακα ενέργειας, $M_G \sim 10^{16}$ GeV, να επιτευχθεί η περιγραφή και των τριών αλληλεπιδράσεων με την ίδια σταθερά ζεύξης. Με την κατάρρευση της αρχικής συμμετρίας μπορεί κανείς να καταλήξει σε αυτή του SM και με την χρήση των εξισώσεων ομάδας επανακανονικοποίησης η αρχική σταθερά ζεύξης μπορεί να εξεληχθεί στις σταθερές ζεύξης του SM.

Η ιδέα αυτή είναι ελκυστική αλλά παρουσιάζει δύο λειτουργικά προβλήματα:

- Το πρόβλημα της ιεράρχησης (hierarchy) των μαζών. Εμφανίζεται ένα ενεργειακό χάσμα ανάμεσα στην κλίμακα του SM $M_Z \sim 10^2$ GeV και στην $M_G$ χωρίς την πρόβλεψη νέων σωματιδίων. Επιπλέον αλληλοδιαγραφές της τάξης των $M_G$ απαιτούνται στο πλαίσιο του μηχανισμού της επανακανονικοποίσης. Και τα δύο αυτά φαινόμενα θεωρούνται αφύσικα.

- Το πρόβλημα της ενοποίησης των σταθερών ζεύξης. Διάφοροι έλεγχοι ακριβείας έδειξαν ότι η εξέλιξη των σταθερών ζεύξης με το σωματιδιακό περιεχόμενο του SM δεν συγκλίνει στο $M_G$.

Και τα δύο αυτά προβλήματα διευθετούνται κατά τρόπο ικανοποιητικό στο πλαίσιο του Ελάχιστου Υπερσυμμετρικού Καθιερωμένου Προτύπου (MSSM). Λέγεται υπερσυμμετρικό γιατί περιέχει σε κάθε σωμάτιο (φερμιόνιο ή μποζόνιο) του SM ένα υπερσυμμετρικό σύντροφο (μποζόνιο ή φερμιόνιο αντίστοιχα) και ελάχιστο γιατί περιέχει τον ελάχιστο αριθμό από τέτοια επιπλέον σωμάτια.

Με βάση τα επιχειρήματα που εκτέθηκαν παραπάνω, υπάρχει μια αρκετά στέραιη αιτιολόγηση της επιλογής του MSSM ως θεωρίας αναφοράς στη διατριβή αυτή. Μια παρουσίαση του προτύπου αυτού θα επιχειρηθεί στο Κεφάλαιο 2. Έμφαση θα δοθεί στο σωματιδιακό φάσμα που θα αποτελέσει και το πεδίο έρευνας στη συνέχεια.

Το δεύτερο αδιευκρίνιστο ερώτημα του τίτλου της διατριβής είναι οι φαινομενολογικοί περιορισμοί που θα επιβληθούν στο επιλεγμένο σωματιδιακό πρότυπο. Αυτοί εκτίθενται στο Κεφάλαιο 3 και προέρχονται από τις διορθώσεις που λαμβάνουν οι μάζες των φερμιονίων και των Higgs της θεωρίας καθώς και από την διαδικασία $b \to s\gamma$. Παρουσιάζεται αναλυτικά το τυπολόγιο που θα εφαρμοστεί προερχόμενο από τη σύγχρονη βιβλιογραφία.

Παράλληλα με το σωματιδιακό υπάρχει και ένα Καθιερωμένο Κοσμολογικό Πρότυπο (: SBB), το οποίο θα αποτελέσει τη βάση των κοσμολογικών αναζητήσεων της διατριβής. Στην περιγραφή του προτύπου αυτού είναι αφιερωμένο το Κεφάλαιο 4. Σε αυτό θεμελιώνεται η συσχέτιση των GUT με την κοσμολογία, καθώς ενέργειες της τάξης της $M_G$ μπορούν να επιτευχθούν στο πολύ πρώιμο Σύμπαν. Η ιδέα του Πληθωρισμού επίσης παρουσιάζεται καθώς η εισαγωγή του θεωρείται αναγκαία για τη θεραπεία κάποιων ατελειών που εμφανίζει η αρχέγονη μορφή του SBB.

Στο Κεφάλαιο 5 βρίσκει απάντηση το τρίτο θέμα του τίτλου της διατριβής. Το κοσμολογικό πρόβλημα που διερευνάται είναι αυτό της Σκοτεινής Ύλης. Η ύπαρξη μια τέτοιας μορφής ύλης είναι αναγκαία για την ερμηνεία της επιπεδότητας του σύγχρονου Σύμπαντος. Το MSSM εφοδιασμένο με την $R$-συμμετρία διαθέτει έναν καλά αιτιολογημένο υποψήφιο για τη λύση του προβλήματος της Σκοτεινής Ύλης. Είναι το ελαφρότατο υπερσυμμετρικό σωμάτιο (: LSP) το οποίο λόγω της προηγούμενης συμμετρίας είναι ευσταθές και ασθενώς αλληλεπιδρόν. Τα σενάρια δομής της Σκοτεινής Ύλης παρέχουν περιοριστικά όρια για την παράμετρο κοσμολογικής πυκνότητας (: CRD) των LSP. Η μέθοδος υπολογισμού της παραμέτρου αυτής, συμπεριλαμβάνοντας διαδικασίες αλληλοκαταστροφής (: ANE) και συγγενικής καταστροφής (: CAE), εκτίθεται λεπτομερειακά, με εφαρμογή σε LSP μορφής Bino. Το σημείο αυτό είναι το πιο κεντρικό της διατριβής, γιατί παρουσιάζονται αναλυτικά αποτελέσματα πρωτότυπων υπολογισμών που συμβάλλουν στον ακριβή προσδιορισμό της CRD.

Η αριθμητική υλοποίηση των ιδεών που αναπτύχθηκαν παραπάνω περιγράφεται στη συνέχεια της διατριβής. Το πλαίσιο μελέτης εφαρμόζεται σε δύο έκδοχές του MSSM με παγκόσμιες αρχικές συνθήκες για τους όρους ασθενούς παραβίασης της υπερσυμμετρίας, ενοποίηση των ζεύξεων Βαθμίδας και παραβίαση της ηλεκτρασθενούς συμμετρίας μέσω κβαντικών διορθώσεων. Στο Κεφάλαιο 6 μελετάται μια εκδοχή του MSSM με παγκόσμιες αρχικές συνθήκες προερχόμενες από την κατάρρευση της Υπερβαρύτητας και ενοποίηση των ζεύξεων Yukawa. Βρίσκεται η περιοχή του παραμετρικού χώρου που ικανοποιεί τους κοσμολογικούς περιορισμούς. Επίσης διερευνάται η συμβατότητα των φαινομενολογικών περιορισμών. Στο Κεφάλαιο 7 μελετάται μια εκδοχή του MSSM με αρχικές συνθήκες προερχόμενες από τη θεωρία Hořava-Witten. Εξετάζεται η ταυτόχρονη ικανοποίηση των φαινομενολογικών και κοσμολογικών περιορισμών με πλήρη, μερική ή μη ενοποίηση των ζεύξεων Yukawa. Τα συμπεράσματα στα οποία οδήγησε η έρευνα των προτύπων αυτών συνοψίζονται στο Κεφάλαιο 8.

Οι αναφορές που ουσιαστικά συνέβαλαν στη μελέτη του θέματος της διατριβής εκτίθενται στη συνέχεια του κειμένου. Προτιμήθηκε η επιλεγμένη παρουσίαση παρά η συνολική καταγραφή της σωρείας των δημοσιεύσεων που κυκλοφορούν σχετικά με το μελετούμενο αντικείμενο.

Για την καλύτερη ταξινόμηση του υλικού της διατριβής εγίνε η επιλογή της διάρθωσης τριών Παραρτημάτων. Στο Παράρτημα Α′ εκτίθενται οι εξισώσεις επανακανονικοποίησης που χρησιμοποιήθηκαν στο αριθμητικό πρόγραμμα. Οι κανόνες Feynman και οι συμβάσεις που συνοδεύουν την εφαρμογή τους, καθώς και τρία παραδείγματα των αναλυτικών υπολογισμών που έγιναν στο πλαίσιο της διατριβής παρουσιάζονται στο Παράρτημα Β′. Τέλος, προς διευκόλυνση του αναγνώστη, θεωρήθηκε σκόπιμη η συνολική καταγραφή των ακρονυμίων που ορίζονται κατά τη διάρκεια ροής του κειμένου, στο Παράρτημα Γ′.

# Κεφάλαιο 2

# Ελάχιστα Υπερσυμμετρικό Καθιερωμένο Πρότυπο

## 2.1 Εισαγωγή

Είναι πέρα από τους σκοπούς αυτής της εργάσιας η μελέτη της Υπερσυμμετρίας. Θα επιχειρηθεί όμως, μια περιεκτίκη εισαγωγή που θα εξοπλίζει τον αναγνώστη με την υποδομή που χρειάζεται, για να παρακολουθήσει πιο ανετα την παρουσίαση του Ελάχιστα Υπερσυμμετρικού Καθιερωμένου Προτύπου (: MSSM) που ειναι ο κυρίως στόχος του κεφαλαίου αυτού. Στην προσπάθεια αυτή πολύτιμη υπήρξε η εμπειρία που αποκομίσθηκε από τη μελέτη των αντίστοιχων κεφαλαίων παλαιότερων διδακτορικών διατριβών των Αν. [1], [2] και [3]. Επίσης, πολλές είναι οι χρήσιμες επισκοπήσεις που κυκλοφορούν στο διαδύκτιο σχετικά με το MSSM από τις οποίες πιο χρήσιμες θεωρούνται οι Αν. [4], [5], [6], [7] και [8]. Ως πιο παιδαγωγική όμως αξιολογείται η Αν. [9]. Οι συμβάσεις, όχι όμως πλήρως και ο συμβολισμός, που υιοθετήθηκαν είναι αυτές της Αν. [11], οι οποίες βρίσκονται σε άμεση συσχέτιση με τις χρησιμοποιούμενες από τις Αν. [12], [13].

Το κεφάλαιο αρχίζει με την καταγραφή των χαρακτηριστικών μιας Υπερσυμμετρικής Θεωρίας Βαθμίδας στο Εδ. 2.2 και στη συνέχεια η μελέτη εξειδικεύεται στην περίπτωση του MSSM. Στο Εδ. 2.3 γίνεται ο καθορισμός του προτύπου, στο Εδ. 2.4 μελετάται ο μηχανισμός Higgs, στο Εδ. 2.5 εξάγεται το SUSY φάσμα του προτύπου και τέλος στο Εδ. 2.6 εξετάζεται η βασική αλληλεπίδραση που θα χρησιμοποιηθεί στη συνέχεια της διατριβής.

## 2.2 Υπερσυμμετρικές Θεωρίες Βαθμίδας

Στόχος δεν είναι η θεωρητική θεμελίωση της υπερσυμμετρίας αλλά η εφαρμογή της προκύπτουσας θεωρίας σε χαμηλή κλίμακα ενέργειας. Θα επιχειρηθεί μια ακροβασία ανάμεσα στην περιγραφική εισαγωγή των απαραίτητων εννοιών και στην ποσοτική έκφραση κάποιων αποτελεσμάτων που εφαρμόζονται εύκολα στην περίπτωση του MSSM. Ειδικότερα στο Εδ. 2.2.1 εισάγονται οι αρχικές έννοιες της SUSY και στη συνέχεια η μελέτη μας προσανατολίζεται σε ένα γενικευμένο SUSY πρότυπο εφοδιασμένο με μια τοπική συμμετρία βαθμίδας. Στο Εδ. 2.2.2 παρουσιάζεται η λαγκραζιανή του προτύπου αυτού, στο Εδ. 2.2.4 εκτίθενται οι δυνατοί όροι παραβίασης της SUSY που δύνανται να καταστίσουν τη λαγκραζιανή αυτή φαινομενολογικά ενδιαφέρουσα και στο Εδ. 2.2.5 δίνεται η μέθοδος εξαγωγής του σωματιδιακού φάσματος. Παρεμβάλλεται το Εδ. 2.2.3, στο οποίο σκιαγραφείται η λύση του προβλήματος της ιεράρχησης.

### 2.2.1 Εισαγωγή της SUSY

Η υπερσυμμετρία (: SUSY) είναι μια συμμετρία που συσχετίζει φερμιόνια και μποζόνια. Ο πιο προωθημένος τρόπος εισαγωγής της είναι με την χρήση της έννοιας του Υπερχώρου. Στο τέλος μιας αρκετά πολύπλοκης διαδικασίας είναι σε θέση κανείς να κατασκευάσει λαγκραζιανές, οι οποίες είναι αναλλοίωτες κάτω από τους SUSY μετασχηματισμούς και περιέχουν τον ίδιο αριθμό μποζονίων και φερμιονίων. Η πορεία αυτή θα επιχειρηθεί να περιγραφεί με τα παρακάτω σημεία:

**α.** Υπερ-άλγεβρα (Superalgebra). Κάθως η SUSY είναι μια συμμετρία θα πρέπει να αναζητηθούν οι γεννήτορες που θα παράγουν τους SUSY μετασχηματισμούς. Επειδή αυτοί θα πρέπει να μετασχηματίζουν σωμάτια με διαφορετικό spin έπεται ότι δεν θα μπορούν να μετατίθενται με τους γεννήτορες των στροφών $M_{\mu\nu}$ της ομάδας Poincaré. Η ομάδα αυτή υλοποιεί τη συμμετρία, που σέβεται η Κ-βαντική Θεωρία Πεδίου και έχει γεννήτορες τους $P_\mu$, $M_{\mu\nu}$, όπου $P_\mu$ ο γενήτορας των χωροχρονικών μετατοπίσεων. Επόμενως, υπάρχει μια ειδοποιός διαφορά ανάμεσα στη SUSY και στις γνωστές συμμετρίες βαθμίδας, οι γεννήτορες των οποίων μετατίθενται με τους $P_\mu$, $M_{\mu\nu}$. Επιπλέον, το θεώρημα Coleman-Mandula απαγορεύει την ύπαρξη μιας μεταθετικής άλγεβρας (Lie) για στοιχεία που δεν μετίθενται με τους $P_\mu$, $M_{\mu\nu}$. Επομένως, οι γεννήτορες της SUSY υπακούουν σε μια αντι-μεταθετική άλγεβρα (και όχι μεταθετική) που ονομάζεται Υπερ-άλγεβρα. Είναι, δηλαδή, σπινοριακά κατασκευάσματα και συνήθως συμβολίζονται $Q_\alpha$ και $\bar{Q}_{\dot{\alpha}}$ (όπου $\alpha = 1, 2$ δείκτες Weyl spinor). Ο $P^2$ συνεχίζει να μετατίθεται με τους γεννήτορες της SUSY πράγμα που σημαίνει ότι η μετατροπή φερμιονίου σε μποζόνιο και το αντίστροφο δεν θα συνοδεύεται με μεταβολή μάζας. Επίσης οι γεννήτορες των συμμετριών βαθμίδας (τοπικών ή ολικών) μετατίθενται με τους $Q_\alpha$, $\bar{Q}_{\dot{\alpha}}$ πράγμα που σημαίνει ότι οι υπερσυμμετρικοί σύντροφοι θα έχουν τους ίδιους κβαντικούς αριθμούς που σχετίζονται με τις συμμετρίες αυτές (ηλεκτρικό φορτίο, ασθενές υπερφορτίο, χρώμα)

**β.** Υπερ-χώρος (Superspace). Οι $Q_\alpha$, $\bar{Q}_{\dot{\alpha}}$ και $P_\mu$ δημιουργούν μια ομάδα συμμετρίας, που δομείται από μετασχηματισμούς με παραμέτρους που ζουν στον υπερχώρο 8 διαστάσεων: $(x^\mu, \theta_\alpha, \bar{\theta}_{\dot{\alpha}})$, όπου $\theta_\alpha$, $\bar{\theta}_{\dot{\alpha}}$ είναι αντιμετατιθέμενες μεταβλητές (Grassmann) που επισυνάπτονται στους $Q_\alpha$, $\bar{Q}_{\dot{\alpha}}$. Αξιοποιώντας τη δομή της ομάδας που διαθέτουν αυτοί οι μετασχηματισμοί μπορεί να βρεθεί η διαφορική αναπαράσταση των $Q_\alpha$ και $\bar{Q}_{\dot{\alpha}}$ και επομένως, μπορεί να επιτευχθεί ο υπολογισμός των SUSY μετασχηματισμών πάνω σε (υπερ-)πεδία.

**γ.** Υπερπεδίο (Superfield). Συναρτήσεις με πεδίο ορισμού τον υπερχώρο ονομάζονται υπερπεδία. Αναπτύσσονται σε πολυώνυμα των $\theta_\alpha$, $\bar{\theta}_{\dot{\alpha}}$ με συντελεστές πεδία, συναρτήσεις του $x^\mu$. Προφανώς το ανάπτυγμα σε δυνάμεις των $\theta_\alpha$, $\bar{\theta}_{\dot{\alpha}}$ σταματά σε τετραγωνικούς όρους, γιατί λόγω της μορφής Grassmann αυτών των μεταβλητών ανώτερης τάξης όροι μηδενίζονται. Επομένως μπορεί να ειπωθεί πιο απλοποιητικά οτι τα υπερπεδία είναι σύνολα από πεδία που αντιπροσωπεύουν σωμάτια διαφορετικού spin. Εισάγονται δύο είδη υπερπεδίων :

- Υπερπεδίο χειραλλότητας (chiral), που συνίσταται από ένα φερμιονικό πεδίο δύο Weyl συνιστωσών και τον SUSY εταίρο του, ένα μιγαδικό βαθμωτό πεδίο, που ονομάζεται sfermion. Επομένως, τα υπερπεδία χειραλλότητας χρησιμοποιούνται για την αναπαράσταση των πεδίων ύλης στις SUSY θεωρίες.
- Διανυσματικό υπερπεδίο, που συνίσταται από ένα διανυσματικό μποζονικό πεδίο και τον SUSY εταίρο του, ένα φερμιονικό πεδίο δύο Weyl συνιστωσών που ονομάζεται gaugino. Επομένως, τα διανυσματικό υπερπεδία χρησιμοποιούνται για την αναπαράσταση των πεδίων βαθμίδας στις SUSY θεωρίες.

Και στις δύο περιπτώσεις που προαναφέρθηκαν εκτός από τις φυσικές συνιστώσες υπάρχουν και κάποιες βοηθητικές. Οι εξισώσεις κίνησης των πεδίων αυτών δεν περιέχουν χρονικές παραγώγους, χρησιμοποιούνται, όμως, για την εξίσωση των μποζονικών και φερμιονικών βαθμών ελευθερίας ενός υπερπεδίου και μετέχουν στις εκφράσεις των SUSY μετασχηματισμών των άλλων συνιστωσών του υπερπεδίου.

**δ.** Υπερδυναμικό (Superpotential). Μια πολυωνιμική συνάρτηση των υπερπεδίων ονομάζεται υπερδυναμικό. Ο χαρακτήρας και οι ιδιότητες της συνάρτησης αυτής καθορίζονται από το φορμαλισμό των υπερπεδίων. Παρακάτω καταγράφονται οι ιδιότητες αυτές χωρίς ιδιαίτερη αιτιολόγηση:

- Είναι αναλυτική συνάρτηση των υπερπεδίων, δεν επιτρέπεται να περιέχει όρους με παραγώγους.
- Είναι συνάρτηση μόνο υπερπεδίων χειραλλότητας και μάλιστα δεν επιτρέπεται να περιέχει ένα υπερπεδίο χειραλλότητας και το μιγαδικό ή ερμητιανό συζυγές του (ολομορφική ιδιότητα κατά την ορολογία της Αν. [6]).

- Περιέχει όρους με δύο και τρία υπερπεδία. Όρος με ένα υπερπεδίο δεν εισάγεται γιατί δεν μπορεί να δώσει ενδιαφέροντες φαινομενολογικά όρους (κινητικής ενέργειας ή αλληλεπίδρασης) ενώ όροι με περισσότερα από τρία υπερπεδία δίνουν θεωρία μη επανακανονικοποιήσιμη.

Όπως θα φανεί και στο Εδ. 2.2.2, από το υπερδυναμικό προκύπτουν οι F όροι για το δυναμικό της θεωρίας και οι όροι των αλληλεπιδράσεων Yukawa των φερμιονίων με τους μποζονικούς τους συντρόφους, πράγμα που συνιστά την πιο σημαντική καινοτομία των SUSY θεωριών. Συνεπώς, το υπερδυναμικό αποτελεί την ταυτότητα μιας SUSY θεωρίας αφού με τον καθορισμό του, καθορίζονται και τα βασικά χαρακτηριστικά της θεωρίας αυτής.

Εξοπλισμένος κανείς με όλα αυτά τα σύνεργα είναι σε θέση να κατασκευάσει SUSY αναλλοίωτες δράσεις. Θα πρέπει να ολοκληρώνει χωρικά, (δηλαδή με όρισμα ολοκλήρωσης το $d^4x$) μία έκφραση για τη λαγκραζιανή, της οποίας ο SUSY μετασχηματισμός να είναι ολική παράγωγος. Τότε η δράση της μετασχηματισμένης λαγκραζιανής θα μηδενίζεται. Η απαίτηση, επίσης της επανακανονικοποιησιμότητας της θεωρίας οδηγεί σχεδόν νομοτελιακά στην κατασκευή συγκεκριμένης μορφής δράσεων. Οι αρχικές αυτές μορφές που είναι εκφρασμένες στη γλώσσα των υπερπεδίων μπορούν να μεταφραστούν στη γλώσσα των συνιστωσών και με τον τρόπο αυτό να προκύψει μια εύληπτη SUSY θεωρία.

Η πορεία που ακολουθεί κανείς στην μελέτη αυτών των θεωριών είναι η εξής. Αρχικά, μελετάται το πρότυπο Wess Zumino το οποίο αποτελείται μόνο από πεδία ύλης. Επομένως, η θεωρία που προκύπτει περιέχει ίσο αριθμό μποζονίων και φερμιονίων με αλληλεπίδραση επαφής χωρίς όμως φορέα αλληλεπίδρασης. Στη συνέχεια μελετάται μια SUSY ηλεκτροδυναμική με την εισαγωγή ενός διανυσματικού υπερπεδίου και της απλούστερης δυνατής συμμετρίας βαθμίδας. Σε αυτή την θεωρία, εκτός από τα πεδία ύλης προκύπτει και ένας άμαζος μποζονικός φορέας αλληλεπίδρασης μαζί με το φερμιονικό του σύντροφο. Τέλος επιτυγχάνεται μια SUSY θεωρία με μια μη αβελιανή συμμετρία βαθμίδας. Η παρουσίαση της μιας τέτοιας θεωρίας στη γλώσσα των συνήθισμένων πεδίων είναι ο σκοπός της επόμενης παραγράφου.

## 2.2.2 Δομή της λαγκραζιανής

Η μορφή του SM βασίζεται στις θεωρίες βαθμίδας. Θεωρίες, δηλαδή, που είναι αναλλοίωτες κάτω από τοπικούς μετασχηματισμούς βαθμίδας, των οποίων οι γεννήτορες $T^a_{ij}$ γενικά σχηματίζουν μια άλγεβρα Lie. Οι χρησιμοποιούμενοι δείκτες $a$, $ij$ τρέχουν ως εξής: $a = 1, ...N$ όπου $N$ η διάσταση της άλγεβρας, $i, j = 1, ...R$ όπου $R$ η διάσταση της αναπαράστασης που έχει επιλεγεί για τα πεδία ύλης που είναι συνήθως η θεμελιώδης. Η απαίτηση αυτής της αναλλοιώτητας εφοδιάζει τη θεωρία με πεδία βαθμίδας $A^a_\mu$ και τις σταθερές ζεύξης τους, $g^a$ με την ύλη. Τα $A^a_\mu$ αρχικά είναι άμαζα. Με το μηχανισμό Higgs μπορεί να δοθεί μάζα σε (κάποια από) αυτά με επανακανονικοποιήσιμο τρόπο.

Η SUSY επέκταση αυτής της θεωρίας γίνεται με την εισαγωγή δύο ειδών υπερπεδίων, σύμφωνα με το Εδ. 2.2.1:

- Υπερπεδία χειραλλότητας που περιέχουν τα πεδία ύλης της θεωρίας, τα φερμιόνια $\psi_i$ και τους μποζονικούς τους συντρόφους, τα sfermions $\varphi_i$.
- Διανυσματικά υπερπεδία που περιέχουν τα πεδία βαθμίδας $A^a_\mu$ και τους φερμιονικούς τους εταίρους, τα gaugino, $\lambda^a$

Η παρουσίαση της μορφής της λαγκραζιανής στη γλώσσα των συνήθων πεδίων μιας τέτοιας θεωρίας που θα μπορούσε να ονομαστεί SUSY Yang-Mills είναι ο σκοπός της παραγράφου αυτής. Για λόγους καλύτερης ταξινόμησης, η λαγκραζιανή μπορεί να τεμαχιστεί στα παρακάτω τμήματα:

$$\mathcal{L}_{\rm SYM} = \mathcal{L}_{\rm KIN} + \mathcal{L}_{\rm INT} - V_{\rm SUSY}, \quad \text{όπου:}$$

**α.** $\mathcal{L}_{\rm KIN}$: Το τμήμα που περιέχει τους κινητικούς όρους των πεδίων της θεωρίας. Υπάρχουν κινητικοί όροι για τα φερμιονικά πεδία $\psi_i$, $\lambda^a$, τα διανυσματικά $A^a_\mu$ και τα βαθμωτά $\varphi_i$. Από αυτούς, το ενδιαφέρον επικεντρώνεται στους κινητικούς όρους των βαθμωτών πεδίων που έχουν τη μορφή:

$$(D^\mu\varphi)^*_i\,(D_\mu\varphi)_i \quad \text{όπου} \quad (D_\mu)_{ij} = \partial_\mu + ig^aT^a_{ij}A^a_\mu \tag{2.1}$$

με $D_\mu$ τη συναλλοίωτη παράγωγο της θεωρίας που περιέχει πεδία βαθμίδας μόνο για την τοπική συμμετρία (και όχι λόγω της SUSY που είναι μια ολική συμμετρία) της θεωρίας. Δε δοθηκε η εκπεφρασμένη μορφή των άλλων κινητικών όρων, γιατί η καταγραφή τους δε θα προσέφερε κάτι σημαντικό στην προσέγγιση που επιχειρείται.

**β.** $\mathcal{L}_{\rm INT}$: Το τμήμα που περιέχει τις αλληλεπιδράσεις Yukawa και fermion-sfermion-gaugino με την εξής μορφή:

$$\mathcal{L}_{\rm INT} = -\frac{1}{2}\frac{\partial^2 W}{\partial\varphi_i\partial\varphi_j}\psi_i\psi_j + i\sqrt{2}g_a\varphi_k^*\lambda^a T_{kl}^a\psi_l + \text{h. c} \tag{2.2}$$

Από τον πρώτο όρο του δεύτερου μέλους της Εξ. (2.2) θα προκύψουν οι όροι μάζας για τα φερμιόνια, τα οποία υπενθυμίζεται ότι αναπαρίστανται με σπίνορες δύο συνιστωσών για αυτό και ο όρος μάζας τους δεν έχει την μορφή που έχουν οι όροι μάζας με Dirac σπίνορες, συμφωνα με τις εξηγήσεις που δίνονται στην Εξ. (Β΄.16). Επίσης το υπερδυναμικό $W(\varphi_j)$ νοείται ως συνάρτηση μόνο των βαθμωτών συνιστωσών των υπερπεδίων που περιέχει.

**γ.** $V_{\rm SUSY}$: Το δυναμικό της πλήρους υπερσυμμετρικής θεωρίας που έχει δύο βασικές συνεισφορές:

$$V_{\rm SUSY} = V_{\rm F} + V_{\rm D}, \quad \text{όπου:}$$

- $V_{\rm F}$, η συνεισφορά που προέρχεται από τα βοηθητικά πεδία $F_i$ που περιέχονται στα υπερπεδία χειραλότητας, γιαυτό και οι όροι αυτοί του δυναμικού ονομάζονται F-όροι:

$$V_{\rm F} = \sum_i F_i^* F_i \quad \text{όπου} \quad F_i = -\frac{\partial W}{\partial\varphi_i} \tag{2.3}$$

- $V_{\rm D}$, η συνεισφορά που προέρχεται από το βοηθητικά πεδία D που περιέχονται στα διανυσματικά υπερπεδία, γιαυτό και οι όροι αυτοί του δυναμικού ονομάζονται D-όροι:

$$V_D = \frac{1}{2}\sum_a D^a D^a \quad \text{όπου} \quad D^a = g^a\varphi_i^* T_{ij}^a\varphi_j \tag{2.4}$$

Αξίζει να σημειωθεί ότι στο τυπολόγιο που παρουσιάστηκε οι σπινοριακοί δείκτες δεν δηλώνονται και υπονοείται άθροιση στους επαναλαμβανόμενους δείκτες της αναπαράστασης της ομάδας βαθμίδας $ij$. Ακόμα οι αφύσικοι βαθμοί ελευθερίας των διανυσματικών υπερπεδίων έχουν αρθεί με την επιλογή της βαθμίδας Wess-Zumino.

### 2.2.3 Το πρόβλημα της ιεράρχησης

Από τη γενική μορφή της λαγκραζιανής που παρουσιαστηκε, μπορεί να κατανοηθεί ότι το πρόβλημα της ιεράρχησης μπαίνει σε πορεία διευθέτησης με την εισαγωγή της SUSY. Κοιτίδα του προβλήματος αυτού είναι η τεράστια ενεργειακή απόσταση που υπάρχει ανάμεσα στο EWS, $M_W \sim 100\,\text{GeV}$ και στην κλίμακα της GUT, $M_G \sim 10^{16}\,\text{GeV}$. Οι παρενέργειες του προβλήματος εκδηλώνονται στις τετραγωνικές αποκλίσεις που παρουσιάζονται σε διαγράμματα της μορφής του Σχ. 2.1(β), τα οποία συνεισφέρουν στη διόρθωση της μάζας, $\delta m_\phi$ ενός βαθμωτού πεδίου $\phi$ που στα πλαίσια του SM είναι το Higgs. Πραγματικά, εισάγωντας ένα $cut-off$ (άνω όριο ορμών) της τάξης των GUT, $M_G$ στο ολοκλήρωμα του βρόχου του Σχ. 2.1(β) , η διόρθωση διαθέτει όρο που αποκλίνει τετραγωνικά:

$$\delta m_\phi \sim \int^{M_G} \frac{d^4k}{k^2} \sim M_G^2 \tag{2.5}$$

Βεβαίως, τέτοιες αποκλίσεις θα μπορούσαν να απορροφηθούν με το πρόγραμμα επανακανονικοποίησης. Όμως αυτό θα απαιτούσε μια αφύσικη λεπτή ρύθμιση των παραμέτρων της θεωρίας, αφού θα έπρεπε να απορροφάται σε κάθε τάξη της θεωρίας διαταραχών μια διόρθωση κατά πολύ μεγαλύτερη της μάζας που διορθώνεται. Οι SUSY θεωρίες προσφέρουν διέξοδο από το πρόβλημα της ιεράρχησης γιατί στο πλαίσιο τους, προστίθενται συνεισφορές στο $\delta m_\phi$ από τους μποζονικούς εταίρους των φερμιονίων, οι οποίες αίρουν τις τετραγωνικές αποκλίσεις της θεωρίας. Αυτό επιτυγχάνεται γιατί οι σταθερές ζεύξεις Yukawa και τεσσάρων βαθμωτών πεδίων βρίσκονται σε συσχέτιση.

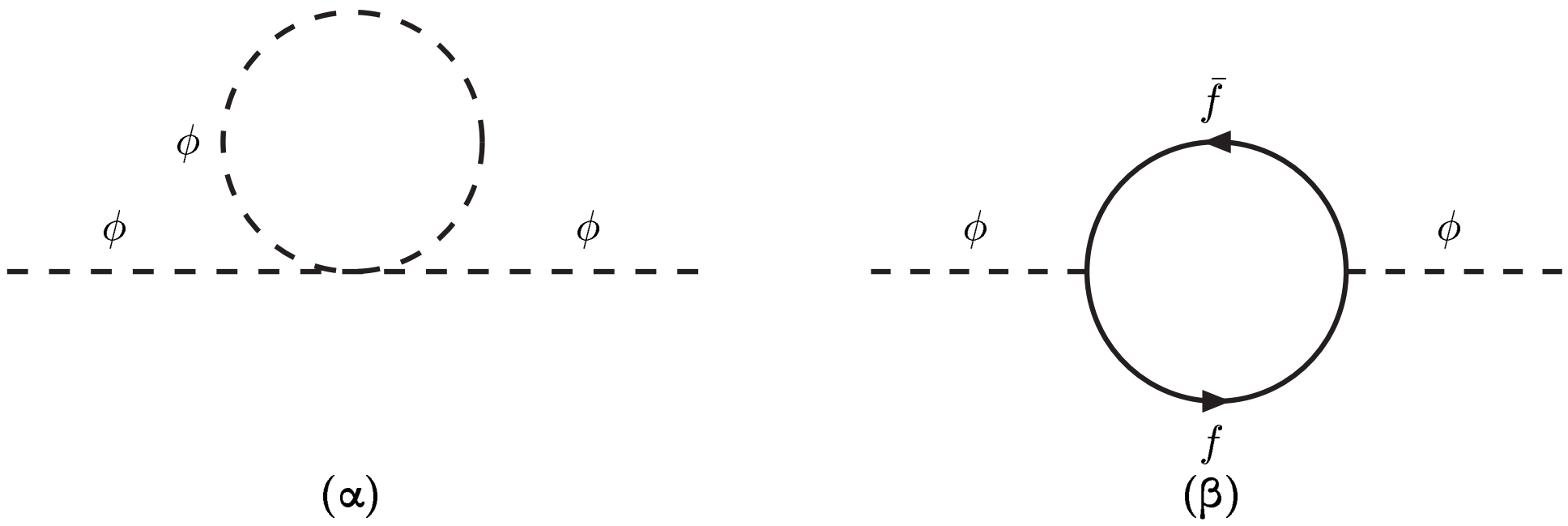

**Σχήμα 2.1:** *Οι βρόχοι των βαθμωτων σωματίων, $\phi$ (α) και των φερμιονίων, $f$ (β) που συνεισφέρουν στην διόρθωση της μάζας του $\phi$.*

Ο μηχανισμός αναίρεσης των τετραγωνικών αποκλίσεων μπορεί να επιδειχθεί μέσω ενός παραδείγματος (πιο αναλυτικές στο σημείο αυτό είναι οι Αν. [1], [5]). Έστω μια στοιχειώδη θεωρία με ένα φερμιόνιο $f$ και ένα βαθμωτό πεδίο $\phi$ στο υπερδυναμικό της οποίας υπάρχει ένας όρος $\lambda\phi^3/3$ γραμμένος υπό μορφή βαθμωτών συνιστωσών. Από τον όρο της λαγκραζιανής στην:

- Εξ. (2.2) προκύπτει μια ζεύξη $g_{\phi ff}\phi ff$ με $g_{\phi ff} = -\lambda$, οπότε η διόρθωση στη μάζα $m_\phi$ από το διάγραμμα (β) του Σχ. 2.1 θα είναι της μορφής:

$$\delta m_\phi^{(\beta)} \sim (-i\lambda)(-i\lambda)ii(-1)M_G^2 = -\lambda^2 M_G^2 \tag{2.6}$$

  όπου τα $i$ προέρχονται από τους διαδότες και το (-1) από τη στατιστική των φερμιονίων.

- Εξ. (2.3) προκύπτει μια ζεύξη $g_{\phi^4}\phi^4$ με $g_{\phi^4} = -\lambda^2$, οπότε η διόρθωση στη μάζα $m_\phi$ από το διάγραμμα (α) του Σχ. 2.1 θα είναι της μορφής:

$$\delta m_\phi^{(\alpha)} \sim (-i\lambda^2)iM_G^2 = \lambda^2 M_G^2 \tag{2.7}$$

Προσθέτωντας τις συνεισφορές των Εξ. (2.6), (2.7) προκύπτει ότι η συνολίκη διόρθωση στη μάζα $m_\phi$ δε περιέχει τετραγωνικές απόκλισεις. Αυτή η αυθόρμητη διευθέτηση του προβλήματος της ιεράρχησης είναι από τους πιο σημαντικούς λόγους εισαγωγής της SUSY στα πλαίσια των GUT.

### 2.2.4 Όροι Ασθενούς Παραβίασης της SUSY

Η διατήρηση της επαφής των θεωρίων με την πραγματικότητα των παρατηρήσεων είναι αναγκαία συνθήκη για τη βιωσιμότητα και την αξιοποίησή τους. Καθώς, λοιπόν, δεν έχουν παρατηρηθεί SUSY σύντροφοι των γνωστών σωματιδίων έπεται ότι αν η SUSY υπάρχει, θα πρέπει να έχει καταρρεύσει με κάποιο τρόπο, ώστε οι SUSY εταίροι να έχουν μάζα σημαντικά υψηλότερη από τα γνωστά σωμάτια. Η κατάρρευση αυτή μπορεί να υλοποιηθεί είτε αυθόρμητα είτε με έξωθεν επέμβαση στη SUSY λαγκραζιανή. Βεβαίως, και με τις δύο τεχνικές επιδιώκεται να μην χαθεί το καθοριστικό κεκτημένο που προσφέρει η εισαγωγή της SUSY, που είναι η λύση του προβλήματος της ιεράρχησης.

Αυθόρμητα σπασμένη λέγεται μια συμμετρία που διαθέτει μία λαγκραζιανή όταν το κενό της θεωρίας δε τη σέβεται εξίσου. Αυτό συμβαίνει όταν η τιμή του δυναμικού στο κενό της θεωρίας, $\langle V_{\rm SUSY}\rangle_0$ δεν είναι μηδέν. Από τις Εξ. (2.3), (2.4) συμπεραίνεται ότι η προηγούμενη συνθήκη μπορεί να εκπληρωθεί όταν $\langle F_i\rangle_0 \neq 0$ ή (και) $\langle D^a\rangle_0 \neq 0$. Αποδεικνύεται στην Αν. [9] ότι η πρώτη περίπτωση δεν παράγει φαινομενολογικά ενδιαφέρουσα μετατόπιση των μαζών των SUSY εταίρων ενώ η δεύτερη περίπτωση είναι ανεφάρμοστη στην περίπτωση της συμμετρίας του SM.

Αναγκαστικά, λοιπόν, απευθύνεται κανείς στη δεύτερη εκδοχή παραβίασης της SUSY που πραγματοποιείται με προσθήκη στη λαγκραζιανή κάποιων όρων ασθενούς (ή ήπιας) παραβίασης της SUSY (: SBT). Τέτοιοι όροι μπορούν να ερμηνευτούν ως προερχόμενοι από από την αυθόρμητη κατάρευση των θεωριών Υπερβαρύτητας. Οι SBT είναι όροι που παραβιάζουν την SUSY αναλοιώτητα της λαγκραζιανής χωρίς την εισαγωγή τετραγωνικών αποκλίσεων στη θεωρία. Αποδεικνύεται στην Αν. [9] ότι όροι που ικανοποιούν αυτή την απαίτηση και δύνανται να προστέθουν στο δυναμικό της θεωρίας είναι οι εξής:

- Όροι μάζας για τα βαθμωτά σωμάτια της μορφής $\frac{1}{2}m^2{}_{ij}\varphi_i\varphi_j^*$
- Όροι μάζας για τα gaugino $-\frac{1}{2}M_\lambda^a\lambda^a\lambda^a$
- Όροι γραμμικοί $C_i\varphi_i + \text{h.c.}$
- Όροι διγραμμικών ζεύξεων $B_{ij}\varphi_i\varphi_j + \text{h.c.}$
- Όροι τριγραμμικών ζεύξεων $A_{ijk}\varphi_i\varphi_j\varphi_k + \text{h.c.}$

Παρατηρείται ότι όροι μάζας φερμιονίων ύλης δεν μπορούν να θεωρηθούν SBT. Επίσης από τους προτεινόμενους παραπάνω SBT επιλέγονται σε κάθε περίπτωση μόνο εκείνοι που σέβονται τη συμμετρία βαθμίδας, την οποία διαθέτει η αρχική λαγκραζιανή. Αυτός ο περιορισμός αφορά τους SBT ζεύξεων, οι οποίοι αναγκαστικά επιλέγεται να έχουν τη μορφή των αντίστοιχων όρων του υπερδυναμικού.

Συμπερασματικά, το δυναμικό μιας ρεαλιστικής SUSY Yang-Mills θεωρίας έχει δύο συνεισφορές:

$$V = V_{\text{SUSY}} + V_{\text{soft}} \tag{2.8}$$

όπου $V_{\text{soft}}$ ένα άθροισμα από τους προτεινόμενους και επιτρεπόμενους από τη συμμετρία SBT.

### 2.2.5 Φάσμα της Θεωρίας

Το διαταρακτικό φάσμα μιας θεωρίας προκύπτει με ανάπτυξη των όρων της λαγκραζιανής της γύρω από το κενό που βρίσκεται από τη συνθήκη ελαχιστοποίησης του δυναμικού:

$$\left\langle \frac{\partial V(\varphi_j)}{\partial \varphi_i} \right\rangle_0 = 0 \tag{2.9}$$

όπου το σύμβολο $\langle ... \rangle_0$ δηλώνει ότι η ποσότητα που περικλείεται έχει υπολογισθεί στο κενό της θεωρίας δηλαδή για $\varphi_i = \langle \varphi_i \rangle_0$. Αναπτύσσοντας τη λαγκραζιανή γύρω από αυτό το κενό λαμβάνονται οι πίνακες μάζας των βαθμωτών, φερμιονικών και διανυσματικών πεδίων της θεωρίας. Συγκεκριμένα:

**α.** Οι όροι μάζας των βαθμωτών μποζονίων αναδύονται αποκλειστικά από την ανάπτυξη του δυναμικού και γράφονται σε πινακική μορφή ως εξής:

$$\frac{1}{2}\begin{pmatrix} \varphi_i^* & \varphi_i \end{pmatrix} \mathcal{M}_{\text{B}}^2 \begin{pmatrix} \varphi_j \\ \varphi_j^* \end{pmatrix} \quad \text{όπου} \quad \mathcal{M}_{\text{B}}^2 = \begin{pmatrix} \left\langle \frac{\partial^2 V}{\partial \varphi_i^* \partial \varphi_j} \right\rangle_0 & \left\langle \frac{\partial^2 V}{\partial \varphi_i^* \partial \varphi_j^*} \right\rangle_0 \\ \left\langle \frac{\partial^2 V}{\partial \varphi_i \partial \varphi_j} \right\rangle_0 & \left\langle \frac{\partial^2 V}{\partial \varphi_i \partial \varphi_j^*} \right\rangle_0 \end{pmatrix} \tag{2.10}$$

**β.** Οι όροι μάζας των φερμιονίων αναδύονται από το τμήμα της Εξ. (2.2) της λαγκραζιανής και γράφονται σε πινακική μορφή ως εξής:

$$-\frac{1}{2}\begin{pmatrix} \lambda^a & \psi_i \end{pmatrix} \mathcal{M}_{\text{F}} \begin{pmatrix} \lambda^b \\ \psi_j \end{pmatrix} \quad \text{όπου} \quad \mathcal{M}_{\text{F}} = \begin{pmatrix} M_\lambda^{ab} & \sqrt{2}i\,\langle D_i^a \rangle_0 \\ \sqrt{2}i\,\langle D_j^b \rangle_0 & \langle f_{ij} \rangle_0 \end{pmatrix} \tag{2.11}$$

$$\text{με} \quad f_{ij} := \left\langle \frac{\partial^2 W}{\partial \varphi_i \partial \varphi_j} \right\rangle_0, \quad D_i^a := \frac{\partial D^a}{\partial \varphi_i} = g\varphi_j^* T_{ji}^a, \quad M_\lambda^{ab} = \frac{\partial^2 V_{\text{soft}}}{\partial \lambda^b \partial \lambda^a} \tag{2.12}$$

**γ.** Οι όροι μάζας των διανυσματικών μποζονίων αναδύονται από το τμήμα της Εξ. (2.1) της λαγκραζιανής και παράγουν τον εξής πίνακα μαζών:

$$\mathcal{M}_{\text{A}}^{ab} = 2\left\langle D_i^a D_{*i}^b \right\rangle_0 \quad \text{με} \quad D_{*i}^a := \frac{\partial D^a}{\partial \varphi_i} = g^a T_{ij}^a \varphi_j^* \tag{2.13}$$

Όπως είναι προφανές από την Εξ. (2.13) τα διανυσματικά μποζόνια της θεωρίας αποκτούν μάζα όταν η θεωρία είναι σπασμένη, δηλαδή όταν $\langle \varphi_i \rangle_0 \neq 0$

Με διαγωνοποιήση των πινάκων αυτών, λαμβάνονται οι ιδιοτιμές (: ΙΔΤ) και ιδιοκαταστάσεις (: ΙΔΚ) μάζας των πεδίων της Θεωρίας. Προκύπτει ίσος αριθμός μποζονικών και φερμιονικών εταίρων με μια διαφορά μάζας μεταξύ τους που οφείλεται στην εισαγωγή των SBT.

## 2.3 Δομή του MSSM

Η εφαρμογή του γενικού πλαισίου που περιγράφηκε στο Εδ. 2.2 στην περίπτωση ενός υπερδυναμικού με τη συμμετρία του SM είναι ο σκοπός αυτής της ενότητας. Ειδικότερα, στο Εδ. 2.3.1 παρουσιάζεται το σωματιδιακό περιεχόμενο του MSSM και στο Εδ. 2.3.2 αναδεικνύονται οι σκέψεις που οδηγούν στη συγκεκριμένη μορφή του υπερδυναμικού. Τέλος στο Εδ. 2.3.3 εφαρμόζοντας το τυπολόγιο του Εδ. 2.2.2 και εξάγονται τα πιο χαρακτηριστικά τμήματα της λαγκραζιανής του MSSM.

### 2.3.1 Συμβολιμοί-Ταξινόμηση-Αναπαραστάσεις Πεδίων

Το πεδιακό περιεχόμενο του MSSM παρουσιάζεται σχηματικά στον Πίνακα 2.1. Κάποιες επιβεβλημένες διευκρινήσεις σχετικά με το χρησιμοποιούμενο συμβολισμό ο οποίος αναγκαστικά, είναι περιεκτικός και σύντομος, δίνονται στην ενότητα αυτή. Τα υπερπεδία που χρησιμοποιούνται για να αναπαραστήσουν και να επεκτείνουν το σωματιδιακό περιεχόμενο του SM μπορούν να κατατάγουν ως εξής:

**α.** Υπερπεδία Ύλης. Πρόκειται για υπερπεδία χειραλότητας στά οποία διευθετούνται τα πεδία ύλης, δηλαδή τα φερμιόνια και τα Sfermions. Απο αυτά τα αριστερόστροφα αναπαρίστανται με διπλέτες ενώ τα δεξιόστροφα είναι μονάδες ως προς την ομάδα συμμετρίας $SU(2)_L$. Ως προς το χαρακτήρα των συνιστωσών τους μπορούν να ταξινομηθούν ως εξής:

- Υπερπεδία λεπτονίων και Sleptons. Συμβολίζονται με $L_l$, $E_l$ όπου ο δείκτης των γενεών τρέχει ως εξής $l = e,\ \mu,\ \tau$. Στα υπερπεδία αυτά διευθετούνται αντίστοιχα:
  - Λεπτόνια, που συμβολίζονται $\psi_{L_l}$, $\psi_{E_l}$ και αναλυτικά παρουσιάζονται παρακάτω:

$$\psi_{L_l} = \begin{pmatrix} \nu_l \\ l \end{pmatrix}_L := \begin{pmatrix} \nu_e \\ e \end{pmatrix}_L, \begin{pmatrix} \nu_\mu \\ \mu \end{pmatrix}_L, \begin{pmatrix} \nu_\tau \\ \tau \end{pmatrix}_L$$

$$\psi_{E_l} = l^c_L := e^c_L, \quad \mu^c_L, \quad \tau^c_L$$

  - Sleptons, που συμβολίζονται $\phi_{L_l}$, $\phi_{E_l}$ και αναλυτικά παρουσιάζονται παρακάτω:

$$\phi_{L_l} = \begin{pmatrix} \tilde{\nu}_l \\ \tilde{l} \end{pmatrix}_L := \begin{pmatrix} \tilde{\nu}_e \\ \tilde{e} \end{pmatrix}_L, \begin{pmatrix} \tilde{\nu}_\mu \\ \tilde{\mu} \end{pmatrix}_L, \begin{pmatrix} \tilde{\nu}_\tau \\ \tilde{\tau} \end{pmatrix}_L$$

$$\phi_{E_l} = \tilde{l}^c_L := \tilde{e}^c_L, \quad \tilde{\mu}^c_L, \quad \tilde{\tau}^c_L$$

- Υπερπεδία Quarks και Squarks. Συμβολίζονται με $Q_q$, $U_q$, $D_q$ όπου ο δείκτης των γενεών τρέχει ως εξής $q = 1\,,2\,,\ 3$. Στα υπερπεδία αυτά διευθετούνται αντίστοιχα:
  - Quarks, που συμβολίζονται $\psi_{Q_q}$, $\psi_{U_q}$, $\psi_{D_q}$ και αναλυτικά παρουσιάζονται παρακάτω:

$$\psi_{Q_q} = \begin{pmatrix} u_q \\ d_q \end{pmatrix}_L := \begin{pmatrix} u \\ d \end{pmatrix}_L, \begin{pmatrix} c \\ s \end{pmatrix}_L, \begin{pmatrix} t \\ b \end{pmatrix}_L$$

$$\psi_{U_q} = u^c_{qL} := u^c_L, \quad c^c_L, \quad t^c_L$$

$$\psi_{D_q} = d^c_{qL} := d^c_L, \quad s^c_L, \quad b^c_L$$

  - Squarks, που συμβολίζονται $\phi_{Q_q}$, $\phi_{U_q}$, $\phi_{D_q}$ και αναλυτικά παρουσιάζονται παρακάτω:

$$\phi_{Q_q} = \begin{pmatrix} \tilde{u}_q \\ \tilde{d}_q \end{pmatrix}_L := \begin{pmatrix} \tilde{u} \\ \tilde{d} \end{pmatrix}_L, \begin{pmatrix} \tilde{c} \\ \tilde{s} \end{pmatrix}_L, \begin{pmatrix} \tilde{t} \\ \tilde{b} \end{pmatrix}_L$$

$$\phi_{U_q} = \tilde{u}^c_{qL} := \tilde{u}^c_L, \quad \tilde{c}^c_L, \quad \tilde{t}^c_L$$

$$\phi_{D_q} = \tilde{d}^c_{qL} := \tilde{d}^c_L, \quad \tilde{s}^c_L, \quad \tilde{b}^c_L$$

**Πίνακας 2.1: Σωματιδιακή Σύσταση του MSSM (ΙΔΚ Βαθμίδας)**

| Superfields | Συνιστώντα Πεδία | | Κβαντικοί αριθμοί | | |
|---|---|---|---|---|---|
| | **Spinors** | **Scalars** | $Q$ | $I_3$ | $Y/2$ |
| | **Πεδία Ύλης** | | | | |
| | **Fermions** | **Sfermions** | | | |
| | **Leptons** | **Sleptons** | | | |
| $L_l$ | $\psi_{L_l} = \begin{pmatrix} \nu_l \\ l \end{pmatrix}_L$ | $\phi_{L_l} = \begin{pmatrix} \tilde{\nu}_l \\ \tilde{l} \end{pmatrix}_L$ | $0$<br>$-1$ | $1/2$<br>$-1/2$ | $-1/2$<br>$-1/2$ |
| $E_l$ | $\psi_{E_l} = l^c_L$ | $\phi_{E_l} = \tilde{l}^c_L$ | $1$ | $0$ | $1$ |
| | $(l = e,\ \mu,\ \tau)$ | | | | |
| | **Quarks** | **Squarks** | | | |
| $Q_q$ | $\psi_{Q_q} = \begin{pmatrix} u_q \\ d_q \end{pmatrix}_L$ | $\phi_{Q_q} = \begin{pmatrix} \tilde{u}_q \\ \tilde{d}_q \end{pmatrix}_L$ | $2/3$<br>$-1/3$ | $1/2$<br>$-1/2$ | $-1/6$<br>$-1/6$ |
| $U_q$ | $\psi_{U_q} = u^c_{qL}$ | $\phi_{U_q} = \tilde{u}^c_{qL}$ | $-2/3$ | $0$ | $-2/3$ |
| $D_q$ | $\psi_{D_q} = d^c_{qL}$ | $\phi_{D_q} = \tilde{d}^c_{qL}$ | $1/3$ | $0$ | $1/3$ |
| | $(q = 1,\ 2,\ 3)$ | | | | |
| | **Διανυσματικά Πεδία** | | | | |
| | **Gauginos** | **Gauge Bosons** | | | |
| $V_1$ | $\tilde{B}$ | $B$ | $0$ | $0$ | $0$ |
| $V_2^a$ | $\begin{pmatrix} \tilde{W}^\pm \\ \tilde{W}^3 \end{pmatrix}$ | $\begin{pmatrix} W^\pm \\ W^3 \end{pmatrix}$ | $\pm 1$<br>$0$ | $0$<br>$0$ | $\pm 1$<br>$0$ |
| $V_3^A$ | $\tilde{g}^A$ | $g^A$ | $0$ | $0$ | $0$ |
| | $(A = 1, ..., 8)$ | | | | |
| | **Πεδία Higgs** | | | | |
| | **Higgsinos** | **Higgs Bosons** | | | |
| $H$ | $\psi_H = \begin{pmatrix} \tilde{H}_0 \\ \tilde{H}_- \end{pmatrix}$ | $\phi_H = \begin{pmatrix} H_0 \\ H_- \end{pmatrix}$ | $0$<br>$-1$ | $+1/2$<br>$-1/2$ | $-1/2$<br>$-1/2$ |
| $\bar{H}$ | $\psi_{\bar{H}} = \begin{pmatrix} \tilde{H}_+ \\ \tilde{\bar{H}}_0 \end{pmatrix}$ | $\phi_{\bar{H}} = \begin{pmatrix} H_+ \\ \bar{H}_0 \end{pmatrix}$ | $+1$<br>$0$ | $+1/2$<br>$-1/2$ | $1/2$<br>$1/2$ |

Είναι προφανές ότι ο φερμιονικός τομέας των υπερπεδίων αυτών, ταυτίζεται με τον αντίστοιχο του SM.

**β.** Διανυσματικά Υπερπεδία. Σε αυτά διευθετούνται τα πεδία άμαζων διανυσματικών μποζονίων και των SUSY φερμιονικών συντρόφων τους, των Gauginos. Η πηγή προέλευσης των μποζονίων αυτών είναι η απαίτηση αναλοιώτητας της λαγκραζιανής κάτω από μετασχηματισμούς που ανήκουν στην ομάδα $SU(3)_c \times SU(2)_L \times U(1)_Y$. Συγκεκριμένα στην ομάδα:

- $U(1)_Y$ με γεννήτορα το υπερφορτίο $Y$, αντιστοιχεί το υπερπεδίο $V_1$ με περιεχόμενα πεδία:

$$B_\mu \;\; \text{και} \;\; \tilde{B} \;\; \text{(Bino)}$$

  και σταθερά ζεύξης $g'$.

- $SU(2)_L$ με γεννήτορες τους γνωστους πίνακες του Pauli της Εξ. (B′.7)

$$\tau^a, \quad \text{όπου} \;\; a = 1, 2, 3$$

  αντιστοιχεί το υπερπεδίο $V_2^a$ με περιεχόμενα πεδία:

$$W_\mu^a \;\; \text{και} \;\; \tilde{W}^a \;\; \text{(Wino)}$$

  και σταθερά ζεύξης $g$.

- $SU(3)_c$ με γεννήτορες τους πίνακες Gell-Mann, αντιστοιχεί το υπερπεδίο $V_3^A$ με $A = 1, ..., 8$ και περιεχόμενα πεδία:

$$g_\mu^A \;\; \text{(gluon)} \;\; \text{και} \;\; \tilde{g}^A \;\; \text{(gluino)}$$

  και σταθερά ζεύξης $g_3$.

Είναι προφανές ότι ο τομέας των διανυσματικών μποζονίων ταυτίζεται με τον αντίστοιχο του SM.

**γ.** Υπερπεδία Higgs. Πρόκειται για υπερπεδία χειραλότητας που αναπαρίστανται με διπλέτες ως προς την ομάδα συμμετρίας $SU(2)_L$ και είναι τα εξής:

- $H$ με την ηλεκτρικά ουδέτερη συνιστώσα στη θέση 11 και περιεχόμενα πεδία:

$$\phi_H := \begin{pmatrix} H_0 \\ H_- \end{pmatrix} , \quad \psi_H := \begin{pmatrix} \tilde{H}_0 \\ \tilde{H}_- \end{pmatrix}$$

- $\bar{H}$ με την ηλεκτρικά ουδέτερη συνιστώσα στη θέση 21 και περιεχόμενα πεδία:

$$\phi_{\bar{H}} := \begin{pmatrix} H_+ \\ \bar{H}_0 \end{pmatrix} , \quad \psi_{\bar{H}} := \begin{pmatrix} \tilde{H}_+ \\ \tilde{\bar{H}}_0 \end{pmatrix}$$

Παρατηρείται ότι εκτός από το χρησιμοποιούμενο στο SM μποζόνιο Higgs υπάρχει και ένα επιπλέον, η ανάγκη εισαγωγής του οποίου θα εξηγηθεί στο Εδ. 2.3.2β.

Οι κβαντικοί αριθμοί που επισυνάπτονται σε κάθε συνιστώσα από τα υπερπεδία που αναφέρθηκαν είναι καταγεγραμμένοι στον Πίνακα 2.1. Με τον όρο ΙΔΚ βαθμίδας νοόυνται οι ΙΔΚ των πεδίων πριν το SSB, όπου κυριαρχεί η συμμετρία βαθμίδας στη θεωρία. Μετά το SSB, οι ΙΔΚ βαθμίδας θα δώσουν τη θέση τους στις ΙΔΚ μάζας όπως θα φανεί στους Πίνακες 2.2, 2.3.

### 2.3.2 Σχεδιασμός του Υπερδυναμικού

Η ανάπτυξη των βασικών επιχειρημάτων, στα οποία εδράζεται η μορφή του υπερδυναμικού του MSSM είναι ο στόχος αυτής της παραγράφου. Οι βασικές προυποθέσεις που επιδιώκεται να ικανοποιεί ένα υπερδυναμικό που θα φιλοδοξούσε να αναπαράγει τα αποτελέσματα του SM με ελάχιστη SUSY επέκταση είναι οι εξής:

**α.** Σεβασμός της συμμετρίας του SM. Οι υποψήφιοι όροι για είσαγωγη στο υπερδυναμικό πρέπει να είναι αναλλοίωτοι υπό τη δράση της $SU(3)_c \times SU(2)_L \times U(1)_Y$. Αυτό εξασφαλίζεται αν το άθροισμα των κβαντικών αριθμών που επισυνάπτονται σε κάθε μέλος ενός υποψήφιου όρου είναι μηδέν. Ενδεικτικά αναφέρεται ότι οι όροι $H^T i\tau_2 L_l E_l$ ή $U_q D_{q'} D_{q''}$ περνούν τον έλεγχο αυτό.

**β.** Παροχή μάζας στα φερμιόνια. Όροι που θα μπορούσαν να προσδώσουν μάζα στα φερμιόνια προέρχονται από τον πρώτο προσθετέο του δεύτερου μέλους της Εξ. (2.2). Κατά τα πρότυπα του SM ένας όρος που θα παρείχε μάζα στα λεπτόνια θα ήταν ο $h_l H^T i\tau_2 L_l E_l$ σε επίπεδο υπερπεδίων ή σε επίπεδο βαθμωτών συνιστωσών $h_l \phi_H^T i\tau_2 \phi_{L_l} \phi_{E_l}$ Λαμβάνοντας τη παράγωγο ποσότητας αυτής και υποθέτωντας ότι το $\phi_H$ αναπτύσσει VEV στην ηλεκτρικά ουδέτερη συνιστώσα του, όπως στην Εξ. (2.37), ο επίμαχος όρος της Εξ. (2.2) γράφεται:

$$-\frac{1}{2} h_l \phi_H^T i\tau_2 \psi_{L_l} \psi_{E_l} + \text{h.c} = -\frac{h_l}{2\sqrt{2}} \begin{pmatrix} v_1 & 0 \end{pmatrix} \begin{pmatrix} 0 & 1 \\ -1 & 0 \end{pmatrix} \begin{pmatrix} \nu_l \\ l \end{pmatrix}_L l_L^c + \text{h.c} = -\frac{1}{2} m_l (l_L l_L^c + \text{h.c}),$$

όπου $m_l = v_1 h_l/\sqrt{2}$. Με τον ίδιο ακριβώς τρόπο εφοδιάζονται με μάζα τα down-quark, οπότε ένας όρος της μορφής $h_{D_q} H^T i\tau_2 Q_q D_q$ είναι επίσης επιβεβλημένος. Είναι, όμως, εμφανές από την προηγούμενη ότι το $\phi_H$ δεν μπορεί να παράσχει μάζα στα up-quark λόγω της θέσης της ηλεκτρικά ουδέτερης συνιστώσας και του υπερφορτίου που αυτή διαθέτει (-1/2). Χρειάζεται ένα Higgs που να έχει την ηλεκτρικά ουδέτερη συνιστώσα στη θέση 21 και υπερφορτίο 1/2. Στο SM αυτή η μετάλλαξη επιτυγχάνεται, λαμβάνοντας το $i\tau_2$ επί το μιγαδικό συζυγές του χρησιμοποιούμενου εκεί πεδίου Higgs. Στην περίπτωση του MSSM, όμως αυτή η συνταγή δεν μπορεί να εφαρμοστεί, λόγω της ολομορφικής ιδιότητας του υπερδυναμικού που έχει ήδη αναφερθεί στο Εδ. 2.2.1. Συνεπώς, αναγκαστική είναι η χρησιμοποίηση ενός δεύτερου υπερπεδίου Higgs $\bar{H}$ με τις απαιτούμενες ιδιότητες, οπότε η διαδικασία παροχής μάζας στα up-quark εξελίσσεται κατά τα γνωστά: Εισάγεται στο υπερδυναμικό ο όρος $-h_{U_q} \bar{H}^T i\tau_2 Q_q U_q$ που σε μορφή βαθωτών συνιστωσών γράφεται $-h_{U_q} \phi_{\bar{H}}^T i\tau_2 \phi_{Q_q} \phi_{U_q}$, οπότε λαμβάνοντας τη παράγωγο της ποσότητας αυτής και υποθέτωντας ότι το $\phi_{\bar{H}}$ αναπτύσσει VEV στην ηλεκτρικά ουδέτερη συνιστώσα του, όπως στην Εξ. (2.37), από τον επίμαχο όρο της Εξ. (2.2) προκύπτει:

$$\frac{1}{2} h_{U_q} \phi_{\bar{H}}^T i\tau_2 \psi_{Q_q} \psi_{U_q} + \text{h.c} = \frac{h_{U_q}}{2\sqrt{2}} \begin{pmatrix} 0 & v_2 \end{pmatrix} \begin{pmatrix} 0 & 1 \\ -1 & 0 \end{pmatrix} \begin{pmatrix} u_q \\ d_q \end{pmatrix}_L u_{qL}^c + \text{h.c} = -\frac{1}{2} m_{u_q} (u_{qL} u_{qL}^c + \text{h.c}),$$

όπου $m_{u_q} = v_2 h_{U_q}/\sqrt{2}$. Με $h_{U_q} > 0$, ένας τέτοιος όρος παροχής μάζας στα up-quark πρέπει να προσημανθεί αντίθετα από τους υπόλοιπους για την επίτευξη πλήρως συμβατού αποτελέσματος.

**γ.** Μη παραβίαση του λεπτονικού και βαρυονικού αριθμού. Υπάρχουν όροι που εκπληρούν τη συνθήκη του σημείου (α), όπως οι παρακάτω:

$$U_q D_{q'} D_{q''}, \quad Q_q^T i\tau_2 L_l D_{q'}, \quad L_l^T i\tau_2 L_{l'} E_{l''}, \quad L_l^T i\tau_2 H$$

οι οποιοί, όμως, είναι ανεπιθύμητοι γιατί παραβιάζουν τη διατήρηση του βαρυονικού, $B$ και λεπτονικού αριθμού, $L$ οι τιμές των οποίων για τα υπερπεδία του MSSM δίνονται παρακάτω:

$$\begin{array}{lllll} B(L_l) = 0, & B(E_l) = 0, & B(Q_q) = 1/3, & B(U_q) = -1/3, & B(D_q) = -1/3 \\ L(L_l) = 1, & L(E_l) = -1, & L(Q_q) = 0, & L(U_q) = 0, & L(D_q) = 0 \end{array} \tag{2.14}$$

Η παραβίαση της διατήρησης αυτών των αριθμών θα επετρέπε τη διάσπαση του πρωτονίου, πράγμα φαινομενολογικά απαράδεκτο.

**δ.** Διατήρηση της ισοτιμίας $R$. Η δια χειρός απόρριψη μερικών όρων, λόγω φαινομενολογικών αντιρρήσεων δεν είναι θεωρητικά υγιής διαδικασία. Πιο ευθύς τρόπος δράσης θα ήταν ο καθορισμός μιας νέας συμμετρίας που θα πρέπει να σέβεται το MSSM. Ορίζεται ένας πολλαπλασιαστικός κβαντικός αριθμός που επισυνάπτεται σε πεδία με την επωνυμία $R$-Parity:

$$R := (-1)^{2S+L+3B} = \begin{cases} +1 & \text{για τα σωμάτια του SM και τα Higgs} \\ -1 & \text{για τα sparticles} \end{cases} \tag{2.15}$$

όπου $S$ το σπιν του πεδίου, και επιβάλλεται η διατήρηση του. Δηλαδή, η $R$-Parity ενός όρου της λαγκραζιανής πρέπει να είναι 1, πράγμα που φαίνεται με μια απλή παρατήρηση στους όρους της Λαγκραζιανής που παρουσιάζονται στο Εδ. 2.3.3. Η επιβολή αυτής της νέας συμμετρίας έχει τις εξής συνέπειες:

- Οι όροι της λαγκραζιανής μπορούν να περιέχουν μόνο άρτιο αριθμό από sparticles.
- Τα sparticles μπορούν να παράγονται μόνο κατά ζεύγη.
- Τα βαρύτερα sparticles διασπώνται στα ελαφρότερα.
- Το LSP είναι ευσταθές γιατί δεν επιτρέπεται να διασπαστεί σε σωμάτια του SM.
- Το LSP είναι ασθενώς αλληλεπιδρών σωμάτιο, γιατί αλληλεπιδρά με την ύλη μόνο με τη μεσολάβηση ενός βαρέως ενδιάμεσου sparticle.

Τα χαρακτηριστικά αυτά που αποκτά το LSP το καθιστούν ένα πολύ ελκυστικό υποψήφιο για τη λύση του προβλήματος της σκοτεινής ύλης, με τρόπο που εκτενώς θα μελετηθεί στο Κεφάλαιο 5.

**ε.** Αποφυγή της ύπαρξης άμαζου μποζονίου Higgs. Όπως θα φανεί στο Εδ. 2.4.5, ο στόχος αυτός επιτυγχάνεται με την εισαγωγή ενός όρου αλληλεπίδρασης, $\mu H^T i\tau_2 \bar{H}$, μεταξύ των υπερπεδίων Higgs.

Η τελική έκφραση του υπερδυναμικού, συμβατή με τους προηγούμενους περιορισμούς, είναι:

$$W_{\rm MSSM} = h_l H^T i\tau_2 L_l E_l + h_{D_q} H^T i\tau_2 Q_q D_q - h_{U_q} \bar{H}^T i\tau_2 Q_q U_q + \mu H^T i\tau_2 \bar{H} \tag{2.16}$$

Σχετικά με τον χρησιμοποιούμενο συμβολισμό, μπορεί να διευκρινιστεί ότι:

- Πινακικοί δείκτες για τη ζεύξη των διπλετών δεν χρησιμοποιούνται γιατί υιοθετήθηκε η παρουσίαση του τυπολόγιου στη γλώσσα των πινάκων και όχι των συνιστωσών.
- Στους επαναλαμβανόμενους δείκτες $l$, $q$ μπορεί να ενοηθεί άθροιση πάνω στις γενεές λεπτονίων και quark αντίστοιχα.
- Αγνοείται η ανάμιξη ανάμεσα στις γενεές, πράγμα που σημαίνει ότι οι ζεύξεις Yukawa είναι αριθμοί και όχι πίνακες.

### 2.3.3 Λανγκραζιανή του MSSM

Η πλήρης μορφή της λαγκραζιανής του MSSM δίνεται στις Αν. [1] και [14]. Η παρουσίαση που θα επιχειρηθεί παρακάτω είναι λιγότερο λεπτομεριακή αλλά ίσως πιο περιεκτική. Χάριν απλότητας παραλείπονται όροι που οφείλονται στη QCD, όροι φαντασμάτων και συγκολιτονίων. Η λανγκραζιανή του MSSM μπορεί να πάρει την γενική μορφή:

$$\mathcal{L}_{\rm MSSM} = \mathcal{L}_{\rm KIN} + \mathcal{L}_{\rm INT} - V_{\rm MSSM}, \quad \text{όπου:}$$

**α.** $\mathcal{L}_{\rm KIN}$: Το τμήμα που περιέχει τους κινητικούς όρους των πεδίων της θεωρίας, των μποζονίων βαθμίδας, των βαθμωτών πεδίων και των φερμιονίων ύλης και gauginos. Από αυτούς, οι χρήσιμοι για την ανάλυση που θα επιχειρηθεί είναι οι κινητικοί όροι των μποζονίων Higgs, οι οποίοι περιέχονται στους όρους των βαθμωτών πεδίων που, με βάση την Εξ. (2.1), είναι:

$$\sum_F \left(D^\mu \phi_F\right)^\dagger D_\mu \phi_F \quad \text{όπου} \quad F := L_l,\, E_l,\, Q_q,\, U_q,\, D_q,\, H,\, \bar{H} \tag{2.17}$$

Η συναλλοίωτη παράγωγος έχει την (πινακική) μορφή:

$$\begin{aligned} D_\mu &= \partial_\mu + ig\frac{\tau^a}{2}W^a_\mu + ig'\frac{Y}{2}B_\mu \\ &= \partial_\mu + i\frac{g}{\sqrt{2}}\left(\tau_- W^+_\mu + \tau_+ W^-_\mu\right) + i\frac{e}{s_W c_W}g_f(I_3, Q)Z_\mu + ieQA_\mu, \end{aligned} \tag{2.18}$$

όπου για την επίτευξη της τελευταίας, πραγματοποιήθηκαν διαδοχικά τα επόμενα βήματα:

- Αναδιατάχτηκαν οι συνιστώσες των διανυσματικών μποζονίων $W^a_\mu$ ως εξής:

$$W^\pm_\mu := \frac{1}{\sqrt{2}}(W^1_\mu \pm iW^2_\mu) \ \text{ και } \ \tau_\pm := \frac{1}{2}(\tau^1 \pm i\tau^2)$$

- Έγινε η στροφή ορισμού των πεδίων $Z_\mu$ και $A_\mu$:

$$\begin{pmatrix} Z_\mu \\ A_\mu \end{pmatrix} = \begin{pmatrix} c_W & -s_W \\ s_W & c_W \end{pmatrix} \begin{pmatrix} W^3_\mu \\ B_\mu \end{pmatrix}. \tag{2.19}$$

όπου χρησιμοποιούνται οι συντμίσεις $s_W := \sin\theta_W$, $c_W := \cos\theta_W$ για τη γωνία Weinberg.

- Έγινε η ταυτοποίηση των σταθερών, σύμφωνα με την Αν. [15]:

$$g s_W = g' c_W = e = \frac{g g'}{\sqrt{g^2 + g'^2}} \tag{2.20}$$

- Ορίστηκε η ποσότητα:

$$g_f(I_3, Q) \ = \ I_3 \ - \ Q s^2_W\,. \tag{2.21}$$

με $Y$ το υπερφορτίο, $Q$ το ηλεκτρικό φορτίο και $I_3$ την τρίτη συνιστώσα του isospin του σωματίου $f$ τα οποία ικανοποιούν τη γνωστή σχέση:

$$Q = I_3 + \frac{Y}{2}$$

**β.** $\mathcal{L}_{\rm INT}$: Το τμήμα που περιέχει τους όρους αλληλεπίδρασης που προκύπτουν από την εισαγωγή της SUSY και βρίσκεται με εφαρμογή της Εξ. (2.2). Επιβεβλημένος μάλιστα, για το λόγο αυτό είναι ο διαχωρισμός του σε δύο τμήματα:

$$\mathcal{L}_{\rm INT} = \mathcal{L}_{\rm Yuk} + \mathcal{L}_{\rm Gau}, \quad \text{όπου:}$$

- $\mathcal{L}_{\rm Yuk}$, το τμήμα στο οποίο ανήκουν οι ζεύξεις Yukawa που προκύπτουν από το υπερδυναμικό:

$$\begin{aligned} \mathcal{L}_{\rm Yuk} \ &= \ -\frac{1}{2}\sum_{F,F'} \frac{\partial^2 W_{\rm MSSM}}{\partial\phi_{Fi}\partial\phi_{F'j}}\psi_{Fi}\psi_{F'j} + {\rm h.\,c} = \\ &- \ \frac{1}{2}\left(h_l \phi^T_H i\tau_2 \psi_{L_l}\psi_{E_l} + h_{D_q}\phi^T_H i\tau_2\psi_{Q_q}\psi_{D_q} - h_{U_q}\phi^T_{\bar H} i\tau_2\psi_{Q_q}\psi_{U_q} + \mu\psi^T_H i\tau_2\psi_{\bar H}\right) + {\rm h.\,c} \\ &- \ \frac{1}{2}\left(h_l \psi^T_H i\tau_2 \psi_{L_l}\phi_{E_l} + h_{D_q}\psi^T_H i\tau_2\psi_{Q_q}\phi_{D_q} - h_{U_q}\psi^T_{\bar H} i\tau_2\psi_{Q_q}\phi_{U_q}\right) + {\rm h.\,c} \\ &- \ \frac{1}{2}\left(h_l \psi^T_H i\tau_2 \phi_{L_l}\psi_{E_l} + h_{D_q}\psi^T_H i\tau_2\phi_{Q_q}\psi_{D_q} - h_{U_q}\psi^T_{\bar H} i\tau_2\phi_{Q_q}\psi_{U_q}\right) + {\rm h.\,c} \end{aligned} \tag{2.22}$$

- $\mathcal{L}_{\rm Gau}$, το τμήμα στο οποίο ανήκουν οροι αλληλεπίδρασης scalars-spinor-gaugino:

$$\mathcal{L}_{\rm Gau} = i\sqrt{2}g' \sum_F \phi^\dagger_F \tilde{B}\frac{Y}{2}\psi_F + i\sqrt{2}g\sum_F \phi^\dagger_F \tilde{W}^a \frac{\tau^a}{2}\psi_F + {\rm h.c} \tag{2.23}$$

Η ανάπτυξη των όρων αυτών, πραγματοποιώντας την άθροιση πάνω στα πεδία $F$, προκύπτει αυθορμήτως με αντικατάσταση των κβαντικών αριθμών των πεδίων από τον Πίνακα 2.1, γιαυτό και δε θα παρουσιαστεί αναλυτικότερη έκφραση αυτού του τμήματος της λαγκραζιανής.

**γ.** $V_{\rm MSSM}$: Το δυναμικό του MSSM που δομείται από δύο διαφορετικής προέλευσης όρους, τον προερχόμενο από την SUSY, $V_{\rm SUSY}$ και τον φέροντα τους SBT, $V_{\rm SBT}$, δηλαδή:

$$V_{\rm MSSM} = V_{\rm SUSY} + V_{\rm SBT}, \ \ \text{με} \ \ V_{\rm SUSY} = V_{\rm F} + V_{\rm D}, \ \ \text{όπου:}$$

- $V_{\rm F}$, το τμήμα που οφείλεται στους F-όρους και βρίσκεται εφαρμόζοντας την Εξ. (2.3):

$$\begin{aligned} V_{\rm F} &= \sum_F \left| \frac{\partial W_{\rm MSSM}}{\partial \phi_F} \right|^2 \\ &= \left| h_l i\tau_2 \phi_{L_l} \phi_{E_l} + h_{D_q} i\tau_2 \phi_{Q_q} \phi_{D_q} + \mu i\tau_2 \phi_{\bar{H}} \right|^2 \\ &+ \left| h_{U_q} i\tau_2 \phi_{Q_q} \phi_{U_q} + \mu \phi_H^T i\tau_2 \right|^2 \\ &+ \left| h_{D_q} \phi_H^T i\tau_2 \phi_{D_q} + h_{U_q} \phi_{\bar{H}}^T i\tau_2 \phi_{U_q} \right|^2 \\ &+ \left| h_l \phi_H^T i\tau_2 \phi_{L_l} \right|^2 + \left| h_{D_q} \phi_H^T i\tau_2 \phi_{Q_q} \right|^2 + \left| h_{U_q} \phi_{\bar{H}}^T i\tau_2 \phi_{Q_q} \right|^2 \end{aligned} \tag{2.24}$$

- $V_{\rm D}$, το τμήμα που οφείλεται στους D-όρους και βρίσκεται εφαρμόζοντας την Εξ. (2.4). Είναι, μάλιστα βολικός ο τεμαχισμός του σε δύο τμήματα:

$$V_{\rm D} = V_{\rm D_Y} + V_{\rm D_L}, \quad \text{όπου:}$$

  – $V_{\rm D_Y}$, το τμήμα που οφείλεται στους D-όρους που προέρχονται από τη συμμετρία $U(1)_Y$:

$$V_{\rm D_Y} = \frac{g'^2}{8}(\sum_F \phi_F^\dagger Y \phi_F)^2 =$$

$$\frac{g'^2}{8}\left(\frac{1}{3}|\phi_{Q_q}|^2 - \frac{4}{3}|\phi_{U_q}|^2 + \frac{2}{3}|\phi_{D_q}|^2 - |\phi_{L_l}|^2 + 2|\phi_{E_l}|^2 + |\phi_{\bar{H}}|^2 - |\phi_H|^2\right)^2 \tag{2.25}$$

  – $V_{\rm D_L}$, το τμήμα που οφείλεται στους D-όρους που προέρχονται από τη συμμετρία $SU(2)_L$:

$$\begin{aligned} V_{\rm D_L} &= \frac{g^2}{8}\sum_a \left(\sum_F \phi_F^\dagger \tau^a \phi_F\right)^2 \\ &= \frac{g^2}{8}\sum_a \left(\phi_H^\dagger \tau^a \phi_H + \phi_{\bar{H}}^\dagger \tau^a \phi_{\bar{H}} + \phi_{Q_q}^\dagger \tau^a \phi_{Q_q} + \phi_{L_l}^\dagger \tau^a \phi_{L_l}\right)^2 \\ &= \frac{g^2}{8}\Big[ \left(|\phi_H|^2\right)^2 + \left(|\phi_{\bar{H}}|^2\right)^2 - 2|\phi_H|^2|\phi_{\bar{H}}|^2 + 4|\phi_H^\dagger \phi_{\bar{H}}|^2 \\ &+ \left(|\phi_{Q_q}|^2\right)^2 + \left(|\phi_{L_l}|^2\right)^2 - 2|\phi_{Q_q}|^2|\phi_{L_l}|^2 + 4|\phi_{Q_q}^\dagger \phi_{L_l}|^2 \\ &- 2|\phi_H|^2|\phi_{Q_q}|^2 + 4|\phi_H^\dagger \phi_{Q_q}|^2 - 2|\phi_{\bar{H}}|^2|\phi_{Q_q}|^2 + 4|\phi_{\bar{H}}^\dagger \phi_{Q_q}|^2 \\ &- 2|\phi_H|^2|\phi_{L_l}|^2 + 4|\phi_H^\dagger \phi_{L_l}|^2 - 2|\phi_{\bar{H}}|^2|\phi_{L_l}|^2 + 4|\phi_{\bar{H}}^\dagger \phi_{L_l}|^2 \Big] \end{aligned} \tag{2.26}$$

  όπου χρησιμοποιήθηκε η ταυτότητα:

$$\sum_a \left(A^\dagger \tau^a A\right)\left(B^\dagger \tau^a B\right) = -|A|^2|B|^2 + |A^\dagger B|^2$$

- $V_{\rm SBT}$, το τμήμα που περιέχει τους SBT, οι οποίοι, με βάση τα εκτεθέντα στο Εδ. 2.2.4, είναι:

$$\begin{aligned} V_{\rm SBT} &= \sum_F m_F |\phi_F|^2 \\ &- \frac{1}{2}M_1 \tilde{B}\tilde{B} - \frac{1}{2}M_2 \sum_a \tilde{W}^a \tilde{W}^a - \frac{1}{2}M_3 \sum_A \tilde{g}^A \tilde{g}^A + \text{h. c} \end{aligned}$$

$$+ h_l A_l \phi_H^T i\tau_2 \phi_{L_l} \phi_{E_l} + h_{D_q} A_{D_q} \phi_H^T i\tau_2 \phi_{Q_q} \phi_{D_q} - h_{U_q} A_{U_q} \phi_{\bar{H}}^T i\tau_2 \phi_{Q_q} \phi_{U_q} + \mu B \phi_H^T i\tau_2 \phi_{\bar{H}} + \text{h. c} \tag{2.27}$$

Από τη λαγκραζιανή που παρουσιάστηκε, μπορεί να ληφθεί το σωματιδιακό φάσμα του MSSM, πράγμα που θα αποτελέσει το επίκεντρο μελέτης των δύο επόμενων ενοτήτων.

## 2.4 Μηχανισμός Higgs στο MSSM

Ο μηχανισμός Higgs είναι από τα πιο χαρακτηριστικά σημεία μιας θεωρίας βαθμίδας. Για την καλύτερη μελέτη του στην περίπτωση του MSSM απομονώνεται το σχετικό τμήμα της λαγκραζιανής στο Εδ. 2.4.1 και βρίσκονται οι συνθήκες λειτουργικότητας του μηχανισμού του SSB στο Εδ. 2.4.2 και ελαχιστοποίησης στο Εδ. 2.4.3. Με αυτή την προετοιμασία, ο μηχανισμός Higgs εξελίσσεται παρέχοντας μάζα σε κάποια μποζόνια βαθμίδας στο Εδ. 2.4.4, αναδεικνύοντας τις ΙΔΚ μάζας των Higgs στο Εδ. 2.4.5 και παρέχοντας μάζα στα φερμιόνια της θεωρίας στο Εδ. 2.4.6.

### 2.4.1 Το τμήμα Higgs της λαγκραζιανής

Το αμιγώς αφιερωμένο στα βαθμωτά πεδία Higgs τμήμα της λαγκραζιανής του MSSM, είναι:

$$\mathcal{L}_{\rm H} = \left(D^{\mu}\phi_H\right)^{\dagger} D_{\mu}\phi_H + \left(D^{\mu}\phi_{\bar H}\right)^{\dagger} D_{\mu}\phi_{\bar H} - V_H, \tag{2.28}$$

όπου οι δύο πρώτοι όροι του δεύτερου μέλους προέρχονται από το τμήμα της Εξ. (2.17), ενώ ο τρίτος λαμβάνει συνεισφορές από τις Εξ. (2.25), (2.26) και (2.27) οπότε η μορφή του είναι:

$$\begin{aligned} V_H &= m_1^2\,|\phi_H|^2 + m_2^2\,|\phi_{\bar H}|^2 + m_{12}^2\,\left(\phi_H^T\, i\tau_2\phi_{\bar H} + {\rm h.\,c}\right) \\ &+ \frac{1}{8}(g^2+g'^2)\,\left(|\phi_H|^2 - |\phi_{\bar H}|^2\right)^2 + \frac{1}{2}\,g^2\,|\phi_H^{\dagger}\,\phi_{\bar H}|^2\,. \end{aligned} \tag{2.29}$$

όπου ορίστηκαν οι ποσότητες:

$$m_1^2 = m_H^2 + \mu^2\,, \tag{2.30}$$

$$m_2^2 = m_{\bar H}^2 + \mu^2\,, \tag{2.31}$$

$$m_{12}^2 = \mu B. \tag{2.32}$$

Αντικαθιστώντας τις οριστικές σχέσεις των $\phi_H$, $\phi_{\bar H}$ στην Εξ. (2.29), αυτή γράφεται ως εξής:

$$\begin{aligned} V_H &= m_1^2\,|H_0|^2 + m_2^2\,|\bar H_0|^2 + m_{12}^2\,\left(H_0\bar H_0 + H_0^*\bar H_0^*\right) \\ &+ \frac{1}{8}(g^2+g'^2)\,\left(|H_0|^2 - |\bar H_0|^2 + |H_-|^2 - |H_+|^2\right)^2 \\ &+ m_1^2\,|H_-|^2 + m_2^2\,|H_+|^2 + m_{12}^2\,\left(-H_+H_- - H_+^*H_-^*\right) \\ &+ \frac{1}{2}\,g^2\,\left(|H_0|^2|H_+|^2 + H_+H_-\bar H_0^*H_0^* + H_+^*H_-^*\bar H_0H_0 + |\bar H_0|^2|H_-|^2\right)\,. \end{aligned} \tag{2.33}$$

Για τον προσδιορισμό του κενού της θεωρίας προφανώς θα χρησιμοποιηθεί μόνο το ηλεκτρικά ουδέτερο τμήμα της Εξ. 2.33 , όπως θα φανεί στα Εδ. 2.4.3, 2.4.2. Αντιθέτως, για τον προσδιορισμό του φάσματος Higgs της θεωρίας, χρειάζεται η πλήρης έκφραση της Εξ. 2.33, όπως θα φανεί στο Εδ. 2.4.5.

### 2.4.2 Συνθήκες επιτυχούς SSB

Το τμήμα των ηλεκτρικά ουδέτερων συνιστωσών των πεδίων Higgs του δυναμικού της Εξ. (2.33) μπορεί να γραφεί ως εξής:

$$V_{H_0} = \frac{1}{8}(g^2+g'^2)\,\left(|H_0|^2 - |\bar H_0|^2\right)^2 + \begin{pmatrix} H_0^* & \bar H_0 \end{pmatrix} \begin{pmatrix} m_1^2 & m_{12}^2 \\ m_{12}^2 & m_2^2 \end{pmatrix} \begin{pmatrix} H_0 \\ \bar H_0^* \end{pmatrix}\,. \tag{2.34}$$

Για να μπορεί να επιτευχθεί το SSB, πρέπει να πληρούνται οι παρακάτω συνθήκες:

- Ο τετραγωνικός πίνακας μάζων να διαθέτει αρνητική ΙΔΤ. Και αυτό, διότι αν δεν υπάρχει αρνητική ΙΔΤ, το $H_0 = \bar H_0 = 0$ θα είναι ένα ευσταθές ελάχιστο του δυναμικού και επομένως το SSB δεν θα ενεργοποιείται. Ισοδύναμα, με βάση την Αν. [8], πρέπει:

$$\det \begin{pmatrix} m_1^2 & m_{12}^2 \\ m_{12}^2 & m_2^2 \end{pmatrix} < 0 \Longrightarrow m_1^2 m_2^2 < m_{12}^4 \tag{2.35}$$

  Η ύπαρξη μιας αρνητικής μάζας, που επιλέγεται να είναι η $m_2^2$, διευκολύνει οπωσδήποτε την εγκαθίδρυση της προηγούμενης και πυροδοτεί (κατά την εύστοχη έκφραση της Αν. [1]) την παραβίαση της ηλεκρασθενούς συμμετρίας.

- Το δυναμικό $V_{H_0}$ να είναι φραγμένο κάτω. Και αυτό, διοτι στην διεύθυνση $H_0 = \bar{H}_0$ υπάρχει ο κίνδυνος ο τετραγωνικός όρος της Εξ. (2.34) να μηδενίζεται. Για να αποφευκτεί αυτό πρέπει σε αυτές τις λεγόμενες D επιπεδες διευθύνσεις να ισχύει:

$$V_{H_0} > 0 \Longrightarrow m_1^2 + m_2^2 > -2m_{12}^2 \tag{2.36}$$

- Να αποφεύγεται η ύπαρξη χρωματιστών ελαχίστων, δηλαδή οι βαθμωτοί σύντροφοι των φερμιονίων ύλης να μην αναπτύσουν VEV και δημιουργούν SSB της $SU(3)_c$. Για να επιτευχθεί αυτό με βάση την Αν. [17], πρέπει:

$$\begin{aligned} A_{U_q}^2 &< 3\,(m_{Q_q}^2 + m_{U_q}^2 + m_{\bar{H}}^2), \\ A_{D_q}^2 &< 3\,(m_{Q_q}^2 + m_{D_q}^2 + m_H^2), \\ A_l^2 &< 3\,(m_{L_l}^2 + m_{E_l}^2 + m_H^2). \end{aligned}$$

Υποθέτωντας ότι ισχύουν οι συνθήκες αυτές, ο μηχανισμός του SSB αναπτύσσεται στις παρακάτω παραγράφους απρόσκοπτα.

### 2.4.3 Συνθήκες ελαχιστοποίησης

Η διαδικασία του SSB αρχίζει με την επιλογή του κενού της θεωρίας που πρέπει να είναι τέτοιο, ώστε να είναι η προκύπτουσα θεωρία αναλοίωτη υπό την $U(1)_{\rm EM}$. Σχηματικά:

$$SU(3)_c \times SU(2)_L \times U(1)_Y \longrightarrow SU(3)_c \times U(1)_{\rm EM}$$

Για την επίτευξη του στόχου αυτού οι VEV αναπτύσσονται στις ηλεκτρικά ουδέτερες συνιστώσες των $\phi_H$, $\phi_{\bar{H}}$ πράγμα που σημαίνει ότι οι γεννήτορας της ομάδας $U(1)_{\rm EM}$ παραμένει άσπαστος (Αν. [16]):

$$\langle \phi_H \rangle_0 = \frac{1}{\sqrt{2}} \begin{pmatrix} v_1 \\ 0 \end{pmatrix} \quad \text{και} \quad \langle \phi_{\bar{H}} \rangle_0 = \frac{1}{\sqrt{2}} \begin{pmatrix} 0 \\ v_2 \end{pmatrix} . \tag{2.37}$$

Έπεται ο επαναορισμός των πεδίων, ώστε η αναμενόμενη τιμή τους να είναι μηδέν, πράγμα που σημαίνει ότι τα χρησιμοποιούμενα πεδία διαταράσσονται ως εξής:

$$\phi_H = \begin{pmatrix} (v_1 + \phi_1 + i\chi_1)/\sqrt{2} \\ H_- \end{pmatrix} \quad \text{και} \quad \phi_{\bar{H}} = \begin{pmatrix} H_+ \\ (v_2 + \phi_2 + i\chi_2)/\sqrt{2} \end{pmatrix} . \tag{2.38}$$

Οι συνθήκες ελαχιστοποίησης της Εξ. (2.9) παρέχουν:

$$\left\langle \frac{\partial V_{H_0}}{\partial \phi_1} \right\rangle_0 = 0 \quad \Longrightarrow \quad 0 = m_1^2 v_1 + m_{12}^2 v_2 + \frac{1}{8}(g^2 + g'^2) v_1 (v_1^2 - v_2^2) \ , \tag{2.39}$$

$$\left\langle \frac{\partial V_{H_0}}{\partial \phi_2} \right\rangle_0 = 0 \quad \Longrightarrow \quad 0 = m_2^2 v_2 + m_{12}^2 v_1 - \frac{1}{8}(g^2 + g'^2) v_2 (v_1^2 - v_2^2) \ . \tag{2.40}$$

Η διαφόριση του $V_{H_0}$ ως προς τις μεταβλητές $\chi_1$, $\chi_2$ παρέχει ταυτότητες. Από τις Εξ. (2.39), (2.40) προκύπτει η σημασία του όρου $m_{12}^2$ και κατα επέκταση του όρου μίξης των υπερπεδίων $H$, $\bar{H}$ στο υπερδυναμικό της Εξ. (2.16), εφόσον ισχύει και σε χαμηλές ενέργειες η Εξ. (2.32). Ακολουθώντας το επιχείρημα της Αν. [9], αν $m_{12}^2 = 0$, τότε οι πιθανές λύσεις των Εξ. (2.39) και (2.40), είναι:

- $v_1 = v_2 = 0$, οπότε δεν συμβαίνει το SSB
- $v_1 = 0$, $v_2^2 = 8m_1^2/(g^2 + g'^2)$, οπότε από τις σκέψεις σχεδιασμού του υπερδυναμικού, του Εδ. 2.3.2, μένουν άμαζα τα down-quark και τα λεπτόνια.
- $v_2 = 0$, $v_1^2 = -8m_2^2/(g^2 + g'^2)$, οπότε πάλι από τα συμπεράσματα του Εδ. 2.3.2, μένουν άμαζα τα up-quark.

Εισάγοντας τη βασική παράμετρο $\beta$ του MSSM μέσω των σχέσεων:

$$v_1 = vc_\beta, \quad v_2 = vs_\beta, \quad \tan\beta := \frac{v_2}{v_1} \tag{2.41}$$

$$v^2 = v_1^2 + v_2^2, \quad 2v_1v_2 = v^2 s_{2\beta}, \quad v_1^2 - v_2^2 = v^2 c_{2\beta} \tag{2.42}$$

με τις εύλογες συντμίσεις $\cos\beta := c_\beta$, $\sin\beta := s_\beta$ $\sin 2\beta := s_{2\beta}$, $\cos 2\beta := c_{2\beta}$ και χρησιμοποιώντας τα μεταγενέστερο αποτελέσμα της Εξ. (2.47), οι Εξ. (2.39), (2.40) γράφονται πιο βολικά:

$$m_1^2 = -m_{12}^2 \tan\beta - \frac{1}{2} M_Z^2 c_{2\beta} \,, \tag{2.43}$$

$$m_2^2 = -m_{12}^2 \cot\beta + \frac{1}{2} M_Z^2 c_{2\beta} \,. \tag{2.44}$$

Χρησιμοποιώντας τους ορισμούς των Εξ. (2.30), (2.31), (2.32), οι προηγούμενες μπορούν να λυθούν ως προς τις παράμετρες $\mu$ (με την απροσδιοριστία ενός προσήμου) και $B$:

$$\mu^2 = \frac{m_H^2 - m_{\bar{H}}^2 \tan^2\beta}{\tan^2\beta - 1} - \frac{1}{2} M_Z^2 \,, \; B = -s_{2\beta} \frac{m_H^2 + m_{\bar{H}}^2 + 2\mu^2}{2\mu} \,. \tag{2.45}$$

Στα αριθμητικά προγράμματα συνήθως υποτίθεται η επιτυχής SSB μέσω κβαντικών διορθώσεων και με δεδομένο το $\tan\beta$ λαμβάνεται το $\mu$ μέσω των Εξ. (2.45), όπως θα εξηγηθεί στα Εδ. 6.4.

### 2.4.4 Τομέας Διανυσματικών Μποζονίων

Οι όροι μάζας για τα διανυσματικά μποζόνια προκύπτουν από τους κινητικούς όρους της Εξ. (2.28). Πιο συγκεκριμένα, δρώντας με τον τελεστή της συναλλοίωτης παραγώγου της Εξ. (2.18) στις Εξ. (2.38), λαμβάνονται εκτός από τους όρους αλληλεπιδράσεων, και οι παρακάτω

$$\frac{g^2}{2}\frac{v_2^2}{2} W_\mu^+ W^{\mu-} + \frac{e^2}{4c_w^2 s_W^2}\frac{v_2^2}{2} Z_\mu Z^\mu + \frac{g^2}{2}\frac{v_1^2}{2} W_\mu^+ W^{\mu-} + \frac{e^2}{4c_w^2 s_W^2}\frac{v_1^2}{2} Z_\mu Z^\mu$$

που μπορούν να ερμηνευτούν ως όροι μάζας. Το φάσμα των διανυσματικών μποζονίων του προτύπου συνίσταται από:

- Δύο διανυσματικά μποζόνια, $W_\mu^+$, $W_\mu^-$ μάζας $M_W$ τέτοιας, ώστε:

$$M_W^2 W_\mu^+ W^{\mu-} = \frac{g^2}{4}(v_1^2 + v_2^2) W_\mu^+ W^{\mu-} \Longrightarrow M_W^2 = \frac{1}{4} g^2 v^2 \tag{2.46}$$

- Ένα ουδέτερο διανυσματικό μποζόνιο, $Z_\mu$ μάζας $M_Z$ τέτοιας, ώστε:

$$\frac{1}{2} M_Z^2 Z_\mu Z^\mu = \frac{e^2}{8c_w^2 s_W^2}(v_1^2 + v_2^2) Z_\mu Z^\mu \Longrightarrow M_Z^2 = \frac{1}{4}(g^2 + g'^2) v^2 \tag{2.47}$$

- Ένα ουδέτερο διανυσματικό μποζόνιο, $A_\mu$ μάζας $M_A = 0$, αφού δεν προκύπτει όρος μάζας για αυτό, πράγμα που ήταν αναμενόμενο αφού επιζητείται ο φορέας του ηλεκτρομαγνητισμού να είναι άμαζος.

Επιπλέον, χρησιμοποιώντας τη σχέση σύνδεσης με τη σταθερά Fermi ($G_F = 1.16639 \times 10^{-5}$ GeV$^{-2}$) της ενεργού θεωρίας των ασθενών αλληλεπιδράσεων, λαμβάνεται η αριθμητική τιμή της VEV:

$$v = \frac{2M_W}{g} = (\sqrt{2} G_F)^{-1/2} = 246 \,\text{GeV}. \tag{2.48}$$

**Πίνακας 2.2: Μηχανισμός Higgs στο MSSM**

| Πριν το SSB | | Μετά το SSB | |
|---|---|---|---|
| Πεδίο | Βαθμοί Ελευθερίας | Πεδίο | Βαθμοί Ελευθερίας |
| $H_0$ | 2 | $h$ | 1 |
| $\bar{H}_0$ | 2 | $H$ | 1 |
| $H_-$ | 2 | $A$ | 1 |
| $H_+$ | 2 | $H^\pm$ | 2 |
| $B_\mu$ | 2 | $A_\mu$ | 2 |
| $W_\mu^3$ | 2 | $Z_\mu$ | 3 |
| $W_\mu^\pm$ | 4 | $W_\mu^\pm$ | 6 |

## 2.4.5 Τομέας Μποζονίων Higgs

Σύμφωνα με την Εξ. (2.10), οι πίνακες μαζών των μποζονίων Higgs προκύπτουν από την ανάπτυξη του δυναμικού της Εξ. (2.33) γύρω από το κενό της θεωρίας

$$V_H = V_0 + \frac{1}{2}\begin{pmatrix}\chi_1 & \chi_2\end{pmatrix}\mathcal{M}_\chi^2\begin{pmatrix}\chi_1\\ \chi_2\end{pmatrix} + \frac{1}{2}\begin{pmatrix}\phi_1 & \phi_2\end{pmatrix}\mathcal{M}_\phi^2\begin{pmatrix}\phi_1\\ \phi_2\end{pmatrix} + \begin{pmatrix}H_- & H_+\end{pmatrix}\mathcal{M}_\pm^2\begin{pmatrix}H_+\\ H_-\end{pmatrix}, \tag{2.49}$$

όπου $V_0 = V_H(v_1, v_2)$. Οι πίνακες των τετραγώνων των μαζών που προκύπτουν, είναι συμμετρικοί και διαγωνοποιούνται με έναν ορθογώνιο μετασχηματισμό

$$R(\theta)^\dagger \mathcal{M}^2_{\chi\,[\phi,\,\pm]} R(\theta) \quad \text{όπου} \quad R(\theta) = \begin{pmatrix}\cos\theta & -\sin\theta\\ \sin\theta & \cos\theta\end{pmatrix}$$

Ειδικότερα οι πίνακες μαζών που εμφανίζονται στην Εξ. (2.49), είναι:

**α.** $\mathcal{M}_\chi^2$ ο πίνακας μαζών των CP-περιττών μποζονίων Higgs

$$\mathcal{M}_\chi^2 = \left(\left\langle \frac{\partial^2 V}{\partial\chi_i\partial\chi_j}\right\rangle_0\right) = -m_{12}^2\begin{pmatrix}\tan\beta & 1\\ 1 & \cot\beta\end{pmatrix}, \tag{2.50}$$

με ΙΔΤ

$$0,\quad m_A^2 = m_1^2 + m_2^2 = -m_{12}^2/s_\beta c_\beta \tag{2.51}$$

και πίνακα διαγωνοποίησης τον $R(-\beta)$. Από την Εξ. (2.51) επαληθεύεται η σημασία του όρου αλληλεπίδρασης των δύο Higgs ώστε να μην προκύπτει άμαζο μποζόνιο Higgs.

**β.** $\mathcal{M}_\phi^2$ ο πίνακας μαζών των CP-άρτιων μποζονίων Higgs

$$\mathcal{M}_\phi^2 = \left(\left\langle \frac{\partial^2 V}{\partial\phi_i\partial\phi_j}\right\rangle_0\right) = \begin{pmatrix} m_A^2 s_\beta^2 + M_Z^2 c_\beta^2 & -(m_A^2+M_Z^2)s_\beta c_\beta\\ -(m_A^2+M_Z^2)s_\beta c_\beta & m_A^2 c_\beta^2 + M_Z^2 s_\beta^2\end{pmatrix} \tag{2.52}$$

με ΙΔΤ

$$m_{H,h}^2 = \frac{1}{2}\left[m_A^2 + M_Z^2 \pm \sqrt{(m_A^2+M_Z^2)^2 - 4m_A^2 M_Z^2 c_{2\beta}^2}\ \right] \tag{2.53}$$

και πίνακα διαγωνοποίησης τον $R(\alpha)$ όπου η $\alpha$ ορίζεται από την επόμενη:

$$\tan 2\alpha = \tan 2\beta\,\frac{m_A^2 + M_Z^2}{m_A^2 - M_Z^2}\,,\quad -\pi/2 \le \alpha \le 0\,. \tag{2.54}$$

Από την Εξ. (2.53), υπάρχει ένα μέγιστο στη μάζα του $h$, που λαμβάνεται για $c_{2\beta}^2 = 1$ και είναι το

$$(m_h^2)_{max} = (m_A^2 + M_Z^2 - |m_A^2 - M_Z^2|)/2 = \begin{cases} m_A^2, & \text{αν } m_A^2 < M_Z^2 \\ M_Z^2, & \text{αν } m_A^2 > M_Z^2 \end{cases} \tag{2.55}$$

Σε κάθε περίπτωση δηλαδή πρέπει $m_h^2 < M_Z^2$. Η πρόβλεψη αυτή τροποποιείται με την προσθήκη των διορθώσεων ενός βρόχου όπως θα φανεί στο Εδ. 3.2.2.

**γ.** $\mathcal{M}_{\pm}^2$ ο πίνακας μαζών των φορτισμένων μποζονίων Higgs

$$\mathcal{M}_{\pm}^2 = (m_A^2 + M_W^2) \begin{pmatrix} c_\beta^2 & s_{2\beta}/2 \\ s_{2\beta}/2 & s_\beta^2 \end{pmatrix}. \tag{2.56}$$

με ΙΔΤ

$$0, \ \ m_{H^\pm}^2 = m_A^2 + M_W^2.$$

και πίνακα διαγωνοποίησης τον $R(\beta)$

Συμπερασματικά, με τον μηχανισμό Higgs οι 8 βαθμοί ελευθερίας των μποζονίων Higgs που διέθετε αρχικά η θεωρία, ανακατανέμονται. Τρεις από αυτούς μετακινούνται στον τομέα των διανυσματικών μποζονίων παρέχοντας μάζα στα $W^+$, $W^-$, $Z$, ενώ οι υπόλοιποι πέντε αναδιατάσσονται στον τομέα των μποζονίων Higgs, παρέχοντας το τελικό φάσμα των βαθμωτών μποζονίων της θεωρίας που είναι:

| | | |
|---|---|---|
| 2 ουδέτερα μποζόνια με CP = 1 | : | $h, H$ |
| 1 ουδέτερο μποζόνιο με CP = -1 | : | $A$ |
| 2 φορτισμένα μποζόνια | : | $H^\pm$ . |

Η καταμέτρηση των βαθμών ελευθερίας των σωματίων που υπεισέρχονται στο μηχανισμό Higgs πριν και μετά το SSB απεικονίζεται στον Πίνακα 2.2.

### 2.4.6 Τομέας Φερμιονίων

Παρεπόμενο του μηχανισμού Higgs είναι ο εφοδιασμός των φερμιονίων με μάζες όπως περιγράφηκε στο Εδ. 2.3.2 . Αξιοποιώντας και τις Εξ. (2.41), (2.47), οι μάζες αυτές λαμβάνουν τη μορφή:

$$m_{l[d_q]} = h_{l[D_q]} v_1 = \frac{h_{l[D_q]}\sqrt{2} M_W c_\beta}{g}, \quad m_{u_q} = h_{U_q} v_2 = \frac{h_{U_q}\sqrt{2} M_W s_\beta}{g}. \tag{2.57}$$

Επισημαίνεται και πάλι ότι αγνοήθηκαν τα φαινόμενα ανάμιξης ανάμεσα στις γενεές που θα απαιτούσε πινακικές ζεύξεις Yukawa και επομένως διαγονοποίηση τους γαι την επίτευξη ΙΔΚ μάζας.

## 2.5 Το SUSY φάσμα του MSSM

Η μέχρι τώρα ανάλυση που επιχειρήθηκε στο MSSM είχε κάποια στοιχεία γενικότητας. Στα προγράμματα αριθμητικής εξομοίωσης χρησιμοποιούνται κάποιες απλοποιητικές προσεγγίσεις. Στόχος της ενότητας αυτής είναι η ανάδειξη των προσεγγίσεων αυτών και η απλοποίηση του χρησιμοποιούμενου συμβολισμού στο Εδ. 2.5.1. Έπεται η παρουσίαση του φάσματος των sparticles του MSSM στα Εδ. 2.5.3 και 2.5.2.

### 2.5.1 Προσαρμογή συμβολισμού

Η βασική παραδοχή που έχει απλοποιητικές επιπτώσεις στο φάσμα του MSSM και στο χρησιμοποιούμενο συμβολισμό είναι ότι τα φερμιόνια των δύο πρώτων οικογενειών θεωρούνται άμαζα. Η υπόθεση αυτή είναι εύλογη αν αναλογισθεί κανείς τη διαφορά μάζας που υπάρχει ανάμεσα στις μάζες των φερμιονίων των δύο πρώτων οικογενειών και σε αυτές της τρίτης. Από τις ζεύξεις Yukawa επιζούν και μετονομάζονται οι παρακάτω:

$$h_\tau, \quad h_{U_3} = h_t, \quad h_{D_3} = h_b \tag{2.58}$$

οπότε ο τύπος μαζών της Εξ. (2.59) μεταφράζεται ως εξής:

$$m_f := \begin{cases} 174\, h_f\, s_\beta \text{ GeV}, & \text{αν } f = t \\ 174\, h_f\, c_\beta \text{ GeV}, & \text{αν } f = b, \tau \\ 0 \text{ GeV}, & \text{αν } f = e,\, \nu_e,\, \nu_\tau,\, u,\, d \end{cases} \tag{2.59}$$

όπου $174 = v/\sqrt{2}$ και χρησιμοποιείται το ίδιο σύμβολο για τα φερμιόνια των δύο πρώτων οικογενειών. Συνέπειες αυτής της απλοποιήσης είναι ότι και οι SBT των δύο πρώτων οικογενειών προκύπτουν εκφυλισμένοι, όπως απορέει από τις RGE του Εδ. Α′.3, οπότε ο συμβολισμός μετεξελίσσεται ως εξής:

$$\begin{array}{lllll} L_\tau = L_L, & E_\tau = \tau_R, & Q_3 = Q_L, & U_3 = t_R, & D_3 = b_R \\ L_{e[\mu]} = l_L, & E_{e[\mu]} = e_R, & Q_{1[2]} = q_L, & U_{1[2]} = u_R, & D_{1[2]} = d_R \end{array} \tag{2.60}$$

Με τις τροποποιήσεις αυτές, ο συμβολισμός εναρμονίζεται με τον χρησιμοποιούμενο στο Παράρτημα Α′.

### 2.5.2 Τομέας Sfermion

Ο πίνακας μαζών των sfermion προκύπτει (Αν. [6]) από την ανάπτυξη κατα Taylor του δυναμικού $V_{MSSM}$ γύρω από το κενό της θεωρίας, σύμφωνα με την Εξ. (2.10). Τα διαγώνια στοιχεία προκύπτουν από τους όρους των Εξ. (2.25), (2.26), τους προσθεταίους της πρώτης γραμμής της Εξ. (2.27) και αυτούς της τελευταίας γραμμής της Εξ. (2.24). Τα μη διαγώνια στοιχεία προκύπτουν από τους όρους της τελευταίας γραμμής της Εξ. (2.27) και τους δύο πρώτους προσθεταίους της Εξ. (2.24). Τελικά ο ενλόγω πίνακας γράφεται:

$$\mathcal{M}^2_{\tilde f} = \begin{pmatrix} m^2_{\tilde f L} & m^2_{\tilde f LR} \\ m^2_{\tilde f LR} & m^2_{\tilde f R} \end{pmatrix}. \tag{2.61}$$

Στον πίνακα αυτό εμφανίζονται:

- Διαγώνια στοιχεία, που μπορούν να γραφούν ως εξής:

$$m^2_{\tilde f L} = m^2_{D_{f_L}} + m_f^2 + M_Z^2 g_{fL} c_{2\beta}, \tag{2.62}$$
$$m^2_{\tilde f R} = m^2_{f_R} + m_f^2 + M_Z^2 g_{fR} c_{2\beta}, \tag{2.63}$$

$$\text{με } f = \begin{cases} \nu_\tau,\, \tau,\, t,\, b & (\text{βαριές γενεές}) \\ \nu_e,\, e,\, u,\, d & (\text{ελαφρές γενεές}) \end{cases} \quad \text{και } D_{f_L} := \begin{cases} L_L\,[Q_L], & \text{αν } f = \nu_\tau,\, \tau\,[t,\, b] \\ l_L\,[q_L], & \text{αν } f = \nu_e,\, e\,[u,\, d] \end{cases} \tag{2.64}$$

Οι $m_{D_{f_L}}$ και $m_{f_R}$ είναι οι SBT που δίνονται στο Παράρτημα Α′.3. Προφανώς, $m^2_{f_R} = 0$, αν $f = \nu_e, \nu_\tau$. Οι ποσότητες $g_{fL[R]}$ υπολογίζονται από την Εξ. (2.21) και ομαδοποιούνται ως εξής:

– Για τα λεπτόνια:

$$\begin{array}{llll} g_{fR} = g_f(0,0), & g_{fL} = g_f(1/2,0) & \text{αν} & f = \nu_e,\, \nu_\tau \\ g_{fL} = g_f(-1/2,-1), & g_{fR} = g_f(0,1) & \text{αν} & f = e,\, \tau \end{array}$$

– Για τα quarks:

$$\begin{array}{llll} g_{fR} = g_f(0,-2/3), & g_{fL} = g_f(1/2,2/3) & \text{αν} & f = u,\, t \\ g_{fL} = g_f(-1/2,-1/3), & g_{fR} = g_f(0,1/3) & \text{αν} & f = d,\, b \end{array}$$

**Πίνακας 2.3: Φάσμα Υπερσυμμετρικών Σωματιδίων**

| Όνομα | ΙΔΚ Βαθμίδας | ΙΔΚ Μάζας | Spin | R- Parity |
|---|---|---|---|---|
| **Sleptons** | $\tilde{\nu}_e,\, \tilde{\nu}_\mu,\, \tilde{\nu}_\tau$<br>$\tilde{e}_L,\, \tilde{e}^c_L,\, \tilde{\mu}^c_L,\, \tilde{\mu}_L$<br>$\tilde{\tau}_L,\, \tilde{\tau}^c_L$ | $\tilde{\nu}_e,\, \tilde{\nu}_\tau$<br>$\tilde{e}_L,\, \tilde{e}^*_R$<br>$\tilde{\tau}_1,\, \tilde{\tau}_2$ | 0 | −1 |
| **Squarks** | $\tilde{u}_L,\, \tilde{u}^c_L,\, \tilde{s}^c_L,\, \tilde{s}_L$<br>$\tilde{d}_L,\, \tilde{d}^c_L,\, \tilde{c}^c_L,\, \tilde{c}_L$<br>$\tilde{t}_L,\, \tilde{t}^c_L,\, \tilde{b}^c_L,\, \tilde{b}_L$ | $\tilde{u}_L,\, \tilde{u}^*_R$<br>$\tilde{d}_L,\, \tilde{d}^*_R$<br>$\tilde{t}_1,\, \tilde{t}_2,\, \tilde{b}_1,\, \tilde{b}_2$ | 0 | −1 |
| **Neutralinos** | $\tilde{B},\, \tilde{W},\, \tilde{H}_0,\, \tilde{\bar{H}}_0$ | $\tilde{\chi},\, \tilde{\chi}^0_2,\, \tilde{\chi}^0_3,\, \tilde{\chi}^0_4$ | 1/2 | −1 |
| **Charginos** | $\tilde{W}^\pm,\, \tilde{H}_-,\, \tilde{H}_+$ | $\tilde{\chi}^\pm_1,\, \tilde{\chi}^\pm_2$ | 1/2 | −1 |
| **Gluinos** | $\tilde{g}^A$ | $\tilde{g}^A$ | 1/2 | −1 |

- Μή διαγώνια στοιχεία, που μπορούν να γραφούν ως εξής:

$$m^2_{\tilde{f}LR} = \begin{cases} m_f \begin{cases} (A_f + \mu \cot\beta), & \text{αν} \;\; f = t \\ (A_f + \mu \tan\beta), & \text{αν} \;\; f = b, \tau \end{cases} \\ \quad 0, \;\; \text{αν} \;\; f = e,\, \nu_e,\, \nu_\tau,\, u,\, d \end{cases} \tag{2.65}$$

Μέσω της Εξ. (2.65), ο πίνακας της Εξ. (2.61), προκύπτει διαγώνιος για τις ελαφρές γενεές ενώ για τις βαριές γενεές, διαγωνοποιείται με έναν ορθογώνιο πίνακα $R_f$ ως εξής:

$$R^\dagger_f \mathcal{M}^2_{\tilde{f}} R_f = \mathrm{diag}(m^2_{\tilde{f}_1}, m^2_{\tilde{f}_1}) \;\; \text{με} \;\; m_{\tilde{f}_1} \geq m_{\tilde{f}_2}\,, \tag{2.66}$$

όπου $m_{\tilde{f}_1}$, $m_{\tilde{f}_2}$ οι ΙΔΤ του πίνακα $\mathcal{M}^2_{\tilde{f}}$ με αντίστοιχες ΙΔΚ τις

$$\begin{pmatrix} \tilde{f}_1 \\ \tilde{f}_2 \end{pmatrix} = R^\dagger_f \begin{pmatrix} \tilde{f}_L \\ \tilde{f}_R \end{pmatrix}, \;\; \text{όπου} \;\; R_f = \begin{pmatrix} c_f & -s_f \\ s_f & c_f \end{pmatrix}. \tag{2.67}$$

και χρησιμοποιήθηκαν οι εύλογες συντμίσεις $s_f := \sin\theta_f$, $c_f := \cos\theta_f$. Προφανώς η διαδικασία διαγωνοποίησης αποκτά βαρύνουσα αξία όταν τα μη διαγώνια στοιχεία του πίνακα της Εξ. (2.61) είναι σημαντικά, πράγμα που συμβαίνει κυρίως όταν η παράμετρος $\tan\beta$ λαμβάνει υψηλές τιμές.

### 2.5.3 Τομέας των Neutralino και Chargino

Με τη διαδικασία του SSB, οι ΙΔΚ βαθμίδας των gauginos και higssinos αναμιγνύονται και παρέχουν τις ΙΔΚ μάζας που ονομάζονται neutralino και chargino. Οι μόνες από τις ΙΔΚ βαθμίδας που δεν επιρεάζονται από την ανάμιξη αυτή και ταυτίζονται με τις ΙΔΚ μάζας είναι αυτές των gluino. Οι ΙΔΤ τους προέρχονται μόνο από τους SBT της δεύτερης γραμμής της Εξ. (2.27) και είναι οι $M_3$.

Οι όροι μάζας των υπολοίπων gauginos και higgsinos ταξινομούνται (Αν. [6]) σχηματίζοντας τους πίνακες των neutralino και chargino, οι οποίοι μελετώνται παρακάτω:

**α.** Ο πίνακας των Neutralino. Προκύπτει από την ομαδοποίηση των ηλεκτρικά ουδέτερων όρων των gauginos και higgsinos, οι οποίοι μπορούν να γραφούν ως εξής:

$$-\frac{1}{2} \begin{pmatrix} -i\tilde{B} & -i\tilde{W}_3 & \tilde{H}_0 & \tilde{\bar{H}}_0 \end{pmatrix} \mathcal{M}_{\tilde{\chi}^0} \begin{pmatrix} -i\tilde{B} \\ -i\tilde{W}_3 \\ \tilde{H}_0 \\ \tilde{\bar{H}}_0 \end{pmatrix} + \mathrm{h.\,c}\,, \tag{2.68}$$

όπου ο πίνακας μαζών των neutralino, είναι:

$$\mathcal{M}_{\tilde{\chi}^0} = \begin{pmatrix} M_1 & 0 & -M_Z s_W c_\beta & M_Z s_W s_\beta \\ 0 & M_2 & M_Z c_W c_\beta & -M_Z c_W s_\beta \\ -M_Z s_W c_\beta & M_Z c_W c_\beta & 0 & \mu \\ M_Z s_W s_\beta & -M_Z c_W s_\beta & \mu & 0 \end{pmatrix}, \tag{2.69}$$

Σύμφωνα με τη γενική Εξ. (2.11), οι όροι του πάνω αριστερου τεταρτημορίου, προέρχονται από την Εξ. (2.27), του κάτω δεξιού από την Εξ. (2.22) και των μη διαγώνιων από την Εξ. (2.23).

0 πίνακας της Εξ. (2.69) είναι συμμετρικός και μπορεί να διγωνοποιηθεί με χρήση ενός ορθογώνιου πίνακα $Z$. Ο προκύπτον πίνακας $Z^T \mathcal{M}_{\tilde{\chi}^0} Z$ είναι διαγώνιος αλλά όχι αναγκαστικά με θετικές ΙΔΤ. Πρέπει να οριστεί ο πίνακας:

$$N = \mathrm{diag}(e^{-i\theta_1^0/2},\, e^{-i\theta_2^0/2},\, e^{-i\theta_3^0/2},\, e^{-i\theta_4^0/2}) Z \tag{2.70}$$

$$\text{όπου} \quad \theta_n^0 = \mathrm{Arg}(Z^T \mathcal{M}_{\tilde{\chi}^0} Z)_{nn}, \quad n = 1,\, 2,\, 3,\, 4 \tag{2.71}$$

Προφανώς δεν εννοείται άθροιση στους δείκτες $n$. Επομένως, η διαγωνοποίηση του πίνακα της Εξ. (2.69) παρέχει τις μάζες των neutralino ως εξής:

$$N^T \mathcal{M}_{\tilde{\chi}^0} N = \mathrm{diag}(m_{\tilde{\chi}},\, m_{\tilde{\chi}_2^0},\, m_{\tilde{\chi}_3^0},\, m_{\tilde{\chi}_4^0}) \tag{2.72}$$

Όπως θα αναφερθεί στο κεφάλαιο 5, το LSP συνήθως προκύπτει να είναι η ελαφρότατη από τις ΙΔΤ του πίνακα της Εξ. (2.69) και γιαυτό συμβολίζεται πιο σύντομα από τις υπόλοιπες ως $m_{\tilde{\chi}}$.

**β.** Ο πίνακας των Chargino. Προκύπτει από την ομαδοποίηση των ηλεκτρικά φορτισμένων όρων των gauginos και higgsinos, οι οποίοι μπορούν να γραφούν ως εξής:

$$-\frac{1}{2}\begin{pmatrix} \psi^{+T} & \psi^{-T} \end{pmatrix} \begin{pmatrix} 0 & \mathcal{M}_{\tilde{\chi}^\pm} \\ \mathcal{M}_{\tilde{\chi}^\pm}^T & 0 \end{pmatrix} \begin{pmatrix} \psi^+ \\ \psi^- \end{pmatrix} + \mathrm{h.\,c} \tag{2.73}$$

όπου ορίστηκαν οι σπίνορες

$$\psi^+ := \begin{pmatrix} -i\tilde{W}^+ \\ \tilde{H}_+ \end{pmatrix} \quad \text{και} \quad \psi^- := \begin{pmatrix} -i\tilde{W}^- \\ \tilde{H}_- \end{pmatrix} \tag{2.74}$$

με $\tilde{W}^\pm := (\tilde{W}^1 \mp i\tilde{W}^2)/\sqrt{2}$ και ο πίνακας μαζών των chargino, είναι:

$$\mathcal{M}_{\tilde{\chi}^\pm} = \begin{pmatrix} M_2 & \sqrt{2} M_W s_\beta \\ \sqrt{2} M_W c_\beta & -\mu \end{pmatrix}. \tag{2.75}$$

Σύμφωνα με τη γενική Εξ. (2.11), ο πάνω αριστερός όρος του προέρχεται από την Εξ. (2.27), ο κάτω δεξιός από την Εξ. (2.22) και οι μη διαγώνιοι από την Εξ. (2.23).

Ο πίνακας $\mathcal{M}_{\tilde{\chi}^\pm}$ δεν είναι ούτε ερμιτιανός ούτε συμμετρικός και επομένως διαγωνοποιείται χρησιμοποιώντας ένα δι-μοναδιαίο μετασχηματισμό $U'^* \mathcal{M}_{\tilde{\chi}^\pm} V^{-1}$, όπου $U'$ και $V$ είναι ερμητιανοί πίνακες που διαγωνοποιούν τον ερμιτιανό πίνακα $\mathcal{M}_{\tilde{\chi}^\pm}^\dagger \mathcal{M}_{\tilde{\chi}^\pm}$ ως εξής:

$$V(\mathcal{M}_{\tilde{\chi}^\pm}^\dagger \mathcal{M}_{\tilde{\chi}^\pm}) V^{-1} = \mathrm{diag}(|m_{\tilde{\chi}_1^+}|^2,\, |m_{\tilde{\chi}_2^+}|^2) = U'^*(\mathcal{M}_{\tilde{\chi}^\pm} \mathcal{M}_{\tilde{\chi}^\pm}^\dagger)(U'^*)^{-1} \tag{2.76}$$

Ο πίνακας $U'^* \mathcal{M}_{\tilde{\chi}^\pm} V^{-1}$ είναι διαγώνιος αλλά όχι αναγκαστικά με θετικές ΙΔΤ. Πρέπει να οριστεί ο πίνακας:

$$U = \mathrm{diag}(e^{-i\theta_1^\pm/2},\, e^{-i\theta_2^\pm/2}) U' \quad \text{όπου} \quad \theta_c^\pm = \mathrm{Arg}(U'^* \mathcal{M}_{\tilde{\chi}^\pm} V^{-1})_{cc}, \quad c = 1,\, 2 \tag{2.77}$$

Προφανώς δεν εννοείται άθροιση στους δείκτες $c$. Επομένως, η διαγωνοποίηση του πίνακα της Εξ. (2.75) παρέχει τις μάζες των chargino ως εξής:

$$U^* \mathcal{M}_{\tilde{\chi}^\pm} V^\dagger = \mathrm{diag}\left(m_{\tilde{\chi}_1^+},\, m_{\tilde{\chi}_2^+}\right). \tag{2.78}$$

Οι ΙΔΚ βαθμίδας (πριν το SSB) και μάζας (μετά το SSB) των sparticles του MSSM καταγράφονται στο συγκεντρωτικό Πίνακα 2.3, μαζί με τους κβανικούς αριθμούς του spin και της $R$-Parity. Παρατηρείται ότι για τις ΙΔΚ μάζας των sfermions των δύο πρώτων οικογενειών χρησιμοποιείται το ίδιο σύμβολο.

## 2.6 Αλληλεπίδραση Fermion-Sfermion-Bino

Η πιο σημαντική αλληλεπίδραση που θα χρησιμοποιηθεί στις ΑΝΕ και στις μικτές CAE ($\tilde{\chi} - \tilde{f}$), όπως θα αναφερθεί στο Εδ. 5.5, είναι η fermion-sfermion-Bino.

Η αλληλεπίδραση αυτή προέρχεται από το τμήμα της λαγκραζιανής της Εξ. (2.23) και δη τον πρώτο όρο της. Ορίζοντας τους σπίνορες με 4 συνιστώσες ως εξής:

$$f = \begin{pmatrix} f_L \\ (f_L^c)^\dagger \end{pmatrix} \quad \text{και} \quad \tilde{\chi} = \begin{pmatrix} -i\tilde{B} \\ i\bar{\tilde{B}} \end{pmatrix}. \tag{2.79}$$

(με το $f$ να παίρνει τις τιμές που φαίνεται στην Εξ. (2.59) ) ο σχετικός όρος Εξ. (2.23) γράφεται με χρήση των ΙΔΚ βαθμίδας (σε αρμονία με τις Εξ. (C.75) της Αν. [13] και (A.23) της Αν. [12])

$$\mathcal{L}_{f\tilde{f}\tilde{\chi}} = \sqrt{2}g\tan\theta_W\left(-Y_{fL}\bar{f}P_R\tilde{f}_L\tilde{\chi} + Y_{fR}\bar{f}P_L\tilde{f}_R\tilde{\chi}\right) + \text{h.c} \tag{2.80}$$

Εκτελώντας τη στροφή ορισμού των ΙΔΚ μάζας της Εξ. (2.67) και χρησιμοποιώντας την Εξ. (2.20), η Εξ. (2.80) γράφεται:

$$\begin{aligned}\mathcal{L}_{f\tilde{f}\tilde{\chi}} = \frac{\sqrt{2}e}{c_W}\Big[&\bar{f}\left(-Y_{fL}P_Rc_f + Y_{fR}P_Ls_f\right)\tilde{f}_1\tilde{\chi} + \bar{\tilde{\chi}}\left(-Y_{fL}P_Lc_f + Y_{fR}P_Rs_f\right)f\tilde{f}_1^* \\ &+\bar{f}\left(Y_{fL}P_Rs_f + Y_{fR}P_Lc_f\right)\tilde{f}_2\tilde{\chi} + \bar{\tilde{\chi}}\left(Y_{fL}P_Ls_f + Y_{fR}P_Rc_f\right)f\tilde{f}_2^*\Big]\end{aligned} \tag{2.81}$$

Οι κόμβοι Feynman που αντλούνται από την Εξ. (2.81) διευθετούνται στο Εδ. Β΄.3.2. Αξίζει να σημειωθεί ότι το $Y_{fL[R]}$ δεν αντιστοιχεί ακριβώς στο υπερφορτίο που σημειώνεται στον Πίνακα 2.1 αλλά στο $Y/2$.

Στο σημείο αυτό ίσως θα έπρεπε να αναφερθούν όλα τα τμήματα της λαγκραζιανής τα οποία παρέχουν αλληλεπιδράσεις που χρησιμοποιήθηκαν στις CAE. Αυτό, όμως, ουσιαστικά θα κατέληγε σε μια αναδημοσίευση των Αν. [12] και [13]. Γιαυτό και έγινε η επιλογή της ενδεικτικής παρουσίασης μόνο μιας αλληλεπίδρασης, της βασικότερης, που είναι η fermion-sfermion-Bino και η διαγραμματική καταγραφή των υπολοίπων στο Εδ. Β΄.3.2.

# Κεφάλαιο 3

# Φαινομενολογικές εφαρμογές-Διορθώσεις

## 3.1 Εισαγωγή

Μια διερεύνηση του MSSM επέκεινα του δενδρικού επιπέδου θα επιχειρηθεί σε αυτό το κεφάλαιο. Το εγχείρημα είναι αναγκαίο κυρίως στις περιπτώσεις μεγάλου $\tan\beta$, το οποίο είναι ύποπτο για την πρόκληση ισχυρών διορθώσεων. Αρχικά μελετώνται οι υπερσυμμετρικές διορθώσεις που λαμβάνουν οι μάζες του CP-even τομέα Higgs του προτύπου στο Εδ. 3.2 και στη συνέχεια, των φερμιονίων της τρίτης οικογένειας στο Εδ. 3.3. Τέλος, στο Εδ. 3.4 παρουσιάζεται το τυπολόγιο που παρέχει τον ανηγμένο λόγο (: branching ratio) της διαδικασίας $b \to s\gamma$. Τα εφόδια του κεφαλαίου αυτού, θα είναι τα βασικά κριτήρια φαινομενολογικής μελέτης διαφόρων ειδικών εκδόσεων του MSSM στο ερευνητικό τμήμα της εργασίας που θα ακολουθήσει.

## 3.2 Διορθώσεις στις μάζες των Μποζονίων Higgs

Κεντρικό ρόλο στο θέμα των διορθώσεων που προστίθενται στο δενδρικού επιπέδου δυναμικό του MSSM παίζει η έννοια της προνομιακής κλίμακας ενέργειας, η οποία εισάγεται στο Εδ. 3.2.1. Ακολουθεί στο Εδ. 3.2.2 το τυπολόγιο που παρέχει τις διορθώσεις στα CP-even Higgs μποζόνια, οι οποίες είναι αξιοσημείωτες για το ελαφρότερο από αυτά, το $h$.

### 3.2.1 Προνομιακή Κλίμακα Ενέργειας

Η έννοια της προνομιακής κλίμακας ενέργειας εισάγεται από τους Drees και Nojiri στην Αν. [18] και εκ τοτε ακολουθείται από πολλές ομάδες ερευνητών. Σύμφωνα με τους εμπνευστές αυτής της ιδέας, υπάρχει κλίμακα ενέργειας, $M_S$, στην οποία οι διορθώσεις στο επανακανονικοποιημένα βελτιωμένο σε επίπεδο ένος βρόχου ενεργό δυναμικό (one-loop renormalization group improved effective potential) περιορίζονται και επομένως, το επανακανονικοποιημένα βελτιωμένο σε δενδρικό επίπεδο ενεργό δυναμικό (tree-level renormalization group improved effective potential) δίνει πιο αξιόπιστα αποτελέσματα σε αυτή την ενεργειακή κλίμακα παρά σε οποιαδήποτε αλλή. Αρα, είναι ωφέλιμο να γίνεται σε αυτήν την ενεργειακή κλίμακα η ελαχιστοποίηση του ενεργού δυναμικού αφού οι διορθώσεις αναμένονται να είναι μικρές. Επιπλέον, σε αυτή την κλίμακα οι διορθώσεις στη μάζα του CP-odd Higgs, $m_A$ είναι εξίσου μικρές.

Οι συνεισφορές που λαμβάνονται υπόψη στις διορθώσεις αυτές προέχονται από βρόχους με sfermions-fermions της τρίτης οικογένειας και δίνονται στο Παράρτημα της Αν. [18]. Επειδή μάλιστα, στις διορθώσεις αυτές, προεξάρχουσα είναι η συνεισφορά των βαρύτερων από τα sfermions, φυσιολογικά των stops η προσεγγιστική ταυτότητα αυτής της ενεργειακής κλίμακας είναι:

$$M_S \simeq \sqrt{m_{\tilde{t}_1} m_{\tilde{t}_2}}\,. \tag{3.1}$$

Είναι δύσκολο να γίνει έλεγχος των προηγουμένων αποφάνσεων. Και αυτό, διότι απαιτείται να υπολογίσει κανείς τις διορθώσεις ανώτερων τάξεων για να πειστεί για την ορθότητα ή μη των ισχυρισμών

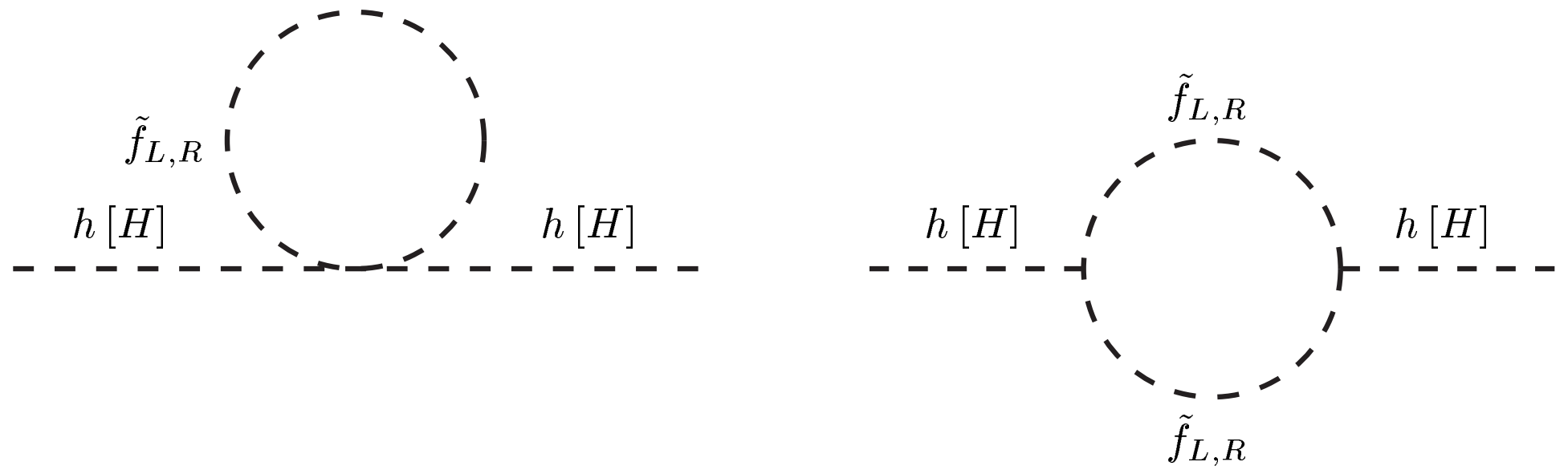

**Σχήμα 3.1:** *Οι βρόχοι των sfermions, $\tilde{f} := \tilde{t}, \tilde{b}$ που συνεισφέρουν στη διόρθωση της μάζας των CP άρτιων Higgs $h$, $H$.*

που διατυπώθηκαν παραπάνω σχετικά με την αξία της ενεργειακής αυτής κλίμακας. Βεβαίως, υπάρχουν νεότερες, βελτιωμένες εκδόσεις των διορθώσεων αυτών, οι πιο πλήρεις των οποίων εκτίθενται στα Παραρτήματα D, E της Αν. [19]. Οι διορθώσεις αυτές έχουν γραφεί και σε αριθμητικό πρόγραμμα που χρησιμοποιήθηκε στον υπολογισμό της Αν. [72]. Εκεί και γίνεται κάποιος σχετικός σχολιασμός των ευρημάτων στην επικίνδυνη περίπτωση της υψηλής $\tan\beta$. Τα συμπεράσματα συνοψίζονται στα παρακάτω σημεία:

- Οι διορθώσεις στην παράμετρο $\mu$ (υπολογισμένη από την ελαχιστοποίηση του επανακανονικοποιημένα βελτιωμένου σε δενδρικό επίπεδο ενεργού δυναμικού σε κλίμακα $M_S$) από την ελαχιστοποίηση του επανακανονικοποιημένα βελτιωμένου σε επίπεδο ένος βρόχου ενεργού δυναμικού είναι σημαντικά μικρές, της τάξεως 1%. Τα αποτελέσματα αυτά, είναι εξαιρετικά ευσταθή με μικρή μετακίνηση της προνομιακής κλίμακας $M_S$.

- Οι συνεισφορές από βρόχους neutralino-neutralino chargino-chargino ενίοτε δεν είναι αμελητέες κυρίως στη περίπτωση του φορτισμένου μποζωνίου Higgs, πράγμα που καθιστά τη διόρθωση αξιοσημείωτη για μικρά $A$ περίπου 20% αλλά ασθενή αργότερα 2%. Επιπλέον, οι διορθώσεις στη μάζα του ελαφρού ουδέτερου μποζονίου Higgs εμφανίζονται μειωμένες κατά 8% περίπου ενώ σε αυτή του βαρύτερου ουδέτερου μποζονίου Higgs εμφανίζονται αυξημένες κατά 10% περίπου. Επιπλέον, παρατηρείται αστάθεια στη συμπεριφορά των αποτελεσμάτων αυτών, με μικρή μετακίνηση από το $M_S$ κυρίως για τις μαζές των $h$, $H$, $H^{\mp}$. Αυτό οφείλεται στις επιπλέον συνεισφορές που σημειώθηκαν προηγουμένως. Γι αυτό και τελικά δεν συμπεριλαμβάνονται στον υπολογισμό πράγμα που καθιστά τα αποτελέσματα σύμφωνα με τα κοινώς κείμενα.

Για τους λόγους αυτούς δεν θεωρήθηκε σκόπιμη η παρουσίαση ενός τόσο σκληρού τυπολόγιου τα αποτελέσματα του οποίου δεν επηρεάζουν αισθήτα τα φαινομενολογικά και κοσμολογικά εξαγόμενά μας. Στα πλαίσια της διατριβής αυτής, οι διορθώσεις αυτές δεν λαμβάνονται υπόψη στο αριθμητικό πρόγραμμα και κατα επέκταση στα σχετικά διαγράμματα. Επιπλέον, αποφασίζεται ότι η προτεινόμενη προνομιακή κλίμακα ενέργειας παρέχει αξιόπιστα αποτελέσματα. Επομένως, σε αυτό το ενεργειακό σημείο γίνεται η ελαχιστοποίηση του επανακανονικοποιημένα βελτιωμένου σε δενδρικό επίπεδο ενεργού δυναμικού και ταυτόχρονα οι RGE του MSSM δίνουν τη θέση τους στις RGE του SM .

### 3.2.2 Διορθώσεις στις μάζες των άρτιων CP Higgs

Οι διορθώσεις στις μάζες των άρτιων CP Higgs μπορούν να ληφθούν αν στη μορφή του δυναμικού του δενδρικού επιπέδου της Εξ. (2.49) προστεθούν οι όροι των κβαντικών διορθώσεων radiative corrections. Τα διαγράμματα που συνεισφέρουν στις διορθώσεις αυτές, περιέχουν βρόχους sfermions και είναι της μορφής των εικονιζόμενων στο Σχ. 3.1. Προεξάρχουσα συμβολή στο τελικό αποτέλεσμα, έχουν τα διαγράμματα που περικλείουν βρόχους από stops, που λογικά είναι πιο βαριά από τα υπόλοιπα sfermions. Στο τυπολόγιο που θα παρουσιαστεί παρακάτω συμεριλαμβάνονται διορθώσεις από βρόχους stops και sbottoms.

Ο διορθωμένος πίνακας των άρτιων CP Higgs μπορεί να γραφεί ως εξής:

$$\left(\mathcal{M}^c_\psi\right)^2 = \begin{pmatrix} M_Z^2 c_\beta^2 + m_A^2 s_\beta^2 + \Delta_{\psi 11} & -(m_A^2 + M_Z^2) s_\beta c_\beta + \Delta_{\psi 12} \\ -(m_A^2 + M_Z^2) s_\beta c_\beta + \Delta_{\psi 21} & M_Z^2 s_\beta^2 + m_A^2 c_\beta^2 + \Delta_{\psi 22} \end{pmatrix} \tag{3.2}$$

Λόγω της συμμετρίας του $\mathcal{M}_\psi^c$ ισχύει $\Delta_{\psi_{ij}} = \Delta_{\psi_{ji}}$ με $i, j = 1, 2$. Οι διορθώσεις $\Delta_{\psi_{11}}$, $\Delta_{\psi_{12}}$ και $\Delta_{\psi_{22}}$ βρίσκονται από την Αν. [18] και υπολογίζονται από τους παρακάτω τύπους:

$$\begin{aligned}\Delta_{\psi_{11}} &= \frac{3g^2}{16\pi^2 M_W^2}\left[\frac{m_b^4}{c_\beta^2}\left(\ln\frac{m_{\tilde{b}_1}^2 m_{\tilde{b}_2}^2}{m_b^4} + 2Z_b \ln\frac{m_{\tilde{b}_1}^2}{m_{\tilde{b}_2}^2}\right)\right. \\ &\left. + \frac{m_b^4}{c_\beta^2} Z_b^2 g(m_{\tilde{b}_1}^2, m_{\tilde{b}_2}^2) + \frac{m_t^4}{s_\beta^2} W_t^2 g(m_{\tilde{t}_1}^2, m_{\tilde{t}_2}^2)\right], \qquad (3.3)\end{aligned}$$

$$\begin{aligned}\Delta_{\psi_{22}} &= \frac{3g^2}{16\pi^2 M_W^2}\left[\frac{m_t^4}{s_\beta^2}\left(\ln\frac{m_{\tilde{t}_1}^2 m_{\tilde{t}_2}^2}{m_t^4} + 2Z_t \ln\frac{m_{\tilde{t}_1}^2}{m_{\tilde{t}_2}^2}\right)\right. \\ &\left. + \frac{m_t^4}{s_\beta^2} Z_t^2 g(m_{\tilde{t}_1}^2, m_{\tilde{t}_2}^2) + \frac{m_b^4}{s_\beta^2} W_b^2 g(m_{\tilde{b}_1}^2, m_{\tilde{b}_2}^2)\right], \qquad (3.4)\end{aligned}$$

$$\begin{aligned}\Delta_{\psi_{12}} &= \frac{3g^2}{16\pi^2 M_W^2}\left[\frac{m_t^4}{s_\beta^2} W_t\left(\ln\frac{m_{\tilde{t}_1}^2}{m_{\tilde{t}_2}^2} + Z_t g(m_{\tilde{t}_1}^2, m_{\tilde{t}_2}^2)\right)\right. \\ &\left. + \frac{m_b^4}{c_\beta^2} W_b\left(\ln\frac{m_{\tilde{b}_1}^2}{m_{\tilde{b}_2}^2} + Z_b g(m_{\tilde{b}_1}^2, m_{\tilde{b}_2}^2)\right)\right], \qquad (3.5)\end{aligned}$$

όπου

$$W_q = \frac{\mu A_q - \mu^2 R_q}{m_{\tilde{q}_2}^2 - m_{\tilde{q}_1}^2}, \qquad (3.6)$$

$$Z_q = \frac{A_q^2 - \mu A_q R_q}{m_{\tilde{q}_2}^2 - m_{\tilde{q}_1}^2}, \qquad (3.7)$$

$$g(m_1^2, m_2^2) = 2 - \frac{m_1^2 + m_2^2}{m_1^2 - m_2^2}\ln\frac{m_1^2}{m_2^2}. \qquad (3.8)$$

με $R_q = \cot\beta\ [\tan\beta]$ για $q = t\ [b]$. Η διορθωμένη γωνία διαγωνοποίησης βρίσκεται από την παρακάτω:

$$\tan 2\alpha = \frac{-(m_A^2 + M_Z^2)s_{2\beta} + 2\Delta_{\psi_{12}}}{-(m_A^2 - M_Z^2)c_{2\beta} + (\Delta_{\psi_{11}} - \Delta_{\psi_{22}})} \qquad (3.9)$$

Το αποτέλεσμα εφαρμογής του τυπόλογιου αυτού δίνει ασήμαντες διορθώσεις στη μάζα του $H$ αλλά σημαντικές στη μάζα του $h$. Με τον συνυπολογισμό των διορθώσεων αυτών, η μέγιστη προβλεπόμενη μάζα του $h$ αυξάνεται από την δενδρικού επιπέδου τιμή της Εξ. (2.55) σε 130 GeV, πράγμα που προστατεύει το MSSM από μια πειραματική απόρριψη. Και αυτό, διοτι ήδη βρίσκονται σε εξέλιξη πειράματα εντοπισμού του $h$ που ανεβάζουν το κάτω επιτρεπτό όριό του ως εξής:

$$m_h > 105\,\text{GeV}. \qquad (3.10)$$

## 3.3 Διορθώσεις στις μάζες των φερμιονίων

Προφανώς απευθυνόμαστε σε μάζες των φερμιονίων της τρίτης γενεάς, αφού σε όλη την έκταση αυτής της διατριβής, τα φερμιόνια των δύο πρώτων γενεών θεωρούνται άμαζα. Οι πλήρεις εξισώσεις που παρέχουν τις διορθώσεις αυτές, με όλες τις δύνατες συνεισφορές που προκύπτουν στα πλαίσια του MSSM και σε επίπεδο ενός βρόχου, δίνονται στο Παράρτημα D της Αν. [19]. Εδώ παρουσιάζονται στα Εδ. 3.3.1, 3.3.2 και 3.3.3 οι προσεγγιστικοί τύποι που κατά τους συγγραφείς της αναφοράς αυτής, αποδίδουν με μεγάλη ακρίβεια και σε κάθε περίπτωση τη σημασία της κάθε διόρθωσης.

### 3.3.1 Μάζα του $b$-quark

Αξιοπρόσεκτες διορθώσεις μπορεί να προκύψουν στη μάζα του $b$-quark από SUSY συνεισφορές σε επίπεδο ενός βρόχου, Οι διορθώσεις αυτές, είναι ιδιαιτέρως αυξημένες στις περιπτώσεις μεγάλων τιμών στην

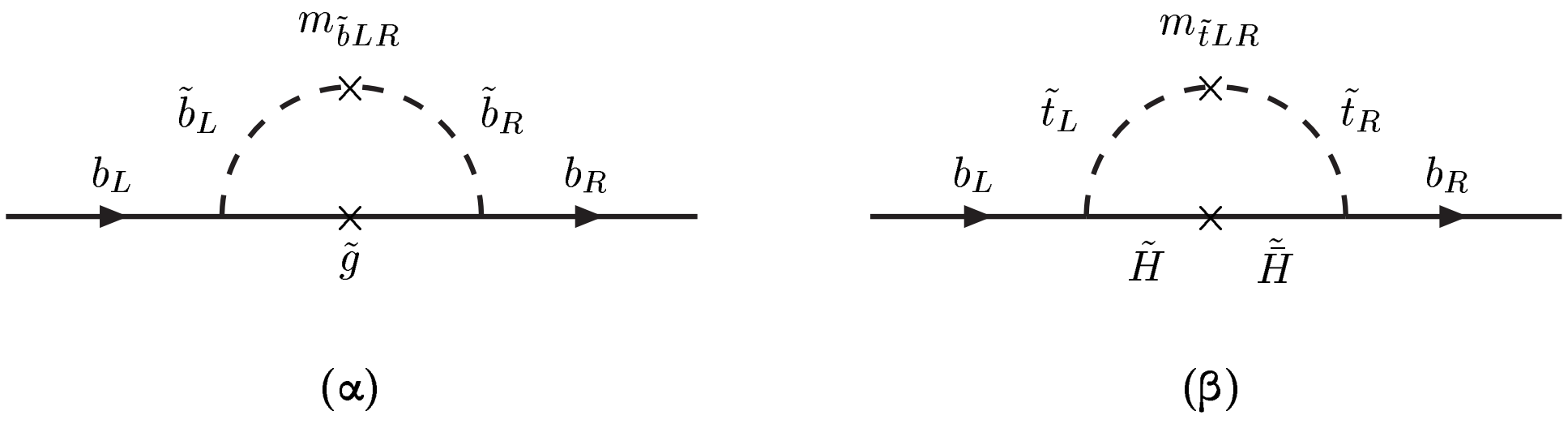

**Σχήμα 3.2:** *Οι βρόχοι sbottom-gluino, (α) και stop-higgsino, (β) που συνεισφέρουν στη διόρθωση της μάζας του b-quark*

παράμετρο $\tan\beta$, σύμφωνα με τις Αν. [20] και [21]. Τα διαγράμματα από τα οποία προέρχονται οι διορθώσεις αυτές φαίνονται στο Σχ. 3.2.

Η διορθωμένη μάζα του $b$-quark δίνεται από τον τύπο:

$$m_b^c = m_b\,(1 + \Delta m_b) \tag{3.11}$$

όπου $m_b$ η δενδρικού επιπέδου μάζα που υπολογίζεται από την Εξ. (2.59) και $\Delta m_b$ η διόρθωση που έχει δύο SUSY συνεισφορές:

$$\Delta m_b = (\Delta m_b)^{\tilde{b}\tilde{g}} + (\Delta m_b)^{\tilde{t}\tilde{\chi}^\pm} \tag{3.12}$$

προερχόμενες από τους βρόχους:

**α.** gluino-sbottom, Σχ. 3.2 (α), με τύπο υπολογισμού τον επόμενο:

$$\begin{aligned}(\Delta m_b)^{\tilde{b}\tilde{g}} &= -\frac{g_3^2}{12\pi^2}\Bigg\{ B_1(M_S, m_{\tilde{g}}, m_{\tilde{b}_1}) + B_1(M_S, m_{\tilde{g}}, m_{\tilde{b}_2}) \\ &\quad - \sin(2\theta_b)\left(\frac{m_{\tilde{g}}}{m_b}\right)\Big[ B_0(M_S, m_{\tilde{g}}, m_{\tilde{b}_1}) - B_0(M_S, m_{\tilde{g}}, m_{\tilde{b}_2})\Big]\Bigg\}\end{aligned} \tag{3.13}$$

όπου $\theta_b$ είναι η γωνία ανάμιξης των sbottom που εμφανίζεται στην Εξ. (2.67) και μπορεί να υπολογισθεί από τον τύπο:

$$\sin 2\theta_b = \frac{2m_b(A_b + \mu\tan\beta)}{m_{\tilde{b}_1}^2 - m_{\tilde{b}_2}^2}\,. \tag{3.14}$$

Αυτή είναι και η προεξάρχουσα συνεισφορά στη διόρθωση του $b$-quark. Έχει ελεγχθεί ότι η φόρμουλα αυτή ταυτίζεται πλήρως τον αντίστοιχο όρο της χρησιμοποιούμενης διόρθωσης από τις Αν. [20], [21] αλλά με αλλαγμένο πρόσημο.

**β.** chargino-stop, Σχ. 3.2 (β), με τύπο υπολογισμού τον επόμενο:

$$\begin{aligned}(\Delta m_b)^{\tilde{t}\tilde{\chi}^\pm} &= \frac{h_t^2}{16\pi^2}\,\mu\,\frac{A_t\tan\beta + \mu}{m_{\tilde{t}_1}^2 - m_{\tilde{t}_2}^2}\Big[ B_0(M_S, \mu, m_{\tilde{t}_1}) - B_0(M_S, \mu, m_{\tilde{t}_2})\Big] \\ &+ \frac{g^2}{16\pi^2}\left\{\frac{\mu M_2\tan\beta}{\mu^2 - M_2^2}\Big[ c_t^2 B_0(M_S, M_2, m_{\tilde{t}_1}) + s_t^2 B_0(M_S, M_2, m_{\tilde{t}_2})\Big] + (\mu \leftrightarrow M_2)\right\}\end{aligned} \tag{3.15}$$

Ο πρώτος όρος της εξίσωσης αυτής παρέχει ίσο κατά απόλυτη τιμή (αλλά με αντίθετο σημείο) αποτέλεσμα με το δεύτερο όρο του χρησιμοποιούμενου από τις Αν. [20], [21] τύπου. Ο δεύτερος όρος της προηγούμενης εξίσωσης δεν περιέχεται στις δύο άλλες αναφορές και δίνει μια μειωμένη όχι όμως αμελητέα συνεισφορά.

Οι συναρτήσεις που υπεισέρχονται στον υπολογισμό είναι:

$$B_0(Q,m_1,m_2) \;=\; -\ \ln\left(\frac{M_{12}^2}{Q^2}\right)+1+\frac{m_{12}^2}{m_{12}^2-M_{12}^2}\ln\left(\frac{M_{12}^2}{m_{12}^2}\right)\ , \qquad (3.16)$$

$$B_1(Q,m_1,m_2) \;=\; \frac{1}{2}\left[-\ \ln\left(\frac{M_{12}^2}{Q^2}\right)+\frac{1}{2}+\frac{1}{1-x_{12}}+\frac{\ln x_{12}}{(1-x_{12})^2}-\theta(1-x_{12})\ln x_{12}\right]\ , \qquad (3.17)$$

με $M_{12}=\max(m_1,m_2)$, $m_{12}=\min(m_1,m_2)$, και $x_{12}=m_2^2/m_1^2$.

Τα πειραματικά όρια για τη μάζα του $b$-quark όπως προκύπτουν από την Αν. [22] είναι τα εξής:

$$m_b(M_Z)=2.67\pm 0.50\,\mathrm{GeV}. \qquad (3.18)$$

### 3.3.2 Μάζα του $t$-quark

Αντίθετα με την περίπτωση του $b$-quark, η μάζα του $t$-quark δεν λαμβάνει σημαντικές SUSY διορθώσεις. Αυτές προέρχονται από το διάγραμμα που περιέχει το gluino-stop βρόχο και υπολογίζονται από την Εξ. (3.13) με την προφανή αντικατάσταση $b\to t$.

Στην περίπτωση του $t$-quark, είναι χρήσιμο να εισαχθεί η έκφραση που παρέχει την φυσική μάζα του $m_t(\mathrm{physical})$ σε κάποια ενεργειακή κλίμακα $M$:

$$m_t(\mathrm{physical})=\ m_t(M)\,[\,1+\ \Delta m_t(M)\,] \qquad (3.19)$$

όπου $m_t(M)$ η δενδρικού επιπέδου μάζα που υπολογίζεται από την Εξ. (2.59) και $\Delta m_t$ η διόρθωση που έχει μια κύρια SM συνεισφορά προερχόμενη από το gluon-top βρόχο που υπολογίζεται σύμφωνα με την Αν. [65] από την:

$$\Delta m_t(M)\;=\;\frac{g_3^2}{12\pi^2}\left[3\ln\left(\frac{M^2}{m_t^2}\right)+4\right]\ . \qquad (3.20)$$

με $g_3$ την ισχυρή ζεύξη βαθμίδας την προερχόμενη από την ομάδα $SU(3)_c$. Το αριθμητικό τμήμα της αγκύλης της Εξ. (3.20) εξαρτάται από το σχήμα επανακανονικοποίησης που έχει επιλεγεί και που στην περίπτωση μας είναι το $\overline{\mathrm{MS}}$.

Συνήθως στην Εξ. (3.20) τίθεται η τρέχουσα τιμή του $t$-quark $m_t(m_t)$, για την οποία τα πειραματικά όρια που χρησιμοποιηθηκαν στη διατριβή (Αν. [19]) είναι τα εξής:

$$m_t(m_t)=166\pm 0.05\,\mathrm{GeV}. \qquad (3.21)$$

οπότε λαμβάνεται η πολική μάζα του $t$-quark που πειραματικά προσδιορίζεται στα $176\pm 8$ GeV.

### 3.3.3 Μάζα του λεπτονίου $\tau$

Οι διορθώσεις στη μάζα του λεπτονίου $\tau$ είναι ασφαλώς μικρότερες σε σχέση με αυτές του $b$-quark όχι όμως και αμεληταίες στην περίπτωση του υψηλού $\tan\beta$. Η διορθωμένη μάζα του λεπτονίου $\tau$ δίνεται από τον τύπο:

$$m_\tau^c=m_\tau(1+\Delta m_\tau) \qquad (3.22)$$

όπου $m_\tau(M)$ η δενδρικού επιπέδου μάζα που υπολογίζεται από την Εξ. (2.59) και $\Delta m_\tau$ η διόρθωση που προέρχεται από το βρόχους chargino-tau sneutrino και δίνεται από την επόμενη φόρμουλα:

$$\Delta m_\tau=\ \frac{g^2}{16\pi^2}\ \frac{\mu M_2\tan\beta}{\mu^2-M_2^2}\left[B_0(M_S,M_2,m_{\tilde\nu_\tau})-B_0(M_S,\mu,m_{\tilde\nu_\tau})\right]\ , \qquad (3.23)$$

όπου η συνάρτηση $B_0(M_S,m_1,m_2)$ δίνεται στην Εξ. (3.16).

Τα πειραματικά όρια για τη μάζα του λεπτονίου $\tau$ όπως προκύπτουν από την Αν. [19] είναι τα εξής:

$$m_\tau(M_Z)=1.74\pm 0.0063\,\mathrm{GeV}. \qquad (3.24)$$

Επισημαίνεται ότι στις Εξ. (3.11) και (3.22) οι SUSY διορθώσεις υπολογίζονται σε ενεργειακή κλίμακα $M_S$ αφού σε αυτό το σημείο σταματά η εξέλιξη των SUSY παραμέτρων.

## 3.4 Το $\mathrm{BR}(b \to s\gamma)$

Η διαδικασία $b \to s\gamma$ υποβάλει σοβαρούς περιορισμούς στο χώρο παραμέτρων του MSSM. Γι αυτό και η θεώρησή της, στα πλαίσια των φαινομενολογικών εφαρμογών, που εξετάζονται, είναι επιβεβλημένη. Παρακάτω παρουσιάζεται το εφαρμοζόμενο στον αριθμητικό υπολογισμό τυπολόγιο, με QCD διορθώσεις κύριας τάξης (: LO) στο Εδ. 3.4.1 και πρώτης μετά την κύρια τάξη (: NLO) στο Εδ. 3.4.2. Κάποια συμπεράσματα σχετικά με τα αποτελέσματα του υπολογισμού αυτού εκτίθενται στο Εδ. 3.4.3 και ακολουθούν τα πειραματικά όρια που πρέπει να ικανοποιούνται στο Εδ. 3.4.4.

### 3.4.1 Τυπολόγιο υπολογισμού με LO QCD διορθώσεις

Ο πρώτος υπολογισμός που αναφέρεται στη βιβλιογραφία είναι αυτός της Αν. [23], ο οποίος όμως, λόγω πολυπλοκότητας, καθίσταται δυσχρηστός. Εύχρηστος, και κοινά αποδεκτός, αν και λίγο απλοποιητικός, είναι ο φορμαλισμός της Αν. [24], ο οποίος και παρουσιάζεται παρακάτω. Ο βασικός τύπος παρέχει τον επίμαχο ανηγμένο λόγο του $b \to s\gamma$, κανονικοποιημένο στο $\mathrm{BR}(b \to ce\bar{\nu})$, μέσω της έκφρασης:

$$\frac{\mathrm{BR}(b \to s\gamma)}{\mathrm{BR}(b \to ce\bar{\nu})} = \frac{|\, V_{ts}^* V_{tb}\,|^2}{|\, V_{cb}\,|^2} \frac{6\alpha_{em}(m_b)}{\pi} \frac{[\eta^{16/23}\mathcal{A}_\gamma + \frac{8}{3}(\eta^{14/23} - \eta^{16/23})\mathcal{A}_g + C_{Wb}]^2}{I(x_{cb})[1 - \frac{2}{3\pi}\alpha_s(m_b) f(x_{cb})]}. \tag{3.25}$$

Διευκρινήσεις σχετικά με την προέλευση των διαφόρων όρων που υπεισέρχονται στον υπολογισμό καθώς και οι αριθμητικές τιμές που χρησιμοποιούνται, δινονται παρακάτω:

- Ο ανηγμένος λόγος της διάσπασης $b \to ce\bar{\nu}$ τίθεται $\mathrm{BR}(b \to ce\bar{\nu}) = .104$.
- Η ηλεκτρομαγνητική σταθερά λεπτής υφής σε κλίμακα ενέργειας $m_b$ έχει τιμή $\alpha_{em}(m_b) = 1/132.5$.
- Ο λόγος των σχετικών στοιχείων του πίνακα CKM έχει κεντρική τιμή που δίνεται:

$$\frac{|V_{ts}^* V_{tb}|^2}{|V_{cb}|^2} \;=\; 0.95 \pm 0.04\;. \tag{3.26}$$

- Η συνάρτηση $I(x)$ προέρχεται από το χώρο των φάσεων $I(x) = 1 - 8x^2 + 8x^6 - x^8 - 24x^4 \ln x$.
- Η προερχόμενη από την QCD διόρθωση για την ημιλεπτονική διαδικασία συμπεριλαμβάνεται μέσω του παράγοντα $f(x_{cb}) = 2.41$ όπου $x_{cb} = \frac{m_c}{m_b} = .316 \pm .013$.
- Το σύμβολο $\eta = \frac{\alpha_s(M_W)}{\alpha_s(m_b)} = .548$, εκτελεί την αναγωγή από κλίμακα ενέργειας $M_W$ σε $m_b$.
- Το σύμβολο $C_{Wb}$ εκφράζει τη LO διόρθωση της QCD στο $b \to s\gamma$ που παίρνει τη μορφή

$$C_{Wb} = \sum_{i=1}^{8} h_i \eta^{d_i}$$

όπου, κατά την Αν. [25], οι τιμές των σταθερών είναι:

$$\begin{array}{l} d = (\ \frac{14}{23},\quad \frac{16}{23},\quad \frac{6}{23},\quad -\frac{12}{23},\quad 0.4086,\quad -0.4230,\quad -0.8994,\quad 0.1456) \\ h = (\ \frac{626126}{272277},\quad -\frac{56281}{51730},\quad -\frac{3}{7},\quad -\frac{1}{14},\quad -0.6494,\quad -0.0380,\quad -0.0186,\quad -0.0057) \end{array} \tag{3.27}$$

Τα πλάτη $\mathcal{A}_\gamma$, $\mathcal{A}_g$ προκύπτουν από τον υπολογισμό διαγραμμάτων (τύπου πινγκουίνου) με τρία εξωτερικά πόδια $b$, $s$ και $\gamma$, $Z$ ή $g$ αντίστοιχα, τα οποία εικονίζονται στο Σχ. 3.3. Προφανώς, η συνεισφορά του πλάτους $\mathcal{A}_\gamma$ είναι εξόχως δεσπόζουσα. Στο μεσολαβούντα βρόχο η αλληλεπίδραση λαμβάνει χώρα με παρουσία δύο σωματίων προέλευσης είτε από το SM, είτε από το MSSM. Γι αυτό, και οι επιμέρους συνεισφορές στα θεωρούμενα πλάτη είναι σκόπιμο να καταταγούν σε τρεις κατηγορίες:

$$\mathcal{A}_{\gamma[g]} = \mathcal{A}_{\gamma[g]}^{SM} + \mathcal{A}_{\gamma[g]}^{H^\pm} + \mathcal{A}_{\gamma[g]}^{\tilde{\chi}^\pm} \tag{3.28}$$

Αναλυτικά, το πλάτος με ενδιάμεσα σωμάτια:

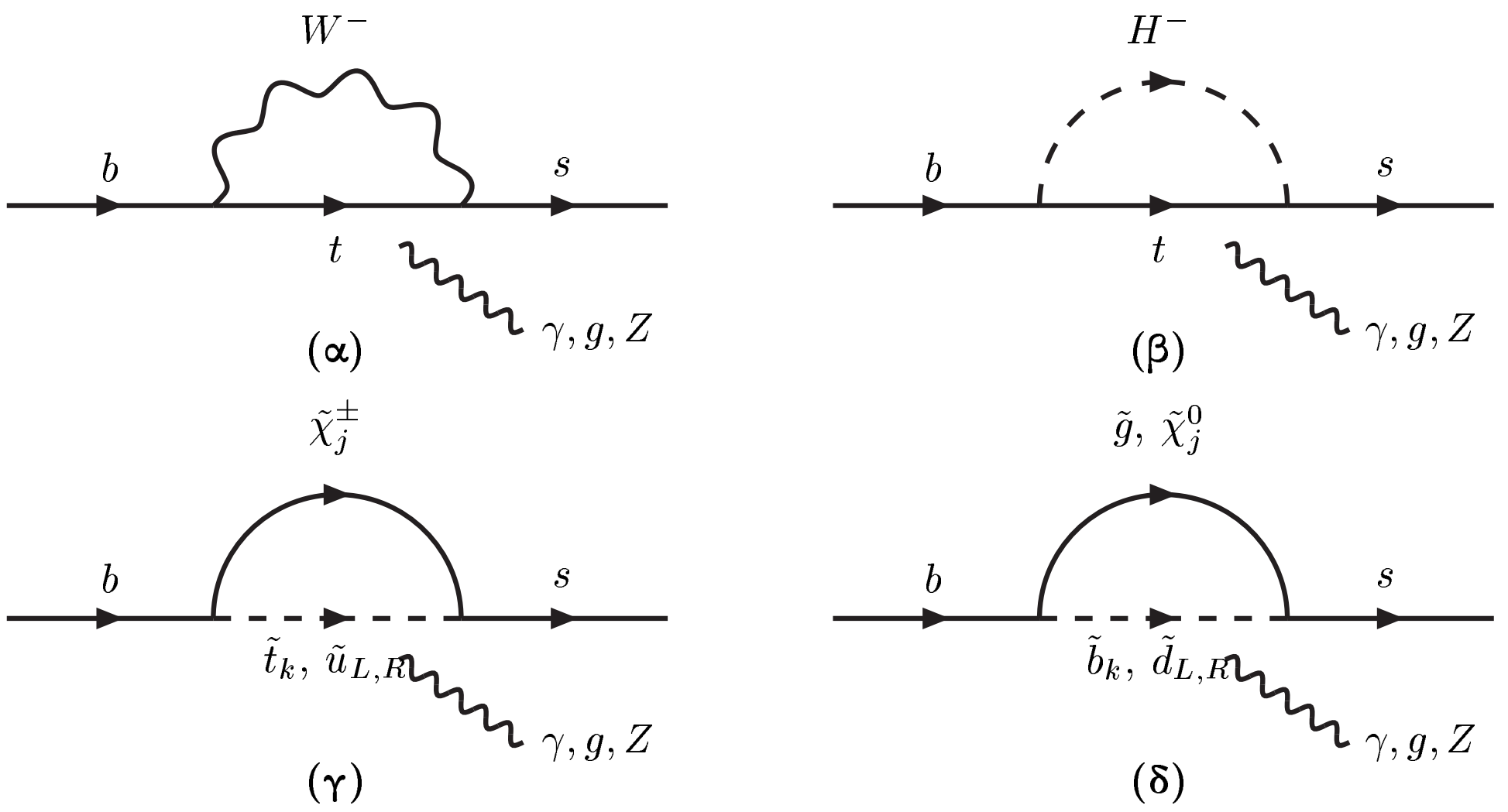

**Σχήμα 3.3:** *Οι βρόχοι $W^-$-top (α), $H^-$-top (β), chargino-stop, sup (γ), gluino, neutralino-sbottom, sdown (δ), που συνεισφέρουν στο* BR($b \to s\gamma$). *Ο εξερχόμενος κλάδος μποζονίου Βαθμίδας, $\gamma$, $Z$, $g$ (χάριν απλότητας και το gluon συμβόλιζεται με τον ίδιο τύπο κλάδου) επισυνάπτεται στο υπόλοιπο διάγραμμα με όλους τους δυνατούς τρόπους*

**α.** $W^{\pm} - u$ (u: up-quark) οφείλεται αποκλειστικά στη Φυσική του SM και δίνεται από τον τύπο:

$$A^{SM}_{\gamma[g]} = \frac{3}{2}\frac{m_t^2}{M_W^2} f^{(1)}_{\gamma[g]}\left(\frac{m_t^2}{M_W^2}\right) \tag{3.29}$$

**β.** $H^{\pm} - u$ δεν οφείλεται αποκλειστικά στη Φυσική MSSM , (λόγω της συνύπαρξης σωματίου του SM, up-quark και του MSSM, $H^{\pm}$ ) γι αυτό και λογίζεται χωριστά. Ο τύπος που το παρέχει είναι:

$$\mathcal{A}^{H^{\pm}}_{\gamma[g]} = \frac{1}{2}\frac{m_t^2}{m_{H^{\pm}}^2}\left[\frac{1}{\tan^2\beta} f^{(1)}_{\gamma[g]}\left(\frac{m_t^2}{m_{H^{\pm}}^2}\right) + f^{(2)}_{\gamma[g]}\left(\frac{m_t^2}{m_{H^{\pm}}^2}\right)\right] \tag{3.30}$$

**γ.** $\tilde{\chi}^{\pm}_{1[2]} - \tilde{u}_{L[R]}$, $\tilde{\chi}^{\pm} - \tilde{t}_{1[2]}$ οφείλεται αποκλειστικά στη Φυσική MSSM και χωρίζεται ως εξής:

$$\mathcal{A}^{\tilde{\chi}^{\pm}}_{\gamma[g]} = \mathcal{A}^{\tilde{\chi}^{\pm}}_{\gamma[g]1} + \mathcal{A}^{\tilde{\chi}^{\pm}}_{\gamma[g]2} + \mathcal{A}^{\tilde{\chi}^{\pm}}_{\gamma[g]3} + \mathcal{A}^{\tilde{\chi}^{\pm}}_{\gamma[g]4}\,, \tag{3.31}$$

όπου οι επιμέρους συνεισφορές δίνονται παρακάτω:

$$\begin{aligned}
\mathcal{A}^{\tilde{\chi}^{\pm}}_{\gamma[g]1} &= \sum_{j=1}^{2}\frac{M_W^2}{m^2_{\tilde{\chi}^{\pm}_j}}|V_{j1}|^2 f^{(1)}_{\gamma[g]}\left(\frac{m^2_{\tilde{u}}}{m^2_{\tilde{\chi}^{\pm}_j}}\right),\\
\mathcal{A}^{\tilde{\chi}^{\pm}}_{\gamma[g]2} &= -\sum_{j,k=1}^{2}\frac{M_W^2}{m^2_{\tilde{\chi}^{\pm}_j}}\left|V_{j1}(R_t)_{k1} - \frac{V_{j2}m_t(R_t)_{k2}}{\sqrt{2}M_W s_\beta}\right|^2 f^{(1)}_{\gamma[g]}\left(\frac{m^2_{\tilde{t}_k}}{m^2_{\tilde{\chi}^{\pm}_j}}\right),\\
\mathcal{A}^{\tilde{\chi}^{\pm}}_{\gamma[g]3} &= -\sum_{j=1}^{2}\frac{M_W}{m_{\tilde{\chi}^{\pm}_j}}\frac{U_{j2}V_{j1}}{\sqrt{2}c_\beta} f^{(3)}_{\gamma[g]}\left(\frac{m^2_{\tilde{u}}}{m^2_{\tilde{\chi}^{\pm}_j}}\right),\\
\mathcal{A}^{\tilde{\chi}^{\pm}}_{\gamma[g]4} &= \sum_{j,k=1}^{2}\frac{M_W}{m_{\tilde{\chi}^{\pm}_j}}\frac{U_{j2}}{\sqrt{2}c_\beta}\left(V_{j1}(R_t)_{k1} - V_{j2}(R_t)_{k2}\frac{m_t}{\sqrt{2}M_W s_\beta}\right)(R_t)^*_{k1} f^{(3)}_{\gamma[g]}\left(\frac{m^2_{\tilde{t}_k}}{m^2_{\tilde{\chi}^{\pm}_j}}\right).
\end{aligned}$$

Όπου $U$, $V$ οι πίνακες διαγωνοποίησης του πίνακα των chargino που ορίζονται στην Εξ. (2.78) και $R_t$ ο πίνακας διαγωνοποίησης του πίνακα των stops που ορίζεται στην Εξ. (2.66). Επίσης, τα squarks των πρώτων οικογενειών, $\tilde{u}_L, \tilde{u}_R$ θεωρήθηκαν εκφυλισμένα μάζας $m_{\tilde{u}}$, υπόθεση που δεν είναι μακριά από την πραγματικότητα, στα φάσματα που ελέγχηκαν.

**δ.** $\tilde{g} - \tilde{d}_{L[R]}$, $\tilde{b}_{1[2]}$ οφείλεται αποκλειστικά στη Φυσική MSSM . Η συνεισφορά του δύναται να αμεληθεί κυρίως όταν έχουν υποτεθεί παγκόσμιες αρχικές συνθήκες και έχει αμεληθεί η αναμιξη μεταξύ γενεών.

**ε.** $\tilde{\chi}$, $\tilde{\chi}^0_{2,\ldots,4} - \tilde{d}_{L[R]}$, $\tilde{b}_{1[2]}$ οφείλεται αποκλειστικά στη Φυσική MSSM. Ομοίως με το προηγούμενο, η συνεισφορά του δύναται να αμεληθεί, σύμφωνα με τα κοινώς κείμενα.

Οι εμπλεκόμενες στον υπολογισμό συναρτήσεις δίνονται παρακάτω:

$$
\begin{aligned}
f^{(1)}_\gamma(x) &= \frac{(7-5x-8x^2)}{36(x-1)^3} + \frac{x(3x-2)}{6(x-1)^4}\ln x\,, \\
f^{(2)}_\gamma(x) &= \frac{(3-5x)}{6(x-1)^2} + \frac{(3x-2)}{3(x-1)^3}\ln x\,, \\
f^{(3)}_\gamma(x) &= (1-x)f^{(1)}_\gamma(x) - \frac{x}{2}f^{(2)}_\gamma(x) - \frac{23}{36}\,, \\
f^{(1)}_g(x) &= \frac{(2+5x-x^2)}{12(x-1)^3} - \frac{x}{2(x-1)^4}\ln x\,, \\
f^{(2)}_g(x) &= \frac{(3-x)}{2(x-1)^2} - \frac{1}{(x-1)^3}\ln x\,, \\
f^{(3)}_g(x) &= (1-x)f^{(1)}_g(x) - \frac{x}{2}f^{(2)}_g(x) - \frac{1}{3}\,.
\end{aligned}
$$

Το τυπολόγιο αυτό βασίζεται σε υπολογισμούς με LO QCD διορθώσεις και ακολουθείται από μια πληθώρα ερευνητών στις Αν. [20], [21], [26] και [30].

### 3.4.2 Τυπολόγιο υπολογισμού με NLO QCD διορθώσεις

Νεότεροι υπολογισμοί της επίμαχης ποσότητας έχουν παρουσιαστεί στις Αν. [27], [28] και [29], λαμβάνοντας υπόψη τους και NLO διορθώσεις προερχόμενες από τη QCD. Οι διορθώσεις αυτές συντείνουν στη μείωση του τελικού αποτελέσματος μέχρι περίπου 30% και στον έλεγχο των σφαλμάτων που προκύπτουν από τις τρεις χρησιμοποιούμενες κλίμακες ενέργειας, στις οποίες γίνεται η συναρμογή των ενεργών (effective) θεωριών που ελέγχουν το φαινόμενο (matching scales).

Ακολουθώντας την απλοποιημένη από πλευράς τυπολογιου και ανεξάρτητη από είδος προτύπου, παρουσίαση της Αν. [29], το BR($b \to s\gamma$) μπορεί να βρεθεί από την σχέση:

$$\mathrm{BR}(b \to s\gamma) = R_\gamma\, \mathrm{BR}^{\mathrm{SM}}(b \to s\gamma)\,, \tag{3.32}$$

όπου η πρόβλεψη του SM με συνυπολογισμό διορθώσεων LO από την QED και NLO από τη QCD, είναι:

$$\mathrm{BR}^{\mathrm{SM}}(b \to s\gamma) = (3.29 \pm 0.33) \times 10^{-4} \tag{3.33}$$

ενώ η θεωρητική πρόβλεψη για το λόγο $R_\gamma$ μπορεί να γραφεί ως εξής:

$$
\begin{aligned}
R_\gamma &= 1 + A_1(\chi)\,(\xi_7 - 1) + A_2(\chi)\left(|\xi_7|^2 - 1\right)\,, && (3.34)\\
\text{όπου}\ \ A_1(\chi) &= \left(B_{27} + \chi B_{28} + (1-\chi)B_{78} + 2\chi(1-\chi)B_{88}\right)/B_T && (3.35)\\
\text{και}\ \ A_2(\chi) &= \left(B_{77} + \chi B_{78} + \chi^2 B_{88}\right)/B_T\,, && (3.36)\\
\text{με}\ \ B_T &= B_{22} + B_{27} + B_{77} + B_{28} + B_{78} + B_{88}. && (3.37)
\end{aligned}
$$

Οι συντελεστές $B$ εξαρτώνται από το σημείο τερματισμού της θεωρίας διαταραχών με NLO QCD διορθώσεις. Οι τιμές τους βρίσκονται από πίνακα της Αν. [29], και επιλέγονται έτσι, ώστε το άθροισμά τους, $B_T$, να ταυτίζεται με την κεντρική τιμή της Εξ. (3.33):

$$B_{22} = 1.258,\quad B_{27} = 1.395,\quad B_{77} = 0.382,\quad B_{28} = 0.161,\quad B_{78} = 0.083,\quad B_{88} = 0.015$$

Τα εμφανιζόμενα στις Εξ. (3.34)-(3.36) σύμβολα ορίζονται ως εξής:

$$\chi = \frac{\xi_8 - 1}{\xi_7 - 1} \qquad \text{όπου} \qquad \xi_{7[8]} = \frac{C_{7[8]}}{C^{\rm SM}_{7[8]}} \tag{3.38}$$

$$\text{με} \quad C^{\rm SM}_{7[8]} = -0.22\,[-0.12] \qquad \text{και} \qquad C_{7[8]} = C^{\rm SM}_{7[8]} + C^{H^\pm}_{7[8]} + C^{\tilde{\chi}^\pm}_{7[8]} \tag{3.39}$$

όπου οι συντελεστές του Wilson $C_{7[8]}$ λαμβάνουν συνεισφορές από τα διαγράμματα του Σχ. 3.3. Για τη συνεισφορά του SM, $C^{\rm SM}_{7[8]}$, τίθενται τα αριθμητικά εξαγώμενα της Αν. [29], ενώ για αυτή από τα φορτισμένα Higgs λαμβάνεται υπόψη το τυπολόγιο της Αν. [28], σύμφωνα με το οποίο:

$$C^{H^\pm}_{7[8]}(M_W) = C^{(0)}_{7[8]}(M_W) + \frac{g_3(M_W)}{16\pi^2} C^{(1)}_{7[8]}(M_W) \tag{3.40}$$

όπου $C^{(0)}_{7[8]}$ και $C^{(1)}_{7[8]}$ οι συνεισφορές με LO και NLO QCD διορθώσεις αντίστοιχα, που είναι:

$$C^{(0)}_{7[8]}(M_W) = \mathcal{A}^{H^\pm}_{\gamma[g]} \quad \text{και} \quad C^{(1)}_{7[8]}(M_W) = G_{7[8]}(y) + \Delta_{7[8]}(y) \ln\frac{M_W^2}{m^2_{H^\pm}} - \frac{4}{9}\left[\frac{1}{6}\right] E(y) \tag{3.41}$$

Οι εμπλεκόμενες συναρτήσεις παρουσιάζονται παρακάτω:

$$\begin{aligned} G_7(y) &= -\frac{4}{3}y\left[\frac{4(-3+7y-2y^2)}{3(y-1)^3}\mathrm{Li}_2\left(1-\frac{1}{y}\right) + \frac{8-14y-3y^2}{3(y-1)^4}\ln^2 y \right.\\ &\left. + \frac{2(-3-y+12y^2-2y^3)}{3(y-1)^4}\ln y + \frac{7-13y+2y^2}{(y-1)^3}\right] \\ &+ \frac{1}{\tan^2\beta}\frac{2}{9}y\left[\frac{y(18-37y+8y^2)}{(y-1)^4}\mathrm{Li}_2\left(1-\frac{1}{y}\right) + \frac{y(-14+23y+3y^2)}{(y-1)^5}\ln^2 y \right.\\ &+ \frac{-50+251y-174y^2-192y^3+21y^4}{9(y-1)^5}\ln y \\ &\left. + \frac{797-5436y+7569y^2-1202y^3}{108(y-1)^4}\right] \end{aligned} \tag{3.42}$$

$$\begin{aligned} \Delta_7(y) &= -\frac{2}{9}y\left[\frac{21-47y+8y^2}{(y-1)^3} + \frac{2(-8+14y+3y^2)}{(y-1)^4}\ln y\right] \\ &+ \frac{1}{\tan^2\beta}\frac{2}{9}y\left[\frac{-31-18y+135y^2-14y^3}{6(y-1)^4} + \frac{y(14-23y-3y^2)}{(y-1)^5}\ln y\right] \end{aligned} \tag{3.43}$$

$$\begin{aligned} G_8(y) &= -\frac{1}{3}y\left[\frac{-36+25y-17y^2)}{2(y-1)^3}\mathrm{Li}_2\left(1-\frac{1}{y}\right) + \frac{19+17y}{(y-1)^4}\ln^2 y \right.\\ &\left. + \frac{-3-187y+12y^2-14y^3)}{4(y-1)^4}\ln y + \frac{3(143-44y+29y^2)}{8(y-1)^3}\right] \\ &+ \frac{1}{\tan^2\beta}\frac{1}{6}y\left[\frac{y(30-17y+13y^2)}{(y-1)^4}\mathrm{Li}_2\left(1-\frac{1}{y}\right) - \frac{y(31+17y)}{(y-1)^5}\ln^2 y \right.\\ &+ \frac{-226+817y+1353y^2+318y^3+42y^4}{36(y-1)^5}\ln y \\ &\left. + \frac{1130-18153y+7650y^2-4451y^3}{216(y-1)^4}\right] \end{aligned} \tag{3.44}$$

$$\begin{aligned} \Delta_8(y) &= -\frac{1}{3}y\left[\frac{81-16y+7y^2}{2(y-1)^3} - \frac{19+17y}{(y-1)^4}\ln y\right] \\ &+ \frac{1}{\tan^2\beta}\frac{1}{6}y\left[\frac{-38-261y+18y^2-7y^3}{6(y-1)^4} + \frac{y(31+17y)}{(y-1)^5}\ln y\right] \end{aligned} \tag{3.45}$$

$$E(y) = \frac{1}{\tan\beta}\left[\frac{y(16-29y+7y^2)}{36(y-1)^3} + \frac{y(3y-2)}{6(y-1)^4}\ln y\right]. \tag{3.46}$$

Τέλος για τη συνεισφορά των charginos $C_{7[8]}^{\tilde{\chi}^\pm}$ χρησιμοποιείται το αποτέλεσμα με LO QCD διορθώσεις γιατί τα αποτελέσματα της Αν. [28] που συνυπολογίζουν NLO QCD διορθώσεις δεν ισχύουν στις εκδόσεις του MSSM που θα εξεταστούν, στις οποίες τα stops δεν είναι σημαντικά ελαφρότερα από τα gluino και squarks. Επομένως,

$$C_{7[8]}^{\tilde{\chi}^\pm} = \mathcal{A}_{\gamma[g]}^{\tilde{\chi}^\pm} \tag{3.47}$$

### 3.4.3 Συμπεριφορά των Αποτελεσμάτων

Δεν θεωρείται σκόπιμη η εμπλοκή στον κυκεώνα των απλουστευτικών προσεγγίσεων, αμφιβόλου περιοχής ισχύος και άδηλης προέλευσης, για να ερμηνευτεί η συμπεριφορά των αποτελεσμάτων που προκύπτουν με εφαρμογή των τυπολογίων των Εδ. 3.4.1 και 3.4.2. Είναι πανθομολογούμενο, ότι η συνεισφορά των φορτισμένων Higgs από την Εξ. (3.30) είναι πάντοτε προσθετική σε αυτήν του SM της Εξ. (3.29). Αντιθέτως, η συνεισφορά των charginos της Εξ. (3.31) είναι άλλοτε επ-οικοδομητική στις δύο προηγούμενες και άλλοτε απ-οικοδομητική. Για τα μελετούμενα σε αυτή την εργασία πρότυπα, και με βάση τις ακολουθούμενες συμβάσεις, όταν $\mu > 0$ η συνεισφορά των charginos καθίσταται προσθετική, ενώ όταν $\mu < 0$ γίνεται αφαιρετική σε σχέση με τη συνεισφορά του SM συν του φορτισμένου Higgs. Προφανώς, η εκκεντρική αυτή συμπεριφορά των αποτελεσμάτων, οφείλεται στους πίνακες διαγωνοποίησης του πίνακα των charginos, $U$, $V$, η μορφή των οποίων εξαρτάται από το πρόσημο της ορίζουσας του πίνακα των charginos.

Ισχυρή συσχέτιση των αποτελεσμάτων αυτών υπάρχει με τις διορθώσεις στη μάζα του $b$-quark. Τα δύο φαινόμενα είναι ουσιαστικά ανταγωνιστικά. Ο κανόνας που διέπει αυτή τη συσχέτιση είναι γενικός και μπορεί να διατυπωθεί με τρόπο ανεξάρτητο από συμβάσεις, σύμφωνα με την Αν. [30]. Αρνητική διόρθωση στη μάζα του $b$-quark συνεπάγεται θετική συνεισφορά των charginos στο $\mathrm{BR}(b \to s\gamma)$. Σαφέστερη εικόνα για την ενδιαφέρουσα αυτή αλληλεπίδραση των δύο φαινομένων δίνεται στα αριθμητικά αποτελέσματα που θα παρουσιαστούν στα κεφάλαια 6 και 7 της διατριβής.

Τα προηγούμενα συμπεράσματα βρίσκονται σε πλήρη συμφωνία με τις διαπιστώσεις των Αν. [20] και [30] με τη σχετική αναγωγή του συμβολισμού στον υιοθετημένο στη διατριβή. Αντίθετα, διαφωνούμε με το διάγραμμα του Σχ. 5 της Αν. [21] ως προς την τοποθέτηση των προσήμων του $\mu$. Προσεκτική εξέταση του συμβολισμού δείχνει ότι υιοθετούνται αντίθετες συμβάσεις για το σημείο του $\mu$ με τις χρησιμοποιούμενες εδώ (ενδεικτικά αναφέρεται η Εξ. (25) ). Κατά συνέπεια, η Εξ. (22) είναι συμβατή με το συμβολισμό, όμως στο σχήμα της εικόνας 5, οι καμπύλες δεν αντιστοιχούν στο αναγραφόμενο πρόσημο του $\mu$ αλλά στο αντίθετο. Επομένως, και τα σχετικά συμπεράσματα είναι αναληθή.

Τέλος κατά την ανάλυση που παρουσιάστηκε η μίξη μεταξύ των οικογένειων αμελήθηκε, τα squarks των ελαφρών οικογένειων θεωρήθηκαν εκφυλισμένα και οι συνεισφορές των βρόχων neutralino, gluino - down-squark, δεν συμπεριλήφθηκαν. Ακόμη δεν διερευνήθηκαν τα σφάλματα που υπεισέρχονται στον υπολογισμό από τις εισαγόμενες στο πρόγραμμα ποσότητες και από τις κλίμακες επανακανονικοποίησης και συναρμογής.

### 3.4.4 Πειραματικά Όρια

Πρόσφατα είναι τα αποτελέσματα από την πειραματική ομάδα ALEPH (Αν. [31])

$$\mathrm{BR}(b \to s\gamma) = (3.11 \pm 0.80 \pm 0.72) \times 10^{-4} \tag{3.48}$$

ενώ η ομάδα CLEO (Αν. [32]), έχει ανακοινώσει:

$$\mathrm{BR}(b \to s\gamma) = (3.15 \pm 0.35 \pm 0.32 \pm 0.26) \times 10^{-4}. \tag{3.49}$$

Επομένως, τα όρια που τίθενται είναι:

$$2 \times 10^{-4} \lesssim \mathrm{BR}(b \to s\gamma) \lesssim 4.5 \times 10^{-4}. \tag{3.50}$$

Η αναμενόμενη βελτίωση των πειραματικών ευρημάτων αναμφίβολα θα βοηθήσει στον ακριβέστερο έλεγχο του MSSM, πάντως σίγουρα θα περιορίσει των επιτρεπτό χώρο παραμέτρων αισθητά.

# Κεφάλαιο 4

# Καθιερωμένο Κοσμολογικό Πρότυπο

## 4.1 Εισαγωγή

Όπως και στην περίπτωση της Φυσικής Στοιχειωδών Σωματιδίων, έτσι και στην Κοσμολογία η απαρχή κάθε αναζήτησης στηρίζεται σε ένα Καθιερωμένο Κοσμολογικό Πρότυπο (: SBB). Το ψηφιδωτό αυτού του προτύπου παρουσιάζεται στο κεφάλαιο αυτό. Έμφαση δίνεται στα έννοιες που θα χρησιμοποιηθούν στο επόμενο κεφάλαιο όπου θα γίνει μια λεπτομεριακή αξιοποίηση του MSSM για την ερμηνεία του κοσμολογικού προβλήματος της Σκοτεινής Ύλης. Στην ανάλυση που θα ακολουθήσει αξιοποιήθηκαν ιεραρχικά οι Αν. [33], [34], [35] και [36].

Στην παρουσίαση του σχετικού τυπολογίου έχει υιοθετηθεί η παρακάτω ταυτοποίηση των σταθέρων

$$\hbar = c = k_B = 1,$$

με $c$ τη σταθέρα του φωτός, $k_B$ τη σταθερά του Boltzman. Για τις αριθμητικές εφαρμογές, χρησιμοποιούνται οι παρακάτω τύποι μετατροπής:

$$\begin{aligned} 1\,\mathrm{GeV}^{-1} &= 1.973\times 10^{-14}\mathrm{cm} = 6.582\times 10^{-25}\mathrm{sec}, \\ 1\,\mathrm{GeV} &= 1.160\times 10^{13}\,{}^0\mathrm{K} = 1.783\times 10^{-24}\mathrm{gr}, \\ 1\,\mathrm{Mpc} &= 3.086\times 10^{24}\mathrm{cm}, \\ 1\,\mathrm{yr} &= 3.156\times 10^{7}\mathrm{sec}. \end{aligned}$$

Επίσης, σε όλη τη διάρκεια του κεφαλαίου αποφεύγεται η χρήση της Σταθεράς του Νεύτωνα ($G_N$) στις εξισώσεις και προτιμάται η πιο οικεία για τους ασχολούμενους με τη Φυσική των Στοιχειωδών Σωματίων, Σταθερά του Plank:

$$M_P = 1/\sqrt{G_N} = 1.22\times 10^{19}\,\mathrm{GeV}.$$

Το κεφάλαιο αρχίζει με τη διατύπωση της Κοσμολογικής Αρχής και των συνεπειών της στο Εδ. 4.2. Ακολουθεί η θεμελίωση των εννοιών που σχετίζονται με τη Δυναμική και τη Θερμοδυναμική του Σύμπαντος στα Εδ. 4.3, 4.4 αντίστοιχα. Εξοπλισμένος κανείς με αυτά τα εφόδια κατανοεί τη Θεωρία της Μεγάλης Εκρηξης που συνοπτικά περιγράφεται στο Εδ. 4.5. Όπως το SM έτσι και το SBB εμφανίζει μερικά μειονεκτήματα. Αυτά εντοπίζονται και λύνονται με την ιδέα του Πληθωρισμού στο Εδ. 4.6.

## 4.2 Κοσμολογική Αρχή

Ακρογωνιαίος λίθος της Σύγχρονης Κοσμολογίας είναι η ανακάλυψη της απόλιθωμένης κοσμικής ακτινοβολίας. Προσεγγίζεται με το φάσμα ενός μελανού σώματος σε θερμοκρασία 2.73 ${}^0$K. Στην ισοτροπία της ακτινοβολίας αυτής (μέχρι τάξεως $10^{-6}$) εδράζεται η Κοσμολογική Αρχή, σύμφωνα με την οποία, σε αρκούντως μεγάλες κλίμακες το σύμπαν είναι Ομογενές και Ισότροπο. Οι άμεσες συνέπειες αυτού του αρχικού αξιώματος περιγράφονται παρακάτω.

### 4.2.1 Συν-κινούμενο Σύστημα Αναφοράς

Σε ένα Ομογενές και Ισότροπο Σύμπαν μπορεί να οριστεί ένα σύστημα αναφοράς στο οποίο η περιεχόμενη ύλη είναι ακίνητη, το οποίο ονομάζεται Συν-κινούμενο (co-moving). Οι θερμοδυναμικές ιδιότητες της ύλης μπορούν να περιγραφούν με θερμοδυναμικά δυναμικά ορισμένα σε αυτό το σύστημα ηρεμίας (: CRF). Παρόλη τη χρήση του συγκεκριμένου συστήματος αναφοράς για τον ορισμό αυτών των ποσοτήτων, οι τιμές που επιτυγχάνονται είναι ανεξάρτητες από σύστημα αναφοράς, πρόκειται, δηλαδή για βαθμωτά (scalar) πεδία, κατά την Αν. [37]. Στο CRF, λοιπόν, ορίζονται:

- Η πυκνότητα αριθμού σωματίων, δηλαδή ο αριθμός σωματίων ανά μονάδα όγκου, $n$
- Η πυκνότητα ενέργειας σωματίων, δηλαδή ο αριθμός σωματίων ανά μονάδα όγκου, $\rho$
- Η πυκνότητα εντροπίας, δηλαδή η εντροπία του ρευστού ανά μονάδα όγκου, $s$
- Η ισοτροπική πίεση, $P$
- Η θερμοκρασία, $T$

Χημικό δυναμικό δεν θα οριστεί, γιατί η παρουσία του θα αγνοηθεί σε όλη τη διάρκεια της διατριβής αυτής. Γενικά όταν δεν υπάρχει ασυμμετρία ανάμεσα στον αριθμό σωματίων και αντι-σωματίων ή αν η υπάρχουσα είναι μικρή, το χημικό δυναμικό μπορεί ασφαλώς να αμεληθεί.

### 4.2.2 Μετρική των Robertson-Walker

Η ομογένεια και η ισοτροπία απλοποιεί σημαντικά το μαθηματικό κατασκεύασμα που πρέπει να επινοηθεί για να περιγράφει το Σύμπαν. Η γεωμετρία του Σύμπαντος προκύπτει από τη μετρική των Robertson-Walker

$$ds^2 = g_{\mu\nu}dx^\mu dx^\nu = dt^2 - R^2(t)\left[\frac{dr^2}{1-kr^2} + r^2(d\theta^2 + \sin^2\theta d\phi^2)\right]. \tag{4.1}$$

Όπου $R(t)$ είναι ο παράγοντας κλίμακας πού εξελίσσεται με το χρόνο, $r, \phi, \theta$ οι συν-κινούμενες χωρικές σφαιρικές συντεταγμένες και η σταθερά $k$ σχετίζεται με την καμπυλότητα του τρισδιάστατου χώρου. Λαμβάνει τιμές $+1$, $0$ ή $-1$ που αντιστοιχούν σε ελλειπτικό (κλειστό), ευκλείδειο (επίπεδο) ή υπερβολικό (ανοικτό) Σύμπαν. Όταν $k = +1$, ο χώρος του Σύμπαντος μπορεί να εκληφθεί ως μία τριών διαστάσεων σφαίρα σε έναν ευκλείδειο χώρο 4 διαστάσεων ακτίνας $R$, με εξίσωση $x_1^2 + x_2^2 + x_3^2 + x_4^2 = R^2$ και όγκο $V = 2\pi^2 R^3$. Γιαυτό και γενικότερα, ως όγκος του σύμπαντος θα λαμβάνεται η ποσότητα $R^3$

### 4.2.3 Ιδανικό Ρευστό

Η επίτευξη της περιγραφής ενός ομογενούς και ισότροπου σύμπαντος απαιτεί και την εισαγωγή ενός είδους ύλης σε αυτό. Το όνομα αυτής, ιδανικό ρευστό, και ο ορισμός της, ένα ρευστό για το οποίο να ισχύουν οι εξισώσεις της Συνέχειας και του Νεύτωνα της Κλασσικής Μηχανικής Ρευστών αλλά υπό σχετικιστική μορφή. Αποδεικνύεται ότι αυτό μπορεί να γίνει τη χρήση του παρακάτω τανυστή ενέργειας-ορμής:

$$T_{\mu\nu} = -Pg_{\mu\nu} + (P+\rho)u_\mu u_\nu \ , \tag{4.2}$$

όπου $u_\mu = dx_\mu/ds$ η τετρα-ταχύτητα ενός σωματίου του ρευστού. Στο συγκινούμενο σύστημα αναφοράς αυτού του σωματίου, η ταχύτητα είναι $(1, \vec{0})$, οπότε η μορφή του τανυστή ενέργειας-ορμής απλοποιείται:

$$T_{\mu\nu} = \mathrm{diag}(\rho, P, P, P). \tag{4.3}$$

### 4.2.4 Ο Νόμος του Hubble

Ο νόμος διαστολής διατυπώθηκε το 1920 από τον Hubble και είναι παρεπόμενο της κοσμολογικής αρχής. Εστω ότι δύο Γαλαξίες απέχουν αρχικά απόσταση $l_0$. Μετά την πάροδο χρόνου $t$ η μεταξύ τους απόσταση θα είναι $l(t) = R(t)l_0$, όπου ο παράγοντας $R(t)$ είναι συνάρτηση μόνο του χρόνου, λόγω της κοσμολογικής αρχής. Η σχετική ταχύτητα των δύο Γαλαξιων θα είναι (ως συνήθως, το σύμβολο ˙ σημαίνει παράγωγο ως προς το χρόνο):

$$v(t) = \frac{dl}{dt} = \dot{R}(t)l_0 = \Big(\frac{\dot{R}(t)}{R(t)}\Big)l(t) = H(t)l(t), \quad \text{με} \quad H(t) = \frac{\dot{R}(t)}{R(t)} \tag{4.4}$$

Η ποσότητα $H(t)$ λέγεται παράμετρος Hubble. Ό νομος του Hubble μας προιδεάζει για μία σχετική κίνηση ανάμεσα στους μακρινούς Γαλαξίες. Η ταυτότητα αυτής της κίνησης δίνεται στο επόμενο εδάφιο.

### 4.2.5 Η Κοσμολογική Υπέρυθρη Μετατόπιση

Η πρώτη επιτυχία της επιλεγμένης μετρικής είναι ότι εξηγεί την παρατηρούμενη υπέρυθρη μετατόπιση του εκπεμπόμενου φάσματος από τους δίαφορους Γαλαξίες και προβλέπει τη διαστολή του σύμπαντος. Βεβαίως, υπέρυθρη μετατόπιση μπορεί να οφείλεται σε βαρυτικά πεδία ή σε φαινόμενο Doppler. Αυτά τα φαινόμενα, όμως είναι αμεληταία σε αποστάσεις κοσμολογικών διαστάσεων.

Έστω ότι μια φωτεινή ακτίνα μήκους κύματος $\lambda(t)$ εκπεμπόμενη τη στιγμή $t_e$ από ένα Γαλαξία σε απόσταση $r_e$, η οποία λαμβάνεται από παρατηρητή τη στιγμή $t_0$ σε απόσταση $r_0 = 0$. Η παρατηρούμενη μετατόπιση, $z$ ορίζεται από την:

$$z := \frac{\lambda(t_0) - \lambda(t_e)}{\lambda(t_e)} \tag{4.5}$$

είναι υπέρυθρη, δηλαδή $z > 0$. Η σύνδεση της παρατήρησης αυτής με τις διαστάσεις του Σύμπαντος, γίνεται με χρήση της μετρικής της Εξ. (4.1). Η θεωρούμενη φωτεινή ακτίνα ταξιδεύει κατα μήκος μιας μηδενικής γεωδαισιακής ($ds^2 = d\theta = d\phi = 0$) στο CRF. Χρησιμοποιώντας την Εξ. (4.1), λαμβάνεται:

$$0 = ds^2 = dt^2 - R^2(t)\Big[\frac{dr^2}{1-kr^2}\Big] \Longrightarrow \frac{dt}{R(t)} = \frac{dr}{\sqrt{1-kr^2}} \tag{4.6}$$

Παρεπιπτώντως, μπορεί να οριστεί ο ορίζοντας ενός σωματίου, δηλαδή η απόσταση που διανύει μια φωτεινή ακτίνα από την αρχή του σύμπαντος $t = 0$ μέχρι μια στιγμή $t$, που είναι προφανώς:

$$d_H(t) := \int_0^t dt = \int_0^{r_H} R(t)\Big[\frac{dr}{1-kr^2}\Big] \Longrightarrow d_H(t) = R(t)\int_0^t \frac{dt'}{R(t')} \tag{4.7}$$

Επιστρέφοντας στον προορισμό μας, και ολοκληρώνοντας την Εξ. (4.6), για δύο γεγονότα που απέχουν πολύ μικρή χρονική απόσταση μεταξύ τους, $\Delta t$ ώστε ο $R(t)$ να μην μεταβάλλεται, λαμβάνεται:

$$\int_{t_e}^{t_0} \frac{dt}{R(t)} \simeq \int_{t_e+\Delta t_e}^{t_0+\Delta t_0} \frac{dt}{R(t)} = \int_0^{r_e} \frac{dr}{\sqrt{1-kr^2}} \tag{4.8}$$

Ορίζοντας τη συνάρτηση $F$, ως εξής:

$$\dot{F}(t) = \frac{1}{R(t)} \Longleftrightarrow F(t) = \int_0^{t_0} \frac{dt}{R(t)}, \tag{4.9}$$

η Εξ. (4.8), γράφεται:

$$F(t_0) - F(t_e) \simeq F(t_0 + \Delta t_0) - F(t_e + \Delta t_e), \tag{4.10}$$

οπότε αναπτύσσοντας κατά Taylor γύρω από τα σημεία $t_e$, $t_0$, λαμβάνεται:

$$\Delta t_0 \dot{F}(t_0) = \Delta t_e \dot{F}(t_e) \Longrightarrow \frac{\Delta t_e}{\Delta t_0} = \frac{R(t_e)}{R(t_0)}, \tag{4.11}$$

Αξιοποιώντας τον ορισμό του μήκους κύματος $\lambda = \Delta t$ και αντικαθιστώντας στην Εξ. (4.5) επιτυγχάνεται τελικά η επιδιωκόμενη:

$$z = \frac{R(t_0)}{R(t_e)} - 1. \tag{4.12}$$

Η υπέρυθρη μετατόπιση, $z > 0$ σημαίνει ότι $R(t_0) > R(t_e)$, δηλαδή η τιμή του παράγοντα κλίμακας τη στιγμή της λήψης της φωτεινής ακτίνας είναι μεγαλύτερη από την τιμή του, τη στιγμή της εκπομπής. Άρα το Σύμπαν διαστέλλεται.

Με βάση την Εξ. (4.12) μπορεί να διατυπωθεί μια σχέση αναγωγής των αποστάσεων $D(t)$ σε διάφορες χρονικές στιγμές:

$$\frac{D(t_0)}{D(t)} = \frac{R(t_0)}{R(t)}. \tag{4.13}$$

Μια εντυπωσιακή εφαρμογή της σχέσης αυτής, δίνεται στην Αν. [36], όπου υπολογίζεται ότι το σημερινό παρατηρούμενο σύμπαν είχε διαστάσεις λίγων χιλιοστών την εποχή του Plank, $(t_P = M_P^{-1})$.

## 4.3 Δυναμική Κοσμολογία

Η Γένική Σχετικότητα είναι η θεωρία πλαίσιο που εφαρμόζεται για την περιγραφή της Δυναμικής του Σύμπαντος. Κεντρικό ρόλο παίζει η εξίσωση πεδίου του Einstein που συσχετίζει τον τένσορα ενέργειας-ορμής $T_{\mu\nu}$ και τη χωροχρονική καμπυλότητα $R_{\lambda\mu\nu\kappa}$:

$$R_{\mu\nu} - \frac{1}{2} g_{\mu\nu}\mathcal{R} = -\frac{8\pi}{M_P^2} T_{\mu\nu} + \Lambda g_{\mu\nu} \ , \tag{4.14}$$

όπου $R_{\mu\nu} := g^{\lambda\kappa} R_{\lambda\mu\kappa\nu}$ ο τανυστής Ricci και $\mathcal{R} = g^{\mu\nu} R_{\mu\nu}$ η βαθμωτή καμπυλότητα Ricci. Αν και η τιμή της κοσμολογικής σταθεράς, $\Lambda$, είναι πολύ μικρή $(\Lambda < 10^{-120} M_P^{-2})$ θεωρείται αναγκαία η διατήρηση της στις εξισώσεις, για λόγους που θα γίνουν φανεροί στην εξέλιξη της εργασίας. Οι επιλογές μετρικής και ύλης εμφυτεύονται στην εξίσωση Einstein, οπότε λαμβάνονται οι εξισώσεις Friedmann-Lemaître που παρουσιάζονται στο Εδ. 4.3.1 Χρήσιμες φυσικές παράμετρες ορίζονται στα Εδ. 4.3.2, 4.3.3 και τέλος η εισαγωγή μιας καταστατικής εξίσωσης στο Εδ. 4.3.4 επιτρέπει την ποιοτική περιγραφή της δυναμικής του Σύμπαντος.

### 4.3.1 Οι εξισώσεις των Friedmann-Lemaître

Η εφαρμογή της εξίσωσης του Einstein στο συν-κινούμενο σύστημα αναφοράς, με τη μετρική της Εξ. (4.1) και τον τανυστή της Εξ. (4.3), απαιτεί τον υπολογισμό του τανυστή Ricci και της καμπυλότητας. Για τους χρονικούς, 00 και χωρικούς, $ij$ δείκτες αντίστοιχα του τανυστή Ricci, λαμβάνεται:

$$R_{00} = -3\frac{\ddot{R}}{R} \tag{4.15}$$

$$R_{ij} = -6\left(\frac{\ddot{R}}{R} + 2\frac{\dot{R}^2}{R^2} + \frac{2k}{R^2}\right). \tag{4.16}$$

ενώ για την καμπυλότητα Ricci

$$\mathcal{R} = 6\left(\frac{\ddot{R}}{R} + \frac{\dot{R}^2}{R^2} + \frac{k}{R^2}\right). \tag{4.17}$$

Αντικαθιστώντας, η Εξ. (4.14) συνεπάγεται για τους χρονικούς, 00 και χωρικούς, $ij$ δείκτες αντίστοιχα

$$\left(\frac{\dot{R}}{R}\right)^2 = \frac{8\pi}{3M_P^2}\rho - \frac{k}{R^2} + \frac{\Lambda}{3} \Longrightarrow H^2 = \frac{8\pi}{3M_P^2}\rho - \frac{k}{R^2} + \frac{\Lambda}{3} \tag{4.18}$$

$$2\frac{\ddot{R}}{R} + \frac{\dot{R}^2}{R^2} + \frac{k}{R^2} = -\frac{8\pi}{3M_P^2} P + \Lambda \Longrightarrow \frac{\ddot{R}}{R} = -\frac{4\pi}{3M_P^2}(\rho + 3P) + \frac{\Lambda}{3} \tag{4.19}$$

Παραγωγίζοντας την Εξ. (4.18) και αντικαθιστώντας στην προκύπτουσα την Εξ. (4.19) επιτυγχάνεται μια εξίσωση που εκφράζει τον πρώτο Θερμοδυναμικό Νόμο για την περίπτωση του κοσμικού ρευστού

$$\dot{\rho} + 3H(\rho + P) + \frac{\Lambda}{3H} = 0 \ . \tag{4.20}$$

ή αλλάζοντας τη μεταβλητή παραγώγησης,

$$\frac{d(\rho R^3)}{dR} = -3pR^2. \tag{4.21}$$

Προφανώς οι Εξ. (4.18), (4.19) και (4.20) δεν είναι ανεξάρτητες μεταξύ τους. Συνήθως ως ανεξάρτητες λαμβάνονται οι (4.18) και (4.20) για να αποφευχθούν οι δεύτερες παράγωγοι. Στον Friedmann αποδίδονται οι Εξ. (4.18) και (4.19) ενώ στον Lemaître η Εξ. (4.20).

### 4.3.2 Παράμετρος Κοσμολογικής Πυκνότητας

Θέτωντας $\Lambda = 0$, ο όρος καμπυλότητας $k/R^2$ στην Εξ. (4.18) είναι θετικός, μηδέν ή αρνητικός, ανάλογα αν η $\rho$ είναι μεγαλύτερη, ίση ή μικρότερη από την κρίσιμη πυκνότητα, $\rho_c$ που ορίζεται ως εξής:

$$\rho_c = \frac{3M_P^2}{8\pi} H^2 . \tag{4.22}$$

Είναι συνεπώς, σκόπιμος ο ορισμός της Παραμέτρου Κοσμολογικής Πυκνότητας, μέσω των σχέσεων:

$$\Omega = \Omega_\rho + \Omega_\Lambda \quad \text{όπου} \quad \Omega_\rho = \frac{\rho}{\rho_c} \quad \text{και} \quad \Omega_\Lambda = \frac{\Lambda}{3H^2}. \tag{4.23}$$

Αντικαθιστώντας τους ορισμούς αυτούς στην Εξ. (4.18) λαμβάνεται η παρακάτω:

$$\Omega_\rho + \Omega_\Lambda - \frac{k}{H^2R^2} = 1 \Longrightarrow \Omega - 1 = \frac{k}{H^2R^2} \tag{4.24}$$

Από την προηγούμενη, συμπεραίνεται ότι $\Omega <[>] 1$ αντιστοιχεί σε ένα ανοικτό [κλειστό] σύμπαν με $k = -1(+1)$ ενώ η τιμή $\Omega = 1$ αντιστοιχεί σε επίπεδο με $k = 0$.

### 4.3.3 Παράμετρος Επιβράδυνσης

Μια εξίσου χρήσιμη παράμετρος που εκφράζει το ρυθμό διαστολής του Σύμπαντος, είναι η Παράμετρος Επιβράδυνσης που ορίζεται ως εξής:

$$q = -\frac{\ddot{R}R}{\dot{R}^2} = \frac{\ddot{R}}{\dot{R}H} \tag{4.25}$$

Από τον ορισμό αυτό, συμπεραίνεται ότι $q >[<] 0$ αντιστοιχεί σε επιβραδυνόμενη [επιταχυνόμενη] διαστολή. Αντικαθιστώντας στην τελευταία την Εξ. (4.19) και χρησιμοποιώντας τις Εξ. (4.22) και (4.23), λαμβάνεται:

$$q = \frac{1}{2}\frac{\rho + 3P}{\rho_c} - \Omega_\Lambda . \tag{4.26}$$

Όπως θα φανεί στη συνέχεια, η διαστολή του Σύμπαντος είναι επιβραδυνόμενη για το SBB, αλλά επιταχυνόμενη για το πληθωριστικό Σύμπαν.

### 4.3.4 Εξισώσεις Κατάστασης

Στις Εξ. (4.18) και (4.20) υπεισέρχονται 3 άγνωστοι, οι $R$, $\rho$, $P$. Επομένως, για τη λύση τους χρειάζεται και μία ακόμα ανεξάρτητη εξίσωση. Αυτή είναι η Καταστατική Εξισώση, $P = P(\rho)$ του κοσμικού ρευστού που παίρνει την συμπαγή μορφή:

$$P = (\omega - 1)\rho, \quad \text{όπου} \quad \omega = \begin{cases} 4/3, & \text{για Επικράτηση Ακτινοβολίας (RD)} \\ 1, & \text{για Επικράτηση Ύλης (MD)} \end{cases} \tag{4.27}$$

Αξιοποιώντας τη νέα αυτή πληροφορία, είναι δυνατόν να εξαχθούν οφέλιμες σχέσεις για τις εμπλεκόμενες μεταβλητές χωρίς την απευθείας λύση των διαφορικών εξισώσεων. Αναλυτικότερα, θέτωντας $\Lambda = 0$ και αντικαθιστώντας την Εξ. (4.27) στην:

- Εξ. (4.20) και ολοκληρώνοντας την προκύπτουσα, λαμβάνεται:

$$\rho R^{3\omega} = \text{cst.} \tag{4.28}$$

**Πίνακας 4.1: Προσεγγιστική Δυναμική του Σύμπαντος**

| Παράμετρος | Είδος Επικράτησης | | |
|---|---|---|---|
| | **RD** | **MD** | **Κενού** |
| $P$ | $\rho/3$ | $0$ | $-\rho$ |
| $\rho$ | $\sim R^{-4}$ | $\sim R^{-3}$ | cst |
| $R$ | $\sim t^{1/2}$ | $\sim t^{2/3}$ | $e^{Ht}$ |
| $T$ | $\sim t^{-1/2}$ | $\sim t^{-2/3}$ | $e^{-Ht}$ |
| $H$ | $1/2t$ | $2/3t$ | cst |
| $d_H$ | $2t$ | $3t$ | $\sim e^{Ht}/H$ |
| $\Omega - 1$ | $t$ | $t^{2/3}$ | $\sim e^{-2Ht}$ |
| $q$ | $\Omega_\rho - \Omega_\Lambda$ | $\Omega_\rho/2 - \Omega_\Lambda$ | $-\Omega_\rho - \Omega_\Lambda$ |

- Εξ. (4.18) και ολοκληρώνοντας την προκύπτουσα, για $k = 0$, λαμβάνεται:

$$Rt^{-2/3\omega} = \text{cst} \Longrightarrow \begin{cases} H(t) & = 2/3\omega t \\ d_H(t) & = 3\omega/(3\omega - 2)t \\ (\Omega(t) - 1) & \sim t^{2-4/3\omega}, \quad (k \neq 0) \end{cases} \tag{4.29}$$

αξιοποιώντας τους ορισμούς των Εξ. (4.4), (4.7) και (4.24).

- Εξ. (4.26) και χρησιμοποιώντας τους ορισμούς της Εξ. (4.24), λαμβάνεται:

$$q = \frac{3\omega - 2}{2}\Omega_\rho - \Omega_\Lambda. \tag{4.30}$$

Ειδικότερα, τα αποτελέσματα εφαρμογής των πιο πάνω σχέσεων στις δύο περιπτώσεις επικράτησης ύλης και ακτινοβολίας καταγράφονται συγκεντρωτικά στον Πίνακα 4.1. Η τρίτη σειρά και η τρίτη στήλη του πίνακα αυτού, περιλαμβάνουν ευρήματα των Εδ. 4.5.1 και 4.6.1, που προτάσσονται χάριν συγκρίσεως και οικονομίας.

## 4.4 Θερμοδυναμική Κοσμολογία

Το κοσμικό ρευστό μπορεί να θεωρηθεί ως σύνολο σωματίων σε κατάσταση θερμοδυναμικής ισοροπίας (: ΘΔΙ) η οποία εγκαθίσταται μέσω αλληλεπιδράσεων καταστροφής και δημιουργίας. Η προσέγγιση του ιδανικού ρευστού είναι επιτυχής εκτός και αν πρόκειται για μετάπτωση φάσεως. Η εμπλοκή της θερμοδυναμικής στην εξέλιξη του σύμπαντος θεμελιώνεται στο Εδ. 4.4.1 και οι θερμοδυναμικές μεταβλήτες που χαρακτηρίζουν αυτό το ρευστό υπολογίζονται στο Εδ. 4.4.3, αφού πρώτα εισαχθούν οι θεμελιώδεις έννοιες από τη Σχετικιστική Στατιστική στο Εδ. 4.4.2.

### 4.4.1 Ισεντροπική Διαστολή

Σε μια περιοχή του συν-κινούμενου συστήματος αναφοράς με όγκο $V = R^3$ το δεύτερο Θερμοδυναμικό αξίωμα έχει τη γνωστή του μορφή:

$$TdS = d(\rho V) + PdV \tag{4.31}$$

Η Εξ. (4.21) αναμορφώνεται ώστε να συγκριθεί με την Εξ. (4.31):

$$\frac{d}{dR}(\rho V) = -P\frac{dV}{dR} \Longrightarrow d(\rho V) + PdV = 0. \tag{4.32}$$

Από τις Εξ. (4.31) και (4.32) προκύπτει αβίαστα ότι $dS = 0$, δηλαδή το σύμπαν διαστέλλεται ισεντροπικά. Επόμενο βήμα, λοιπόν θα είναι ο συσχετισμός της εντροπίας με τις άλλες μεταβλητές. Από την Εξ. (4.31), λαμβάνεται:

$$dS = \frac{1}{T}\left[Vd\rho + (P+\rho)dV\right]. \tag{4.33}$$

Η Εξ. (4.31) αξιοποιείται με δύο διαφορετικούς τρόπους:

- Θεωρώντας την εντροπία ως συνάρτηση της μορφής $S(V,T)$, η Εξ. (4.33) συνεπάγεται:

$$\begin{aligned} \frac{\partial S}{\partial V} = \frac{\rho+P}{T} \quad \text{και} \quad \frac{\partial S}{\partial T} &= \frac{V}{T}\frac{d\rho}{dT} \Longrightarrow \\ \frac{d}{dT}\left(\frac{\rho+P}{T}\right) &= \frac{1}{T}\frac{d\rho}{dT} \Longrightarrow \\ \frac{dP}{dT} &= \frac{(\rho+P)}{T}. \end{aligned} \tag{4.34}$$

- Προσπαθώντας να δημιουργηθεί τέλειο διαφορικό, λαμβάνεται

$$dS = \frac{1}{T}d\left[V(P+\rho)\right] - \frac{1}{T}VdP \tag{4.35}$$

Αντικαθιστώντας την Εξ. (4.34) στο τελευταίο όρο της Εξ. (4.35) επιτυγχάνεται:

$$dS = d\left[\frac{(P+\rho)V}{T}\right] \Longrightarrow S = \left[\frac{(P+\rho)V}{T} + \text{cst}\right] \tag{4.36}$$

Συνεπώς, η εντροπία του σύμπαντος παραμένει σταθερή και δίνεται από την Εξ. (4.36) με απροσδιοριστία μιας αυθαίρετης σταθεράς. Πηγή παραβίασης της ισεντροπικότητας αυτής, είναι οι μεταπτώσεις φάσεως (phase transition) κατα τη διάρκεια, των οποίων μπορεί να ελευθερωθεί θερμότητα αυξάνοντας την εντροπία. Σε τέτοιες περιπτώσεις η προσέγγιση του ιδανικού ρευστού δεν είναι πλέον εφαρμόσιμη και επιβάλλεται η χρήση της Θεωρίας Πεδίου καθορισμένης θερμοκρασίας, πράγμα που είναι πέρα από τους σκοπούς της εργασίας αυτής.

### 4.4.2 Έννοιες Σχετικιστικής Στατιστικής

Για ιδανικό ρεύστο, η συνάρτηση κατανομής, $D(p,T)$ στο χώρο των φάσεων για σωμάτιο μάζας $m$ με $g$ βαθμούς ελευθερίας σε ΘΔΙ θερμοκρασίας $T$, ορίζεται ως εξής:

$$\frac{1}{(2\pi)}gD(p,T) = \frac{d\mathcal{N}}{d^3xd^3p} \tag{4.37}$$

και δίνεται, κατά την Αν. [39], από την παρακάτω αναλλοίωτη κατά Lorentz έκφραση:

$$D(p,T) = \left[e^{(p^\mu u_\mu/T)} + \text{sf}\right]^{-1}, \tag{4.38}$$

όπου $E(|\vec{p}|) := \sqrt{m^2 + \vec{p}^{\,2}}$, η ενέργεια του ενλόγω σωματίου και sf ένας φάκτορας που επιλέγεται ανάλογα με τη στατιστική φύση του σωματίου

$$\text{sf} = \begin{cases} -1, & \text{Στατιστική Bose – Einsein (: BES)} \\ +1, & \text{Στατιστική Fermi – Dirac (: FDS)} \\ 0, & \text{Στατιστική Maxwell – Boltsmann (: MBS)} \end{cases} \tag{4.39}$$

Οι δύο πρώτες επιλογές ισχύουν για σχετικιστικά μποζόνια ή φερμιόνια ενώ η τρίτη εφαρμόζεται για μη σχετικιστικά μποζόνια και φερμιόνια. Υπενθυμίζεται οτι το χημικό δυναμικό μπορεί ασφαλώς να αγνοηθεί για τις ανάγκες της διατριβής. Οι ποσότητες που πρέπει να υπολογισθούν με τη χρήση της παραπάνω κατανομής είναι οι θερμοδυναμικές μεταβλητές $n$, $\rho$, $P$. Αυτές σε ένα τυχαίο σύστημα αναφοράς μπορούν να οριστούν με σχετικιστικά συναλλοίωτη μορφή ως εξής:

- Το $n$ αποτελεί μέλος του τετραδιανύσματος $N^\mu = (n, \vec{j}\,)$, όπου $\vec{j}$ είναι η ροή των σωματίων

$$N^\mu = g\int \frac{d^3p}{(2\pi)^3}\frac{p^\mu}{E(p)}\,D(p,T)\,. \tag{4.40}$$

- Τα $\rho$, $P$ προκύπτουν από τον τανυστή Ενέργειας-Ορμής του ρευστού που ορίζεται από τη σχέση:

$$T^{\mu\nu} = g\int \frac{d^3p}{(2\pi)^3}\frac{p^\mu p^\nu}{E(p)}\,D(p,T)\,. \tag{4.41}$$

Αποδεικνύεται ότι το όρισμα ολοκλήρωσης $d^3p/E(p)$ είναι κατά Lorentz αναλλοίωτο, επομένως οι προηγούμενες ποσότητες είναι ορισμένες κατά Lorentz συναλλοίωτα. Στο CRF, όπου $u^\mu = (1, \vec{0})$, οι παραπάνω γενικοί ορισμοί εξειδικεύονται και, αφού εκτελεστεί η σχετική σφαιρική ολοκλήρωση,

$$\frac{1}{(2\pi)^3}d^3p = \frac{1}{(2\pi)^3}(4\pi)|\vec{p}\,|^2 d|\vec{p}\,|, \tag{4.42}$$

γράφονται ως εξής :

$$n(T) = N^0 = \frac{g}{2\pi^2}\int_0^\infty D(|\vec{p}\,|,T)|\vec{p}\,|^2 d|\vec{p}\,| \tag{4.43}$$

$$\rho(T) = T^{00} = \frac{g}{2\pi^2}\int_0^\infty D(|\vec{p}\,|,T)E(|\vec{p}\,|)|\vec{p}\,|^2 d|\vec{p}\,| \tag{4.44}$$

$$P(T) = \frac{1}{3}T^{ii} = \frac{g}{2\pi^2}\int_0^\infty D(|\vec{p}\,|,T)|\vec{p}\,|^4 d|\vec{p}\,|. \tag{4.45}$$

Ο υπολογισμός των ολοκληρωμάτων αυτών γίνεται εύκολα με αξιοποίηση των προσεγγίσεων που γίνονται σε κάθεμια από τις παρακάτω περιπτώσεις:

**α.** Σχετικιστικό όριο ($T \gg m$). Σε αυτή την περίπτωση, η μάζα του θεώρουμενου σωματίου μπορεί να αμεληθεί, $E \simeq |\vec{p}\,|$ οπότε η συνάρτηση κατανομής, γράφεται:

$$D(|\vec{p}\,|,T) = \left[e^{|\vec{p}\,|/T} \pm 1\right]^{-1} \tag{4.46}$$

και εφαρμόζοντας τους παρακάτω ολοκληρωτικούς τύπους:

$$\int_0^\infty dt\frac{t^z-1}{e^t-1} = (z-1)!\zeta(z), \tag{4.47}$$

$$\int_0^\infty dt\frac{t^z-1}{e^t+1} = (1-2^{1-z})(z-1)!\zeta(z), \tag{4.48}$$

$$\text{με}\ \ \zeta(2) = \frac{\pi^2}{6},\ \ \zeta(3) = 1.202,\ \ \zeta(4) = \frac{\pi^4}{90} \tag{4.49}$$

(όπου $\zeta$ η Zeta συνάρτηση Riemann) λαμβάνεται αντίστοιχα για τα:

- Μποζόνια

$$n(T) = g\frac{\zeta(3)}{\pi^2}T^3, \tag{4.50}$$

$$\rho(T) = g\frac{\pi^2}{30}T^4, \tag{4.51}$$

$$P(T) = \frac{1}{3}\rho(T). \tag{4.52}$$

- Φερμιόνια

$$n(T) = \frac{3}{4} g \frac{\zeta(3)}{\pi^2} T^3, \tag{4.53}$$

$$\rho(T) = \frac{7}{8} g \frac{\pi^2}{30} T^4, \tag{4.54}$$

$$P(T) = \frac{1}{3} \rho(T). \tag{4.55}$$

Η πυκνότητα εντροπίας και στις δύο προηγούμενες περιπτώσεις υπολογίζεται από την Εξ. (4.36) που με αντικατάσταση των Εξ. (4.52), (4.55) παίρνει τη μορφή:

$$s(T) = \frac{4}{3} \frac{\rho}{T}, \tag{4.56}$$

**β.** Μη σχετικιστικό όριο ($T \ll m$). Σε αυτή την περίπτωση, ο όρος μάζας γίνεται κυρίαρχος, $|\vec{p}|^2 \ll m^2$ οπότε:

$$E(|\vec{p}\,|) = (|\vec{p}\,|^2 + m^2)^{1/2} = \left[ m^2 \left( 1 + \frac{|\vec{p}\,|^2}{m^2} \right) \right]^{1/2} \simeq m + \frac{|\vec{p}\,|^2}{2m} \tag{4.57}$$

και η συνάρτηση κατανομής, γράφεται:

$$D(p,T) = e^{\left(-m/T - |\vec{p}\,|^2/2mT\right)}, \tag{4.58}$$

Ο τύπος ολοκληρωμάτων που αξιοποιείται είναι:

$$I_z(A) = \int_0^\infty dt\, t^z e^{-At^2} \quad \text{με} \quad A > 0 \quad \text{και επαγωγική σχέση:}$$

$$\frac{d}{dA} I_z(A) = -I_{z+2}(A) \quad \text{όπου} \quad I_0 = \frac{1}{2}\sqrt{\frac{\pi}{A}} \quad \text{και} \quad I_1 = \frac{1}{2A}$$

Τα αποτελέσματα καταγράφονται παρακάτω:

$$n(T) = g \left( \frac{mT}{2\pi} \right)^{3/2} e^{-m/T} \tag{4.59}$$

$$\rho(T) = nm + \frac{3}{2} nT \tag{4.60}$$

$$P(T) = nT \tag{4.61}$$

Στην Εξ. (4.61) με νοσταλγία αναγνωρίζουμε τη γυμνασιακή καταστατική εξίσωση των τελείων αερίων. Επίσης, στην Εξ. (4.60) ο τελευταίος όρος συνήθως αγνοείται γιατί προέρχεται από ολοκλήρωση παράγοντα τάξεως $|\vec{p}\,|^4$.

### 4.4.3 Θερμοδυναμικές Μεταβλητές σε RD

Αφού τα φωτόνια βρίσκονται πάντα σε ΘΔΙ στο κοσμικό ρευστό, είναι χρήσιμο η μεταβλητή $T$ να λογίζεται ως η θερμοκρασία των φωτονίων και οι θερμοδυναμικές μεταβλήτες κάθε άλλου συστατικού να υπολογίζονται αναφορικά με τις αντίστοιχες του φωτονικού υποβάθρου. Εφαρμόζοντας τις Εξ. (4.50)-(4.51) για φωτονικό ρευστό με $g_\gamma = 2$, λαμβάνεται:

$$n_\gamma = 2 \frac{\zeta(3)}{\pi^2} T^3 \tag{4.62}$$

$$\rho_\gamma = 3P_\gamma = 2 \frac{\pi^2}{30} T^4. \tag{4.63}$$

Συνακόλουθα, η πυκνότητα αριθμού σωματιδίων, ενέργειας και εντροπίας ενός ρευστού που αποτελείται από ακρέως σχετικιστικά σωμάτια (μποζόνια και φερμιόνια) θερμοκρασίας $T_i$ το καθένα, είναι:

$$n_{RD} = g_{N*}\frac{\zeta(3)}{\pi^2}T^3, \tag{4.64}$$

$$\rho_{RD} = 3P_{RD} = g_*\frac{\pi^2}{30}T^4, \tag{4.65}$$

$$s_{RD} = g_{S*}\frac{2\pi^2}{45}T^3. \tag{4.66}$$

όπου οι ενεργοί αριθμοί βαθμών ελευθερίας ορίζονται αντίστοιχα:

$$g_{N*} = \sum_{i=bosons} g_i\left(\frac{T_i}{T}\right)^3 + \frac{3}{4}\sum_{i=fermions} g_i\left(\frac{T_i}{T}\right)^3, \tag{4.67}$$

$$g_* = \sum_{i=bosons} g_i\left(\frac{T_i}{T}\right)^4 + \frac{7}{8}\sum_{i=fermions} g_i\left(\frac{T_i}{T}\right)^4, \tag{4.68}$$

$$g_{S*} = \sum_{i=bosons} g_i\left(\frac{T_i}{T}\right)^3 + \frac{7}{8}\sum_{i=fermions} g_i\left(\frac{T_i}{T}\right)^3. \tag{4.69}$$

Προφανώς οι Εξ. (4.64), (4.65) ισχύουν σε εποχή RD. Σε εποχή μεγάλης θερμοκρασίας (νωρίτερα από 1 sec) πρακτικά $T_i = T$, οπότε οι ενεργοί αριθμοί βαθμών ελευθερίας απλοποιούνται ως εξής:

$$g_* = g_{S*} = \sum_{i=bosons} g_i + \frac{7}{8}\sum_{i=fermions} g_i \quad \text{και} \quad g_{N*} = \sum_{i=bosons} g_i + \frac{3}{4}\sum_{i=fermions} g_i \tag{4.70}$$

Αυτή η έκφραση χρησιμοποιείται στον Πίνακα 4.2 για τον υπολογισμό του $g_*(T)$ στην περίπτωση των σωματίων του SM και MSSM. Στον πίνακα αυτό τα sfermions των δύο πρώτων γενεών παριστάνονται με το ίδιο σύμβολο. Επίσης, οι βαθμοί ελευθερίας ενός:

- Πραγματικού βαθμωτού πεδίου είναι 1.
- Μιγαδικού βαθμωτού πεδίου είναι 2, όσες και οι πραγματικές συνιστώσες του.
- Λεπτονίου είναι 2, όσες και οι ΙΔΚ ορισμένου spin.
- Quark είναι 6, 2 για κάθε ΙΔΚ καθορισμένου χρώματος.
- Άμαζου διανυσματικού μποζονίου είναι 2, όσες και οι ΙΔΚ ορισμένης ελικότητας.
- Έμμαζου διανυσματικού μποζονίου είναι 3, όσες και οι ΙΔΚ ορισμένης ελικότητας.

Εξοπλισμένοι με τις παραπάνω απαραίτητες έννοιες, είναι δυνατόν να θεμελιωθεί εναργέστερα η θεωρία εξέλιξης του σύμπαντος στην επόμενη παράγραφο.

## 4.5 Η Θεωρία της Μεγάλης Έκρηξης

Η Θεωρία της Μεγάλης Έκρηξης (Big-Bag) είναι, κατα γενική ομολογία, μια επιτυχημένη προσπάθεια ερμηνείας του Θαύματος της Δημιουργίας. Οι προβλέψεις της θεωρίας για τα αρχικά στάδια του Σύμπαντος αναφέρονται στο Εδ. 4.5.1 Ο βασικός μηχανισμός της αποδέσμευσης περιγράφεται στο Εδ. 4.5.2 και στο Εδ. 4.5.3 αποδεικνύεται μια χρήσιμη σχέση σύνδεσης της θερμοκρασίας με το χρόνο. Εφαρμογή των εννοίων ακολουθεί στο Εδ. 4.5.4. Τέλος στο Εδ. 4.5.5 γίνεται μια σύντομη καταγραφή της ιστορίας του Σύμπαντος από φυσικής απόψεως και στο Εδ. 4.5.6 καταγράφονται οι τιμές διάφορων μετρήσιμων μεγεθών.

**Πίνακας 4.2: Ενεργός Αριθμός Βαθμών Ελευθερίας**

| $T <$ | **Σωμάτια σε ΘΔΙ** | $g_*(T)$ |
|---|---|---|
| $m_e$ | $\gamma$ | 2 |
| $T_D(\nu)$ | $+\,e^{\pm}$ | $+2\times2\times7/8$ |
| $m_\mu$ | $+\nu_e,\,\nu_\mu,\,\nu_\tau$ | $+3\times2\times7/8$ |
| $m_s$ | $+\mu^{\pm}$ | $+2\times2\times7/8$ |
| | $+u,\,\bar{u},\,d,\,\bar{d}$ | $+4\times3\times2\times7/8$ |
| | $+g$ | $+8\times2$ |
| $m_c$ | $+s,\,\bar{s}$ | $+3\times2\times2\times7/8$ |
| $m_\tau$ | $+c,\,\bar{c}$ | $+3\times2\times2\times7/8$ |
| $m_b$ | $+\tau,\,\bar{\tau}$ | $+2\times2\times7/8$ |
| $M_W$ | $+b,\,\bar{b}$ | $+3\times2\times2\times7/8$ |
| $m_t$ | $+W^{\pm},\,Z$ | $+3\times3$ |
| $M_S$ | $+t,\,\bar{t}$ | $+3\times2\times2\times7/8$ |
| | $+h$ | $+1\ (=106.75)$ |
| $M_G$ | $+\tilde{\nu}_e,\,\tilde{e}_L,\,\tilde{e}_R^*$ | $+6\times2$ |
| | $+\tilde{\nu}_\tau,\,\tilde{\tau}_1,\,\tilde{\tau}_2$ | $+3\times2$ |
| | $+\tilde{u}_L,\,\tilde{u}_R^*,\,\tilde{d}_L,\,\tilde{d}_R^*$ | $+8\times3\times2$ |
| | $+\tilde{t}_1,\,\tilde{t}_2,\,\tilde{b}_1,\,\tilde{b}_2$ | $+4\times3\times2$ |
| | $+\tilde{\chi},\,\tilde{\chi}_2^0,\,\tilde{\chi}_3^0,\,\tilde{\chi}_4^0$ | $+4\times2$ |
| | $+\tilde{\chi}_1^{\pm},\,\tilde{\chi}_2^{\pm}$ | $+4\times2$ |
| | $+g^A$ | $+8\times2$ |
| | $+H,\,H^{\pm},\,A$ | $+4\ (=228.75)$ |

### 4.5.1 Αρχικά στάδια

Στα αρχικά στάδια εξέλιξης του, το σύμπαν ήταν RD. Το συμπέρασμα αυτό μπορεί να αιτιολογηθεί αν υπολογισθεί ο λόγος των πυκνοτήτων ενέργειας $\rho_{MD}$, $\rho_{RD}$ σε MD και RD αντίστοιχα και ληφθεί το όριο $R\to0$ που αντιστοιχεί στην αρχή του σύμπαντος. Από τις Εξ. (4.28) με $\omega=4/3$ και $\omega=1$ διαιρώντας κατά μέλη, λαμβάνεται:

$$\frac{\rho_{MD}}{\rho_{RD}}\sim R\to0\ \ \text{όταν}\ \ R\to0. \tag{4.71}$$

Επομένως, το πρώιμο Σύμπαν κυριαρχείται από ακρέως σχετικιστική ύλη. Λόγω της ισεντροπικότητας ισχύει: $s_{RD}R^3=\text{cst}$. Σε αυτή τη σχέση, αντικαθιστώντας την Εξ. (4.56) και λύνοντας ως προς τη θερμοκρασία, λαμβάνεται:

$$T\sim g_{S*}^{-1/3}R^{-1}. \tag{4.72}$$

Θεωρώντας το $g_{S*}$ σταθερό, συμπεραίνεται οτι η θερμοκρασία, και κατά συνέπεια και η ενέργεια, του σύμπαντος για $R\to0$ ήταν πολύ μεγάλη. Αντιθέτως, καθώς το σύμπαν διαστέλλεται η θερμοκρασία ελαττώνεται προοδευτικά, σε μια εξάρτηση από το χρόνο που καταγράφεται στην τρίτη σειρά του συγκεντρωτικού πίνακα 4.1.

Τα προηγούμενα συμπεράσματα επιτρέπουν την υπόθεση ότι η εξέλιξη του σύμπαντος ξεκίνησε από μια μεγάλη έκρηξη. Αμέσως μετά, το σύμπαν βρισκόταν σε μια κατάσταση υψηλής θερμοκρασίας και πυκνότητας όπου ακτινοβολία και ύλη ήταν συνδεδεμένες. Επομένως, οι GUT αποκτούν λόγο ύπαρξης και πεδίο ελέγχου, αφού καλούνται να εξηγήσουν την κατάσταση του σύμπαντος στα πρώτα του στάδια και να προβλέψουν ή μάλλον να ερμηνεύσουν σημερινές καταστάσεις.

### 4.5.2 Μηχανισμός Αποσύνδεσης

Τα σωμάτια που συνιστούν το κοσμικό ρευστό παραμένουν στη ΘΔΙ μέσω αλληλεπιδράσεων σκέδασης, αλληλοκαταστροφής και δημιουργίας. Καθώς το Σύμπαν ψύχεται και διαστέλεται, οι ακτίνες αλληλεπίδρασης, Γ εξασθενούν συγκρινόμενες με ακτίνα διαστολής του Σύμπαντος $H$, οπότε έρχεται στιγμή, που εξαρτάται από τις μάζες των σωματίων και από τις σταθερές ζεύξης των αλληλεπιδράσεων, κατά την οποία τα σωμάτια αποσυνδέονται από το κοσμικό ρευστό και δίνουν μια εναπομένουσα πυκνότητα.

Όπως διαφάνηκε από την εισαγωγή αυτής της ενότητας, κεντρικό ρόλο στο μηχανισμό αποσύνδεσης, παίζει η έννοια της ακτίνας αλληλεπίδρασης που ορίζεται κατά τρόπο σχετικιστικά αναλλοίωτο, βάση της Αν. [38], ως εξής:

$$\Gamma(T) := n^{\rm eq}(T)\, \langle \sigma v_{\rm rel} \rangle\, (T)\,, \tag{4.73}$$

όπου $n^{\rm eq}(T)$ η πυκνότητα αριθμού των σωματίων στην ΘΔΙ, $\sigma$ η ενεργός διατομή της αλληλεπίδρασης, $v_{\rm rel}$ η σχετική ταχύτητα των αλληλεπιδρώντων σωματίων και $\langle ... \rangle$ δηλώνει τη θερμική μέση τιμή της περιεχόμενης ποσότητας. Ο υπολογισμός της ποσότητας αυτής, γίνεται, συνήθως, με διαστατικά επιχειρήματα και βεβαίως πολύ προσεγγιστικά. Στο κεφάλαιο 5, της διατριβής αυτής, ο ακριβής υπολογισμός της ποσότητας αυτής, στην περίπτωση του LSP θα είναι ένας από τους βασικούς στόχους μας. Επομένως, μετατίθενται σε αυτό το σημείο και οι λεπτομερειακοί ορισμοί των μεγεθών που υπεισέρχονται στην Εξ. (4.73).

Ένα σωμάτιο $i$, μάζας $m_i$ αποσυνδεέται από το κοσμικό λουτρό σε θερμοκρασία $T_D$, όταν ισχύει η συνθήκη αποσύνδεσης ή αποδέσμευσης:

$$\Gamma(T_D) \simeq H(T_D) \tag{4.74}$$

Όταν $\Gamma(T_D) \gg H(T_D)$, ο χρόνος αλληλεπίδρασης των σωματίων είναι μικρότερος από την ηλικία του σύμπαντος οπότε η αλληλεπίδραση προλαβαίνει να επισυμβεί, ενώ όταν $\Gamma(T_D) \ll H(T_D)$, δεν προλαβαίνει.

Ανάλογα με την στατιστική του συμπεριφορά τη στιγμή της αποσύνδεσης, το λείψανο ή πιο ήπια το υπόλειμμα ή απομεινάρι του σωματίου χαρακτηρίζεται:

- Θερμό, όταν $m_i \ll T_D$. Από τα σωμάτια του SM, αυτή η συνθήκη ισχύει μόνο για τα νετρίνο τα οποία είναι και άμαζα και ασθενώς αλληλεπιδρώντα.

- Ψυχρό, όταν $m_i \gg T_D$. Από τα σωμάτια του SM, όλα τα υπόλοιπα είναι και έμμαζα και ισχυρά ή (και) ηλεκτρομαγνητικά αλληλεπιδρώντα και επόμενως, όταν $m_i < T_D$ η πυκνότητα αριθμού τους, φθίνει ταχέως με απότελεσμα την ολοκληρωτική τους καταστροφή μέχρι την εποχή της αποσύνδεσης που είναι μεταγενέστερη. Αντιθέτως, από τα σωματία του MSSM, υπάρχει ασθενώς αλληλεπιδρών σωμάτιο, το LSP που επιζεί φθίνοντας, βεβαίως, μέχρι την εποχή της αποσύνδεσης, και στη συνέχεια η πυκνότητα αριθμού του εξελίσσεται μέχρι την εποχή μας. Σε αυτή την τελευταία εκδοχή είναι αφιερωμένο το επόμενο κεφάλαιο της εργασίας.

Η διατήρηση της εντροπίας είναι ο άλλος νόμος που ελέγχει την διαδικασία αποσύνδεσης και μας επιτρέπει να εξάγουμε συμπεράσματα για την πορεία του σύμπαντος πριν και μετά την αποσύνδεση κάθε σωματιού.

### 4.5.3 Συσχετισμός Χρόνου Θερμοκρασίας

Στην καταγραφή της αλληλουχίας των κοσμικών γεγονότων προφανώς πολύ χρήσιμη θα ήταν η ύπαρξη μιας εξίσωσης συσχετισμού των δύο βασικών παραμέτρων της θεωρίας, του χρόνου και της Θερμοκρασίας. Αυτός είναι ο σκοπός της παραγράφου αυτής. Σημείο επαφής των δύο μεγεθών είναι η πυκνότητα ενέργειας, για την οποία ισχυεί η Εξ. (4.65) σε RD και μπορεί να αποδειχθεί ότι:

$$\rho_{RD} = \frac{3M_P^2}{32\pi}\, t^{-2} \tag{4.75}$$

Η Εξ. (4.75) προκύπτει εύκολα διαφορίζοντας την Εξ. (4.28) με $\omega = 4/3$ και ολοκληρώνοντας την προκύπτουσα

$$\frac{\dot\rho_{RD}}{\rho_{RD}} = -4H \quad \text{όπου } H = \sqrt{\frac{8\pi\,\rho_{RD}}{3\,M_P^2}} \tag{4.76}$$

οπότε έπεται η αποδεικτέα Εξ. (4.75). Εξισώνοντας τις Εξ. (4.65) και (4.75) και λύνοντας ως προς $t$, επιτυγχάνεται:

$$t = \sqrt{\frac{45}{16\pi^3}} M_P \; g_*(T)^{-1/2} \, T^{-2} = 2.41846 \; g_*(T)^{-1/2} \left(\frac{T}{\mathrm{MeV}}\right)^{-2} \mathrm{sec.} \tag{4.77}$$

όπου αξιοποίηθηκε ο συσχετισμός των μονάδων του Εδ. 4.1. Η χρηστικότητα της Εξ. (4.77) είναι προφανής στην περιγραφή της εξέλιξης του σύμπαντος. Ο ενεργός αριθμός βαθμών ελευθερίας $g_*(T)$ που πρέπει να τεθεί στην Εξ. (4.77) εξαρτάται από τη σύσταση του κοσμικού ρευστού σε δεδομένη θερμοκρασία $T$. Στον Πίνακα 4.2 δίνονται οι τιμές του $g_*(T)$ για διάφορες θερμοκρασίες $T$ οι οποίες συγκρίνονται με τις μάζες των σωματίων του SM και του MSSM. Σε ενέργειες της τάξης του EWS $g_*(T) = 106.75$ ενώ της τάξης GUT είναι $g_*(T) = 228.75$.

### 4.5.4 Αποσύνδεση νετρίνων

Εφαρμογή των προηγούμενων εννοιών μπορεί να γίνει στην περίπτωση αποσύνδεσης των νετρίνο. Εξάλλου, είναι χρήσιμος αυτός ο υπολογισμός, γιατί μας εξασφαλίζει μια πίο ακριβή εκτίμηση των ενεργών βαθμών ελευθερίας της σημερινής περιόδου.

Δεδομένου ότι οι αλληλεπιδράσεις των νετρίνο είναι ασθενείς η θερμική μέση τιμή της ενεργού διατομής είναι: $\langle\sigma v_{\rm rel}\rangle \sim G_F^2 T^2$, οπότε η ακτίνα αλληλεπίδρασης είναι $\Gamma = n\langle\sigma v_{\rm rel}\rangle \sim G_F^2 T^5$ (αφού $n \approx T^3$). Η ακτίνα διαστολής, λόγω της Εξ. (4.76) έχει τη μορφή $H \sim T^2/M_P$. Εφαρμόζοντας την Εξ. (4.74), η θερμοκρασία αποσύνδεσης των νετρίνο, εκτιμάται ότι είναι:

$$T_\nu \sim (G_F^2 M_P)^{-1/3} \sim 1\,\mathrm{MeV}. \tag{4.78}$$

αφού $G_F \sim 10^{-5}\,\mathrm{GeV}^{-2}$. Μέσω της Εξ. (4.77), υπολογίζεται οτι το γεγονός αυτό συμβαίνει σε χρόνο $t = 1.03$ sec, με $g_*(T > m_e) = 11/2$ και εντροπία $S(T \gtrsim m_e) \sim 11(RT_\nu)^3/2$, Εξ. (4.56). Όταν $T \lesssim m_e$, σε χρόνο $t = 6.57$ sec, με $g_*(T \lesssim m_e) = 2$ η εντροπία γίνεται $S(T \lesssim m_e) \sim 2(RT)^3$. Η διατήρηση της εντροπίας συνεπάγεται:

$$S(T \lesssim m_e) = S(T \gtrsim m_e) \Longrightarrow \left(\frac{T_\nu}{T}\right)^3 = \frac{4}{11}. \tag{4.79}$$

Εφαρμόζοντας τις Εξ. (4.69), (4.68) οι σημερινές τιμές ενεργών βαθμών ελευθερίας, είναι:

$$g_{S*0} = g_\gamma + \frac{7}{8} N_\nu \, g_\nu \left(\frac{T_\nu}{T}\right)^3 = \frac{43}{11}\,, \tag{4.80}$$

$$g_{*0} = g_\gamma + \frac{7}{8} N_\nu \, g_\nu \left(\frac{T_\nu}{T}\right)^4 = 3.36\,, \tag{4.81}$$

όπου θεωρήθηκαν 3 άμαζα νετρίνο ($N_\nu$=3) με $g_\nu = 2$.

### 4.5.5 Περίληψη της Ιστορίας του Σύμπαντος

Σε αδρές γραμμές η ιστορία του Σύμπαντος καταγράφεται στον Πίνακα 4.3. Παρακάτω δίνονται κάποιες σύντομες επεξηγήσεις για τα κομβικά σημεία της διαδρομής, προτάσοντας τη χρονική απόσταση από τη στιγμή της Μεγάλης Έκρηξης και την αντίστοιχη θερμοκρασία οι οποίες συνδέονται μέσω της Εξ. (4.77):

- $t_P \sim 10^{-44}$ sec με $T_P \sim 10^{19}$ GeV: Η εποχή Plank, στην οποία κυριαρχεί η Κβαντική Βαρύτητα.
- $t_G \sim 10^{-39}$ sec με $T_G \sim 10^{16}$ GeV: Η εποχή SUSY-GUT στην όποια εντάσονται η μετάπτωση φάσης που συνδέεται με το Πληθωρισμό, και η πραγματοποίηση της Βαρυογένεσης. Είναι το σημείο στο οποίο η σύγχρονη Φυσική Στοιχειωδών Σωματίων εφαρμόζει και ελέγχει τις θεωρίες της.
- $t_N \sim 1\,\mathrm{sec} - 3\,\mathrm{min}$ με $T_N \sim 1 - 0.1\,\mathrm{MeV}$: Λαμβάνει χώρα η κατασκευή των νουκλεονίων, και των ελαφρων στοιχείων στη συνέχεια.

**Πίνακας 4.3: Ιστορία του Σύμπαντος**

| **Εποχή** | $t$ (sec) | $T$ (GeV) | **Χαρακτηριστικά** |
|---|---|---|---|
| Υπερχορδών | $10^{-44}$ | $10^{19}$ | Κβαντική |
| (Plank) | | | Βαρύτητα |
| Υπερσυμμετρικής | $10^{-36}$ | $10^{15}$ | Ενοποιημένη |
| Ενοποίησης | | | Μετάπτωση Φάσεως |
| (SUSY-GUT) | $10^{-32}$ | $10^{10}$ | Πληθωρισμός |
| | | | Βαρυογένεση |
| Ηλεκτρασθενής | $10^{-10}$ | $10^{2}$ | Ηλεκτρασθενής |
| | | | Μετάπτωση Φάσεως |
| Αδρονική | $10^{-5}$ | $3\times10^{-1}$ | Σύνθεση Αδρονίων |
| | $4\times10^{-5}$ | $1.4\times10^{-1}$ | Καταστροφή $\pi^{\pm}$ |
| Λεπτονική | $8\times10^{-5}$ | $1\times10^{-1}$ | Καταστροφή $\mu^{\pm}$ |
| | $8\times10^{-3}$ | $1\times10^{-2}$ | Αποδέσμευση $\nu_\mu$ |
| | 0.7 | $1\times10^{-1}$ | Αποδέσμευση $\nu_e$ |
| | 6 | $.5\times10^{-1}$ | Καταστροφή $e^{\pm}$ |
| Φωτονική | 60 | $.1\times10^{-3}$ | Νουκλεοσύνθεση |
| | $6\times10^{12}$ | $.4\times10^{-9}$ | Αποσύνδεση |
| | | | Φωτονίων |
| | $2\times10^{17}$ | $10^{-12}$ | Παρών |

- $t_d \sim 10^{11}$ sec και $T_d \sim 10^{-9}$ GeV: Η ενεργειακή πυκνότητα της ακτινοβολίας γίνεται ίση με αυτήν της ύλης η οποία σταδιακά γίνεται κυρίαρχη. Στη συνέχεια, τα $e^-$ συνδέονται με τους πυρήνες δημιουργώντας τα πρώτα άτομα. Όταν η ενέργεια των φωτονίων του υποβάθρου, γίνει μικρότερη από την ενέργεια σύνδεσης των $e^-$ στο άτομο του $H$ (13.6 eV), γίνεται η αποσύνδεση των φωτονίων από την ύλη. Τα άτομα αποκτούν ευστάθεια αφού τα φωτόνια του υποβάθρου δεν έχουν την απαιτούμενη ενέργεια για να τα διεγείρουν. Επομένως, τα φωτόνια παύουν να αλληλεπιδρούν με την ύλη και συνεχίζουν ελεύθερα την πορεία τους διαμέσου των αιώνων συνιστώντας την CBR.
- $t_0 \sim 10^{17}$ sec με $T_0 \sim 10^{-12}$ GeV: Η σύγχρονη εποχή των παρατηρήσεων και των αναζητήσεων.

### 4.5.6 Τρέχουσες τιμές Κοσμολογικών Παραμέτρων

Τα μετρήσιμα παρατηρησιακά δεδομένα στα οποία στηρίζεται η σύγχρονη Κοσμολογία είναι δύο (ο δείκτης 0 σε ένα μέγεθος δηλώνει σημερινή τιμή):

- Η θερμοκρασία της CBR, που μετρίεται από τον Cosmic Background Explorer (: COBE):

$$T_0 = (2.726 \pm 0.01)\,^0\mathrm{K}, \tag{4.82}$$

με μια ανισοτροπία που επίσης, μετράται με μεγάλη ακρίβεια:

$$\frac{\delta T_0}{T_0} \simeq 6.6 \times 10^{-6}. \tag{4.83}$$

- Η παράμετρος του Hubble, που δίνεται με χρήση μιας νέας παραμέτρου αναγωγής $h$,

$$H_0 = 100\, h\,\text{km sec}^{-1}\,\text{Mpc}^{-1}, \quad \text{όπου} \quad h = 0.65 \pm 0.15. \tag{4.84}$$

Το μεγαλύτερο παρατηρησιακό πρόβλημα στον προσδιορισμό του $H_0$ είναι η αβεβαιότητα στον καθορισμό των κοσμολογικών αποστάσεων.

- Ο λόγος της αριθμητικής πυκνότητας των βαρυονίων προς αυτή των φωτονίων, που

$$1.9 \times 10^{-10} < \eta_0 < 5.8 \times 10^{-10} \quad \text{όπου} \quad \eta_0 = \frac{n_{B_0}}{n_{\gamma_0}} \tag{4.85}$$

Η μέτρηση αυτή, όπως φαίνεται, είναι εξαιρετικά ακριβής.

- Η παράμετρος κοσμολογικής πυκνότητας φράσεται κάτωθεν από υπολογισμούς σε σμήνοι γαλαξιών και άνωθεν από τη συμπεριφορά της γαλαξιακής αριθμητικής πυκνότητας ως εξής:

$$0.1 \leq \Omega_0 \leq 2. \tag{4.86}$$

Με τα δεδομένα αυτά μπορούν να υπολογισθούν:

- Η κρίσιμη πυκνότητα χρησιμοποιώντας τις Εξ. (4.22) και (4.84):

$$\rho_{c_0} = (2.99863 \times 10^{-12} h^{1/2}\,\text{GeV})^4 \simeq 1.0523 \times 10^{-5} h^2\,\text{GeVcm}^{-3}. \tag{4.87}$$

- Το άνω όριο της ηλικίας του Σύμπαντος:

$$t_0 < H_0^{-1} = 9.778 \times 10^9\ h^{-1}\,\text{yr}. \tag{4.88}$$

- Οι τρέχουσες τιμές για τις θερμοδυναμικές μεταβλητές που χαρακτηρίζουν το υπόβαθρο ακτινοβολίας:

$$\rho_{R_0} = g_{*0}\frac{\pi^2}{30}T_0^4 \simeq 4.54 \times 10^{-10}\,\text{GeV cm}^{-3}, \tag{4.89}$$

$$\Omega_{R_0} h^2 = 4.18 \times 10^{-5}, \tag{4.90}$$

$$s_0 = g_{S*0}\frac{2\pi^2}{45}T_0^3 \simeq 2970\,\text{cm}^{-3}. \tag{4.91}$$

Ειδικά για τα φωτόνια του υποβάθρου χαρακτηριστικό νούμερο είναι αυτό που προκύπτει με εφαρμογή της Εξ. (4.62), $n_{\gamma_0} \simeq 422\,\text{cm}^{-3}$.

- Η παράμετρος κοσμολογικής πυκνότητας των βαρυονίων, χρησιμοποιώντας τις Εξ. (4.23), (4.87), (4.60), (4.85) τη μάζα του πρωτονίου $m_B := .937\,\text{GeV}$ και το $n_{\gamma_0}$:

$$0.007 \leq \Omega_{B_0} h^2 \leq 0.021. \tag{4.92}$$

Στις ποσότητες που παρουσιάστηκαν σε αυτό το εδάφιο θα στηρηχθεί η αξιολόγηση του κοσμολογικού προτύπου στη συνέχεια.

## 4.6 Πληθωρισμός

Ο πληθωρισμός (inflation) είναι ένα στάδιο στην εξέλιξη του Σύμπαντος κατά το οποίο ο παράγοντας κλίμακας αυξάνεται εκθετικά. Η εμφύτευση μίας τέτοιας διαδικασίας στην πορεία εξέλιξης του Σύμπαντος κρίθηκε αναγκαία, γιατί το SBB παρουσίαζε κάποια βασικά μειονεκτήματα που ισχυρά απειλούσαν το οικοδόμημά του. Η υλοποιήση της ιδέας του πληθωρισμού εκτίθεται στο Εδ. 4.6.1 οι συνέπειες της στο Εδ. 4.6.2 και η διευθέτηση των προβλημάτων στο Εδ. 4.6.3. Επιλέγεται μια προσεγγιστική παρουσίαση του θέματος, γιατί αυτό δε συνδέεται άμεσα με το αντικείμενο, στο οποίο είναι αφιερωμένη η εργασία αυτή. Πιο εκτεταμένη ανάλυση γίνεται στις ειδικές Αν. [40], [41] και [42].

### 4.6.1 Σχεδιάζοντας τον Πληθωρισμό

Υποτίθεται η ύπαρξη ενός κοσμολογικού βαθμωτού πεδίου $\phi$ (του inflaton) το οποίο είναι μόνο του χρόνου συνάρτηση $\phi(t)$ και ελέγχεται από τις κλασικές εξισώσεις κίνησης. Το κτίσιμο ενός συγκεκριμένου πληθωριστικού προτύπου απαιτεί επίσης την υιοθέτηση ενός δυναμικού χωρίς αυτό να αλλοιώνει τα γένικα χαρακτηριστικά του πληθωρισμού που θα αναφερθούν παρακάτω. Συνήθως στα πλαίσια μιας GUT, ο πληθωρισμός εξελίσσεται κατά τη διάρκεια μιας μετάπτωσης φάσης. Το δυναμικό εξαρτάται από τη θερμοκρασία $T$ με κρίσιμο και εξελικτικό τρόπο. Δηλαδή υπάρχει θερμοκρασία κρίσιμη $T_c$, που συναρτάται άμεσα με το κενό της θεωρίας $\langle\phi\rangle$. Όταν $T > T_c$, το $\langle\phi\rangle = 0$ είναι το ολικό ελάχιστο του δυναμικού. Όταν $T = T_c$, το $\langle\phi\rangle = 0$ και $\pm M_G$ είναι ελάχιστα. Όταν $T < T_c$, το $\langle\phi\rangle = 0$ είναι ένα τοπικό ελάχιστο του δυναμικού ενώ το ολικό ελάχιστο είναι στο $\langle\phi\rangle = M_G$. Εάν το inflaton βρισκόταν κατά λάθος στο απατηλό ελάχιστο $\langle\phi\rangle = 0$ θα κινηθεί προς το πραγματικό $\langle\phi\rangle = M_G$ είτε με θερμικές διακυμάνσεις, είτε με το φαινόμενο της σήραγγας. Αν επιπλέον, το δυναμικό είναι αρκούντως ομαλό κατά τη διάρκεια της κίνησης του inflaton προς το πραγματικό κενό, ώστε η κινητική του ενέργεια να μπορεί να αμεληθεί σε σχέση με τη δυναμική, τότε υπάρχουν συνθήκες ευδοκίμησης του πληθωρισμού.

Η μαθηματική μελέτη ξεκινά με τη λανγκραζιανή πυκνότητα του πεδίου $\phi$ στο CRF, που είναι:

$$\mathcal{L}_\phi = \frac{1}{2}\partial_\mu\phi\partial^\mu\phi - V(\phi) \tag{4.93}$$

από την οποία λαμβάνεται ο τανυστής ενέργειας-ορμής:

$$T^{\mu\nu}_\phi = \frac{\partial\mathcal{L}_\phi}{\partial(\partial_\mu\phi)}\partial^\nu\phi - g^{\mu\nu}\mathcal{L}_\phi = \partial^\mu\phi\partial^\nu\phi - g^{\mu\nu}\left(\frac{1}{2}\partial_\lambda\phi\partial^\lambda\phi - V(\phi)\right), \tag{4.94}$$

Από τον τανυστή αυτό, προκύπτουν:

- Η πυκνότητα ενέργειας και η πίεση του $\phi$:

$$\rho_\phi = T^{00}_\phi = \frac{1}{2}\dot{\phi}^2 + V(\phi) \tag{4.95}$$

$$P_\phi = T^{ii}_\phi = \frac{1}{2}\dot{\phi}^2 - V(\phi) \tag{4.96}$$

- Το πρώτο θερμοδυναμικο αξίωμα, Εξ. (4.20), με μορφή:

$$\dot{\rho_\phi} = -3H(\rho_\phi + P_\phi) \tag{4.97}$$

Με αντικατάσταση των Εξ. (4.95), (4.96) στην Εξ. (4.97), λαμβάνεται η εξίσωση κίνησης του $\phi$,

$$\ddot{\phi} + 3H\dot{\phi} + V'(\phi) = 0. \tag{4.98}$$

Βεβαίως, η ίδια εξίσωση λαμβάνεται και με ελαχιστοποιήση της δράσης, όμως είναι πιο χρονοβόρα η παρουσίαση και δεν υιοθετήθηκε. Σύνδυασμός των Εξ. (4.18), (4.95) και (4.98) παρέχει την παρακάτω χρήσιμη:

$$H^2 = \frac{8\pi}{3M_P^2}\rho_\phi = \frac{8\pi}{3M_P^2}\left(\frac{1}{2}\dot{\phi}^2 + V(\phi)\right) \Longrightarrow \dot{H} = -\frac{4\pi}{M_P^2}\dot{\phi}^2. \tag{4.99}$$

Βασική παραδοχή του πληθωρισμού είναι ότι κατά τη διάρκεια της πορείας του προς το πραγματικό κενό, το $\phi$ ολισθαίνει αργά, δηλαδή ο όρος τριβής υπερισχυεί και η κινητική ενέργεια είναι υποτονική σε σχέση με την δυναμική. Ποσοτικά εκφραζόμενοι:

$$\dot{\phi} \gg \ddot{\phi} \;\; \text{και} \;\; V(\phi) \gg \dot{\phi}^2 \tag{4.100}$$

Αξιοποιώντας τις προηγούμενες, προκύπτει:

$$(4.98) \Longrightarrow 3H\dot{\phi} = -V'(\phi) \tag{4.101}$$

$$(4.95) \text{ και } (4.96) \Longrightarrow P_\phi \simeq -\rho_\phi, \tag{4.102}$$

$$(4.102) \text{ και } (4.97) \Longrightarrow \rho_\phi \simeq \text{cst}, \tag{4.103}$$

$$(4.103) \text{ και } (4.99) \Longrightarrow H(\phi) \simeq \text{cst}. \tag{4.104}$$

Με αξιοποίηση της τελευταίας η Εξ. (4.18) ολοκληρώνεται άμεσα. Μάλιστα, ο όρος καμπυλότητας, $k/R^2$, μπορεί ασφαλώς να παραληφθεί γιατί, όπως προκύπτει, ο $R(t)$ αυξάνει ταχέως και καθιστά την προηγούμενη ποσότητα αμεληταία. Επομένως,

$$H_\phi^2 := \left(\frac{\dot{R}}{R}\right)^2 = \frac{8\pi}{3M_P^2}V_0, \tag{4.105}$$

όπου για μια τυπική GUT $V_0 := V(\pm\langle\phi\rangle) \sim M_G^4$. Συμβολίζοντας με $t_i, t_f$ την αρχική και τελική στιγμή του πληθωρισμού, συμπεραίνεται οτι ο παραγοντας κλίμακας εχει ισχυρά ενισχυθεί στο τέλος του πληθωρισμού ενώ η θερμοκρασία έχει εξίσου ελλατωθεί σύμφωνα με την Εξ. (4.72):

$$R(t_f) = R(t_i)e^{H_\phi(t_f-t_i)} := R(t_i)e^{N_e}, \tag{4.106}$$

$$T(t_f) = T(t_i)e^{-H_\phi(t_f-t_i)} \tag{4.107}$$

Με κατάλληλη επιλογή του αριθμού $N_e$ των $e$-πτυχών είναι δυνατόν να λυθούν όλα τα προβλήματα του SBB, πράγμα που εξηγείται στο επόμενο εδάφιο. Ειδικότερα, με αντικατάσταση της Εξ. (4.106), υπολογίζεται:

- Το μέγεθος του ορίζοντα, χρησιμοποιώντας την οριστική Εξ. (4.7):

$$d_H(t) = e^{H_\phi t}\int_0^t \frac{dt'}{e^{H_\phi t'}} = e^{H_\phi t}(1-e^{-H_\phi t})/H_\phi \sim H_\phi^{-1}e^{H_\phi t}, \tag{4.108}$$

  όπου ο όρος $e^{-H_\phi t}$ μπορεί να παραλειφθεί, γιατί οι χρόνοι αναφοράς είναι της τάξης του $t_G$.

- Η συμπεριφορά της παραμέτρου κοσμολογικής πυκνότητας, χρησιμοποιώντας τη Εξ. (4.24):

$$\Omega(t) - 1 \sim 1/R(t)^2 = e^{-2H_\phi t} \tag{4.109}$$

Επιπλέον, με αντικατάσταση της Εξ. (4.102) στην Εξ. (4.26), λαμβάνεται ότι η παράμετρος επιβράδυνσης είναι αρνητική κατά τον πληθωρισμό, δηλαδή η διαστολή γίνεται επιταχυνόμενα, πράγμα που αποτελεί μια ακόμα ειδοποιό διαφορά της πληθωριστικής διαστολής από αυτή του SBB.

Τα ευρήματα της παραγράφου αυτής σχετικά με τη συμπεριφορά των κοσμολογικών παραμέτρων κατά τη διάρκεια του πληθωρισμού, έχουν ήδη καταγραφεί στην τρίτη στήλη του συγκεντρωτικού Πίνακα 4.1.

### 4.6.2 Παρεπόμενα του Πληθωρισμού

Ακροθιγώς δίνεται ένα περίγραμμα της θεωρίας του πληθωρισμού. Η διαδικασία που περιγράφηκε στο Εδ. 4.6.1 είναι μια απλοποιημένη μορφή του πλήρους μηχανισμού. Αυτός εξελίσσεται στα παρακάτω στάδια:

**α.** Της αργής ολίσθησης. Σε αυτό το στάδιο ήταν αφιερωμένο το Εδ. 4.6.1. Κατά τη διαρκεία του, η κίνηση του $\phi$ ελέγχεται από την Εξ. (4.101), οπότε αυτή μοιάζει με μια απλή διαφορική εξίσωση που περιγράφει κίνηση με παρουσία τριβής, από όπου και η ονομασία του στάδιου αυτού. Πιο ακριβείς συνθήκες από την Εξ. (4.100) μπορούν να διατυπωθουν, οι οποίες δίνουν την ελευθερία η παράμετρος του Hubble και ο αριθμός των $e$-πτυχών να εξαρτώνται από το inflaton.

**β.** Των συναφών ταλαντώσεων. Είναι στάδιο επακόλουθο της αργής ολίσθησης. Κατά τη διάρκεια του, ο όρος $\ddot{\phi}$ γίνεται υπολογίσιμος οπότε η Εξ. (4.98) θυμίζει σύστημα που ταλαντώνεται. Το inflaton διασπάται σε άλλα πολύ πιο ελαφρά από αυτό σωμάτια.

**γ.** Της Επαναθέρμανσης. Είναι στάδιο παράλληλο των συναφών ταλαντώσεων. Το inflaton καθώς διασπάται, ακτινοβολεί, οπότε το Σύμπαν περνά από την κυριαρχία της ενέργειας του κένου στην κυριαρχία της ακτινοβολίας. Η θερμοκρασία στην οποία αποκαθίσταται ισότητα ανάμεσα στις δύο πυκνότητες ενέργειας, (equidensity point) λέγεται θερμοκρασία επαναθέρμανσης και είναι της τάξης $T_r \sim 10^9$ GeV. Επειδή κατά τη διάρκεια του πληθωρισμού η θερμοκρασία ελλατώνεται, κατά την Εξ. (4.107), πολύ πιο απότομα από ό,τι την εποχή της ακτινοβολίας, η πάροδος από τη μία φάση στην άλλη, μπορεί να θεωρηθεί ως επαναθέρμανση του σύμπαντος.

Το σενάριο του πληθωρισμού, ελέγχεται από τις επιπτώσεις του στη CBR. Μπορεί να θεμελιωθεί ένας φορμαλισμός που συνδέει τις διακυμάνσεις της CBR, που παρατήρουνται από τον COBE, Εξ. (4.83), με τη μορφή του χρησιμοποιούμενου για τον πληθωρισμό δυναμικού και τον αριθμό των $e$-πτυχών. Επομένως, ο πληθωρισμός παύει να αποτελεί ένα αφηρημένο μαθηματικό κατασκεύασμα, αφού τα σημάδια του ψηλαφώνται στο σύγχρονο σύμπαν.

### 4.6.3 Τα προβλήματα του SBB και η λύση τους

Η εφαρμογή της SBB έκτος από τις σημαντικές επιτυχίες που έχει στην ερμηνεία των κοσμολογικών παρατηρήσεωνν οδηγεί και σε κάποιες αντιφάσεις, οι οποίες αίρονται ταυτόχρονα με χρήση της ιδέας του Πληθωρισμού. Παρακάτω επισημαίνονται τα προβλήματα και σκιαγραφείται η λύση τους:

**α.** Το πρόβλημα του Ορίζοντα. Ο ορίζοντας ενός σωματίου εκφράζει το μέγεθος του σύμπαντος που μπορεί να είναι σε επικοινωνία με το σωμάτιο αυτό, μέσω φωτεινών σημάτων. Η συσχέτιση του ορίζοντα $d_H$ και του σημερινού μεγέθους του σύμπαντος $D$ κάποια στιγμή $t$ δίνεται από την παρακάτω σχέση:

$$\frac{D(t)}{d_H(t)} = \frac{R(t)}{R(t_0)}\frac{t_0}{t} \tag{4.110}$$

όπου χρησιμοποιήθηκαν οι Εξ. (4.13) και (4.29) και θεωρήθηκε οτι $D(t_0) = d_H(t_0) = 3t_0$. Λαμβάνοντας υπόψη, την Εξ. (4.29), εφαρμογή της Εξ. (4.110) παρέχει:

$$\frac{D(t)}{d_H(t)} = \frac{R(t)}{R(t_d)}\frac{R(t_d)}{R(t_0)}\frac{t_0}{t} = \left(\frac{t}{t_d}\right)^{1/2}\left(\frac{t_d}{t_0}\right)^{2/3}\frac{t_0}{t} \Longrightarrow \begin{cases} D(t_d)/d_H(t_d) & \sim 10^2 \\ D(t_G)/d_H(t_G) & \sim 10^{28} \end{cases} \tag{4.111}$$

Ώστε, ο σωματιδιακός ορίζοντας είναι κατά πολύ μικρότερος από το μέγεθος του σύμπαντος κάθε χρονική στιγμή. Εγείρεται, λοιπόν το ερώτημα, πώς η CBR είναι τόσο πολύ ομογενής αφού τα φωτόνια δεν είναι δυνατόν να επικοινωνούν μεταξύ τους. Με τη μεσολάβηση του πληθωρισμού, η παραπάνω ποσότητα γράφεται:

$$\begin{aligned} \frac{D(t)}{d_H(t)} &= \frac{R(t)}{R(t_f)}\frac{R(t_f)}{R(t_d)}\frac{R(t_d)}{R(t_0)}\frac{t_0}{t} \\ &= e^{-H(t_f-t)}\left(\frac{t_f}{t_d}\right)^{1/2}\left(\frac{t_d}{t_0}\right)^{2/3}\frac{t_0}{t} \Longrightarrow D(t_G)/d_H(t_G) \sim 10^{-1} \end{aligned} \tag{4.112}$$

όπου χρησιμοποιήθηκε $N_e \sim 65$ και $t_f = t_G + N_e/H_\phi$. Συμπεραίνεται, λοιπόν, ότι η συμβολή του πληθωρισμού είναι καθοριστική στην άρση του παράδοξου του ορίζοντα.

**β.** Το πρόβλημα της Επιπεδότητας. Τα παρατηρησιακά ευρήματα της Εξ. (4.86) για την παράμετρο $\Omega$ βρίσκονται πολύ κοντά στην οριακή τιμή που αρμόζει σε ένα επίπεδο Σύμπαν, πράγμα που κατα αρχή χρήζει εξηγήσεως. Συνακόλουθα, αναζητώντας την τιμή της παραμέτρου αυτής στο αρχέγονο σύμπαν βρισκόμαστε προ εκπλήξεως:

$$\frac{\Omega(t_0)-1}{\Omega(t_G)-1} = \left(\frac{t_0}{t_d}\right)^{2/3}\left(\frac{t_d}{t_G}\right) \sim 10^{54} \Longrightarrow \Omega(t_G) \sim 1 \pm 10^{-54} \tag{4.113}$$

όπου χρησιμοποιήθηκαν τα αποτελέσματα του Πίνακα 4.1. Η παράδοξη αυτή ακρίβεια είναι ένδειξη παθογένειας του SBB. Μια πιο λογική πιθανότητα θα ήταν η παράμετρος $\Omega$ να βρίσκεται κοντά στη μονάδα με την ίδια πάντα ακρίβεια. Πραγματικά, με την παρουσία του πληθωρισμού, ο λόγος της Εξ. (4.113) με χρήση και της Εξ. (4.106) γράφεται:

$$\frac{\Omega(t_0)-1}{\Omega(t_G)-1} = \left(\frac{t_0}{t_d}\right)^{2/3}\left(\frac{t_d}{t_f}\right)e^{2N_e} \Longrightarrow \Omega(t_G) \sim 1 \tag{4.114}$$

με επιλογή $N_e \geq 60$. Συνεπώς, ο πληθωρισμός προβλέπει ένα επίπεδο Σύμπαν.

**γ.** Το πρόβλημα των ανεπιθύμητων λειψάνων. Αυτό το πρόβλημα ανακύπτει από τη σύνδεση του SBB με GUT. Κατά τις μεταπτώσεις φάσεως που είναι αναπόφευκτες ακόμα και με απουσία του πληθωρισμού, αφού χωρίς αυτές δεν μπορεί να μεταπηδήσει κανείς στην συμμετρία του SM, δημιουργούνται βαριά σωμάτια, λόγω τοπολογικών ανωμαλιών. Τα πιο ανεπιθύμητα από αυτά είναι τα μαγνητικά μονόπολα. Επειδή οι αλληλεπιδράσεις τους είναι ασθενείς αυτά τα σωμάτια αποσυνδέονται εύκολα από το κοσμικό ρευστό και εξελίσσονται μέχρι σήμερα, η αριθμητική τους πυκνότητα φθίνει πιο αργά, σαν $R^{-3}$ από

ό,τι η ακτινοβολία (που φθίνει σαν $R^{-4}$). Θα μπορούσαν, λοιπόν να γίνουν το κυρίαρχο υλικό του Σύμπαντος, πράγμα που αντικρούεται από τα παρατηρησιακά δεδομένα. Η εισαγωγή του πληθωρισμού συντείνει στην απαλλαγή της θεωρίας από τέτοια ανεπιθύμητα κατασκευάσματα. Και αυτό, διότι η ενεργειακή πυκνότητα του inflaton, είναι κατα πολύ μεγαλύτερη και πιο ανθεκτική στην επίδραση της διαστολής από την αριθμητική πυκνότητα των μονοπόλων, η οποία γρήγορα καθίσταται αμελητaία. Στην υλοποιήση του μηχανισμού αυτού χρειάζεται προσοχή στο σημείο της επαναθέρμανσης, ώστε η θερμοκρασία να μην ανυψωθεί τόσο όσο θα επέτρεπε την επαναδημιουργία των μονοπόλων. Πρέπει λοιπόν, το $T_r$ να είναι σημαντικά μικρότερη από την μάζα των μονοπόλων.

**δ.** Το πρόβλημα των μεγάλης κλίμακας διακυμάνσεων. Το SBB δεν διαθέτει μηχανισμό παραγωγής αρχέγονων διακυμάνσεων στην πυκνότητα ενέργειας, $(\delta\rho/\rho)$ ούτε και στην θερμοκρασία της CBR. Με την προσθήκη του πληθωρισμού, όπως ήδη αναφέρθηκε, αναπτύσσεται ένας φορμαλισμός σύνδεσης των διακυμάνσεων αυτών με τον αριθμό των $e$ πτυχών, που επιτρέπει την εξήγηση της μικρής ανισοτροπίας, Εξ. (4.83), στην CBR.

Συμπερασματικά, ο πληθωρισμός είναι ένα φαινόμενο που πιστεύεται οτι χαρακτηρίζει την εξέλιξη του πολύ πρώιμου σύμπαντος μετά το τέλος, του οποίου η ροή της ιστορίας συνεχίζεται κατά τα προβλεπόμενα από το SBB. Συνεπώς, ο πληθωρισμός δεν υποκαθιστά το SBB, απλά το συμπληρώνει, θεραπεύει τις ατέλειές του και βελτιώνει τις προβλέψεις του.

## 4.7 Κριτική του Κοσμολογικού Προτύπου

Μια αποτίμηση του Κοσμολογικού Προτύπου που παρουσιάστηκε στο κεφάλαιο αυτό είναι ο στόχος αυτού του εδαφίου βάση των Αν. [41] και [43]. Παρακάτω καταγράφονται:

**α.** Οι επιτυχίες του SBB. Στις επιτυχίες του SBB συγκαταλέγονται τα εξής:

- Η εξήγηση της διαστολής του Σύμπαντος.
- Η εξήγηση της ύπαρξης και της σύστασης της CBR.
- Η εξήγηση της εναπομένουσας πυκνότητας των ελαφρών στοιχείων του Σύμπαντος.

Επιπλέον με τη συνδρομή του πληθωρισμού, αίρονται τα παράδοξα που εμφάνιζε το συγκεκριμένο πρότυπο και εξηγείται η παρατηρούμενη ανισοτροπία, στην CBR.

**β.** Μερικά αδιευκρίνιστα θέματα. Κάποια σημεία της θεωρίας παραμένουν αδιευκρίνιστα, δηλαδή, δεν έχει καταστεί δυνατόν να διατυπωθεί με ακρίβεια μια κοινά αποδεκτή ερμηνεία για αυτά. Αποτελούν, επομένως, αντικείμενα έρευνας. Δύο είναι τα βασικά ανοικτά θέματα της Σύγχρονης Κοσμολογίας:

- Η Βαρυογένεση. Δύο είναι τα ερωτήματα που πρέπει να κατανοηθούν. Που οφείλεται η προφανής ασυμμετρία μεταξύ ύλης και αντιύλης στο σύμπαν και ποιά είναι η προέλευση του παρατηρούμενου λόγου $\eta_0$ στην Εξ. (4.85). Οι ιδέες που προτείνονται μπορούν να ταξινομηθούν σε δύο κατηγορίες. Μία που παράγει την βαρυονική ασυμμετρία σε κλίμακα ενέργειας GUT και μία σε EWS. Η παραγωγή μπορεί να γίνει είτε άμεσα είτε μέσω λεπτογένεσης. Αυτό σε πρότυπα $SO(10)$ γίνεται με διάσπαση των δεξιόστροφων νετρίνο που παράγονται στο τέλος του πληθωρισμού από το inflaton.
- Η Σκοτεινή Ύλη. Ο πληθωρισμός προβλέπει ένα επίπεδο Σύμπαν πράγμα που σημαίνει ότι $\Omega = 1$. Από τις μέχρι τώρα παρατηρήσεις, οι συνισταμένες του $\Omega$ στις Εξ. (4.90) και (4.92), απέχουν πολύ από την επαλήθευση της πληθωριστικής πρόβλεψης. Είναι, όμως γενικά αποδεκτό ότι εκτός από τη φωτεινή ύλη, η οποία έχει καταμετρηθεί, υπάρχει και σκοτεινή η οποία είναι δύσκολο να ανιχνευτεί και γιαυτό παραμένει απαρατήρητη. Με αυτή την αρχική υπόθεση, έχουν επινοηθεί διάφορα σωματιδιακά σενάρια που καταλήγουν στη θεμελίωση ενός επίπεδου σύμπαντος.

Όπως φαίνεται από τα προηγούμενα, η συνδρομή της Φυσικής Στοιχειωδών Σωματιδίων είναι αποφασιστικής σημασίας για τη διερεύνηση των ανοικτών πεδίων της Κοσμολογίας. Σε μια τέτοια σύζευξη αντικειμένων για τη μελέτη του θέματος της Σκοτεινής Ύλης είναι αφιερωμένο το επόμενο κεφάλαιο της διατριβής αυτής.

# Κεφάλαιο 5

# Υπερσυμμετρική Σκοτεινή Ύλη

## 5.1 Εισαγωγή

Ο βασικός φορμαλισμός που θα επιτρέψει την σύνδεση της Φυσικής Στοιχειωδών Σωματιδίων με τις αναζητήσεις της Σύγχρονης Κοσμολογίας για τη Σκοτεινή Ύλη (: DM), αναπτύσσεται σε αυτό το κεφάλαιο. Ειδικότερα, στο Εδ. 5.2 εξηγείται διεξοδικά το θέμα της DM, και επιλέγεται το σενάριο που θα τεθεί σε υλοποίηση. Το τυπολόγιο που θα χρησιμοποιηθεί κατά την πραγματοποίηση των υπολογισμών μας θεμελιώνεται στα Εδ. 5.3, 5.4. Τέλος στο Εδ. 5.5 γίνεται εφαρμογή του τυπολόγιου στην κατεύθυνση που έχει επιλέγει και παρουσιάζονται τα αναλυτικά αποτελέσματα των Αν. [44] και [45].

## 5.2 Η υπόθεση της Σκοτεινής Ύλης

Η πειστική εισαγωγή της ιδέας της DM είναι ο στόχος αυτής της ενότητας. Η αναγκαιότητα εισαγωγής της θεμελιώνεται θεωρητικά στο Εδ. 5.2.1, ενώ τα παρατηρησιακά επιχειρήματα γύρω από το θέμα εκτίθενται περιληπτικά στο Εδ. 5.2.2. Σενάρια δομής της παραμέτρου πυκνότητας και σύστασης της DM ακολουθούν στα Εδ. 5.2.3, 5.2.4, 5.2.5. Τέλος, μια ποιοτική παρουσίαση της υποψηφιότητας που θα υποστηριχθεί κατά τη διάρκεια του ερευνητικού τμήματος αυτής της εργασίας δίνεται στο Εδ. 5.2.6.

### 5.2.1 Αναγκαιότητα της υπόθεσης

Όπως αναφέρθηκε στο Εδ. 4.6.3, στο πρόβλημα της επιπεδότητας, η τιμή $\Omega = 1$ είναι η πιο φυσική τιμή που μπορεί να λάβει η παράμετρος πυκνότητας. Επιπρόσθετα, αυτή είναι η τιμή που προβλέπει ο Πληθωρισμός. Παρατηρησιακά, όμως η μόνη μαρτυρία που υπάρχει για την επίμαχη ποσότητα, καταγράφεται στην Εξ. (4.92) και είναι πολύ πιο χαμηλά από τις θεωρητικές προσδοκίες. Για τη γεφύρωση του χάσματος αυτού, έχει γίνει αποδεκτή στην επιστημονική κοινότητα η ύπαρξη μιας νέας μορφής Ύλης που ονομάζεται Σκοτεινή. Η ονομασία της οφείλεται στο ότι ούτε εκπέμπεται ούτε απορροφάται ηλεκτρομαγνητική ακτινοβολία στα γνωστά μήκη κύματος από αυτή. Γιαυτό και παραμένει απαρατήρητη.

Η σύσταση αυτής της μορφής ύλης και ο τρόπος οργάνωσης των διάφορων συνεισφορών της για την επίτευξη της προβλεπόμενης τιμής $\Omega = 1$ είναι αντικείμενο θεωρητικής και πειραματικής έρευνας. Σημαντικό στην κατεύθυνση αυτή θα είναι η ακόμη πιο ακριβής μέτρηση της ανισοτροπίας της CBR που θα κρίνει τελεσίδικα την εγκυρότητα ή μη του Πληθωριστικού προτύπου.

### 5.2.2 Παρατηρησιακά δεδομένα

Από τη φύση της, η υπόθεση της DM δεν είναι επιδεκτική παρατηρησιακών αποδείξεων. Παρόλη τη δυσκολία ψηλάφησης της DM, υπάρχουν παρατηρησιακές ενδείξεις που καταφάσκουν στην ύπαρξή της, σύμφωνα με τις Αν. [47], [50]. Οι βασικότερες πηγές άντλησης πληροφοριών για το υπό εξέταση θέμα είναι οι εξής:

**α.** Η ηλικία του σύμπαντος. Υπάρχει τύπος (Αν. [51]) άμεσης συσχέτισης της ηλικία του Σύμπαντος, $t_0$, με την παράμετρο του Hubble και τις διάφορές συνεισφορές στην $\Omega$. Αν για παράδειγμα, τεθεί $h \simeq 0.7$, υπολογίζεται $t_0 \sim 11.5$ Gyr που απαιτεί ή $\Omega_M \lesssim 0.3$ ή $\Omega_M \sim 0.5$, ανάλογα με το αν επιτρέπεται ή όχι η χρήση της $\Lambda$. Φαίνεται, λοιπόν οτι οι προβλέψεις για το $\Omega$ είναι πολύ ανώτερες από τις παρατηρούμενες, Εξ. 4.92, 4.90.

**β.** Η φωτεινότητα ουράνιων αντικειμένων. Η παράμετρος πυκνότητας που οφείλεται στη φωτεινή ύλη του σύμπαντος μπορεί να προκύψει από μέτρηση της μέσης τιμής του λόγου της μάζας προς το φως διάφορων αστροφυσικών αντικειμένων. Τα απότελέσματα τέτοιων μετρήσεων δίνουν τα παρακάτω όρια για την ορατή παράμετρο πυκνότητας:

$$0.003 \leq \Omega_{vis} \leq 0.017\,. \tag{5.1}$$

Το αποτέλεσμα αυτό είναι χαμηλότερο από αυτό της Εξ. (4.92) πράγμα που δείχνει ότι ακόμα και σε επίπεδο βαρυονικής ύλης μπορεί να υπάρχει DM.

**γ.** Η καμπύλες περιστροφής γύρω από γαλαξίες. Σε τέτοιες παρατηρήσεις κανείς μετράει την ταχύτητα με την οποία νέφοι αερίων, μικροί γαλαξίες ή σμήνοι αστεροειδών κινούνται γύρω από κάποιον γαλαξία. Όπως είναι γνωστό από τη γυμνασιακή Φυσική, η γραμμική ταχύτητα που προκύπτει από την έκφραση της βαρυτικής έλξης ως κεντρομόλας, ελαττώνεται σαν $\sqrt{G_N M/r}$, όπου $r$ η απόσταση του περιστρεφόμενου αντικειμένου από το κέντρο του εν λόγω γαλαξία μάζας $M$. Οι παρατηρήσεις δείχνουν ότι η ταχύτητα παραμένει προσεγγιστικά σταθερή, που σημαίνει ότι η μάζα των γαλαξιών αυτών δεν είναι συγκεντρωμένη στο φωτεινά τους μέρη μόνο. Μάλιστα, για να είναι η ταχύτητα σταθερή θα πρέπει η μάζα του γαλαξία να αυξάνεται γραμμικά με την απόσταση $r$. Μελέτες αυτού του τύπου δηλώνουν ότι το 90% της μάζας αυτών των γαλαξιών είναι DM κατά την Αν. [46]. Εκτιμήσεις της Αν. [47] από μετρήσεις αυτού του τύπου δίνουν

$$\Omega_M \simeq 0.1^{+0.1}_{-0.05}\,. \tag{5.2}$$

Λαμβάνοντας υπόψη τα δεδομένα των Εξ. (4.92), (5.1) και (5.2), μπορεί κανείς να καταλήξει στα παρακάτω συμπεράσματα:

- Ένα σημαντικό ποσό της ύλης του Σύμπαντος είναι μη ορατή.
- Ένα πολύ μικρό ποσό της DM είναι βαρυονική.
- Ένα σημαντικό ποσό της DM είναι μη βαρυονική.

### 5.2.3 Σύσταση της DM

Ένα σωμάτιο θεωρείται αξιόλογος υποψήφιος για την ερμηνεία της σύστασης της DM, πρέπει να είναι ασθενώς αλληλεπιδρών και ευσταθές, ή τουλάχιστον να διαθέτει μακροβιότητα μετρούμενη βεβαίως, σε χρόνους κοσμολογικής κλίμακας. Οι μέχρι τώρα προτεινόμενοι υποψήφιοι μπορούν να ταξινομηθούν ως εξής:

**α.** Υποψήφιοι για βαρυονική DM. Οι εκπρόσωποι αυτής της κατηγορίας εκφέρονται με τη γενική ονομασία MACHOs που είναι ακρονύμιο της Massive Compact Halo Objects. Αντικείμενα προερχόμενα από την αστροφυσική ανήκουν σε αυτή την ομάδα, λευκοί νάνοι, αστέρες νετρονίων, υπολείματα μελανών οπών.

**β.** Υποψήφιοι για μη βαρυονική DM. Οι εκπρόσωποι αυτής της κατηγορίας προέρχονται από τον μικρόκοσμο και ταξινομούνται στις εξής συνομοταξίες:

- Υποψήφιοι για θερμή DM (: HDM). Στην κατηγορία αυτή εγκλείωνται σωμάτια που μπορούν να θεωρηθούν σχετικιστικά την στιγμή της αποσύνδεσης, σύμφωνα με τα εκτεθέντα στο Εδ. 4.74. Ισχυρή υποψηφιότητα στην κατεύθυνση αυτή υποβάλουν τα νετρίνο. Ενδείξεις για μη μηδενικές μάζες των νετρίνο προέρχονται από μετρήσεις στις ταλαντώσεις των ηλιακών και ατμοσφερικών νετρίνο. Μετρώνται οι διαφορές τετραγώνων μάζας των σωματίων αυτών. Εναλλακτική υποψηφιότητα είναι αυτή του axion που η αναγκαιότητα ύπαρξής του σχετίζεται με το ισχυρό CP-πρόβλημα της κβαντικής χρωμοδυναμικής.

**Πίνακας 5.1: Τα DM Σενάρια**

| Σενάριο | Συνεισφορές στην $\Omega$ | | | | $h$ | $\Omega_{LSP}h^2$ |
|---|---|---|---|---|---|---|
| | $\Omega_M$ | | | $\Omega_\Lambda$ | | |
| | $\Omega_{DM}$ | | $\Omega_B$ | | | |
| | $\Omega_{CDM}$ | $\Omega_{HDM}$ | | | | |
| **CHDM** | $0.7 \pm 0.05$ | $0.25 \pm 0.05$ | $0.05 \pm 0.005$ | $0$ | $0.5 \pm 0.05$ | $0.175 \pm 0.045$ |
| **ΛCDM** | $0.35 \pm 0.1$ | $0$ | $0.05 \pm 0.005$ | $0.65 \pm 0.5$ | $0.65 \pm 0.05$ | $0.155 \pm 0.065$ |

- Υποψήφιοι για ψυχρή DM (: CDM). Στην κατηγορία αυτή εγκλείωνται σωμάτια που μπορούν να θεωρηθούν μη σχετικιστικά την στιγμή της αποσύνδεσης, σύμφωνα με τα εκτεθέντα στο Εδ. 4.74. Πληθώρα υποψηφιοτήτων υπάρχει στην περίπτωση αυτή. Προεξάρχουσα αυτών είναι η υποψιότητα του Ελαφρότατου Υπερσυμμετρικού Σωματιού (: LSP) που μπορεί να είναι neutralino, sneutrino, gravitino ή axino. Εναλλακτικά σενάρια κτίζονται με μη τοπολογικά σολιτόνια ή βαριά μη θερμικά λείψανα.

Προφανώς, κάποιο είδος DM μπορεί να συντίθεται από περισσότερα του ενός είδους σωματίων. Στη διατριβή αυτή, υποτίθεται οτι η CDM δομείται μόνο από LSP, η ταυτότητα του οποίου θα καθοριστεί στο Εδ. 5.5.

### 5.2.4 Πιθανές Συνεισφορές στην $\Omega$

Από την αρχέγονη σχέση της Εξ. (4.23) φαίνεται ότι η παράμετρος κοσμολογικής πυκνότητας δομείται από δύο συνιστώσες:

**α.** Την οφειλόμενη στην πυκνότητα ύλης $\Omega_M$, η οποία σύμφωνα με όλες τις ενδείξεις, συμπεριλαμβάνει τις εξής υποδιαιρέσεις:

- Την οφειλόμενη στην πυκνότητα βαρυονικής ύλης $\Omega_B$.
- Την οφειλόμενη στην πυκνότητα DM, $\Omega_{DM}$, η οποία υποδιαιρείται περαιτέρω σε τμήμα που οφείλεται σε πυκνότητα θερμής DM, $\Omega_{HDM}$ και ψυχρής DM, $\Omega_{CDM}$.

Συμπερασματικά, μπορεί να γραφεί η παρακάτω:

$$\Omega_M = \Omega_B + \Omega_{DM} \quad \text{όπου} \quad \Omega_{DM} = \Omega_{HDM} + \Omega_{CDM}. \tag{5.3}$$

**β.** Την οφειλόμενη στη κοσμολογική σταθερά, $\Omega_\Lambda$.

### 5.2.5 Τα CDM Σενάρια

Υπάρχουν δύο βασικά σενάρια διευθέτησης των διάφορων συνεισφορών της $\Omega$, ώστε η προκύπτουσα τιμή να είναι συμβατή με πληθωριστική πρόγνωση ($\Omega = 1$). Αυτά παρουσιάζονται συγκεντρωτικά στο συγκριτικό Πίνακα 5.1 και παρακάτω δίνονται τα βασικά τους χαρακτηριστικά χωρίς εμμονή στα πεδία απροσδιοριστίας:

**α.** Το HCDM Σενάριο. Σύμφωνα με το σενάριο αυτό, οι συνεισφορές στην $\Omega$ διευθετούνται ως εξής:

$$\Omega_B \simeq 0.1, \quad \Omega_{CDM} \simeq 0.7, \quad \Omega_{HDM} \simeq 0.2$$

Το σενάριο αυτό, Αν. [48], έχει τα εξής χαρακτηριστικά:

- Η ανηγμένη παράμετρος Hubble, λαμβάνεται στα χαμηλότερα επιτρεπτά όριά της, $h \simeq 0.5$.

- Δεν επιτρέπεται συνεισφορά από τον όρο που οφείλεται στη Λ.
- Η κεντρική τιμή για την επίμαχη ποσότητα είναι: $\Omega_{CDM}\ h^2 \simeq 0.175$
- Η θερμή συνιστώσα της CDM, μπορεί να συνίσταται από ελαφρά εκφυλισμένα νετρίνο, πράγμα που είναι συμβατό με τις ταλαντώσεις των ατμοσφερικών και ηλιακών νετρίνο μέσα στα πλάσια ενός υπερσυμμετρικού πληθωριστικού προτύπου σύμφωνα με την Αν. [52].

**β.** Το ΛCDM Σενάριο. Σύμφωνα με το σενάριο αυτό οι συνείσφορές στην Ω διευθετούνται ως εξής:

$$\Omega_B \simeq 0.05, \quad \Omega_{CDM} \simeq 0.3, \quad \Omega_\Lambda \simeq 0.65$$

Το σενάριο αυτό, Αν. [53], έχει τα εξής χαρακτηριστικά:

- Η ανηγμένη παράμετρος Hubble, λαμβάνεται στα υψηλότερα επιτρεπτά όριά της, $h \simeq 0.65$.
- Δεν προβλέπεται συνεισφορά από τη θερμή συνιστώσα της CDM. Επόμενως, το σενάριο είναι συμβατό με ιεράρχηση στις μάζες των νετρίνο, σύμφωνα με την Αν. [54].
- Η κεντρική τιμή για την επίμαχη ποσότητα είναι: $\Omega_{CDM}\ h^2 \simeq 0.155$.
- Είναι συνεπές με πρόσφατες μετρήσεις της μεγάλης υπέρυθρης μετατόπισης στον υπερκαινοφανή τύπου Ia (Αν. [55]), οι οποίες διαγνώσκουν επιταγχυνόμενη διαστολή του σύμπαντος, πράγμα που συνεπάγεται ότι μια υπολογίσιμη συνεισφορά στην Ω θα οφείλεται στην Λ, με σχετικά ελαττωμένη συνεισφορά από τη $\Omega_M$, όπως επιβεβαιώνεται από τα αναγραφόμενα στον Πίνακα 4.1. Αυτό σημαίνει ότι το σύμπαν είσερχεται σε μια νέα φάση πληθωρισμού, όπου η Λ θα γίνει κυρίαρχη, όποτε η ολοκλήρωση του βιολογικού κύκλου του σύμπαντος έχει αρχίσει. Βεβαίως, πολλά από τα στοιχεία των παρατηρήσεων χρήζουν διασαφηνίσεων.

Έλεγχος των σεναρίων αυτών, γίνεται με σύγκριση του φάσματος των διακύμανσεων πυκνότητας ($\delta\rho/\rho$) που παρέχουν με αυτό που προκύπτει από τις μετρήσεις της ανισοτροπίας της CBR σύμφωνα με την Αν. [49]. Συμπεραίνεται ότι και τα δύο είναι εξίσου αποδεκτές ερμηνείες για τη σύνθεση της κοσμικής δομής.

Συμπερασματικά, υποθέτωντας ότι η CDM συντίθεται εξολοκλήρου από ένα συστατικό, το LSP, σε όλη την έκταση της διατριβής αυτής θα ληφθούν τα παρακάτω όρια (Αν. [44], [53]) που είναι συμβατά και με τα δύο σενάρια που εκτέθηκαν:

$$0.09 \lesssim \Omega_{LSP} h^2 \lesssim 0.22. \tag{5.4}$$

### 5.2.6 Η υποψηφιότητα του LSP

Ανάμεσα στους διάφορους υποψήφιους για σκοτεινή ύλη, ο πιο διάσημος και πολλά υποσχόμενος είναι το LSP. Μια ποιοτική περιγραφή του σεναρίου που καλούμαστε να υλοποιήσουμε, θα επιχειρηθεί σε αυτή την παράγραφο, ακολουθώντας την Αν. [56]. Βασίζεται στην ύπαρξη ενός σωματίου με τις παρακάτω ιδιότητες:

- Ελαφρότατο. Με τη συνθήκη αυτή εξασφαλίζεται ότι με το δίαβα του χρόνου, όλα τα σωμάτια του MSSM θα διασπαστούν σε LSP χωρίς αυτό να μπορεί να διασπαστεί σε άλλα σωμάτια του MSSM.
- Ευσταθές. Δεν μπορεί να διασπαστεί από μόνο του σε σωμάτια του SM. Εγγύηση για αυτήν την υπόθεση μπορεί να αποτελέσει η υιοθέτηση της ισοτιμίας-$R$ που παραβιάζεται σε διαδικασίες διάσπασης ενός sparticle σε σωμάτια αμιγώς του SM, όπως εξηγείται στο Εδ. 2.3.2δ.
- Ασθενώς αλληλεπιδρόν. Οι διαδικασίες αλληλοκαταστροφής που ελλατώνουν την αριθμητική πυκνότητα του σωματίου γίνονται πολύ αργά.

Εν αρχή, υπήρχε το πυκνό και θερμό Σύμπαν. Το LSP βρισκόταν σε ΘΔΙ με τα άλλα σωμάτια του κοσμικού ρευστού αλληλεπιδρώντας μαζί τους με αμφίδρομες αντιδράσεις:

$$\text{LSP} + \text{LSP} \leftrightarrow X + Y \tag{5.5}$$

στις οποίες η $R$-ισοτιμία διατηρούνταν. Σε αυτές τις θερμοκρασίες, το LSP ήταν ένα ισχυρά σχετικιστικό σωμάτιο με πυκνότητα αριθμού να δίνεται από την Εξ. (4.53). Καθώς το Σύμπαν ψύχεται, η θερμοκρασία προοδευτικά ελαττώνεται και κάποια στιγμή πέφτει κάτω από τη μάζα του LSP, οπότε η πυκνότητα αριθμού των LSP αρχίζει να φθίνει εκθετικά, βάση της Εξ. (4.59). Οι αμφίδρομες αντιδράσεις της Εξ. (5.5) μετατρέπονται σε μονόδρομες καταστροφικές. Εποπτικά τα προηγούμενα συνοψίζονται στην επόμενη:

$$n_{\tilde{\chi}} \sim \begin{cases} T^3, & T \gg m_{\tilde{\chi}} \\ (m_{\tilde{\chi}} T)^{3/2} e^{-m_{\tilde{\chi}}/T}, & T \ll m_{\tilde{\chi}} \end{cases} \tag{5.6}$$

Αν τα δεδομένα του σεναρίου ήταν αυτά, το LSP θα ολοκλήρωνε το ζωτικό του κύκλο σβήνοντας αργά υπό την επίρροια του παράγοντα της MBS μέσα στο κοσμικό λουτρό. Από το άσημο αυτό τέλος, το LSP σώζεται για δύο λόγους:

- Είναι ασθενώς αλληλεπιδρόν σωμάτιο, οπότε οι αντιδράσεις καταστροφής οι οποίες το φθείρουν πραγματοποιούνται πολύ αργά.

- Κάποια στιγμή δραπετεύει από το κοσμικό λουτρό. Αυτό γίνεται όταν η συνθήκη αποσύνδεσης, Εξ. (4.74) αποκατασταθεί, σε θερμοκράσια $T_{\rm F}$, όπου $(m_{\tilde{\chi}}/T_{\rm F} \sim 25)$.

Μετά από την περιπέτεια αυτή, η αριθμητική πυκνότητα των LSP αρχίζει να φθίνει πολύ πιο αργά ($n_{\tilde{\chi}} \sim 1/R^3$). Επιπλέον η υπόθεση ευστάθειάς του, το καθιστά ανθεκτικό στο πέρασμα των αιώνων. Ψυχρό και σκοτεινό, όπως είναι, περνά απαρατήρητο. Αν μάλιστα, η CRD των απολιθωμάτων του μπορεί να επαληθεύσει τα CDM κοσμολογικά όρια, τότε το LSP γίνεται ένας επιτυχής CDM υποψήφιος. Αυτό, όμως δεν είναι πάντα ελεγχόμενο και μπορεί εύκολα η υποψηφιότητα να αποβεί ατελέσφορη, είτε γιατί υπερπηδά είτε γιατί δεν προσεγγίζει τα CDM κοσμολογικά όρια.

Πιο συναρπαστικό γίνεται το σενάριο, αν το LSP αποκτήσει συνεργό (ή και συνεργούς) στην επιχείρηση απόδρασής του από το κοσμικό λουτρό. Το ρόλο αυτό καλείται να διαδραμάτισει το NLSP. Μάλιστα, η συμμετοχή του γίνεται πρωταγωνιστική, αν αυτό διαθέτει συγγένεια μάζας με το LSP, αν δηλαδή η μάζα του $m_{NLSP}$ είναι περίπου ίση με την $m_{LSP}$. Η δραστηριότητα του NLSP εξελίσσεται σε δύο στάδια:

- Συνεισφέρει στην ανταλλαγή αριθμού σωματιδίων ανάμεσα στα σωμάτια του SM και του MSSM μέσω διαδικασιών του τύπου:
$$\mathrm{LSP} + X \leftrightarrow \mathrm{NLSP} + Y, \tag{5.7}$$
όπου τα $X, Y$ είναι σωμάτια του SM. Οι διαδικασίες αυτές συμβαίνουν συχνά σε θερμοκρασίες $T \sim T_{\rm F}$, διότι η ακτίνα αντίδρασης είναι ανάλογη μιας δύναμης του παράγοντα της MBS $\exp(-m_{\rm LSP}/T_{\rm F}) \simeq \exp(-25)$ για $m_{\rm LSP} \simeq m_{\rm NLSP}$ (τα περισσότερα $X$, $Y$ είναι σχετικιστικά σε θερμοκρασίες $T \sim T_{\rm F}$). Αντίθετα οι καταστροφές LSP-LSP συμβαίνουν λιγότερο συχνά, αφού η ακτίνα αντίδρασης είναι ανάλογη δύο δυνάμεων του ίδιου φάκτορα. Οι αλληλεπιδράσεις του τύπου της Εξ. (5.7) διατηρούν σχετική ισορροπία ανάμεσα στο LSP και στο NLSP για αρκετό χρόνο μετά την απόδραση όλων των MSSM σωματιδίων από το κοσμικό ρευστό.

- Συνεισφέρει στην εξέλιξη της αριθμητικής πυκνότητας των LSP μέσω των διαδικασιών συγγενικής καταστροφής (: CAE):

$$\mathrm{LSP} + \mathrm{NLSP}^{(*)} \leftrightarrow X + Y \;\; \text{και} \;\; \mathrm{NLSP} + \mathrm{NLSP}^{(*)} \leftrightarrow X + Y. \tag{5.8}$$

Προοδευτικά όλα τα NLSP, NLSP$^{(*)}$ θα διασπαστούν σε LSP (συν σωμάτια του SM). Η σημερινή αριθμητική πυκνότητα των LSP βρίσκεται με λύση της εξίσωσης Boltzmann. Αυτό θα είναι θέμα της επόμενης παραγράφου. Μια πρόγευση του απότελέσματος μπορεί να δοθεί στο σημείο αυτό ακολουθώντας την Αν. [57]. Κεντρικό ρόλο στην εξέλιξη της αριθμητικής πυκνότητας των LSP, παίζει η δρώσα ενεργός διατομή:

$$\sigma_{\rm eff} \sim \sigma(\mathrm{LSP} - \mathrm{LSP}) + g_{\rm NLSP}\sigma(\mathrm{LSP} - \mathrm{NLSP}) + (g_{\rm NLSP})^2 \sigma(\mathrm{NLSP} - \mathrm{NLSP}^{(*)}). \tag{5.9}$$

όπου έχουν διατηρηθεί μόνο οι ουσιώδεις παράγοντες της MBS με:

$$g_{\rm NLSP} \sim (1 + \Delta_{\rm NLSP})^{3/2} e^{-\Delta_{\rm NLSP} m_{\rm LSP}/T} \tag{5.10}$$

Επομένως, ρυθμιστικό ρόλο στην ένταση των CAE παίζει η σχετική διαφορά μάζας των δύο πρωταγωνιστών του σεναρίου:

$$\Delta_{NLSP} = (m_{NLSP} - m_{LSP})/m_{LSP} \tag{5.11}$$

και βεβαίως, οι εμπλεκόμενες ενεργές διατομές των CAE. Ώστε, η συνέργεια του NLSP στην απόδραση του LSP από το κοσμικό ρευστό είναι δραστική, όταν ισχύουν οι εξής συνθήκες (Αν. [57]):

$$\Delta_{\rm NLSP} \ll 1 \ \ \text{και} \ \ \sigma({\rm LSP}-{\rm NLSP})+\sigma({\rm NLSP}-{\rm NLSP}^{(*)}) \gg \sigma({\rm LSP}-{\rm LSP}). \tag{5.12}$$

Αυστηρότερη μαθηματική επεξεργασία και επέκταση των εννοιών που εισήχθηκαν εντελώς περιγραφικά στην παράγραφο αυτή γίνεται στο Εδ. 5.3. Το πλεονέκτημα μιας τέτοιας προκαταρκτικής παρουσίασης είναι η αποφυγή εμπλοκής σε μαθηματικά κατασκευάσματα που μερικές φορές συντείνουν στην απώλεια του νήματος.

## 5.3 Εναπομένουσα Πυκνότητα των LSP

Η θεμελίωση του βασικού τυπολόγιου που χρησιμοποιείται για τον υπολογισμό της εναπομένουσας πυκνότητας (: CRD) των LSP είναι ο σκοπός αυτής της ενότητας. Στο Εδ. 5.3.1 παρουσιάζεται η βασική εξίσωση που διέπει το πρόβλημα. Ακολουθείται η προσέγγιση των συγγραφέων της Αν. [58], στην οποία είσαγονται για πρώτη φορά οι CAE. Λαμβάνεται, επίσης, υπόψη η ανα-καίνιση του φορμαλισμού που επιχειρήθηκε στην Αν. [59]. Στο Εδ. 5.3.2 εισάγεται η έννοια της δρώσας ενεργού διατομής και στο Εδ. 5.3.3 αναδεικνύεται η εξίσωση που παρέχει τη θερμοκρασία ψύξης του LSP. Μια εύχρηστη τελική λύση επιτυγχάνεται στο Εδ. 5.3.5.

### 5.3.1 Η εξίσωση Boltzmann

Θεωρούμε την αλληλοκαταστροφή $N$ υπερσυμμετρικών σωματίων $\tilde{\chi}_i$ $(i=1,\ldots,N)$ με μάζες $m_i$, διατεταγμένες ως εξής $m_1 \leq m_2 \leq \cdots \leq m_{N-1} \leq m_N$ και εσωτερικούς βαθμούς ελευθερίας (στατιστικά βάρη) $g_i$. Προς αποφυγή συγχύσεως, επισημαίνεται ότι για τη μάζα του LSP θα χρησιμοποιείται το σύμβολο $m_1$ σε όλη τη διάρκεια της παρούσας ενότητας 5.3.

Η εξέλιξη της αριθμητικής πυκνότητας $n_i$ του σωματίου $i$ δίνεται από την εξίσωση Boltzmann που με βάση την Αν. [36], γράφεται:

$$\begin{aligned}\frac{dn_i}{dt} &= -3Hn_i-\sum_{j=1}^{N}\langle\sigma_{ij}v_{ij}\rangle\left(n_in_j-n_i^{\rm eq}n_j^{\rm eq}\right)\\ &-\sum_{j\neq i}\left[\langle\sigma'_{Xij}v_{ij}\rangle\left(n_in_X-n_i^{\rm eq}n_X^{\rm eq}\right)-\langle\sigma'_{Xji}v_{ij}\rangle\left(n_jn_X-n_j^{\rm eq}n_X^{\rm eq}\right)\right]\\ &-\sum_{j\neq i}\left[\Gamma_{ij}\left(n_i-n_i^{\rm eq}\right)-\Gamma_{ji}\left(n_j-n_j^{\rm eq}\right)\right].\end{aligned} \tag{5.13}$$

Ο πρώτος όρος στο δεύτερο μέλος οφείλεται στη διαστολή του σύμπαντος, ο δεύτερος περιγράφει τις αλληλοκαταστροφές και τις συγγενικές καταστροφές $\tilde{\chi}_i\tilde{\chi}_j$, των οποίων η ολική ενεργός διατομή είναι:

$$\sigma_{ij} = \sum_X \sigma(\tilde{\chi}_i\tilde{\chi}_j \to X). \tag{5.14}$$

Ο τρίτος όρος περιγράφει τις μεταλλαγές $\tilde{\chi}_i \to \tilde{\chi}_j$ έξω από το κοσμικό υπόβαθρο, με ενεργές διατομές:

$$\sigma'_{Xij} = \sum_Y \sigma(\tilde{\chi}_iX \to \tilde{\chi}_jY) \tag{5.15}$$

Ο τελευταίος όρος συμπεριλαμβάνει τις διασπάσεις των $\tilde{\chi}_i$ με ακτίνες διάσπασης:

$$\Gamma_{ij} = \sum_X \Gamma(\tilde{\chi}_i \to \tilde{\chi}_jX). \tag{5.16}$$

Στις προηγούμενες εκφράσεις, $X$ και $Y$ είναι σωμάτια (ή σύνολα από σωμάτια) του SM που μετέχουν στις διαδικασίες, $v_{ij}$ είναι η σχετική ταχύτητα των $i$ και $j$ και $n_i^{\rm eq}$ η αριθμητική πυκνότητα των σωματίων $\tilde{\chi}_i$ που λαμβάνεται απο την Εξ. (4.59) αφού σε θερμοκρασίες $T < m_i$ τα σωματία $\tilde{\chi}_i$ υπόκεινται σε MBS.

Κανονικά, οι διασπάσεις όλων των σωματίων, $\tilde{\chi}_i$ με $i \neq 1$ (αφού το 1 θεωρήθηκε ευσταθές) γίνονται γρηγορότερα από την ηλικία του σύμπαντος. Συνεπώς, έχοντας υποθέσει διατήρηση της $R$-parity, όλα αυτά τα σωμάτια διασπώνται στο LSP λόγω των διαδικασιών της Εξ. (5.16) . Επομένως, η τελική αριθμητική πυκνότητα του LSP θα εξαρτάται από το άθροισμα των πυκνοτήτων όλων των σωματίων του MSSM,

$$n = \sum_{i=1}^{N} n_i. \tag{5.17}$$

Για το $n$ λαμβάνεται η επόμενη εξίσωση εξέλιξης:

$$\frac{dn}{dt} = -3Hn - \sum_{i,j=1}^{N} \langle \sigma_{ij} v_{ij} \rangle \left( n_i n_j - n_i^{\rm eq} n_j^{\rm eq} \right) \tag{5.18}$$

όπου οι όροι στη δεύτερη και τρίτη γραμμή της Εξ. (5.13) αλληλοδιαγράφονται αθροιζόμενοι. Επόμενως, οι διαδικασίες στις Εξ. (5.15), (5.16), δεν συνεισφέρουν στην εξέλιξη της αριθμητικής πυκνότητας του LSP.

### 5.3.2 Δρώσα Ενεργός Διατομή

Η έννοια της δρώσας ενεργού διατομής συντείνει σε σημαντική απλοποίηση της εξίσωσης Boltzmann και μας ωθεί σε πορεία τελικής λύσης της. Σημαντικό ρόλο στο σημείο αυτό παίζουν οι διαδικασίες της Εξ. (5.15). Λόγω των διαδικασιών αυτών, όλα τα ελαφρώς βαρύτερα από το LSP σωμάτια του MSSM διατήρουνται σε σχετική ΘΔΙ με το LSP για αρκετό χρόνο μετά την αποσύνδεση του LSP από το κοσμικό ρευστό. Ποσοτικά αλλά προσεγγιστικά εκφραζόμενοι, επικαλούμενοι την Εξ. (4.59) για τα $\tilde{\chi}_i$ και τις Εξ. (4.50), (4.53) για τα $X$, επιτυγχάνουμε:

$$\begin{aligned} \sigma_{ij} n_i n_j &\sim T^3 m_i^{3/2} m_j^{3/2} \sigma_{ij} e^{-(m_i+m_j)/T}, & (5.19)\\ \sigma'_{Xij} n_i n_X &\sim T^{9/2} m_i^{3/2} \sigma'_{Xij} e^{-m_i/T}, & (5.20)\\ \text{οπότε } n_X/n_j &\sim (T/m_j)^{3/2} e^{-m_j/T} \sim 10^9 & (5.21) \end{aligned}$$

υποθέτωντας ότι οι $\sigma_{ij}$, $\sigma'_{Xij}$ δεν είναι δραματικά διαφορετικές. Αφού την θερμοκρασία αποσύνδεσης την ελέγχουν οι διαδικασίες στην Εξ. (5.14) επιτρέπεται να υποθέσουμε ότι οι λόγοι των αριθμητικών πυκνοτήτων ισουνται με τους αντιστοιχους σε ΘΔΙ:

$$\frac{n_i}{n} \simeq \frac{n_i^{\rm eq}}{n^{\rm eq}} \Longrightarrow n_i n_j - n_i^{\rm eq} n_j^{\rm eq} \simeq \frac{n_i^{\rm eq} n_j^{\rm eq}}{(n^{\rm eq})^2} \left( n^2 - (n^{\rm eq})^2 \right). \tag{5.22}$$

Αντικαθιστώντας στην Εξ. (5.22), λαμβάνεται η κατά πολύ απλούστερη έκφραση

$$\frac{dn}{dt} = -3Hn - \langle \sigma_{\rm eff} v \rangle \left( n^2 - (n^{\rm eq})^2 \right) \tag{5.23}$$

από την οποία ο ορισμός της δρώσας ενεργού διατομής αναδύεται αυθόρμητα

$$\sigma_{\rm eff} v = \sum_{i,j} \sigma_{ij} v_{ij} r_i(x) r_j(x) , \tag{5.24}$$

με τα νέα βολικά σύμβολα $r_i(x)$ οριζόμενα ως εξής:

$$r_i(x) := \frac{n_i^{\rm eq}}{n^{\rm eq}} \tag{5.25}$$

και υπολογιζόμενα κομψά από την επόμενη:

$$r_i(x) = \frac{g_i (1+\Delta_i)^{3/2} e^{-\Delta_i x}}{g_{\rm eff}} \quad \text{με} \quad x = m_1/T. \tag{5.26}$$

Στην Εξ.(5.26) έχουν εισαχθεί τα σύμβολα της σχετικής διαφοράς μάζας,

$$\Delta_i = (m_i - m_1)/m_1 \tag{5.27}$$

και του δρώντος στατιστικού βάρους,

$$g_{\rm eff}(x) = \sum_i g_i(1+\Delta_i)^{3/2} e^{-\Delta_i x}. \tag{5.28}$$

Η απόδειξη της εύχρηστης Εξ.(5.26) γίνεται αντικαθιστώντας στην Εξ.(5.25) τις εκφράσεις για τις πυκνότητες αριθμού σωματίων σε κατάσταση ισορροπίας στο κοσμικό ρευστό που εδράζονται στη μη σχετικιστική προσέγγιση. Επομένως, για κάθε σωμάτιο $i$, ισχύει:

$$n_i^{\rm eq}(x) = \frac{g_i}{(2\pi)^{3/2}} m_i^{3/2} m_1^{3/2} x^{-3/2} e^{-x m_i/m_1} \tag{5.29}$$

και διαιρώντας με την αντίστοιχη έκφραση για το σωμάτιο αναφοράς 1, λαμβάνεται:

$$\frac{n_i^{\rm eq}(x)}{n_1^{\rm eq}(x)} = \frac{g_i}{g_1}(\frac{m_i}{m_1})^{3/2} x^{-3/2} e^{-x\Delta_i} \Longrightarrow$$

$$n_i^{\rm eq}(x) = \frac{g_i}{(2\pi)^{3/2}}(1+\Delta_i)^{3/2} e^{-\Delta_i x} m_1^3 x^{-3/2} e^{-x} \tag{5.30}$$

Αθροίζοντας τις Εξ. (5.30), λαμβάνεται η πυκνότητα αριθμού όλων των σωματίων:

$$n^{\rm eq}(x) = \frac{g_{\rm eff}}{(2\pi)^{3/2}} m_1^3 x^{-3/2} e^{-x} \tag{5.31}$$

οπότε διαιρώντας την Εξ. (5.30) με την Εξ. (5.31), επιτυγχάνεται η αποδεικτέα Εξ. (5.26).

### 5.3.3 Θερμοκρασία Αποσύνδεσης των LSP

Η θερμοκρασία αποσύνδεσης (ή ψύξης, αφού πρόκειται για ψυχρά απομεινάρια) των LSP, $T_{\rm F}$ ορίζεται ως εκείνη στην οποία τα LSP απομακρύνονται από τη κοσμικό ρευστό. Στη συνθήκη αποδεύσμεσης, Εξ. (4.74)

$$\Gamma(x_{\rm F}) \simeq H(x_{\rm F}) \quad \text{όπου} \quad x_{\rm F} = m_1/T_{\rm F} \tag{5.32}$$

αντικαθίστανται:

- Η οριστική έκφραση της ακτίνας αλληλεπίδρασης, Εξ. (4.73) που παίρνει τη μορφή:

  $$\Gamma(x_{\rm F}) = n^{\rm eq}(x_{\rm F})\langle\sigma_{\rm eff} v\rangle(x_{\rm F}), \tag{5.33}$$

  όπου η αριθμητική πυκνότητα των LSP στην ΘΔΙ βρίσκεται εφαρμόζοντας την Εξ. (4.59):

  $$n^{\rm eq}(x_{\rm F}) = \frac{g_{\rm eff}}{(2\pi)^{3/2}} m_1^3 x_{\rm F}^{-3/2} e^{-x_{\rm F}}. \tag{5.34}$$

- Η παράμετρος Hubble, που για ένα κυριαρχούμενο από ακτινοβολία σύμπαν, γράφεται συνδυάζοντας τις Εξ. (4.76) και (4.65):

  $$H(x_{\rm F}) = \sqrt{\frac{8\pi\rho_{RD}}{3M_P^2}} = 2\pi\sqrt{\frac{\pi}{45}}\frac{g_*^{1/2}}{M_P}\frac{m_1^2}{x_{\rm F}^2} \tag{5.35}$$

Η κατάληξη των αντικαταστάσεων καταγράφεται αδρά στις παρακάτω γραμμές:

$$\frac{2^{5/2}\pi^3}{\sqrt{45}}\frac{g_*^{1/2}}{M_P} = m_1 x_{\rm F}^{1/2} e^{-x}\langle\sigma_{\rm eff} v\rangle(x_{\rm F}) \tag{5.36}$$

$$\Longrightarrow x_{\rm F} = \ln\frac{0.0382435\, g_{\rm eff}(x_{\rm F})\, M_P\, m_1\, \langle\sigma_{\rm eff} v\rangle(x_{\rm F})}{g_*^{1/2} x_{\rm F}^{1/2}}\,. \tag{5.37}$$

Η τελική εξίσωση μπορεί να λυθεί αριθμητικά. Οι τιμές που επιτυγχάνονται για τη θερμοκρασία αποδέσμευσης είναι της τάξης $m_1/25$ και επομένως, η μη σχετικιστική προσέγγιση που επιλέχθηκε στην Εξ.(5.26) είναι τουλάχιστον αυτοσυνεπής.

### 5.3.4 Προσεγγιστική λύση της Boltzmann

Η επίτευξη προσεγγιστικής λύσης απαιτεί τον ανασχηματισμό της μορφής της εξίσωσης Boltzmann, Εξ. (5.23), ώστε ως ανεξάρτητη μεταβλήτη να τεθεί η θερμοκρασία. Με αυτόν ως απώτερο σκοπό, ορίζεται:

$$Y = \frac{n}{s} \tag{5.38}$$

οπότε η διατήρηση της εντροπίας $\frac{d}{dt}(s(T)R^3) = 0$ συνεπάγεται δύο νέες εξισώσεις:

- Μία που συνδέει τη νέα παράμετρο με την τρέχουσα πυκνότητα αριθμού σωματίων

$$\frac{d}{dt}(\frac{n}{Y}R^3) = 0 \Longrightarrow \dot{n} + 3Hn = s\dot{Y} \tag{5.39}$$

Με βάση την προηγούμενη η Εξ. (5.23) μπορεί να γραφει στη μορφή:

$$\dot{Y} = -s\langle\sigma_{\rm eff}v\rangle \left(Y^2 - Y_{\rm eq}^2\right). \tag{5.40}$$

- Μία που εξασφαλίζει την εισαγωγή της θερμοκρασίας στην αλληλουχία των εξισώσεων

$$\dot{T} = -3\frac{\dot{R}}{R}\frac{s(T)}{s'(T)}. \tag{5.41}$$

Εκμεταλευόμενοι αυτή τη δεύτερη συνέπεια της ισεντροπικής κοσμικής διαστολής,

$$\dot{Y} = \frac{dY}{dx}\frac{dx}{dt} = -\frac{dY}{dx}x\frac{\dot{T}}{T} = 3H\frac{dY}{dx}\frac{x^2}{m_{\tilde{\chi}}}\frac{s(T)}{s'(T)} \tag{5.42}$$

Συνδυάζοντας τα προηγούμενα κεκτημένα, επιτυγχάνεται

$$\frac{dY}{dx} = -\frac{m_1}{x^2}\frac{s'}{3H}\langle\sigma_{\rm eff}v\rangle \left(Y^2 - Y_{\rm eq}^2\right). \tag{5.43}$$

Σε ένα σύμπαν κυριαρχούμενο από ακτινοβολία, ισχύουν οι Εξ. (4.76) και (4.66) οπότε αντικαθιστώντας τη ποσότητα $s'/3H$, η Εξ. (5.43) τελικά γράφεται:

$$\frac{dY}{dx} = -\sqrt{\frac{\pi}{45}}\frac{g_*^{1/2}m_1}{x^2}M_P\langle\sigma_{\rm eff}v\rangle \left(Y^2 - Y_{\rm eq}^2\right) \tag{5.44}$$

Μετά την αποδέσμευση των LSP, η τρέχουσα πυκνότητα αριθμού σωματίων είναι αυξημένη σε σχέση με αυτή που είναι σε ισορροπία η οποία για $T >> m_1$ φθίνει ταχέως λόγω του εκθετικού που οφείλεται στην MBS. Επομένως, η προσέγγιση

$$n^2 - (n^{eq})^2 \simeq n^2 \Longrightarrow Y^2 - (Y_{eq})^2 \simeq Y^2$$

είναι καλά αιτιολογημένη. Επίσης, λαμβάνεται το $g_*$ να είναι πρακτικά σταθερό στην τιμή που αποκτά τη στιγμή της αποδέσμευσης. Οπότε η ολοκλήρωση της Εξ. (5.44) από $x_{\rm F}$ μέχρι $x_0 = m_1/T_0 \simeq \infty$ όπου $T_0$ είναι η σημερινή θερμοκρασία των φωτονίων στο σύμπαν, καθίσταται ένα τετριμμένο εγχείρημα με αποτέλεσμα την σημερινή τιμή του λόγου $Y_0 = Y(\infty)$, που είναι:

$$Y_0 = -\left(\sqrt{\frac{\pi}{45}}g_*^{1/2}m_1 M_P x_{\rm F}^{-1}\, J_{\rm eff}\right)^{-1}, \tag{5.45}$$

όπου χρησιμοποιήθηκε η προσέγγιση $Y(\infty) - Y(x_{\rm F}) \simeq Y(\infty)$ (και πάλι λόγω της MBS) και ορίστηκε η ποσότητα:

$$J_{\rm eff} := x_{\rm F}\int_{x_{\rm F}}^{\infty}\langle\sigma_{\rm eff}v\rangle x^{-2}dx\,, \tag{5.46}$$

η χρησιμότητα της οποίας φαίνεται στην αριθμητική επεξεργασία των αποτελεσμάτων.

### 5.3.5 Τύπος Υπολογισμού της CRD των LSP

Από τον ορισμό της παραμέτρου πυκνότητας, στην Εξ. (4.23) και της βοηθητικής ποσότητας $Y$ στην Εξ. (5.38), λαμβάνεται:

$$\Omega_{LSP} = \rho_{1_0}/\rho_{c_0} \;\; \mu\varepsilon \;\; \rho_{1_0} = m_1 s_0 Y_0 \tag{5.47}$$

όπου χρησιμοποιήθηκε προφανώς η Εξ. (4.60). Αντικαθιστώντας τις σημερινές τιμές για την κρίσιμη πυκνότητα, $\rho_{c_0}$ και την πυκνότητα εντροπίας $s_0$ από τις Εξ. (4.87), (4.91) και το αποτέλεσμα για το $Y_0$ από την Εξ. (5.45), προκύπτει η τελική έκφραση:

$$\Omega_{LSP}\, h^2 \simeq \frac{1.06647 \times 10^9 \text{ GeV}^{-1}}{g_*^{1/2}(x_{\rm F})\, M_P\, x_{\rm F}^{-1}\, J_{\rm eff}} \tag{5.48}$$

Για ένα τυπικό LSP μάζας 200 GeV και για $x_{\rm F} \simeq 20$, προκύπτει ότι $m_b < T_{\rm F} < M_W$, οπότε από τον Πίνακα 4.2 συνάγεται οτι $g_*(x_{\rm F}) \simeq 86.25$ που αντιστοιχεί σε χρόνο $t_{\rm F} \simeq 2.6 \times 10^{-9}$ sec, μέσω της Εξ. (4.77).

Η απλή Εξ. (5.48) που επιτεύχθηκε δεν παύει να είναι προσεγγιστική και ισχύει με μια ακρίβεια περίπου 5% όπως ισχυρίζονται οι εμπνευστές της στην Αν. [36]

## 5.4 Μέθοδος υπολογισμού της $\langle\sigma_{ij}v_{ij}\rangle$

Ο φορμαλισμός που αναπτύχθηκε στην προηγούμενη παράγραφο θα απέβαινε ανενεργός, αν δεν συνοδευόταν με έναν αποτελεσματικό και εύχρηστο αλγόριθμο υπολογισμού της ποσοτήτων $< \sigma_{ij}v_{ij} >$ που θα μπορούσε να εφαρμοστεί σε μια πληθώρα διαδικασίων Συγγενικής Καταστροφής, κατά βάση. Υποβοηθητικό στοιχείο στην κατεύθυνση αυτή είναι η χαμηλή θερμοκρασία αποσύνδεσης των LSP που επιτρέπει την ανάπτυξη της ποσότητας $\sigma_{ij}v_{ij}$ σε δυνάμεις της $v_{ij}$ ως εξής:

$$\sigma v_{ij} = a_{ij} + b_{ij} v_{ij}^2 + \cdots \tag{5.49}$$

Η ανάπτυξη αυτή βεβαίως διευκολύνει και την εύρεση της θερμικής μέσης τιμής της επίμαχης ποσότητας. Επομένως, το πρόβλημα μετατίθεται στην ανάδειξη μεθόδου εξαγωγής των συντελεστών $a_{ij}$ και $b_{ij}$. Σε αυτό το θέμα είναι αφιερώμενη η επόμενη υποπαράγραφος. Στη μεθεπόμενη, εξάγεται η θερμική μέση τιμή.

Χάριν απλότητας, κατα τη διάρκεια θεμελίωσης της μεθόδου, θα αποσυρθούν οι δείκτες από την ενεργό διατομή $\sigma_{ij}$, την ταχύτητα $v_{ij}$ και στους συντελεστές $a_{ij}$ και $b_{ij}$. Εννοείται φυσικά, ότι τα αποτελέσματα χρησιμοποιούνται για κάθε διαδικασία αλληλοκαταστροφής ή συγγενικής καταστροφής σωματίων, με ορισμένες αρχικές και τελικές καταστάσεις. Οι συντελεστές $a_{ij}$ και $b_{ij}$ προκύπτουν με συνυπολογισμό όλων των διαδικασιών που αντιστοιχούν στις ίδιες αρχικές καταστάσεις $ij$.

### 5.4.1 Ανάπτυξη του $\sigma v_{\rm rel}$ σε δυνάμεις του $v_{\rm rel}$

Ακολουθείται η μέθοδος που αναπτύσσεται από το J. Wells στην Αν. [60]. Επεκτείνεται, όμως για να συμπεριλάβει και διαδικασίες στις οποίες τα δύο αλληλεπιδρώντα αρχικά σωμάτια δεν έχουν την ίδια μάζα.

Έστω $m_1$, $m_2$ οι μάζες των εισερχομένων σωματίων ($m_1 \leq m_2$) με τετραορμές $p_1, p_2$ και ενέργειες $E_1$, $E_2$ και $m_3$, $m_4$ οι μάζες των εξερχομένων σωματίων με τετραορμές $p_3$, $p_4$ και ενέργειες $E_3$, $E_4$ αντίστοιχα. Στο σύστημα αναφόρας κέντρου μάζας, ισχύει:

$$p_1(E_1, \vec{p}_{\rm i}) \;,\; p_2(E_2, -\vec{p}_{\rm i}) \tag{5.50}$$

$$p_3(E_3, \vec{p}_{\rm o}) \;,\; p_4(E_4, -\vec{p}_{\rm o}) \tag{5.51}$$

όπου τα μέτρα των εισερχομένων $|\vec{p}_{\rm i}|$ και εξερχομένων $|\vec{p}_{\rm o}|$ ορμών είναι:

$$|\vec{p}_{\rm i}| = \frac{m_1 v_{\rm cm}}{\sqrt{1 - v_{\rm cm}^2}} \;\; \text{και} \;\; |\vec{p}_{\rm o}| = \frac{\sqrt{(s - (m_3 + m_4)^2)(s - (m_3 - m_4)^2)}}{2\sqrt{s}} \tag{5.52}$$

με $v_{\rm cm}$ τη ταχύτητα του σωματίου μάζας $m_1$ ως προς το σύστημα αναφόρας κέντρου μάζας, $v_{\rm cm} = |\vec{p}_{\rm i}|/E_1$ και $s$, $t$, $u$ οι μεταβλητές Maldestam ($s + t + u = m_1^2 + m_2^2 + m_3^2 + m_4^2$). Η διαφορική ενεργός διατομή στο ενλόγω σύστημα αναφόρας, είναι:

$$\frac{d\sigma}{d\Omega} = \frac{|\mathcal{A}|^2}{64\pi^2 s} \frac{|\vec{p}_{\rm o}|}{|\vec{p}_{\rm i}|}. \tag{5.53}$$

Θέτωντας

$$J(s,t) = |\mathcal{A}|^2 \quad \text{και} \quad K(s) = \frac{|\vec{p}_o|}{16\pi m_1 s}, \tag{5.54}$$

η επίμαχη ποσότητα μπόρει να γραφεί:

$$v_{\rm cm}\frac{d\sigma}{d\Omega} = \frac{J(s,t)}{4\pi}K(s)\sqrt{1-v_{\rm cm}^2} \tag{5.55}$$

με $\theta_c$ τη γωνία σκέδασης των σωματίων στο θεωρούμενο σύστημα αναφοράς. Όπως φαίνεται από την προηγούμενη, η εύρεση των συντελεστών $a$ και $b$ απαιτεί την κατά Taylor ανάπτυξη των ποσοτήτων $J(s,t)$, $K(s)$ γύρω από τα $s_0$, $t_0$.

Αρχικά, γίνεται η ανάπτυξη των δύο μεταβλητών Mandelstam:

$$s := (p_1+p_2)^2 = (E_1+E_2)^2 = s_0 + s_2 v_{\rm cm}^2 + \mathcal{O}(v_{\rm cm}^4) \tag{5.56}$$

όπου

$$s_0 = (m_1+m_2)^2 \quad \text{και} \quad s_2 = 2m_1^2 + \frac{m_1}{m_2}(m_1^2+m_2^2). \tag{5.57}$$

Επίσης,

$$\begin{aligned} t: &= (p_1-p_3)^2 = m_1^2 + m_3^2 - 2E_1E_3 + |\vec{p}_i||\vec{p}_o|\cos\theta_c && (5.58)\\ &= t_0 + t_1\cos\theta_c v_{\rm cm} + t_2 v_{\rm cm}^2 + \mathcal{O}(v_{\rm cm}^3) && (5.59)\end{aligned}$$

όπου

$$t_0 = \frac{-m_1^2 m_2 - m_1 m_2^2 + m_2 m_3^2 + m_1 m_4^2}{m_1+m_2} \tag{5.60}$$

$$t_1 = \frac{m_1}{m_1+m_2}\mu_1^2 \tag{5.61}$$

$$t_2 = \frac{m_1}{2m_2(m_1+m_2)}\mu_2^2 \tag{5.62}$$

με

$$\mu_1^2 = \sqrt{(m_1+m_2)^2-(m_3-m_4)^2}\sqrt{(m_1+m_2)^2-(m_3+m_4)^2} \tag{5.63}$$

$$\mu_2^2 = (m_1+m_2)^3 + (m_1-m_2)(m_4^2-m_3^2). \tag{5.64}$$

Από την Εξ. (5.63) προκύπτει ο κινηματικός περιορισμός που πρέπει να ικανοποιείται για την πραγματοποίηση μιας διαδικασίας: $(m_1+m_2) \geq (m_3+m_4)$.

Ακολούθως, αναπτύσσεται το $K(s)$ κατά Taylor:

$$K(s) = K_0 + K_2 v_{\rm cm}^2 + \mathcal{O}(v_{\rm cm}^4) \tag{5.65}$$

όπου

$$K_0 = \frac{\mu_1^2}{32\pi m_1(m_1+m_2)^3} \tag{5.66}$$

$$K_2 = \frac{-(m_1+m_2)^4 - 3(m_3^2-m_4^2)^2 + 4(m_1+m_2)^2(m_3^2+m_4^2)}{64\pi m_2(m_1+m_2)^3\mu_1^2}. \tag{5.67}$$

Αναπτύσσοντας και το $J(s,t)$ γύρω από το $s=s_0$ και $t=t_0$, λαμβάνεται:

$$\begin{aligned} J(s,t) &= J(s_0,t_0) + \frac{\partial J(s_0,t_0)}{\partial s}(s-s_0) + \frac{\partial J(s_0,t_0)}{\partial t}(t-t_0)\\ &+\frac{1}{2}\frac{\partial^2 J(s_0,t_0)}{\partial s^2}(s-s_0)^2 + \frac{1}{2}\frac{\partial^2 J(s_0,t_0)}{\partial t^2}(t-t_0)^2 + \frac{\partial^2 J(s_0,t_0)}{\partial s\partial t}(s-s_0)(t-t_0) + \cdots \end{aligned} \tag{5.68}$$

Το κρίσιμο σημείο εδώ είναι ότι το $J(s_0,t_0)$ και οι παράγωγοι $\partial^{m+n}J_0/\partial s^m \partial t^n$ δεν εξαρτώνται από τη γωνία $\cos\theta_c$. Μόνο οι δυνάμεις $(t-t_0)^n$ εξαρτώνται από τη γωνία $\cos\theta_c$. Η εξάρτηση από τις περιττές δυνάμεις καταπνίγεται, αφού $\int d\Omega \cos\theta_c = \int d\Omega \cos^3\theta_c = 0$, οπότε απομένει η εξάρτηση από τη δεύτερη δύναμη που δίνει $\int d\Omega \cos^2\theta_c = 4\pi/3$. Το αποτέλεσμα της ολοκλήρωσης της ποσότητας $J(s,t)/4\pi$ στο όριο χαμηλής ταχύτητας είναι:

$$\int d\Omega \frac{J(s,t)}{4\pi} = J_0 + J_2 v_{\rm cm}^2 + \mathcal{O}(v_{\rm cm}^4) \tag{5.69}$$

όπου

$$J_0 = J(s_0,t_0), \tag{5.70}$$

$$J_2 = \frac{\partial J(s_0,t_0)}{\partial s}s_2 + \frac{\partial J(s_0,t_0)}{\partial t}t_2 + \frac{1}{6}\frac{\partial^2 J(s_0,t_0)}{\partial t^2}t_1^2. \tag{5.71}$$

Τελικά, εισάγοντας τις αναπτυγμένες μορφές στην εξ.(5.55) και κρατώντας όρους μεχρι δεύτερης τάξης ως προς την $v_{\rm cm}$ λαμβάνεται:

$$\sigma v_{\rm cm} = a_{\rm cm} + b_{\rm cm} v_{\rm cm}^2 \tag{5.72}$$

όπου

$$a_{\rm cm} = J_0 K_0, \tag{5.73}$$

$$b_{\rm cm} = J_0 K_2 - \frac{1}{2} J_0 K_0 + J_2 K_0. \tag{5.74}$$

Η μετάβαση από την ταχύτητα του σωματίου μάζας $m_1$ ως προς το σύστημα αναφόρας κέντρου μάζας, $v_{\rm cm}$ στη σχετική ταχύτητα $v_{\rm rel}$ γίνεται αξιοποιώντας τον ορισμό της Αν. [38]:

$$v_{\rm rel} = \frac{\sqrt{(p_1p_2)^2 - m_1^2 m_2^2}}{E_1 E_2} \tag{5.75}$$

οπότε αντικαθιστώντας

$$(p_1p_2)^2 - m_1^2 m_2^2 = (E_1 + E_2)^2 |\vec{p}_{\rm i}|^2$$

και αναπτύσσοντας κατά Taylor , λαμβάνεται:

$$v_{\rm rel} = (\frac{m_1+m_2}{m_2}) v_{\rm cm} + \frac{m_1(-m_1^2+m_2^2)}{2m_2^3} v_{\rm cm}^3 + \mathcal{O}(v_{\rm cm}^5) \tag{5.76}$$

Συνεπώς, οι συντελεστές του αναπτύγματος:

$$\sigma v_{\rm rel} = a + b v_{\rm rel}^2 \tag{5.77}$$

είναι:

$$a = \frac{m_1+m_2}{m_2} a_{\rm cm} \quad \text{και} \quad b = \frac{m_2}{m_1+m_2} b_{\rm cm}. \tag{5.78}$$

Το μόνο αδιευκρίνιστο βήμα, λοιπόν, στον υπολογισμό των $a$ και $b$ είναι η εύρεση του πλάτους $|\mathcal{A}|$. Η διαδικασία αυτή βασίζεται σε γνώσεις εφαρμογής των κανόνων Feynman και άλγεβρας αντικειμένων σπινοριακού χαρακτήρα. Μέθοδος, όμως, δεν μπορεί να δοθεί αφού οι τεχνικές ποικίλουν κατα περίπτωση. Περισσότερο φως σε αυτό το σημείο του υπολογισμού επιχειρείται να δοθεί στο Παράρτημα Β′ όπου αναφέρονται οι σχετικοί κανόνες Feynman και εκτίθενται τρία απλά αλλά αντιπροσωπευτικά παραδείγματα.

### 5.4.2 Εξαγωγή της θερμικής μέσης τιμής του $\sigma v_{\rm rel}$

Η χρησιμότητα των συντελεστών $a$ και $b$ φαίνεται από την εμπλοκή τους στην έκφραση της θερμικής μέσης τιμής του $\sigma v_{\rm rel}$. Αναλυτικότερα, από τον ορισμό της θερμικής μέσης τιμής, ισχύει (αποφεύγεται για λίγο ακόμα η χρήση δεικτών)

$$< \sigma v_{\rm rel} > (T) = \frac{\int \sigma v_{\rm rel} e^{-E_1/T} e^{-E_2/T} d^3p_1 d^3p_2}{\int e^{-E_1/T} e^{-E_2/T} d^3p_1 d^3p_2}, \tag{5.79}$$

όπου $p_1$ ($E_1$) και $p_2$ ($E_2$) οι τετραορμές και οι ενέργειες των συγκρουόμενων σωματιδίων στο κοσμικό σύστημα αναφοράς ηρεμίας και $T$ η θερμοκρασία. Χρησιμοποιώντας τη μη σχετικιστική προσέγγιση της Εξ. (4.57), η παραπάνω γράφεται σε πιο χρηστική μορφή

$$< \sigma v_{\rm rel} > (T) = \frac{\int \sigma v_{\rm rel} e^{-p_1^2/2m_1T} e^{-p_2^2/2m_2T} d^3p_1 d^3p_2}{\int e^{-p_1^2/2m_1T} e^{-p_2^2/2m_2T} d^3p_1 d^3p_2}. \tag{5.80}$$

Αντικαθιστώντας στη συνέχεια, την ανεπτυγμένη μορφή της $\sigma v_{\rm rel}$ αναπτύσσοντας και τη σχετική ταχύτητα σε δυνάμεις των $p_1$ και $p_2$, ($v_{\rm rel}^2 \simeq (p_1/m_1 - p_2/m_2)^2$) και αλλάζοντας μεταβλητές από τις $p_1$ και $p_2$ στις $p_{\rm rel} = p_1 - p_2$, $p_T = p_1 + p_2$ το αποτέλεσμα της ολοκλήρωσης προκύπτει εύκολα. Εισάγοντας και δείκτες στην τελική έκφραση αυτή παίρνει την εξής μορφή:

$$< \sigma_{ij} v_{ij} > (x) = a_{ij} + \left(6 + (-12 + \frac{8}{\pi})\Delta_{ij} + \mathcal{O}(\Delta_{ij}^2)\right) b_{ij}/x, \tag{5.81}$$

όπου $\Delta_{ij} = (m_j - m_i)/m_i$ η ανηγμένη διαφορά μάζας των δύο αλληλεπιδρώντων σωματίων, $i$,$j$. Προφανώς, οι όροι της προηγούμενης εξίσωσης που είναι ανάλογοι του $\Delta_{ij}$ μπορούν να αμεληθούν, γιατί οι τιμές που λαμβάνει το $\Delta_{ij}$ είναι μικρότερες του 0.25, πράγμα που απαιτείται για να ειναι σημαντική και η συγγενική καταστροφή. Επομένως, το τελικό αποτέλεσμα παίρνει τη εξής σύντομη μορφή:

$$< \sigma_{ij} v_{ij} >= a_{ij} + 6b_{ij}/x \tag{5.82}$$

### 5.4.3 Αξιολόγηση των συνεισφορών στη δρώσα ενεργό διατομή

Η συνεισφορά των διαφόρων διαδικασιών και ξεχωριστά των συντελεστών $a_{ij}$ και $b_{ij}$ στο τελικό αποτέλεσμα εκτιμάται καλύτερα με ομαδοποίηση αυτών που έχουν σταθερό ολοκληρωτικό παράγοντα στην ποσότητα $J_{\rm eff}$. Πιο συγκεκριμένα, χρησιμοποιώντας τις Εξ. (5.24), (5.46) και (5.82), επιτυγχάνεται:

$$J_{\rm eff} = \sum_{(ij)} J_{(ij)}\ , \quad \text{όπου} \quad J_{(ij)} := \sum_{(ij)} (J^a_{(ij)} a_{ij} + J^b_{(ij)} b_{ij}) \tag{5.83}$$

και η άθροιση λογίζεται σε ομάδες αρχικών καταστάσεων $(ij)$ με ίδια τιμή στους ολοκληρωτικούς παράγοντες:

$$J^a_{(ij)} = c_{(ij)} x_{\rm F} \int_{x_{\rm F}}^{\infty} \frac{dx}{x^2} r_i(x) r_j(x)\ , \tag{5.84}$$

$$J^b_{(ij)} = 6c_{(ij)} x_{\rm F} \int_{x_{\rm F}}^{\infty} \frac{dx}{x^3} r_i(x) r_j(x) \cdot \tag{5.85}$$

Τα $c_{(ij)}$ είναι σταθερές που εκφράζουν τον αριθμό που εμφανίζεται η ενεργός διατομή μιας διαδικασίας τύπου $(ij)$ στο άθροισμα της Εξ. (5.24). Είναι αξιοσημείωτο ότι για $\Delta_{1j} = 0$ ο συσχετισμός των ολοκληρωτικών παραγόντων λαμβάνει την εξής απλή μορφή (αφού τα $r_i(x)$ γίνονται ανεξάρτητα του $x$):

$$J^b_{(ij)} = 3J^a_{(ij)}/x_{\rm F} \tag{5.86}$$

Από τις εκφράσεις που επιτεύχθηκάν πιο πάνω μπορούν να γίνουν οι παρακάτω παρατηρήσεις σχετικά με τις επιμέρους συνεισφορές στο $J_{\rm eff}$:

- Η συνεισφορά των διαδικασίων συγγενικής καταστροφής είναι σημαντική μόνο για μικρά $\Delta_{1j} < 0.25$ λόγω των υπαρχόντων εκθετικών στα ολοκληρώματα των Εξ. (5.85) που φθίνουν ταχέως.
- Από τις διαδικασίες συγγενικής καταστροφής, οι πιο σημαντικές ποσοτικά είναι οι τύπου $(1i)$, με $i > 1$ λόγω της ύπαρξης μόνο ενός εκθετικού στα ολοκληρώματα των Εξ. (5.85)
- Η συνεισφορά των διαδικασίων συγγενικής καταστροφής εφράζεται κυρίως μέσω των συντελεστών $a_{ij}$ αφού η συνεισφορά των $b_{ij}$ είναι συμπιεσμένη κατά ένα σχετικό παράγοντα τουλάχιστον 1/7, όπως εναργώς προκύπτει από την Εξ. (5.86).
- Όταν οι διαδικασίες συγγενικής καταστροφής παύουν να δρουν αποικοδομητικά, $\Delta_{1j} > 0.25$, η διαδικασία αλληλοκαταστροφής, 11 γίνεται κυρίαρχη για το τελικό αποτέλεσμα.

## 5.5 Εφαρμογή σε LSP μορφής Bino

Η ελαφρότατη ΙΔΚ μάζας του πίνακα των neutralino μπορεί να γραφεί ως γραμμικός συνδυασμός των ΙΔΚ βαθμίδας από τις οποίες προέρχεται ως εξής:

$$\tilde{\chi} = N_{11}(-i\tilde{B}) + N_{21}(-i\tilde{W}_3) + N_{31}\tilde{H}_0 + N_{41}\tilde{\bar{H}}_0. \tag{5.87}$$

Οι συντελεστές $N_{11} - N_{14}$ είναι τα στοιχεία του πίνακα $N$ που διαγωνοποιεί τον πίνακα των neutralino κατά την Εξ. (2.72). Η ταξινόμηση της φύσης ενός neutralino γίνεται μέσω του μεγέθους της καθαροτητάς του που ορίζεται ως εξής:

$$P := N_{11}^2 + N_{21}^2.$$

Αν $P > 0.9$, το neutralino είναι μορφής gaugino, ενώ αν $P < 0.1$ είναι μορφής higgsino και αν $0.1 \leq P \leq 0.9$ είναι μικτής μορφής. Από αυτή την ποικιλία περιπτώσεων, το neutralino στο οποίο είναι αφιερωμένη η μελέτη της διατριβής αυτής είναι όχι μόνο μορφής gaugino αλλά ακόμα ειδικότερα, μορφής Bino, που σημαίνει ότι:

$$N_{11} \simeq 1, \quad N_{21} \simeq N_{31} \simeq N_{41} \simeq 0. \tag{5.88}$$

Δεν είναι τυχαία η επιλογή αυτή. Αρχικά, πολλές είναι οι μαρτυρίες (Αν. [46]) που συγκλίνουν στο συμπέρασμα οτι δεν είναι επιτυχής υποψήφιος για τη λύση του προβλήματος της CDM ένα LSP μορφής:

- Higgsino, γιατί η μείωση που υφίσταται η CRD του, λόγω των ΑΝΕ σε ζεύγη $W$, $Z$ είναι κατακλυσμιαία, με αποτέλεσμα να μην επιτυγχάνονται τα CDM όρια για μάζες του LSP μικρότερες του .5TeV, ακόμα και χωρίς την παρουσία των CAE σύμφωνα με την Αν. [61].
- $W_3$, γιατί η CRD υφίσταται δραστική ταπείνωση λόγω των αναπόφευκτων CAE με το σχεδόν εκφυλισμένο ελαφρότατο chargino, οπότε η μάζα του LSP πρέπει να γίνει μεγαλύτερη του 1TeV για να γίνει ενδιαφέρον για τα CDM σενάρια.

Επιπλέον, η χρησιμοποίηση παγκόσμιων αρχικών συνθηκών για τους όρους ασθένους παραβίασης της υπερσυμμετρίας, σαν και αυτές που θα χρησιμοποιηθούν στα επόμενα κεφάλαια, οδηγεί με μεγάλη πιθανότητα σε LSP μορφής Bino σύμφωνα με την Αν. [62]. Κυρίως όταν και η παράμετρος $\tan\beta$ παίρνει μεγάλες τιμές, η υπόθεση μας είναι ιδιαιτέρως δικαιολογημένη. Για μικρότερες τιμές της $\tan\beta$ παρατηρήθηκαν αποκλίσεις από την συμπεριφορά αυτή οπότε και θα γίνεται άμεση αναφορά στο μειονέκτημα της υπόθεσής μας.

Μία ακόμα συνθήκη που πρέπει να ισχύει ώστε η αρχική μας υπόθεση στην Εξ. (5.88) να μην αλλοιώνει τη γενικότητα των αποτελεσμάτων μας, είναι να φυλάσσεται αρκετή απόσταση της μάζας του LSP από το διπλάσιο των μαζών που εμφανίζονται στους διαδότες των διαδικασιών μέσω του διαύλου $s$. Ποσοτικά εκφραζόμενοι, πρέπει:

$$m_{\tilde{\chi}} \gg [\ll] \; M_Z/2, \, m_h/2, \, m_H/2, \, m_A/2 \tag{5.89}$$

Αυτές οι περιοχές των πόλων γενικά είναι απομονωμένες στον παραμετρικό χώρο των προτύπων που θα εξεταστούν και για αυτό, ο δεύτερος αυτός περιορισμός που πρέπει να ισχύει για να είναι εφαρμόσιμα τα αποτελεσματά μας δεν επιρρεάζει αισθητά την εξαγωγή γενικών συμπερασμάτων.

Όπως φαίνεται από την Εξ. (5.48) για τον υπολογισμό της CRD των LSP απαιτείται ο υπολογισμός των Ενεργών Διατομών των ΑΝΕ και CAE. Σε αυτά τα θέματα είναι αφιερωμένα τα Εδ. 5.5.1, 5.5.2, 5.5.3. Τέλος διαφοροποιήσεις, διαφωνίες και διασαφηνήσεις σχετικά με τα αποτελέσματα της μελέτης που παρουσιάζεται στην Αν. [64] διευθετούνται στο Εδ. 5.5.4.

Στους εκτεθήμενους Πίνακες 5.2, 5.3 και 5.6 των Εδ. 5.5.1, 5.5.2 και 5.5.3 περιγράφονται συμβολικά όλα τα διαγράμματα Feynman που συνεισφέρουν στη δρώσα ενεργό διατομή για κάθε περίπτωση. Εφαρμόζοντας το γενικό φορμαλισμό του Εδ. 5.3, οι εκεί αριθμητικοί δείκτες $ij$, ενσαρκώνονται σε s-particles συγκεκριμένης ταυτότητας. Προφανώς $1 := \tilde{\chi}$. Η ταυτοποίηση των άλλων αριθμών γίνεται κατά περίπτωση στα αντίστοιχα εδάφια. Τα σωμάτια που εναλλάσονται σε κάθε ζεύγος αρχικών $(ij)$ και τελικών καταστάσεων δηλώνονται ως ορίσματα μέσα στα κανάλια αλληλεπίδρασης την οποία πραγματοποιούν. Συγκεκριμένα, τα σύμβολα $s(p)$, $t(p)$ και $u(p)$ δηλώνουν δενδρικού επιπέδου διαγράματα, στα οποία το σωμάτιο $p$ εναλλάσεται μέσω του $s$-, $t$- ή $u$-διαύλου. Το σύμβολο $c$, προερχόμενο από τη λέξη contact, δηλώνει διαγράμματα στα οποία τα τέσσερα εξωτερικά πόδια τους βρίσκονται σε επαφή. Αναλυτικά αποτελέσματα για τις σημαντικές συνεισφορές στη δρώσα ενεργό διατομή παρουσιάζονται στους Πίνακες 5.4 και 5.7 και στις Εξ. (5.90), (5.91) και (5.95).

### 5.5.1 Ενεργός Διατομή Αλληλοκαταστροφής

**Πίνακας 5.2: Διαγράμματα Αλληλοκαταστροφής**

| Καταστάσεις | | Κανάλια |
|---|---|---|
| Αρχική | Τελική | Αλληλεπίδρασης |
| $\tilde{\chi}\tilde{\chi}$ | $f\bar{f}$ | $s(h)$, $s(H)$, $s(A)$, $s(Z)$ |
| | | $t(\tilde{f}_{1[2]})$, $u(\tilde{f}_{1[2]})$ |
| | $hh$, $hH$, $HH$ | $s(h)$, $s(H)$, $t(\tilde{\chi}_i^0)$, $u(\tilde{\chi}_i^0)$ |
| | $AA$, $ZA$ | $s(h)$, $s(H)$, $t(\tilde{\chi}_i^0)$, $u(\tilde{\chi}_i^0)$ |
| | $hA$, $HA$ | $s(Z)$, $s(A)$, $t(\tilde{\chi}_i^0)$, $u(\tilde{\chi}_i^0)$ |
| | $W^+W^-$, $H^+H^-$ | $s(h)$, $s(H)$, $s(Z)$, $t(\tilde{\chi}_i^\pm)$, $u(\tilde{\chi}_i^\pm)$ |
| | $H^+W^-$ | $s(h)$, $s(H)$, $s(A)$, $t(\tilde{\chi}_i^\pm)$, $u(\tilde{\chi}_i^\pm)$ |
| | $ZZ$ | $s(h)$, $s(H)$, $t(\tilde{\chi}_i^0)$, $u(\tilde{\chi}_i^0)$ |
| | $Zh$, $ZH$ | $s(A)$, $s(Z)$, $t(\tilde{\chi}_i^0)$, $u(\tilde{\chi}_i^0)$ |

Οι αρχικές μας υποθέσεις των Εξ. (5.88) και (5.89) έχουν δραστική επίπτωση στον υπολογισμό της Ενεργού Διατομής Αλληλοκαταστροφής. Και αυτό, διότι αν δεν ίσχυε η Εξ. (5.88) θα έπρεπε στον υπολογισμό της Ενεργού Διατομής Αλληλοκαταστροφής να συμπεριληφθούν όλα τα κανάλια αλληλεπίδρασης που καταγράφονται στον Πίνακα 5.2. Στην εξεταζόμενη περίπτωση, όμως, μόνο τα διαγράμματα με τελικές καταστάσεις $f\bar{f}$, $H^+H^-$ επιζούν. Πραγματικά, αντικαθιστώντας τις Εξ. (5.88) στις εκφράσεις των σχετικών κορυφών Feynman από την Αν. [12], προκύπτει ότι:

- Οι ζεύξεις neutralino-neutralino-higgs[gauge boson], $\tilde{\chi}-\tilde{\chi}_i^0-h, H, A, Z$ μηδενίζονται. Επομένως όλες οι δυνατές αλληλεπιδράσεις μέσω του διαύλου $s$ αποκλείονται. Το συμπέρασμα αυτό θεμελιώνεται ασφαλέστερα με επίκλιση και της δεύτερης αρχικής παραδοχής μας στην Εξ. (5.89). Και αυτό, διότι ο μικρός παρονομαστής που θα προέκυπτε σε αντίθετη περίπτωση, ακόμα και με παρουσία μικρών ζευξέων θα είχε ως αποτέλεσμα την απότομη αύξηση της ενεργού διατομής. Ακόμα, λόγω του μηδενισμού των ίδιων ζεύξεων, όλα τα διαγράμματα με εναλλαγή neutralino στο δίαυλο $t$ ή $u$ επίσης αποκλείονται.

- Οι ζεύξεις neutralino-chargino-gauge boson, $\tilde{\chi}-\tilde{\chi}_i^\pm-W^\mp$ μηδενίζονται. Επομένως όλα τα διαγράμματα με εναλλαγή chargino στο δίαυλο $t$ ή $u$ αποκλείονται εκτός από τα διαγράμματα με τελική κατάσταση $H^+H^-$.

- Οι ζεύξεις neutralino-chargino-charge higgs, $\tilde{\chi}-\tilde{\chi}_i^\pm-H^\mp$ επιζούν. Τα διαγράμματα που κατασκευάζονται με χρήση της κορυφής αυτής προκύπτει ότι έχουν μειωμένη συνεισφορά σε σχέση με αυτή που προκύπτει από τα διαγράμματα που βασίζονται στην κορυφή neutralino-sfermion-fermion. Επιπλέον σε περιπτώσεις βαρύ τομέα Higgs (για μικρές η ενδιάμεσες τιμές του $\tan\beta$) τα διαγράμματα με τελική κατάσταση $H^+H^-$ είναι κινηματικά εξαιρεταία. Συνεπώς, δεν περικλείονται στον υπολογισμό.

- Οι ζεύξεις neutralino-sfermion-fermion, $\tilde{\chi}-\tilde{f}-f$ είναι κυρίαρχες. Η δεσπόζουσα συνεισφορά στην Ενεργό Διατομή Αλληλοκαταστροφής προέρχεται από την εναλλαγή ενός sfermion $\tilde{f}$, μέσω των $t$ και $u$ διαύλων που οδηγεί σε τελικές καταστάσεις $f\bar{f}$ όπου $f$ κάποιο quark ή lepton. Από όλα τα δυνατά διαγράμματα, αυτά που οδηγούν σε quark είναι συμπιεσμένα σε σχέση με αυτά που οδηγούν σε κάποιο lepton λόγω των βαρύτερων μαζών των squark στους διαδότες και λόγω του μικρότερου υπερφορτίου των squark, το οποίο υψώνεται στην τετάρτη δύναμη όπως φαίνεται στην Εξ. (5.90) και (5.91).

Συμπερασματικά, τα διαγράμματα που περιλαμβάνονται στον υπολογισμό μας διαμορφώνονται με αρχικές, τελικές καταστάσεις και κανάλια αλληλεπίδρασης αντίστοιχα ως εξής:

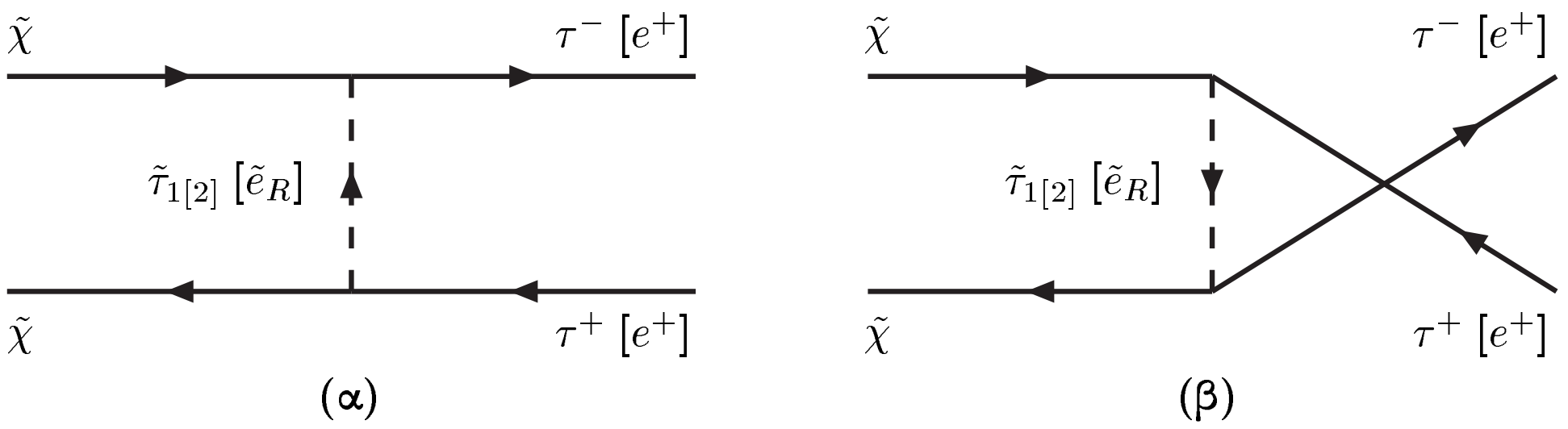

**Σχήμα 5.1:** *Τα διαγράμματα Feynman που συνεισφέρουν στην ενεργό διατομή αλληλοκαταστροφής $\tilde{\chi}\tilde{\chi} \to \tau\bar{\tau}\ [e\bar{e}]$ μέσω των διαύλων αλληλεπίδρασης $t$, (α) και $u$, (β)*

$$\tilde{\chi}\tilde{\chi} \quad \tau\bar{\tau} \quad t(\tilde{\tau}_{1[2]}),\ u(\tilde{\tau}_{1[2]}) \quad \text{και}$$
$$\tilde{\chi}\tilde{\chi} \quad e\bar{e} \quad t(\tilde{e}_R),\ \ u(\tilde{e}_R)$$

Η διαγραμματική μορφή αυτών των διαδικασιών φαίνεται στο Σχ. 5.1

Η διαδικάσια εξαγωγής των συντελεστών $a$ και $b$ του Εδ. 5.4.1 δεν χρειάζεται να ενεργοποιηθεί για την περίπτωση των διαδικασιών αλληλοκαταστροφής, γιατί οι συντελέστες αυτοί μπορούν να εξαχθούν απευθείας από το τυπολόγιο του παραρτήματος της Αν. [63]. Αμελώντας τις μάζες των λεπτονίων της τελικής κατάστασης οι συντελεστές $a_{\tilde{\chi}\tilde{\chi}}$ και $b_{\tilde{\chi}\tilde{\chi}}$ της Εξ. (5.82) βρίσκονται:

$$a_{\tilde{\chi}\tilde{\chi}} = \frac{e^4}{2\pi c_W^4}\, s_\tau^2 c_\tau^2 Y_L^2 Y_R^2 m_{\tilde{\chi}}^2 \left(\frac{1}{\Sigma_2} - \frac{1}{\Sigma_1}\right)^2 , \tag{5.90}$$

$$\begin{aligned} b_{\tilde{\chi}\tilde{\chi}} \;=\;& \frac{e^4 m_{\tilde{\chi}}^2}{24\Sigma_1^4\Sigma_2^4\pi c_W^4}\Big[2\Big((2m_{\tilde{\chi}}^4 - 2m_{\tilde{\chi}}^2\Sigma_1 + \Sigma_1^2)\Sigma_2^2 Y_L^4 + (1\leftrightarrow 2)\Sigma_1^2 Y_R^4\Big)c_\tau^4 + 2(1\leftrightarrow 2)s_\tau^4 \\ +\;& \Big(4\Sigma_1^2\Sigma_2^2\left(2m_{\tilde{\chi}}^4 + \Sigma_1\Sigma_2 - m_{\tilde{\chi}}^2(\Sigma_1+\Sigma_2)\right)(Y_R^2+Y_L^2) + \big((\Sigma_1-\Sigma_2)^2\left[9\Sigma_1^2\Sigma_2^2\right. \\ -\;& \left.20m_{\tilde{\chi}}^2\Sigma_1\Sigma_2(\Sigma_1+\Sigma_2) + 4m_{\tilde{\chi}}^4(3\Sigma_1^2 + 4\Sigma_1\Sigma_2 + 3\Sigma_2^2)\right]\big)\,Y_L^2Y_R^2\Big)c_\tau^2 s_\tau^2\Big] \\ +\;& 2\,\frac{e^4}{12\pi c_W^4}\,\frac{m_{\tilde{\chi}}^2(m_{\tilde{\chi}}^4 + m_{\tilde{e}_R}^4)}{\Sigma_e^4} , \end{aligned} \tag{5.91}$$

όπου $Y_{L(R)} = -1/2(-1)$ το υπερφορτίο του $\tau_{L(R)}$, $\Sigma_{1,2} = m_{\tilde{\chi}}^2 + m_{\tilde{\tau}_{1,2}}^2$ και $\Sigma_e = m_{\tilde{\chi}}^2 + m_{\tilde{e}_R}^2$ με $m_{\tilde{e}_R}$ να είναι η κοινή μάζα των δεξιόστροφων slepton $\tilde{e}_R$, $\tilde{\mu}_R$ των ελαφρότερων γενέων. Μερικά σχόλια είναι τώρα απαραίτητα:

- Ο συντελεστής $a_{\tilde{\chi}\tilde{\chi}}$ επιζεί μόνο στις περιπτώσεις μεγάλου $\tan\beta$ γιατί είναι ανάλογος του $s_\tau^2$, το οποίο μηδενίζεται σε περιπτώσεις μικρού $\tan\beta$ και συνεπώς ελάχιστης ανάμιξης $\tilde{\tau}_L - \tilde{\tau}_R$.
- Η κύρια συνεισφορά στο $b_{\tilde{\chi}\tilde{\chi}}$ προέρχεται από το πρώτο όρο στην αγγύλη του δεύτερου μέλους της Εξ. (5.91). Οι άλλοι όροι στην ίδια αγκύλη οφείλονται στην ανάμιξη $\tilde{\tau}_L - \tilde{\tau}_R$.
- Ο τελευταίος όρος στο δεξί μέρος της Εξ. (5.91) παριστάνει τη συνεισφορά των δύο ελαφρότερων γενεών. Τα δεξιόστροφα sleptons θεωρούνται εκφυλισμένα με μάζα $m_{\tilde{e}_R}$. Σε περιπτώσεις μεγάλου $\tan\beta$ οι τιμές του $m_{\tilde{e}_R}$ είναι κατα πολύ μεγαλύτερες από αυτές του $m_{\tilde{\tau}_2}$ και επομένως οι αντίστοιχες συνεισφορές στο $b_{\tilde{\chi}\tilde{\chi}}$ είναι μικρότερες από εκείνες που προέρχονται από την εναλλαγή του $\tilde{\tau}_2$. Αντίθετα σε περιπτώσεις μικρού $\tan\beta$ όλες οι επιμέρους συνεισφορές είναι παρόμοιες.
- Η συνεισφορά στο $b_{\tilde{\chi}\tilde{\chi}}$ από το διάγραμμα με εναλλαγή $\tilde{\tau}_1$ είναι μικρή, μολονότι παρουσιάζεται στην Εξ. (5.91). Βρίσκεται ότι η συνεισφορά του είναι μειωμένη κατά $1/6 - 1/8$ συγκρινόμενη από τη

**Πίνακας 5.3: Διαγράμματα Συγγενικής Καταστροφής Ι**

| Καταστάσεις | | Κανάλια |
|---|---|---|
| Αρχική | Τελική | Αλληλεπίδρασης |
| $\tilde{\chi}\tilde{\tau}_2$ | $\tau h$, $\tau H$, $\tau Z$ | $s(\tau)$, $t(\tilde{\tau}_{1,2})$ |
| | $\tau A$ | $s(\tau)$, $t(\tilde{\tau}_1)$ |
| | $\tau\gamma$ | $s(\tau)$, $t(\tilde{\tau}_2)$ |
| $\tilde{\tau}_2\tilde{\tau}_2$ | $\tau\tau$ | $t(\tilde{\chi})$, $u(\tilde{\chi})$ |
| $\tilde{\tau}_2\tilde{\tau}_2^*$ | $hh$, $hH$, $HH$, $ZZ$ | $s(h)$, $s(H)$, $t(\tilde{\tau}_{1,2})$, $u(\tilde{\tau}_{1,2})$, $c$ |
| | $AA$ | $s(h)$, $s(H)$, $t(\tilde{\tau}_1)$, $u(\tilde{\tau}_1)$, $c$ |
| | $hZ$, $HZ$ | $s(Z)$, $t(\tilde{\tau}_{1,2})$, $u(\tilde{\tau}_{1,2})$ |
| | $h\gamma$, $H\gamma$ | $t(\tilde{\tau}_2)$, $u(\tilde{\tau}_2)$ |
| | $hA$, $HA$ | $s(Z)$, $t(\tilde{\tau}_1)$, $u(\tilde{\tau}_1)$ |
| | $AZ$ | $s(h)$, $s(H)$, $t(\tilde{\tau}_1)$, $u(\tilde{\tau}_1)$ |
| | $H^+H^-$, $W^+W^-$ | $s(h)$, $s(H)$, $s(\gamma)$, $s(Z)$, $c$ |
| | $H^+W^-$ | $s(h)$, $s(H)$ |
| | $\gamma\gamma$, $\gamma Z$ | $t(\tilde{\tau}_2)$, $u(\tilde{\tau}_2)$, $c$ |
| | $t\bar{t}$, $b\bar{b}$ | $s(h)$, $s(H)$, $s(\gamma)$, $s(Z)$ |
| | $\tau\bar{\tau}$ | $s(h)$, $s(H)$, $s(\gamma)$, $s(Z)$, $t(\tilde{\chi})$ |
| | $u\bar{u}$, $d\bar{d}$, $e\bar{e}$ | $s(\gamma)$, $s(Z)$ |
| | $\nu\bar{\nu}$ | $s(Z)$ |

συνεισφορά από καθεμία από τις ελαφρότερες γενέες. Αυτό μπορεί να κατανοηθεί από την ακόλουθη παρατήρηση. Παρά το γεγονός οτι η μάζα στο διαδότη αυτού του διαγράμματος, $m_{\tilde{\tau}_1}$, δεν είναι πολύ υψηλότερα από τη $m_{\tilde{e}_R}$, η κύρια συνεισφορά της περιέχει ένα παράγοντα της μορφής $c_\tau^4 Y_L^4$ όπως φαίνεται . Ένας πιο εύχρηστος και πιο σύντομος τύπος μπορεί να ληφθεί αν αμεληθεί η συνεισφορά του διαγράμματος με εναλλαγή του $\tilde{\tau}_1$ που είναι:

$$b_{\tilde{\chi}\tilde{\chi}}^{\tilde{\tau}_2,\tilde{e}_R} = \frac{e^4}{12\pi c_W^4}\frac{m_{\tilde{\chi}}^2}{\Sigma_2^4}\left[(s_\tau^4 Y_L^4 + c_\tau^4 Y_R^4)(m_{\tilde{\chi}}^4 + m_{\tilde{\tau}_2}^4) + \frac{s_\tau^2 c_\tau^2 Y_L^2 Y_R^2}{2}(m_{\tilde{\chi}}^4 + 9m_{\tilde{\tau}_2}^4 - 2m_{\tilde{\chi}}^2 m_{\tilde{\tau}_2}^2)\right]$$
$$+2\,\frac{e^4}{12\pi c_W^4}\,\frac{m_{\tilde{\chi}}^2(m_{\tilde{\chi}}^4 + m_{\tilde{e}_R}^4)}{\Sigma_e^4}\,. \qquad (5.92)$$

## 5.5.2 Ενεργές Διατομές Συγγενικής Καταστροφής Bino-Stau

Σε περιοχές του παραμετρικού χώρου, όπου λαμβάνεται $\tan\beta \geq 15$, η ανάμιξη των $\tilde{\tau}_L - \tilde{\tau}_R$ παίζει ρόλο ιδιαζόντως σημαντικό, γιατί ταπεινώνει τη μάζα του ελαφρότερου από αυτά, $\tilde{\tau}_2$ πλέον, χαμηλότερα από τα right sleptons των δύο πρώτων οικογενειών, στα οποία η ανάμιξη μπορεί ασφαλώς να αμεληθεί λόγω των μικρών μαζών των αντίστοιχων φερμιονίων. Σε τέτοιες περιπτώσεις, ο μηχανισμός της Συγγενικής Καταστροφής ενεργοποιείται ανάμεσα στα δύο, μακράν των άλλων, ελαφρότερα σωμάτια του SUSY φάσματος. Επομένως, στην εξίσωση Boltzmann η πυκνότητα αριθμού μετεχόντων σωματιδίων, είναι:

$$n = n_{\tilde{\chi}} + n_{\tilde{\tau}_2} + n_{\tilde{\tau}_2^*}\,, \qquad (5.93)$$

**Πίνακας 5.4: Συνεισφορές στους Συντελεστές $a_{ij}$ $(ij \neq \tilde{\chi}\tilde{\chi})$**

| **Διαδικασία** | **Συνεισφορά στο Συντελεστή $a_{ij}$** |
|---|---|
| $\tilde{\chi}\tilde{\tau}_2 \to \tau h$ | $e^2(1-\bar{m}_h^2)^2\{2Y_LY_Rg_{h\tau\tau}[2s_\tau c_\tau g_h/(m_{\tilde{\tau}_2}-\bar{m}_h^2 m_{\tilde{\chi}}) - c_{2\tau}\bar{g}_{h1}/$<br>$(\bar{m}_{\tilde{\tau}_1}^2+\bar{m}_{\tilde{\chi}}(\bar{m}_{\tilde{\tau}_2}-\bar{m}_h^2))] + [g_{h\tau\tau}^2 + g_h^2/(m_{\tilde{\chi}}\bar{m}_h^2 - m_{\tilde{\tau}_2})^2](s_\tau^2Y_L^2+c_\tau^2Y_R^2)$<br>$+\bar{g}_{h1}^2(c_\tau^2Y_L^2+s_\tau^2Y_R^2)/(\bar{m}_{\tilde{\tau}_1}^2+\bar{m}_{\tilde{\chi}}(\bar{m}_{\tilde{\tau}_2}-\bar{m}_h^2))^2 - 2s_\tau c_\tau(Y_L^2-Y_R^2)$<br>$\bar{g}_h\bar{g}_{h1}/(\bar{m}_{\tilde{\tau}_2}-\bar{m}_{\tilde{\chi}}\bar{m}_h^2)(\bar{m}_{\tilde{\tau}_1}^2+\bar{m}_{\tilde{\chi}}(\bar{m}_{\tilde{\tau}_2}-\bar{m}_h^2))\}/32\pi c_W^2 m_{\tilde{\tau}_2}(m_{\tilde{\tau}_2}+m_{\tilde{\chi}})$ |
| $\tilde{\chi}\tilde{\tau}_2 \to \tau\gamma$ | $e^4(s_\tau^2Y_L^2+c_\tau^2Y_R^2)/16\pi c_W^2 m_{\tilde{\tau}_2}(m_{\tilde{\chi}}+m_{\tilde{\tau}_2})$ |
| $\tilde{\chi}\tilde{\tau}_2 \to \tau Z$ | $e^2(1-\bar{M}_Z^2)\{\bar{m}_{\tilde{\tau}_2}(1-\bar{M}_Z^2)^3[g_{\tilde{\tau}_2\tilde{\tau}_2 Z}^2(s_\tau^2Y_L^2+c_\tau^2Y_R^2)/(\bar{M}_Z^2\bar{m}_{\tilde{\chi}}-\bar{m}_{\tilde{\tau}_2})^2$<br>$+g_{\tilde{\tau}_1\tilde{\tau}_2 Z}^2(c_\tau^2Y_L^2+s_\tau^2Y_R^2)/(\bar{m}_{\tilde{\tau}_1}^2+\bar{m}_{\tilde{\chi}}(\bar{m}_{\tilde{\tau}_2}-\bar{M}_Z^2))^2$<br>$-2g_{\tilde{\tau}_1\tilde{\tau}_2 Z}g_{\tilde{\tau}_2\tilde{\tau}_2 Z}s_\tau c_\tau(Y_L^2-Y_R^2)/(\bar{m}_{\tilde{\tau}_2}-\bar{M}_Z^2\bar{m}_{\tilde{\chi}})(\bar{m}_{\tilde{\tau}_1}^2+\bar{m}_{\tilde{\chi}}(\bar{m}_{\tilde{\tau}_2}-\bar{M}_Z^2))]$<br>$-2g_Z(\bar{M}_Z^2-1)^2[g_{\tilde{\tau}_2\tilde{\tau}_2 Z}(L_\tau s_\tau^2Y_L^2+R_\tau c_\tau^2Y_R^2)/(\bar{m}_{\tilde{\tau}_2}-\bar{M}_Z^2\bar{m}_{\tilde{\chi}})$<br>$-g_{\tilde{\tau}_1\tilde{\tau}_2 Z}s_\tau c_\tau(L_\tau Y_L^2-R_\tau Y_R^2)/(\bar{m}_{\tilde{\tau}_1}^2+\bar{m}_{\tilde{\chi}}(\bar{m}_{\tilde{\tau}_2}-\bar{M}_Z^2))]$<br>$+g_Z^2(L_\tau^2s_\tau^2Y_L^2+R_\tau^2c_\tau^2Y_R^2)(1+\bar{M}_Z^2-2\bar{M}_Z^4)(1+\hat{m}_{\tilde{\chi}})\}/32\pi c_W^2 M_Z^2$ |
| $\tilde{\tau}_2\tilde{\tau}_2 \to \tau\tau$ | $e^4(s_\tau^4Y_L^4+c_\tau^4Y_R^4)m_{\tilde{\chi}}^2/\pi c_W^4\Sigma_2^2$ |
| $\tilde{\tau}_2\tilde{\tau}_2^* \to \gamma\gamma$ | $e^4/8\pi m_{\tilde{\tau}_2}^2$ |
| $\tilde{\tau}_2\tilde{\tau}_2^* \to \gamma Z$ | $-e^2g_{\tilde{\tau}_2\tilde{\tau}_2 Z}^2(\hat{M}_Z^2-4)/16\pi m_{\tilde{\tau}_2}^2$ |
| $\tilde{\tau}_2\tilde{\tau}_2^* \to ZZ$ | $(1-\hat{M}_Z^2)^{1/2}\{[(g_h^2g_{hZZ}^2P_1/(\hat{m}_h^2-4)+12g_hg_{hZZ}g_{\tilde{\tau}_2\tilde{\tau}_2 Z}^2M_Z^2\hat{M}_Z^2)/(\hat{m}_h^2-4)$<br>$-4g_hg_{hZZ}g_{\tilde{\tau}_1\tilde{\tau}_2 Z}^2m_{\tilde{\tau}_2}^2(P_4-\hat{m}_{\tilde{\tau}_1}^2P_1)/(1+\hat{m}_{\tilde{\tau}_1}^2-\hat{M}_Z^2)(\hat{m}_h^2-4)+(h\leftrightarrow H)]$<br>$+g_{\tilde{\tau}_2\tilde{\tau}_2 Z}^4M_Z^4P_3/(\hat{M}_Z^2-2)^2+2g_hg_{hZZ}g_Hg_{HZZ}P_1/(\hat{m}_h^2-4)(\hat{m}_H^2-4)$<br>$-8g_{\tilde{\tau}_2\tilde{\tau}_2 Z}^2g_{\tilde{\tau}_1\tilde{\tau}_2 Z}^2M_Z^4[P_5-3\hat{m}_{\tilde{\tau}_1}^2(\hat{M}_Z^2-2)]/(1+\hat{m}_{\tilde{\tau}_1}^2-\hat{M}_Z^2)(\hat{M}_Z^2-2)$<br>$+4g_{\tilde{\tau}_1\tilde{\tau}_2 Z}^4m_{\tilde{\tau}_2}^4[\hat{m}_{\tilde{\tau}_1}^4P_1+(1-\hat{M}_Z^2)^2P_2-2\hat{m}_{\tilde{\tau}_1}^2P_4]/(1+\hat{m}_{\tilde{\tau}_1}^2-\hat{M}_Z^2)^2\}$<br>$/64\pi M_Z^4m_{\tilde{\tau}_2}^2$ |
| $\tilde{\tau}_2\tilde{\tau}_2^* \to W^+W^-$ | $(1-\hat{M}_W^2)^{1/2}(4-4\hat{M}_W^2+3\hat{M}_W^4)[g_hg_{hW^+W^-}/(\hat{m}_h^2-4)$<br>$+g_Hg_{HW^+W^-}/(\hat{m}_H^2-4)+g_{\tilde{\tau}_2\tilde{\tau}_2W^+W^-}m_{\tilde{\tau}_2}^2]^2/32\pi M_W^4m_{\tilde{\tau}_2}^2$ |
| $\tilde{\tau}_2\tilde{\tau}_2^* \to t\bar{t}$ | $3(1-\hat{m}_t^2)^{3/2}[g_hg_{htt}/(\hat{m}_h^2-4)+g_Hg_{Htt}/(\hat{m}_H^2-4)]^2/4\pi m_{\tilde{\tau}_2}^4$ |

και η δρώσα ενεργός διατομή στην Εξ. (5.24) για $v \simeq v_{ij}$, γράφεται:

$$\sigma_{\rm eff} = \sigma_{\tilde{\chi}\tilde{\chi}}r_{\tilde{\chi}}r_{\tilde{\chi}} + 4\sigma_{\tilde{\chi}\tilde{\tau}_2}r_{\tilde{\chi}}r_{\tilde{\tau}_2} + 2(\sigma_{\tilde{\tau}_2\tilde{\tau}_2}+\sigma_{\tilde{\tau}_2\tilde{\tau}_2^*})r_{\tilde{\tau}_2}r_{\tilde{\tau}_2}\,. \qquad (5.94)$$

όπου τα $r_i$ δίνονται από την Εξ. (5.26), αντικαθιστώντας $i = \tilde{\chi}, \tilde{\tau}_2, \tilde{\tau}_2^*$ και $g_i = 2,\ 1,\ 1$, αντίστοιχα. Προφανής επίσης είναι η αριθμητική ταυτοποίηση στο άθροισμα της Εξ. (5.24), $1 := \tilde{\chi}, 2 := \tilde{\tau}_2, 3 := \tilde{\tau}_2^*$. Οι παράγοντας 4, (2) στην Εξ. (5.94) εξηγείται από το γεγονός ότι οι παράγοντες 12, 13, 21, 31 (23, 32) στο άθροισμα της Εξ. (5.24) είναι ταυτόσημοι.

Η ποσότητα $J_{\rm eff}$ βρίσκεται από την Εξ. (5.83) αθροίζοντας στις εξής ομάδες $(ij) = (\tilde{\chi}\tilde{\chi})$, $(\tilde{\chi}\tilde{\tau}_2)$ και $(\tilde{\tau}_2\tilde{\tau}_2^{(*)})$ με

$$a_{\tilde{\tau}_2\tilde{\tau}_2^{(*)}} = a_{\tilde{\tau}_2\tilde{\tau}_2} + a_{\tilde{\tau}_2\tilde{\tau}_2^*},$$
$$b_{\tilde{\tau}_2\tilde{\tau}_2^{(*)}} = b_{\tilde{\tau}_2\tilde{\tau}_2} + b_{\tilde{\tau}_2\tilde{\tau}_2^*}$$

και οι ποσότητες $J^a_{(ij)}$, $J^b_{(ij)}$ υπολογίζονται χρησιμοποιώντας τις Εξ. (5.85), (5.84) με τις σχετικές σταθερές

**Πίνακας 5.5: Τα Σύμβολα $\lambda$**

| **Διαδικασία** | $\lambda_h$ | $\lambda_H$ | $\lambda_1$ | $\lambda_2$ | $\lambda_c$ |
|---|---|---|---|---|---|
| $\tilde\tau_2\tilde\tau_2^* \to hh$ | $g_h g_{hhh}$ | $g_H g_{Hhh}$ | $g_{h1}^2$ | $g_h^2$ | $g_{\tilde\tau_2\tilde\tau_2 hh}$ |
| $\tilde\tau_2\tilde\tau_2^* \to hH$ | $g_h g_{Hhh}$ | $g_H g_{hHH}$ | $g_{h1} g_{H1}$ | $g_h g_H$ | $g_{\tilde\tau_2\tilde\tau_2 hH}$ |
| $\tilde\tau_2\tilde\tau_2^* \to HH$ | $g_h g_{hHH}$ | $g_H g_{HHH}$ | $g_{H1}^2$ | $g_H^2$ | $g_{\tilde\tau_2\tilde\tau_2 HH}$ |
| $\tilde\tau_2\tilde\tau_2^* \to AA$ | $g_h g_{hAA}$ | $g_H g_{HAA}$ | $-g_{A1}^2$ | 0 | $g_{\tilde\tau_2\tilde\tau_2 AA}$ |
| $\tilde\tau_2\tilde\tau_2^* \to H^+H^-$ | $g_h g_{hH^+H^-}$ | $g_H g_{HH^+H^-}$ | 0 | 0 | $g_{\tilde\tau_2\tilde\tau_2 H^+H^-}$ |

$c_{(ij)} = 1,\, 4,\, 2$ για $(ij) = (\tilde\chi\tilde\chi)$, $(\tilde\chi\tilde\tau_2)$ και $(\tilde\tau_2\tilde\tau_2^{(*)})$. Όταν $\Delta_{\tilde\tau_2} = 0$, $J^a_{(ij)} = 1/4,\, 1/2,\, 1/8$ αντίστοιχα.

Στον Πίνακα 5.3 περιγράφονται όλα τα διαγράματα Feynmann που συνεισφέρουν στη δρώσα ενεργό διατομή. Γενικά, από τα αναγραφόμενα διαγράμματα, τα παρακάτω:

$$\tilde\tau_2\tilde\tau_2^* \to h[H]A,\ h[H]\gamma,\ h[H]Z,\ AZ,\ H^-W^+,\ \nu\bar\nu$$

προκύπτει ότι έχουν ασθενέστατη συνεισφορά στη δρώσα ενεργό διατομή, λόγω είτε των ασθενών ζεύξεων είτε των μεγάλων μαζών, οπότε μπορούν, εκ του ασφαλούς, να αμεληθούν από τον τελικό υπολογισμό. Οι συνεισφορές από τις δίαφορες CAE που καταγράφονται στον Πίνακα 5.3 στους συντελεστές $a_{ij}$ και $b_{ij}$ $(ij \neq \tilde\chi\tilde\chi)$ της Εξ.(5.82) υπολογίζονται εφαρμόζοντας τη μέθοδο που περιγράφηκε στο Εδ. 5.4.1. Οι μάζες των leptons και quarks (εκτός από το $t$-quark) στις τελικές καταστάσεις ή στους διαδότες λαμβάνονται μηδέν. Αντιθέτως, οι $b$ και $\tau$ ζεύξεις Yukawa δεν αγνοούνται γιατί, στις περιπτώσεις μεγάλου $\tan\beta$, η επίδρασή τους προκύπτει να είναι πολύ σημαντική. Οι κυρίαρχες συνεισφορές στην $J_{\rm eff}$ της Εξ.(5.83) προέρχονται από τους συντελεστές $a_{ij}$ στην περίπτωση της CAE.

Στον Πίνακα 5.4 εκτίθενται μερικές από τις διαδικασίες που συνεισφέρουν στα $a_{ij}$ $(ij \neq \tilde\chi\tilde\chi)$ μαζί με τις αναλυτικές τους εκφράσεις για αυτές τους τις συνεισφορές. Τα σύμβολα $\hat{}$ $[\bar{}]$ πάνω από μια ποσότητα δηλώνουν ότι αυτή η ποσότητα μετριέται σε μονάδες της μάζας $m_{\tilde\tau_2}$ $[\,m_{\tilde\chi} + m_{\tilde\tau_2}]$. Τα σύμβολα $c_\tau$, $s_\tau$ είναι τα στοιχεία του πίνακα διαγωνοποιεί τον πίνακα μαζών των stau και ορίζονται στην Εξ. (2.67). Επίσης,

$$P_{1(2)} = 3\hat M_Z^4 - (+)4\hat M_Z^2 + 4\ ,\ P_3 = 3\hat M_Z^4 - 8\hat M_Z^2 + 8\ ,$$
$$P_4 = 3\hat M_Z^6 - 3\hat M_Z^4 - 4\hat M_Z^2 + 4\ ,\ P_5 = 3\hat M_Z^4 - 5\hat M_Z^2 + 2\ .$$

Η συνεισφορά από τη διαδικασία $\tilde\chi\tilde\tau_2 \to \tau H$ (ή $\tau A$) στο συντελεστή $a_{\tilde\chi\tilde\tau_2}$ λαμβάνεται από την έκφραση για $\tilde\chi\tilde\tau_2 \to \tau h$ στον Πίνακα 5.4, αντικαθιστώντας $h$ με $H$ (ή $A$ και $\cos 2\tau$ με 1). Για τη συνεισφορά στη $a_{\tilde\tau_2\tilde\tau_2^*}$ από κάθεμία από τις πέντε διαδικασίες με δύο higgs στην τελική κατάσταση (στον Πίνακα 5.3), ένας γενικευμένος τύπος μπορεί να δοθεί:

$$a_{\tilde\tau_2\tilde\tau_2^*\to H_pH_q} = (\frac{1}{2})\frac{1}{128\pi m^6_{\tilde\tau_2}}(4-(\hat m_{H_p}-\hat m_{H_q})^2)^{1/2}(4-(\hat m_{H_p}+\hat m_{H_q})^2)^{1/2}$$

$$\left(\frac{\lambda_h}{4-\hat m_h^2} + \frac{\lambda_H}{4-\hat m_H^2} + \frac{4\lambda_1}{\hat m^2_{H_p} + \hat m^2_{H_q} - 2\hat m^2_{\tilde\tau_1} - 2} + \frac{4\lambda_2}{\hat m^2_{H_p} + \hat m^2_{H_q} - 4} - \lambda_c m^2_{\tilde\tau_2}\right)^2, \qquad (5.95)$$

όπου τα $H_p$, $H_q$ δηλώνουν τα $h$, $H$, $A$, $H^+$, $H^-$, ο παράγοντας $1/2$ παραμένει μόνο όταν υπάρχουν ταυτοτικά σωμάτια στην τελική κατάσταση και τα σύμβολα $\lambda_h$, $\lambda_H$, $\lambda_1$, $\lambda_2$, $\lambda_c$ αντιστοιχούν σε διαγράμματα εναλλαγής $s(h)$, $s(H)$, $t(\tilde\tau_{1,2})$ (ή $u(\tilde\tau_{1,2})$), $c$ στον Πίνακα 5.3 και δίνονται στον Πίνακα 5.5.

Τα εμφανιζόμενα σύμβολα $g$ στους Πίνακες 5.4 και 5.5 αντιστοιχούν σε διάφορους κόμβους αλληλεπίδρασης με τα αλληλεπιδρώντα σωμάτια δηλωμένα ως δείκτες. Οι εκφράσεις τους δίνονται στο Εδ. Β′.3.2 του Παραρτήματος Β′, όπου εκτίθενται σε διαγραμματική μορφή και οι χρησιμοποιούμενοι κανόνες Feynman.

**Πίνακας 5.6: Διαγράμματα Συγγενικής Καταστροφής ΙΙ**

| Καταστάσεις | | Κανάλια Αλληλεπίδρασης |
|---|---|---|
| Αρχική | Τελική | |
| $\tilde{\chi}\tilde{e}_R$ | $eh$, $eH$, $eZ$ | $s(e)$, $t(\tilde{e}_R)$ |
| | $e\gamma$ | $s(e)$, $t(\tilde{e}_R)$ |
| $\tilde{e}_R\tilde{e}_R$ | $ee$ | $t(\tilde{\chi})$, $u(\tilde{\chi})$ |
| $\tilde{e}_R\tilde{e}_R^*$ | $hh$, $hH$, $HH$, $ZZ$ | $s(h)$, $s(H)$, $t(\tilde{e}_R)$, $u(\tilde{e}_R)$, $c$ |
| | $hZ$, $HZ$ | $s(Z)$, $t(\tilde{e}_R)$, $u(\tilde{e}_R)$ |
| | $h\gamma$, $H\gamma$ | $t(\tilde{e}_R)$, $u(\tilde{e}_R)$ |
| | $H^+H^-$ | $s(h)$, $s(H)$, $s(\gamma)$, $s(Z)$, $c$ |
| | $W^+W^-$ | $s(h)$, $s(H)$, $s(\gamma)$, $s(Z)$ |
| | $H^+W^-$ | $s(h)$, $s(H)$ |
| | $\gamma\gamma$, $\gamma Z$ | $t(\tilde{e}_R)$, $u(\tilde{e}_R)$, $c$ |
| | $t\bar{t}$, $b\bar{b}$ | $s(h)$, $s(H)$, $s(\gamma)$, $s(Z)$ |
| | $\tau\bar{\tau}$ | $s(h)$, $s(H)$, $s(\gamma)$, $s(Z)$, $t(\tilde{\chi})$ |
| | $u\bar{u}$, $d\bar{d}$, $e\bar{e}$ | $s(\gamma)$, $s(Z)$ |
| | $\nu\bar{\nu}$ | $s(Z)$ |
| $\tilde{\tau}_2\tilde{e}_R$ | $\tau e$ | $t(\tilde{\chi})$ |
| $\tilde{\tau}_2\tilde{e}_R^*$ | $\tau\bar{e}$ | $t(\tilde{\chi})$ |
| $\tilde{e}_R\tilde{\mu}_R$ | $e\mu$ | $t(\tilde{\chi})$ |
| $\tilde{e}_R\tilde{\mu}_R^*$ | $\tau\bar{\mu}$ | $t(\tilde{\chi})$ |

Δεν δίνονται οι αναλυτικές εκφράσεις των συνεισφορών στο συντελεστή $a_{\tilde{\tau}_2\tilde{\tau}_2^*}$ από τις διαδικασίες με $b\bar{b}$ και $\tau\bar{\tau}$ στις τελικές καταστάσεις επειδή είναι πολύ μικρές. Εντούτοις, συμπεριλαμβάνονται στο αριθμητικό πρόγραμμα. Αξιοσημειωτο είναι ότι οι συνεισφορές από τις διαδικασίες με τελικές καταστάσεις $u\bar{u}$, $d\bar{d}$ και $e\bar{e}$ στο συντελεστή $a_{\tilde{\tau}_2\tilde{\tau}_2^*}$ μηδενίζονται. Οι διαδικασίες αυτές συνεισφέρουν μόνο στο συντελεστή $b_{\tilde{\tau}_2\tilde{\tau}_2^*}$. Επίσης, για τους συντελεστές $b_{ij}$ ($ij \neq \tilde{\chi}\tilde{\chi}$), οι οποίοι συμπεριλαμβάνονται στο αριθμητικό πρόγραμμα, δεν παρέχονται αναλυτικές εκφράσεις γιατί οι συνεισφορές τους στην $J_{\rm eff}$ είναι, γενικά, αμεληταίες.

### 5.5.3 Ενεργές Διατομές Συγγενικής Καταστροφής Bino-Stau-Selectron

Σε περιοχές του παραμετρικού χώρου, όπου λαμβάνεται $\tan\beta \leq 15$, φαινόμενα Συγγενικής Καταστροφής με τα sleptons των ελαφρών οικογενειών $\tilde{e}_R$, $\tilde{e}_R^*$, $\tilde{\mu}_R$, $\tilde{\mu}_R^*$ κάνουν την εμφάνισή τους και γίνονται πιό αισθητά όσο μειώνεται η τιμή της $\tan\beta$. Και αυτό, διότι η μίξη των $\tilde{\tau}_1 - \tilde{\tau}_2$ γίνεται ασθενέστερη και δεν είναι ικανή να ελαττώσει τη μάζα του $\tilde{\tau}_2$ αρκούντως, ώστε να βρίσκεται σε απόσταση από τα down sleptons των πρώτων οικογενειών, η οποία να εξασφαλίζει ότι αυτά δεν συμμετέχουν στο μηχανισμό της Συγγενικής Καταστροφής. Επομένως, στην εξίσωση Boltzmann η πυκνότητα αριθμού μετεχώντων σωματιδίων, είναι:

$$n = n_{\tilde{\chi}} + n_{\tilde{\tau}_2} + n_{\tilde{\tau}_2^*} + n_{\tilde{e}_R} + n_{\tilde{\mu}_R} + n_{\tilde{e}_R^*} + n_{\tilde{\mu}_R^*} \,,$$

και η δρώσα ενεργός διατομή στην Εξ. (5.24) για $v \simeq v_{ij}$, γράφεται:

$$\sigma_{\rm eff} \;=\; \sigma_{\tilde{\chi}\tilde{\chi}} r_{\tilde{\chi}} r_{\tilde{\chi}} + 4\sigma_{\tilde{\chi}\tilde{\tau}_2} r_{\tilde{\chi}} r_{\tilde{\tau}_2} + 2\times 4\sigma_{\tilde{\chi}\tilde{e}_R} r_{\tilde{\chi}} r_{\tilde{e}_R}$$

$$
\begin{aligned}
&+ \quad 2(\sigma_{\tilde{\tau}_2\tilde{\tau}_2} + \sigma_{\tilde{\tau}_2\tilde{\tau}_2^*}) r_{\tilde{\tau}_2} r_{\tilde{\tau}_2} + 2\times 2(\sigma_{\tilde{e}_R\tilde{e}_R} + \sigma_{\tilde{e}_R\tilde{e}_R^*}) r_{\tilde{e}_R} r_{\tilde{e}_R} \\
&+ \quad 2\times 2\times 2(\sigma_{\tilde{\tau}_2\tilde{e}_R} + \sigma_{\tilde{\tau}_2\tilde{e}_R^*}) r_{\tilde{\tau}_2} r_{\tilde{e}_R} + 2\times 2(\sigma_{\tilde{e}_R\tilde{\mu}_R} + \sigma_{\tilde{e}_R\tilde{\mu}_R^*}) r_{\tilde{e}_R} r_{\tilde{e}_R}. \qquad (5.96)
\end{aligned}
$$

όπου τα $r_i$ δίνονται από την Εξ. 5.26, αντικαθιστώντας $i = \tilde{\chi}, \tilde{\tau}_2, \tilde{\tau}_2^*, \tilde{e}_R, \tilde{e}_R^*, \tilde{\mu}_R, \tilde{\mu}_R^*$ και $g_i = 2, 1, 1, 1, 1, 1, 1$ αντίστοιχα. Οι επιπλέον παράγοντες 2 που τίθενται στην προηγούμενη οφείλονται στον εκφυλισμό των δύο πρώτων γενέων.

Η ποσότητα $J_{\rm eff}$ βρίσκεται από την Εξ. (5.83) αθροίζοντας στις εξής ομάδες $(ij) = (\tilde{\chi}\tilde{\chi})$, $(\tilde{\chi}\tilde{l})$ και $(\tilde{l}\tilde{l}^{(*)})$ με

$$
\begin{aligned}
a_{\tilde{\chi}\tilde{l}} &= a_{\tilde{\chi}\tilde{\tau}_2} + a_{\tilde{\chi}\tilde{e}_R} \\
a_{\tilde{l}\tilde{l}^{(*)}} &= a_{\tilde{\tau}_2\tilde{\tau}_2} + a_{\tilde{\tau}_2\tilde{\tau}_2^*} + a_{\tilde{e}_R\tilde{e}_R} + a_{\tilde{e}_R\tilde{e}_R^*} + a_{\tilde{\tau}_2\tilde{e}_R} + a_{\tilde{\tau}_2\tilde{e}_R^*} + a_{\tilde{e}_R\tilde{\mu}_R} a_{\tilde{e}_R\tilde{\mu}_R^*}
\end{aligned}
$$

και αντίστοιχα

$$
\begin{aligned}
b_{\tilde{\chi}\tilde{l}} &= b_{\tilde{\chi}\tilde{\tau}_2} + b_{\tilde{\chi}\tilde{e}_R} \\
b_{\tilde{l}\tilde{l}^{(*)}} &= b_{\tilde{\tau}_2\tilde{\tau}_2} + b_{\tilde{\tau}_2\tilde{\tau}_2^*} + b_{\tilde{e}_R\tilde{e}_R} + b_{\tilde{e}_R\tilde{e}_R^*} + b_{\tilde{\tau}_2\tilde{e}_R} + b_{\tilde{\tau}_2\tilde{e}_R^*} + b_{\tilde{e}_R\tilde{\mu}_R} b_{\tilde{e}_R\tilde{\mu}_R^*}
\end{aligned}
$$

Οι ποσότητες $J^a_{(ij)}$, $J^b_{(ij)}$ υπολογίζονται χρησιμοποιώντας τις Εξ. (5.85), (5.84) με τις σχετικές σταθερές $c_{(ij)} = 1, 4, 2$ για $(ij) = (\tilde{\chi}\tilde{\chi})$, $(\tilde{\chi}\tilde{l})$ και $(\tilde{l}\tilde{l}^{(*)})$.

Τα διαγράμματα Feynmann που περιλαμβάνονται στον υπολογισμό για τις ενεργές διατομές με αρχικές καταστάσεις:

- $\tilde{\chi}\tilde{\chi}$, $\tilde{\chi}\tilde{\tau}_2$, $\tilde{\tau}_2\tilde{\tau}_2$, $\tilde{\tau}_2\tilde{\tau}_2^*$ αναγράφονται στον Πίνακα 5.3.
- $\tilde{\chi}\tilde{e}_R$, $\tilde{e}_R\tilde{e}_R$, $\tilde{e}_R\tilde{e}_R^*$ λαμβάνονται από τα προηγούμενα κάνοντας τις αντικατάστασεις $\tilde{\tau}_2 \to \tilde{e}_R$, $\tau \to e$ και αμελώντας την εναλλαγή του $\tilde{\tau}_1$ στις αντίστοιχες διαδικασίες.
- $\tilde{\tau}_2\tilde{e}_R$, $\tilde{\tau}_2\tilde{e}_R^*$, $\tilde{e}_R\tilde{\mu}_R$, $\tilde{e}_R\tilde{\mu}_R^*$ και τελικές $\tau e$, $\tau\bar{e}$, $e\mu$, $e\bar{\mu}$ αντίστοιχα, πραγματοποιούνται μέσω μιας $\tilde{\chi}$ εναλλαγής στο $t$-κανάλι.

Τα επιπλέον διαγράμματα αναγράφονται στον Πίνακα 5.6. Λόγω των επιπρόσθετων αυτών διαγραμμάτων, προκύπτουν και επιπλέον συνεισφορές στους συντελεστές $a_{ij}$. Ειδικότερα, οι συνεισφορές από διαδικασίες με αρχικές καταστάσεις:

- $\tilde{\chi}\tilde{\tau}_2$, $\tilde{\tau}_2\tilde{\tau}_2$, $\tilde{\tau}_2\tilde{\tau}_2^*$ αναγράφονται στον Πίνακα 5.4.
- $\tilde{\chi}\tilde{e}_R, \tilde{e}_R\tilde{e}_R, \tilde{e}_R\tilde{e}_R^*$ λαμβάνονται κάνοντας την αντικατάσταση $\tilde{\tau}_2 \to \tilde{e}_R$ και αμελώντας τη μίξη των staus και τη μάζα του $\tau$ στο τυπολόγιο του Πίνακα 5.4 και στα εμφανιζόμενα εκεί $g$ σύμβολα. Δηλαδή, πρέπει $c_\tau = 1$, $s_\tau = 0$, $m_\tau = 0$ στους τύπους του Πίνακα 5.4 και του Εδ. Β$'$.3.2.
- $\tilde{\tau}_2\tilde{e}_R$, $\tilde{\tau}_2\tilde{e}_R^*$, $\tilde{e}_R\tilde{\mu}_R$, $\tilde{e}_R\tilde{\mu}_R^*$ αναγράφονται στον παρακάτω Πίνακα:

**Πίνακας 5.7: Συνεισφορές στους Συντελεστές $a_{ij}$**

| Διαδικασία | Συνεισφορά στο Συντελεστή $a_{ij}$ |
|---|---|
| $\tilde{\tau}_2\tilde{e}_R \to \tau e$ | $e^4 Y_R^4 c_\tau^2 m_{\tilde{\chi}}^2 (m_{\tilde{e}_R} + m_{\tilde{\tau}_2})^2 /$ $8\pi c_W^4 m_{\tilde{e}_R} m_{\tilde{\tau}_2} (m_{\tilde{\chi}}^2 + m_{\tilde{e}_R} m_{\tilde{\tau}_2})$ |
| $\tilde{\tau}_2\tilde{e}_R^* \to \tau\bar{e}$ | $e^4 Y_L^2 Y_R^2 s_\tau^2 m_{\tilde{\chi}}^2 (m_{\tilde{e}_R} + m_{\tilde{\tau}_2})^2 /$ $8\pi c_W^4 m_{\tilde{e}_R} m_{\tilde{\tau}_2} (m_{\tilde{\chi}}^2 + m_{\tilde{e}_R} m_{\tilde{\tau}_2})$ |
| $\tilde{e}_R\tilde{\mu}_R \to e\mu$ | $e^4 Y_R^4 m_{\tilde{\chi}}^2 / 2\pi c_W^4 \Sigma_e^2$ |
| $\tilde{e}_R\tilde{\mu}_R^* \to e\bar{\mu}$ | $e^4 Y_R^4 m_{\tilde{e}_R}^2 / 12\pi c_W^4 \Sigma_e^2$ |

όπου $\Sigma_e = m_{\tilde{\chi}}^2 + m_{\tilde{e}_R}^2$ με $m_{\tilde{e}_R}$ να είναι η κοινή μάζα των δεξιόστροφων sleptons $\tilde{e}_R$, $\tilde{\mu}_R$ των πρώτων γενεών και $Y_{L(R)} = -1/2(-1)$ το υπερφορτίο των αριστερόστροφων και δεξιόστροφων λεπτονίων.

### 5.5.4 Σύγκριση Αποτελεσμάτων

Τα αποτελέσματα που παρουσιάστηκαν στα Εδ. 5.5.2 και 5.5.3 προήλθαν από ενδελεχή μελέτη και αξιοποίηση του παραρτήματος της Αν. [64]. Συγκρίνοντας τα αποτελέσματα αυτά, με τα αντίστοιχα της Αν. [64], μπορούν να επισημανθούν τα ακόλουθα:

**α.** Ουσιώδης διαφορά της ανάλυσής μας με αυτή της Αν. [64] είναι ότι συμπεριλαμβάνεται η ανάμιξη των $\tilde{\tau}_1 - \tilde{\tau}_2$ που ενίοτε (για $\tan\beta \geq 15$) συμβαίνει να είναι πολύ σημαντική. Βεβαίως, η ανάμιξη των ελαφρότερων οικογενειών αμελείται και στις δύο προσεγγισεις. Λόγω της ανάμιξης των $\tilde{\tau}_1 - \tilde{\tau}_2$ ο υπολογισμός μας (σε αντίθεση με αυτόν της Αν. [64]) συμπεριλαμβάνει:

- Νέους δίαυλους αλληλεπίδρασης με εναλλαγή του $\tilde{\tau}_1$ σωματίου στις περισσότερες από τις θεωρούμενες διαδικασίες, οι οποίοι πραγματοποιούνται μέσω των αντίστοιχων $g$ συμβόλων ($g_{\tilde{\tau}_1\tilde{\tau}_2 Z}$, $g_{h[H]1}$, $g_{A1}$).
- Το διάγραμμα επαφής (c) στη διαδικασία $\tilde{\tau}_2\tilde{\tau}_2^* \to W^+W^-$ μέσω του συμβόλου $g_{\tilde{\tau}_2\tilde{\tau}_2 W^+W^-}$.
- Όρους ανάλογους των $c_\tau$, $s_\tau$ σε διάφορα $g$ σύμβολα ($g_{h[H]}$, $g_{\tilde{\tau}_2\tilde{\tau}_2 Z}$, $g_{\tilde{\tau}_2\tilde{\tau}_2 hh[HH]}$, $g_{\tilde{\tau}_2\tilde{\tau}_2 hH}$, $g_{\tilde{\tau}_2\tilde{\tau}_2 AA}$, $g_{\tilde{\tau}_2\tilde{\tau}_2 H^+H^-}$).

**β.** Λόγω των ενίοτε ($\tan\beta \geq 15$) υψηλών ζεύξεων Yukawa ο υπολογισμός μας (σε αντίθεση με αυτόν της Αν. [64] ) συμπεριλαμβάνει:

- Σύμβολα $g$ ανάλογα των μαζών των φερμιονίων της τρίτης γενεάς, $m_\tau$, $m_t$, ($g_{A1}$, $g_{h[H]\tau\tau}$, $g_{A\tau\tau}$).
- Όρους ανάλογους των μαζών των φερμιονίων της τρίτης γενεάς, σε κάποια άλλα $g$ σύμβολα ($g_{h[H]1}$, $g_{h[H]}$, $g_{\tilde{\tau}_2\tilde{\tau}_2 hh[HH]}$, $g_{\tilde{\tau}_2\tilde{\tau}_2 hH}$, $g_{\tilde{\tau}_2\tilde{\tau}_2 AA}$, $g_{\tilde{\tau}_2\tilde{\tau}_2 H^+H^-}$).
- Το δίαυλο αλληλεπίδρασης $s(\tau)$ στις διαδικασίες $\tilde{\chi}\tilde{\tau}_2 \to \tau A$, $\tau h$ και $\tau H$.
- Τους δίαυλους αλληλεπίδρασης $s(h)$, $s(H)$ στις διαδικασίες $\tilde{\tau}_2\tilde{\tau}_2^* \to b\bar{b}$ και $\tau\bar{\tau}$.

**γ.** Τυπογραφικά λαθη και παραλήψεις στο παράρτημα της Αν. [64] εχουν προσεκτικά επισημανθεί και δεν συμπεριλαμβάνονται στο τυπολόγιο που παρουσιάστηκε και στο αριθμητικό προγραμμα που εφαρμόζεται. Ενδεικτικά (χρησιμοποιώντας και το συμβολισμό του ενλόγω παραρτήματος) αναφέρεται ότι στις διαδικασίες:

- $\tilde{\tau}\tilde{\tau}^* \to \tau\bar{\tau}$, $f\bar{f}$, $t\bar{t}$ ο δεύτερος πολλαπλασιαστικός παράγοντας στα σύμβολα $f_{3c}$ και $f_{3d}$ που αναδύεται από τη ζεύξη $Z$- fermion-fermion πρέπει να διαθέτει στον παρονομαστή τον όρο $4\cos\theta_W$ και όχι τον αναγραφόμενο $4\cos^2\theta_W$ όπως επιβεβαιώνεται και στην Εξ. (C.62) της Αν. [13].
- $\tilde{\tau}\tilde{\tau}^* \to hH$ δεν συμπεριλαμβάνεται το $u$-κανάλι χωρίς αυτό να οφείλεται στην μη θεώρηση της ανάμιξης $\tilde{\tau}_1 - \tilde{\tau}_2$.
- $\tilde{\tau}\tilde{\tau}^* \to hh$ ο πρώτος πολλαπλασιαστικός παράγοντας στο σύμβολο $f_2$, που αντιστοιχεί στο οικείο σύμβολο $g_H$ πρέπει να περιέχει $\cos(\alpha+\beta)$ και όχι $\sin(\alpha+\beta)$.
- $\tilde{\tau}\tilde{\tau}^* \to HH$, $H^+H^-$ ο πρώτος πολλαπλασιαστικός παράγοντας στο σύμβολο $f_1$, που αντιστοιχεί στο οικείο σύμβολο $g_H$ πρέπει να περιέχει $\cos(\alpha+\beta)$ και όχι $\sin(\alpha+\beta)$ και να προσημανθεί αντιθέτως.
- $\tilde{\tau}\tilde{\tau}^* \to HH$, $H^+H^-$ ο πρώτος πολλαπλασιαστικός παράγοντας στο σύμβολο $f_2$, που αντιστοιχεί στο οικείο σύμβολο $g_h$ πρέπει να προσημανθεί αντιθέτως.
- $\chi\tilde{\tau} \to \tau\gamma$, $\tau h[H]$, $\tau Z$ διαθέτουν την average των αρχικών σπιν για το $\chi$ ενώ η $\tilde{\tau}\tilde{\tau}^* \to \tau\tau$ δε διαθέτει τον παράγοντα 1/2 που πρέπει να συμπεριληφθεί λόγω των ταυτοτικών τελικών καταστάσεων.

Συμπερασματικά, ο υπολογισμός που παρουσίαστηκε στα δύο προηγούμενα εδάφια έχει πολλαπλώς ελεγχθεί και πιστεύεται ότι παρέχει αξιόπιστα αποτελέσματα για όλες τις δυνατές τιμές της παραμέτρου $\tan\beta$ και για LSP μορφής bino, βεβαίως.

# Κεφάλαιο 6

# MSSM με ενοποίηση Yukawa

## 6.1 Εισαγωγή

Από την πληθώρα εκδόσεων του MSSM, επιλέγεται και μελετάται μία που προέρχεται από τη συζεύξη της SUSY με μία ομάδα μεγάλης Ενοποίησης, που προβλέπει ενοποίηση των ζεύξεων Βαθμίδας και Yukawa. Μια σύντομη αιτιολόγηση της επιλογής μας αυτή γίνεται στο Εδ. 6.2 και έπονται οι αριθμητικές εφαρμογές. Στο Εδ. 6.3 καταδεικνύεται η προιούσα ενοποίηση και στα Εδ. 6.4 και 6.5 μελετώνται τα παραμετρικά και φαινομενολογικά χαρακτηριστικά του προτύπου. Ακολουθεί η εφαρμογή του στο πρόβλημα της Σκοτεινής Ύλης, στο Εδ. 6.6. Συγκεντρωτικά, τα συμπεράσματα της μελέτης του προτύπου αυτού, που προέρχονται από τα αριθμητικά ευρήματα των Αν. [44] και [72], καταγράφονται στο Εδ. 6.7. Στο τελευταίο Εδ. 6.8 γίνεται μια διερεύνηση των δυνατοτήτων που διανοίγονται αν η απαίτηση της ενοποίησης Yukawa εγκαταλειφθεί.

## 6.2 SUSY-GUT με ενοποίηση Yukawa

Μια αιτιολόγηση της επιλογής μελέτης ενός SUSY-GUT με Μεγάλη ομάδα συμμετρίας $G$ την $SO(10)$ ή την $E_6$ επιχειρείται παρακάτω. Σύμφωνα με τις Αν. [65], [66] και [67] σε τέτοια πρότυπα :

- Και οι τρεις αλληλεπιδράσεις του SM ενοποιούνται με μια απλή ομάδα Lie. Επιπλέον, η SUSY λύνει το πρόβλημα της ιεραρχίας ανάμεσα στην EWS και στην GUT κλίμακα ενέργειας.

- Τα 15 φερμιόνια κάθε γενέας του SM συν ένα δεξιόστροφο νετρίνο διευθετούνται οικονομικά στην 16-διαστάσεων σπινοριακή αναπαράσταση της $SO(10)$. Παρενθετίκα επισημαίνεται ότι η χρήση του όρου $SO(10)$ γίνεται καταχρηστικά και αναφέρεται στην καλύπτουσα (covering) ομάδα της $SO(10)$, την $spin(10)$ ομάδα που διαθέτει τη σπινοριακή αναπαράσταση.

- Το singlet νετρίνο superfield(s) $\hat{N}^c$ μπορεί να αναπτύξει μια Majorana μάζα σε πολύ μεγάλες ενέργειες $\sim 10^{11} - 10^{16}$GeV. Συνδυάζοντας αυτή με τη συνηθισμένη Dirac μάζα που μπορεί να αναπτύξει το νετρίνο, οδηγείται κανείς σε μάζες για τα αριστερόστροφα νετρίνο της τάξης των eV, ενώ τα δεξιόστροφα νετρίνο παραμένουν απαρατήρητα, μέσω του μηχανισμού της τραμπάλας see-saw. Οι λαμβανόμενες μάζες για τα νετρίνο μπορούν εύκολα να βρεθούν σε συμφωνία με τα δεδομένα από τα πειράματα για τα ηλιακά και ατμοσφαιρικά νετρίνο.

- Υπάρχει μόνο μία ζεύξη Yukawa ανά γενεά οπότε προβλέπεται ενοποίηση Yukawa σε GUT κλίμακα ενέργειας. Ειδικά αυτό αφορά την τρίτη γενεά, αλλά μπορεί να επεκταθεί και στις ελαφρότερες γενεές.

- Η παράμετρος $\tan\beta$ δεν είναι ελεύθερη, δεν εισάγεται αυθαιρέτως στο αριθμητικό πρόγραμμα προσομοίωσης, αλλά προσδιορίζεται από την απαίτηση της ενοποίησης Yukawa. Επομένως, τα πρότυπα αυτά έχουν το προνόμιο μιας λιγότερης ανεξάρτητης μεταβλητής.

- Η διατήρηση της $R$-parity είναι μια φυσική συνέπεια της συμμετρίας βαθμίδας του προτύπου σύμφωνα με την Αν. [68]. Όταν η $SO(10)$ σπάει, μέσω μιας ορισμένης αναπαράστασης των πεδίων Higgs, η $R$-parity παραμένει διατηρήσιμη ακόμα και στην ηλεκτρασθενή κλίμακα ενέργειας.

- Η βαρυογένεση στο πρώιμο Σύμπαν μπορεί να εξηγηθεί ως συνέπεια των διασπάσεων των δεξιόστροφών νετρίνων που έχουν δραπετεύσει από τη ΘΔΙ του κοσμικού ρευστού.

Συμπερασματικά, τα πρότυπα SUSY-GUT με ενοποίηση Yukawa έχουν αυξημένη προβλεψιμότητα και ισχυρό καθορισμό. Στην μελέτη που θα ακολούθησει γίνονται οι ακόλουθες υποθέσεις:

- Ένα πρότυπο SUSY GUT με μία ζεύξη Yukawa ανα γενεά είναι ισχυρό σε κλίμακα ενέργειας $M > M_G$.
- Η ομάδα συμμετρίας $G$ καταρρέει μέσω κάποιου μηχανισμού στην ομάδα του MSSM σε κλίμακα ενέργειας $M = M_G$. Υποτίθεται παγκοσμιότητα των SBT, όπως θα εξηγηθεί στο Εδ. 6.4.
- Η ηλεκτρασθενής συμμετρία καταρρέει μέσω κβαντικών διορθώσεων σε κλίμακα ενέργειας $M = M_S$.

Με τις υποθέσεις αυτές, το φάσμα της θεωρίας και οι φαινομενολογικές και κοσμολογικές παράμετρες του προτύπου μελετώνται στη συνέχεια του κεφαλαίου.

## 6.3 Ενοποίηση ζεύξεων Βαθμίδας και Yukawa

Κάποια χαρακτηριστικά του αριθμητικού πρόγράμματος που χρησιμοποιείται για την υλοποίηση της ενοποίησης των ζεύξεων Βαθμίδας και Yukawa δίνονται αντίστοιχα στα Εδ. 6.3.1, 6.3.2.

### 6.3.1 Ενοποίηση ζεύξεων Βαθμίδας

Μία από τις σημαντικές επιτυχίες της υπερσυμμετρίας είναι η επίτευξη πιο ακριβούς ενοποίησης των ζεύξεων Βαθμίδας, από αυτήν που το μη υπερσυμμετρικό SM επιτύγχανε. Υπάρχει, δηλαδή, σημείο στό οποίο οι ζεύξεις Βαθμίδας των ισχυρών, ασθενών και ηλεκτρομαγνητικών ενοποιούνται με πολύ καλή ακρίβεια καθώς εξελίσσονται με τις εξισώσεις επανακανονικοποίησης (: RGE) από κλίμακα χαμηλής ενέργειας σε κλίμακα υψηλότερης ενεργειας. Το σημείο αυτό καθορίζεται από την τιμή της ενέργειας του, $M_G$ και την τιμή της (ενοποιημένης) ζεύξης, $g_G$.

Πρακτικά, στο αριθμητικό πρόγραμμα που κατασκευάστηκε, οι RGE τρέχουν από κλίμακα υψηλής ενέργειας με αρχικές συνθήκες υποθετικές:

$$g_1(M_G) = g_2(M_G) = g_3(M_G) := g_G\ . \tag{6.1}$$

όπου οι σταθερές ζεύξης σχετίζονται με τις αντίστοιχες που παρουσιάστηκαν στο Κεφάλαιο 2 ως εξής:

$$g'(t) = \sqrt{\frac{3}{5}}g_1(t),\quad g(t) = g_2(t)\quad g_3(t) = g_3(t)\quad \text{όπου}\quad t = \ln M \tag{6.2}$$

από τις οποίες υπολογίζονται οι αντίστοιχες σταθερές λεπτής υφής:

$$\alpha_i(t) = \frac{g_i^2(t)}{4\pi}\quad \text{όπου}\quad i = 1,\, 2,\, 3 \tag{6.3}$$

με $M$, την κλίμακα ενέργειας που είναι και η ανεξάρτητη μεταβλητή στις RGE.

Οι RGE του MSSM σε επίπεδο 2 βρόχων εξελίσσονται από την κλίμακα ενοποίησης, $M_G$ μέχρι μια ενδιάμεση κλίμακα $M_S$ συγκρίσιμη με την προνομιακή που έχει οριστεί στην Εξ. (3.1) και στη συνέχεια, δίνουν τη θέση τους στις RGE του SM σε επίπεδο 2 βρόχων που συνεχίζουν την εξέλιξη μέχρι την κλίμακα $M_Z$. Σε αυτή την κλίμακα υπολογίζονται οι επίμαχες ποσότητες

$$s_W^2(M_Z) = \frac{3\alpha_1(M_Z)/5}{3\alpha_1(M_Z)/5 + \alpha_2(M_Z)},\quad \alpha_{em}(M_Z) = \alpha_2(M_Z)s_W^2(M_Z),\quad \alpha_S(M_Z) = \alpha_3(M_Z) \tag{6.4}$$

χρησιμοποιώντας τις Εξ. (2.20). Επιδιώκοντας να επιτυγχάνονται στην ενέργεια $M_Z$ οι πειραματικές τιμές:

$$\alpha_{em}(M_Z) \simeq 1/128,\quad s_W^2(M_Z) \simeq 0.232, \tag{6.5}$$

$$\alpha_S(M_Z) \simeq 0.12,\quad \text{όπου}\quad M_Z \simeq 91.18\,\text{GeV}, \tag{6.6}$$

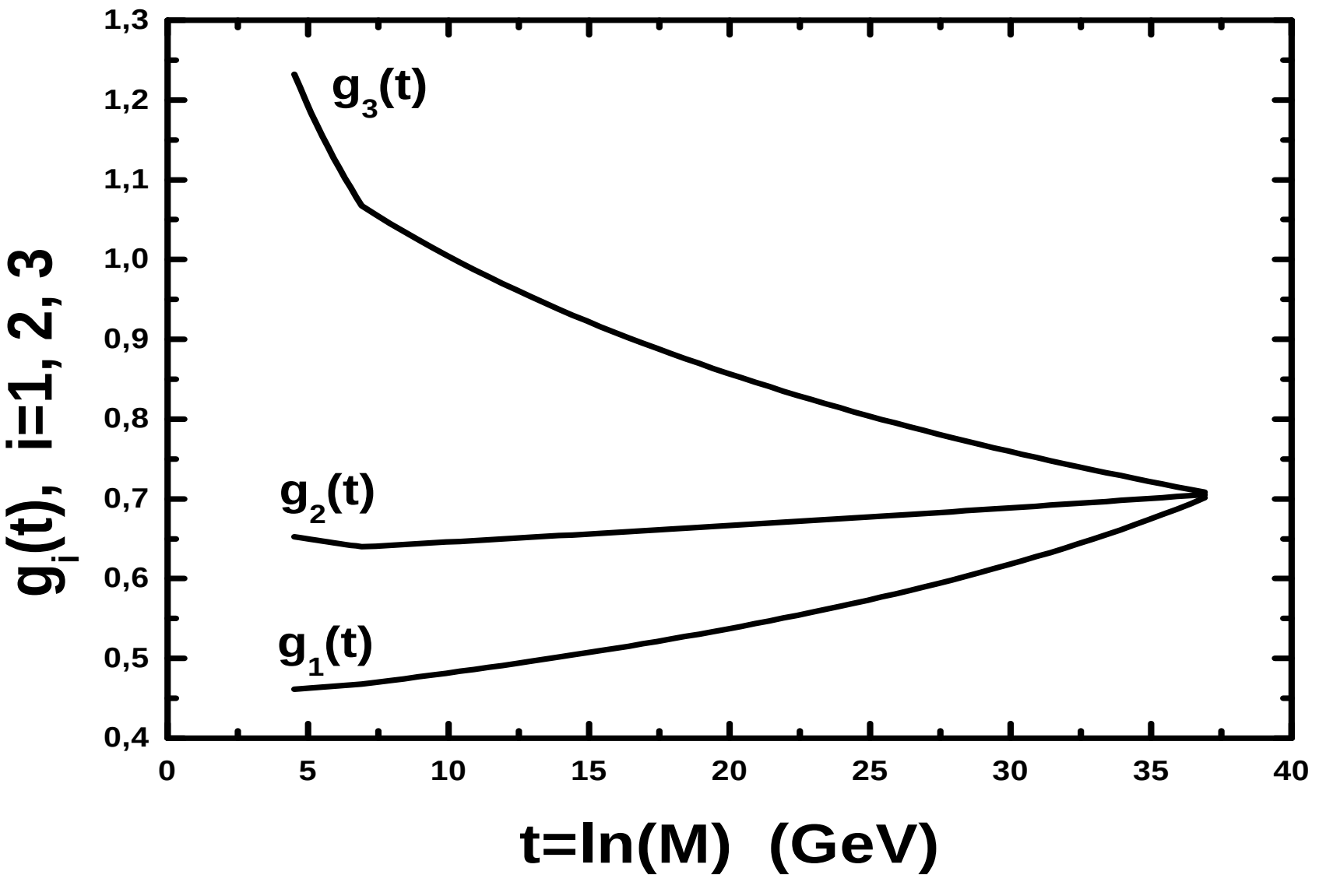

**Σχήμα 6.1:** *Η εξέλιξη των ζεύξεων Βαθμίδας για* $M_S = 1\,\text{TeV}$.

καθορίζεται η κλίμακα GUT, $M_G \sim 1.5 \times 10^{16}$ GeV και η τιμή της ενοποιημένης ζεύξης $\alpha_G \sim 1/25.2$.

Οι χρησιμοποιούμενες RGE στο αριθμητικό πρόγραμμα δίνονται στο Εδ. Α΄.2. Στόχος του προγράμματος δεν είναι η εκτέλεση ελέγχου ακριβείας της ενοποίησης, αλλά η χρησιμοποίηση των αποτελεσμάτων για την διερεύνηση φαινομενολογικών και κοσμολογικών συνεπειών του MSSM. Μερικά ενδεικτικά αποτελέσματα των επιδόσεων μας παρουσιάζονται στον Πίνακα 6.1, ενώ μια αντιπροσωπευτική εικόνα ενοποίησης δίνεται στο Σχ. 6.1.

### 6.3.2 Ενοποίηση ζεύξεων Yukawa

Ενοποίηση Yukawa σημαίνει ότι οι ασυμπτωτικές τιμές (στην κλίμακα $M_G$) των ζεύξεων Yukawa λαμβάνουν τήν ίδια τιμή, $h_0$. Συμβολικά:

$$h_t(M_G) = h_b(M_G) = h_\tau(M_G) := h_0\ . \tag{6.7}$$

Οι χρησιμοποιούμενες στο αριθμητικό πρόγραμμα RGE του MSSM σε επίπεδο 2 βρόχων δίνονται στο Εδ. Α΄.2. Οι RGE του SM, όμως χρησιμοποιούνται σε επίπεδο 1 βρόχου. Ο προσδιορισμός της απαιτούμενης αρχικής συνθήκης $h_0$ γίνεται ως εξής:

- Δίνεται αρχικά μια δοκιμαστική τιμη $0 < h_0 < 1$
- Με την τιμή αυτή ως αρχική συνθήκη το πρόγραμμα ρέει ως το $M_S$ λύνοντας τις RGE του MSSM και στη συνέχεια μέχρι το $M_Z$ λύνοντας τις RGE του SM.
- Εισάγοντας στο $M_S$, τη μάζα $m_\tau(M_S) = 1.78\,\text{GeV}$, λαμβάνεται μια τιμή για την $\beta$ μέσω της Εξ. (2.59), έχοντας δεδομένο από την εξέλιξη των RGE το $h_\tau(M_S)$.
- Με την τιμή αυτή της $\beta$ προσδιορίζεται η μάζα $m_t(m_t)$ μέσω της Εξ. (2.59), χρησιμοποιώντας μια επαναληπτική συνθήκη στο πρόγραμμα που θα εξασφαλίζει την ταυτότητα συνάρτησης και ορίσματος.
- Αν το εξαγώμενο για τη μάζα $m_t(m_t)$ είναι μέσα στα όρια της Εξ. (3.21) η διαδικασία σταματά, ειδάλλως επαναλαμβάνεται.

**Πίνακας 6.1: Ενοποίηση ζεύξεων Βαθμίδας και Yukawa**

| | | | | |
|---|---|---|---|---|
| sign$\mu$ | $-1$ | $-1$ | $-1$ | $+1$ |
| $M_S$ (GeV) | 850 | 1130 | 1380 | 2550 |
| **Μεταβλητές Εισόδου** | | | | |
| $M_G$ ($10^{16}$ GeV) | 1.55 | 1.394 | 1.123 | 8.92 |
| $\alpha_G^{-1}$ | 25.13 | 25.23 | 25.28 | 25.61 |
| $h_G$ | .0668 | .660 | .658 | .65 |
| $m_\tau(M_S)$ (GeV) | 1.99 | 2.01 | 2.018 | 1.76 |
| **Μεταβλητές Εξόδου** | | | | |
| $\alpha_3(M_Z)$ | .1203 | .1208 | .1207 | .12059 |
| $s^2_W(M_Z)$ | .23106 | .2307 | .2308 | .230 |
| $\alpha^{-1}_{em}(M_Z)$ | 128.016 | 127.7 | 127.12 | 126.69 |
| $\tan\beta$ | 46.28 | 45.96 | 45.89 | 52.8 |
| $m_t(m_t)$ (GeV) | 166.006 | 166.026 | 166.038 | 166.006 |
| $m^c_b(M_Z)$ (GeV) | 4.42 | 4.42 | 4.40 | 2.38 |
| $m^c_\tau(M_Z)$ (GeV) | 1.743 | 1.744 | 1.742 | 1.741 |

Η διαδικασία μπορεί να βελτιωθεί αν συμπεριληφθούν οι διορθώσεις στο $m_\tau$ με βάση το τυπολόγιο του Εδ. 3.3.3. Όποτε η μάζα που θα τεθεί σε κλίμακα $M_S$ δεν θα είναι 1.78GeV αλλά τόση όση χρειάζεται, ώστε μετά την προσθήκη των διορθώσεων από την Εξ. (3.23) να γίνεται ίση με την πειραματικά προβλεπόμενη τιμή της Εξ. (3.24). Συνακόλουθα μεταβάλλεται και η τιμή που επιτυγχάνεται για την $\tan\beta$, αφού οι διορθώσεις της (3.23) έχουν το πρόσημο του $\mu$. Συγκεκριμένα:

- Όταν $\mu < 0$, λαμβάνεται $46.3 \geq \tan\beta \geq 45$
- Όταν $\mu > 0$, λαμβάνεται $53.5 \geq \tan\beta \geq 51.7$

καθώς το $m_A$ μεταβάλλεται στο διάστημα $100\text{GeV} \leq m_A \leq 700\text{GeV}$. Χωρίς την εισαγωγή των διορθώσεων του $\tau$, και στις δύο προηγούμενες περιπτώσεις λαμβάνεται: $\tan\beta \simeq 52$. Παραδείγματα υλοποίησης των ιδεών αυτών καταγράφονται στον Πίνακα 6.1 όπου εκτίθενται μερικά από τα εισαγόμενα και τα εξαγόμενα του αριθμητικού προγράμματος. Μερικά από τα φάσματα που χρησιμοποιήθηκαν για τον υπολογισμό των $m^c_b(M_Z)$ και $m^c_\tau(M_Z)$ παρέχονται στον Πίνακα 6.4 με σημείο αναφοράς την τιμή της $\tan\beta$ και το sign$\mu$. Μια αντιπροσωπευτική εικόνα ενοποίησης των ζεύξεων Yukawa δίνεται στο Σχ. 6.2 με $M_S = 1$ TeV

Το πλεονέκτημα αυτής της προσέγγισης είναι ότι η παράμετρος $\tan\beta$ είναι αποτέλεσμα της ασυμπτωτικής συνθήκης που ικανοποιούν οι ζεύξεις Yukawa και όχι μια ελεύθερη παράμετρος του προβλήματος. Μειονέκτημα είναι οτι υπάρχει πρόβλεψη για τη μάζα του $b$ quark, η οποία κατα κανονα εξωθείται κοντά στα ανώτερα πειραματικά της όρια της Εξ. (3.18). Για το θέμα αυτό θα γίνει διεξοδική αναλυση στο Εδ. 6.5.1.

## 6.4 Παραμετρική μελέτη

Όπως φάνηκε από το Εδ. 6.3, η απαίτηση της ενοποίησης των ζεύξεων βαθμίδας και Yukawa συνεπικουρούμενη από τα πειραματικά δεδομένα για τα $\alpha_{em}(M_Z)$, $s^2_W$ και τις μάζες των φερμιονίων $t$, $\tau$, προσδιορίζει πλήρως την εξέλιξη του πρώτου συνόλου των RGE. Το δεύτερο σύνολο των RGE ελέγχει την εξέλιξη των SBT. Οι χρησιμοποιούμενες στο αριθμητικό πρόγραμμα RGE του MSSM σε επίπεδο ενός βρόχου δίνονται στο Εδ. Α΄.3. Χρησιμοποιούνται παγκόσμιες (universal) αρχικές συνθήκες και δη:

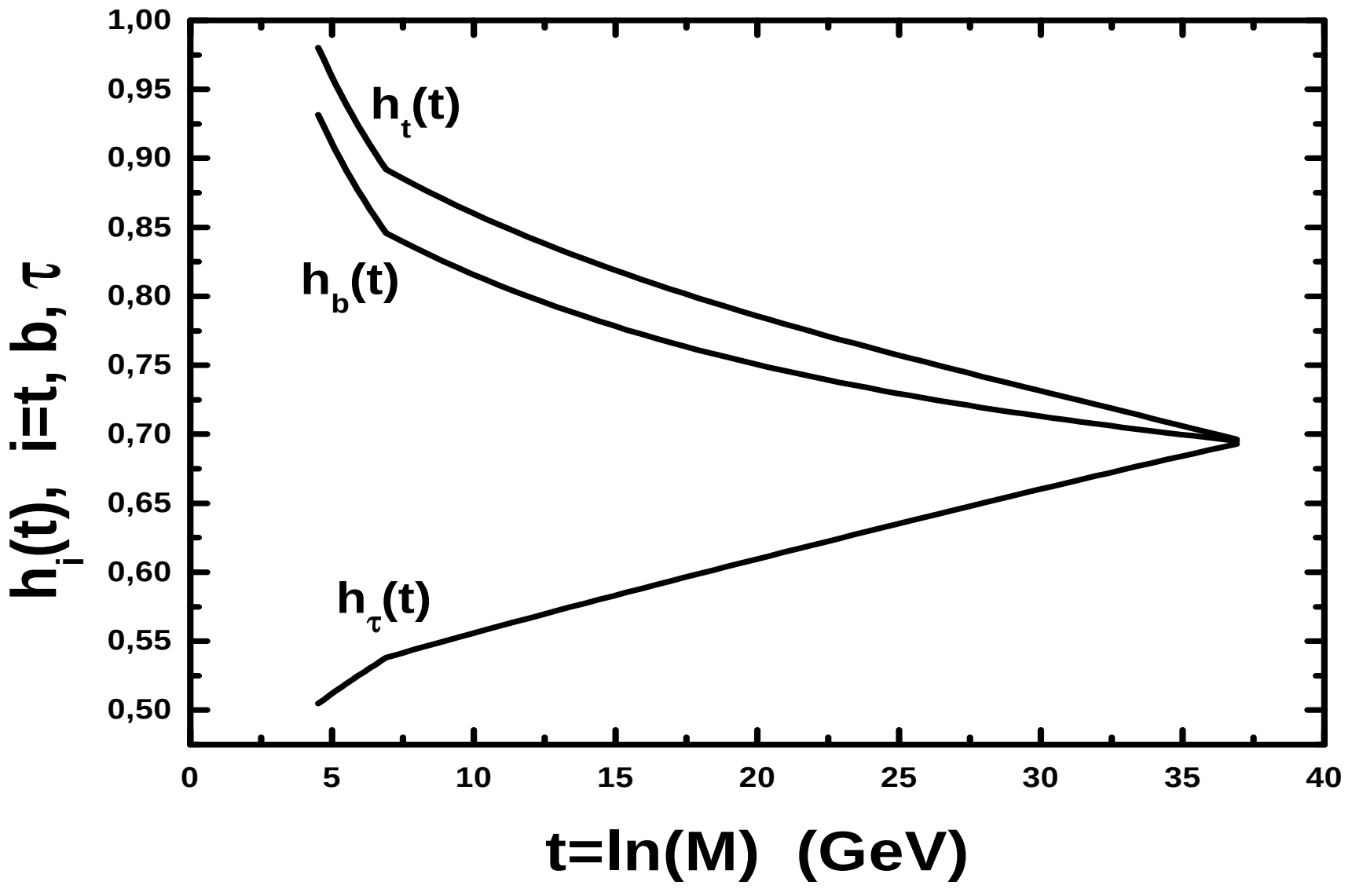

**Σχήμα 6.2:** *Η εξέλιξη των ζεύξεων Yukawa για* $M_S = 1$ TeV.

- Μια κοινή μάζα για τα gaugino, $M_{1/2}$. Συγκεκριμένα,

$$M_1(M_G) = M_2(M_G) = M_3(M_G) = M_{1/2}. \tag{6.8}$$

- Μια κοινή μάζα για τα βαθμωτά πεδία, $m_0$. Συγκεκριμένα, οι ασυμπτωτικές συνθήκες:
  - Για τις βαριές γενεές, είναι:

$$m_{L_L}(M_G) = m_{Q_L}(M_G) = m_{\tau_R}(M_G) = m_{t_R}(M_G) = m_{b_R}(M_G) = m_0. \tag{6.9}$$

  - Για τις ελαφρές γενεές, είναι:

$$m_{l_L}(M_G) = m_{q_L}(M_G) = m_{e_R}(M_G) = m_{u_R}(M_G) = m_{d_R}(M_G) = m_0. \tag{6.10}$$

  - Για τα Higgs, είναι:

$$m_H(M_G) = m_{\bar{H}}(M_G) = m_0. \tag{6.11}$$

- Μια κοινή τριγραμμική ζεύξη, $A_0$. Για τις βαριές γενεές, οι ασυμπτωτικές συνθήκες είναι:

$$A_t(M_G) = A_b(M_G) = A_\tau(M_G) = A_0. \tag{6.12}$$

Για τις ελαφρές γενεές, οι ασυμπτωτικές συνθήκες είναι προφανώς παρόμοιες, αλλά στην προσέγγιση που εφαρμόζεται, με τα φερμιόνια των ελαφρών οικογενειών να θεωρούνται άμαζα, η εξέλιξη των τριγραμμικών ζεύξεων των ελαφρών γενεών καθίσταται άχρηστη.

Η λύση των διαφορικών εξισώσεων γίνεται αριθμητικά με χρήση της *Mathematica* (Αν. [69]). Η ολοκλήρωση του συστήματος των εξισώσεων αυτών γίνεται από κλίμακα ενέργειας $M_G$ μέχρι $M_S$. Σε αυτή την κλίμακα ενέργειας βρίσκεται η παράμετρος $\mu$ χρησιμοποιώντας την Εξ. (2.45) στα πλαίσια της προσέγγισης του tree-level renormalization group improved effective potential σύμφωνα με τα αναφερόμενα στο Εδ.

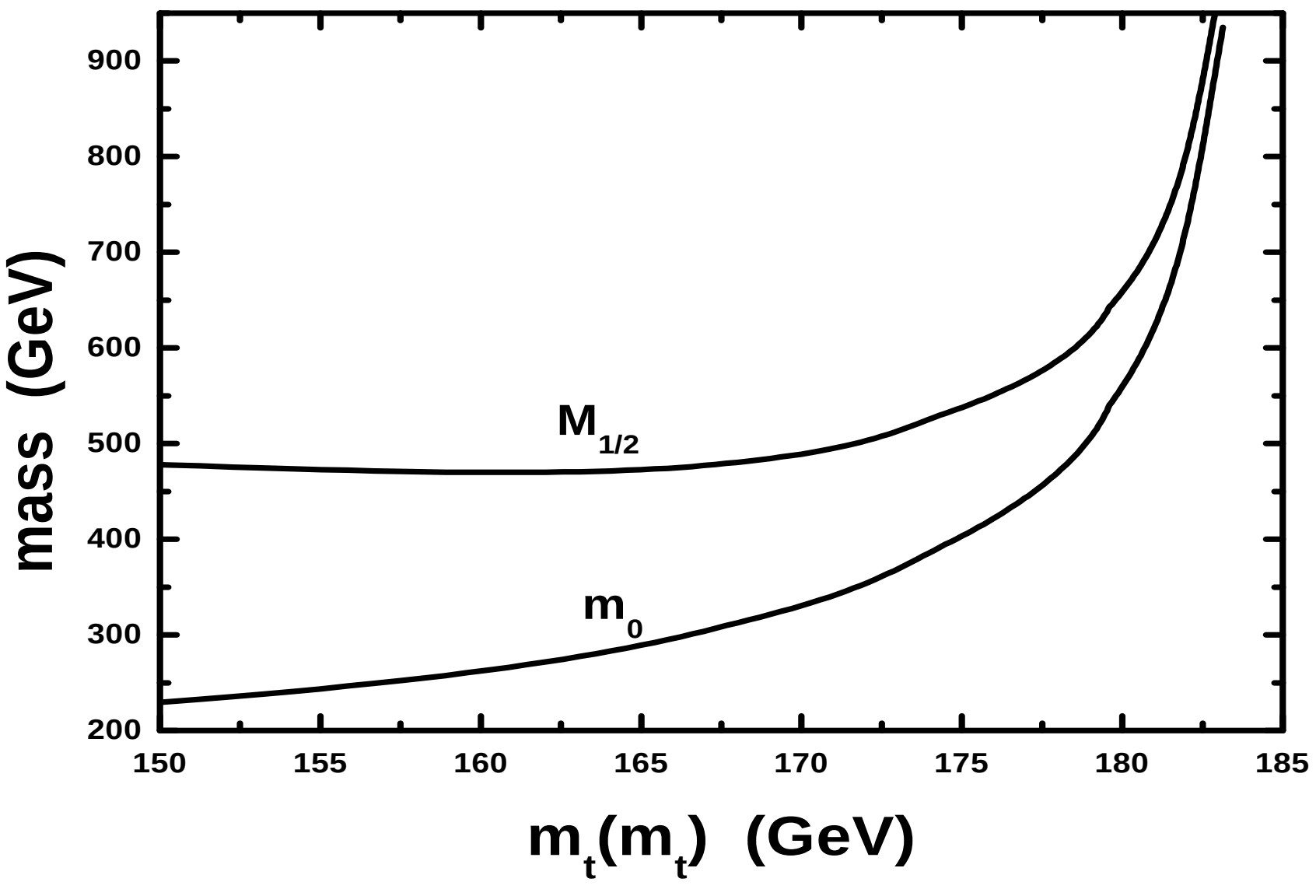

**Σχήμα 6.3:** *Οι ασυμπτωτικές τιμές των ασθενών μαζών για* $\Delta_{NLSP} = 0, m_A = M_Z$ *και* $M_S = 1$ TeV.

3.2.1. Ακολούθως, υπολογίζεται το SUSY φάσμα της θεωρίας σε κλίμακα ενέργειας $M_S$ ακολουθώντας το τυπολόγιο του Εδ. 2.5. Το LSP της θεωρίας προκύπτει ότι είναι μορφής bino με καθαρότητα 98% ενώ το NLSP είναι το ελαφρότερο stau μορφής δεξιόστροφης. Λόγω της υψηλής τιμής της παραμέτρου $\tan\beta$ οι ΙΔΤ των άλλων sfermions είναι σημαντικά υψηλότερα από τις μάζες των προηγούμενων δύο σωματίων. Μετά τον υπολογισμό του φάσματος γίνεται μια σύγκριση της δοκιμαστικής τιμής του $M_S$ που επιλέγει με αυτή της Εξ. (3.1), οπότε η όλη διαδικασία επαναλαμβάνεται μέχρι την επίτευξη ενός αυτοσυνεπούς αποτελέσματος. Συνήθως, αυτό συμβαίνει μετά δύο ή τρεις επαναλήψεις.

Η επεξεργασία και η παρουσίαση των απότελεσμάτων σε αυτό το προκαταρκτικό στάδιο του υπολογισμού δίνονται στα Εδ. 6.4.1 και 6.4.2.

## 6.4.1 Μετάθεση παραμέτρων

Οι αρχικά ελεύθερες παράμετροι του προτύπου είναι:

$$m_0,\ M_{1/2},\ A_0,\ \mathrm{sign}\mu\ .$$

Τιθέται $A_0 = 0$. Η επιλογή αυτή είναι σίγουρα αυθαίρετη, βοηθάει όμως, στον καλύτερο προσδιορισμό του προτύπου απαλλάσοντάς το από μια παράμετρο. Τα αποτελέσματα δεν επηρεάζονται κρίσιμα από αυτή την επιλογή, παρότι οι ζεύξεις αυτές υπεισέρχονται στον υπολογίσμο αρκετών $g$-συμβόλων.

Στην Αν. [20], επισημαίνεται ότι μπορεί να θεμελιωθεί αριθμητικά μια σχέση ανάμεσα στις παράμετρες $m_A$, $m_0$ και $M_{1/2}$. Πραγματικά, βασιζόμενοι σε τεχνίκες Fit αποδεικνύουμε τη σχέση αυτή:

$$m_A^2 \simeq c_M M_{1/2}^2 + c_m m_0^2 - M_Z^2\ , \tag{6.13}$$

με συντελέστες $c_M \sim 0.1$ και $c_m \sim -0.1$, καθοριζόμενους από το αριθμητικό πρόγραμμα για κάθε τιμή των $m_t(m_t)$, $M_S$. Η εξαγωγή της προηγούμενης πολύ χρήσιμης σχέσης γίνεται ως εξής:

- Με σταθέρο $M_{1/2} = M_c$ βρίσκεται ένα σύνολο τιμών, $m_A^2 - m_0^2$, που ικανοποιούν την γραμμική σχέση

$$m_A^2 \simeq c_m m_0^2 + \mathrm{cst}_m \tag{6.14}$$

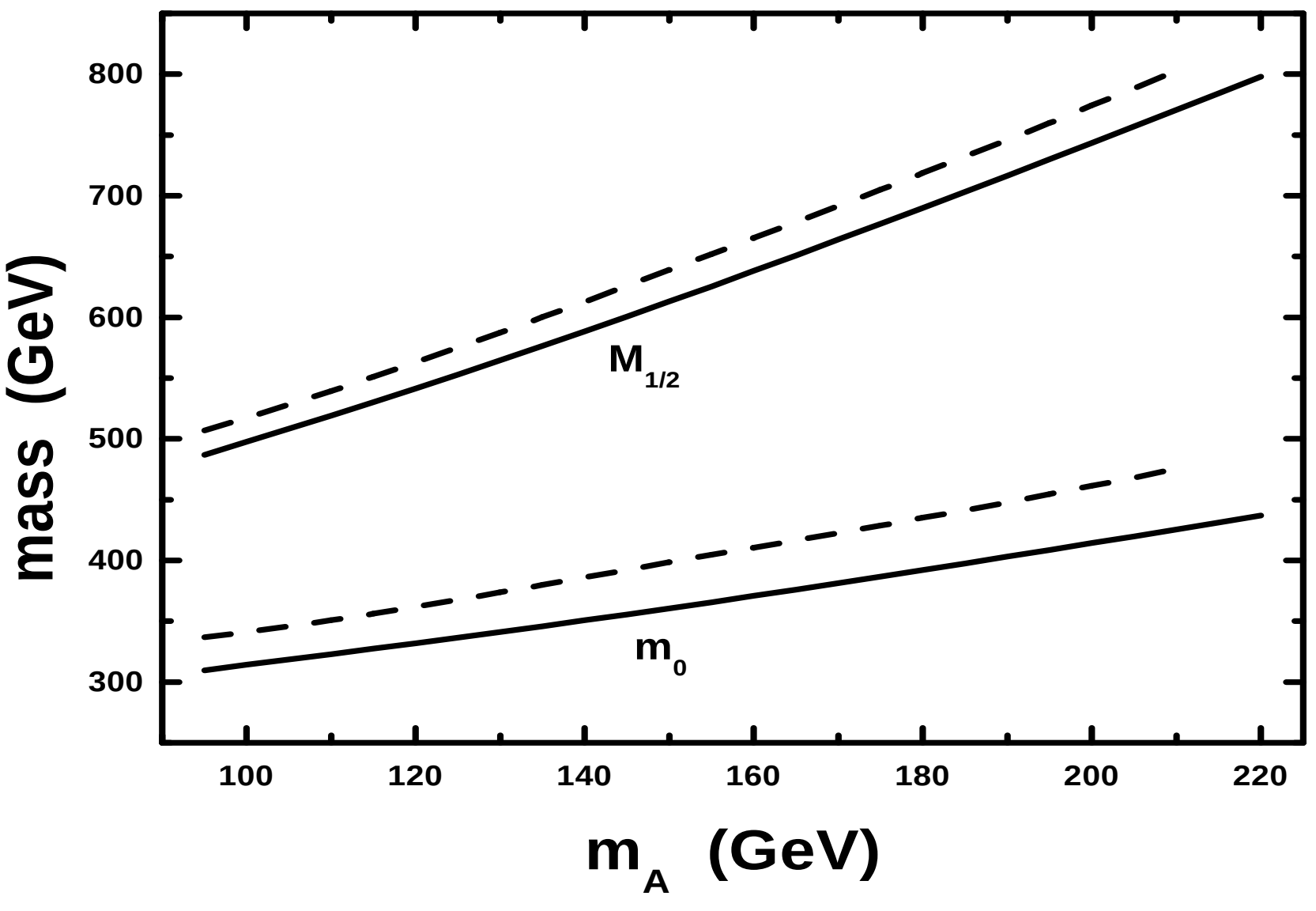

**Σχήμα 6.4:** *Οι ασυμπτωτικές τιμές των ασθενών μαζών, $M_{1/2}$ και $m_0$, ως συναρτήσεις της $m_A$ για $\Delta_{NLSP} = 0.2$ (συνεχής γραμμή) και 0.8 (διακεκομένη γραμμή) με $M_S = 1$ TeV και $\mu < 0$.*

- Με σταθέρο $m_0 = m_c$ βρίσκεται ένα σύνολο τιμών, $m_A^2 - M_{1/2}^2$, που ικανοποιούν την γραμμική σχέση

$$m_A^2 \simeq c_M M_{1/2}^2 + \mathrm{cst}_M \tag{6.15}$$

- Προφανώς, αν ισχύει μία σχέση της μορφής

$$m_A^2 \simeq c_m m_0^2 + c_M M_{1/2}^2 + \mathrm{cst}\ , \tag{6.16}$$

  πρέπει να ισχύουν ταυτόχρονα οι σχέσεις

$$\begin{cases} \mathrm{cst}_M = c_m m_c^2 + \mathrm{cst} \\ \mathrm{cst}_m = c_M M_c^2 + \mathrm{cst} \end{cases}$$

  Επαληθεύεται πολλαπλώς ότι cst $\simeq -M_Z^2$. Το γεγονός αυτό μπορεί να εξηγηθεί από τη Εξ. (2.51) ορισμού του $m_A$ αν σε αυτή αντικατασταθεί το $\mu$ από την Εξ. (2.45). Προκύπτει αυτός ο όρος που δεν αλλοιώνεται με την εξέλιξη των $m_H, m_{\bar{H}}$, τα οποία εξαρτώνται ομαλά από τις αρχικές συνθήκες $m_0$, $M_{1/2}$.

Η διαδικασία που περιγράφηκε πιο πάνω, είναι από τα πιο πρωτότυπα σημεία της παρούσας διατριβής.

Εξοπλισμένοι με τη σχέση αυτή, επιτύγχανουμε μια πολύ σημαντική μετατόπιση παραμέτρων. Μπορούν να ληφθούν ως μεταβλητές εισόδου στο πρόγραμμα οι $m_A$, $M_{1/2}$ και το $m_0$ να προσδιορίζεται από αυτές μέσω της Εξ. (6.13).

Όπως έχει τονιστεί στο Εδ. 5.3, κεντρικό ρόλο στην υλοποίηση του μηχανισμού της CAE, παίζει η παράμετρος $\Delta_{NLSP}$, που ρυθμίζει τη σχετική διαφορά μάζας ανάμεσα στα δύο ελαφρότατα σωμάτια του φάσματος, που ορίζεται από την Εξ. (5.11). Το επιθυμητό $\Delta_{NLSP}$ επιτυγχάνεται για κάθε τμή του $m_A$ με δοκιμαστική αλλαγή του $M_{1/2}$. Με αύξηση του $M_{1/2}$ αυξάνεται το $\Delta_{NLSP}$.

Καταληκτικά, οι ελεύθερες παράμετροι του σωματιδιακού προτύπου είναι:

$$m_A,\ \Delta_{NLSP},\ \mathrm{sign}\mu\ .$$

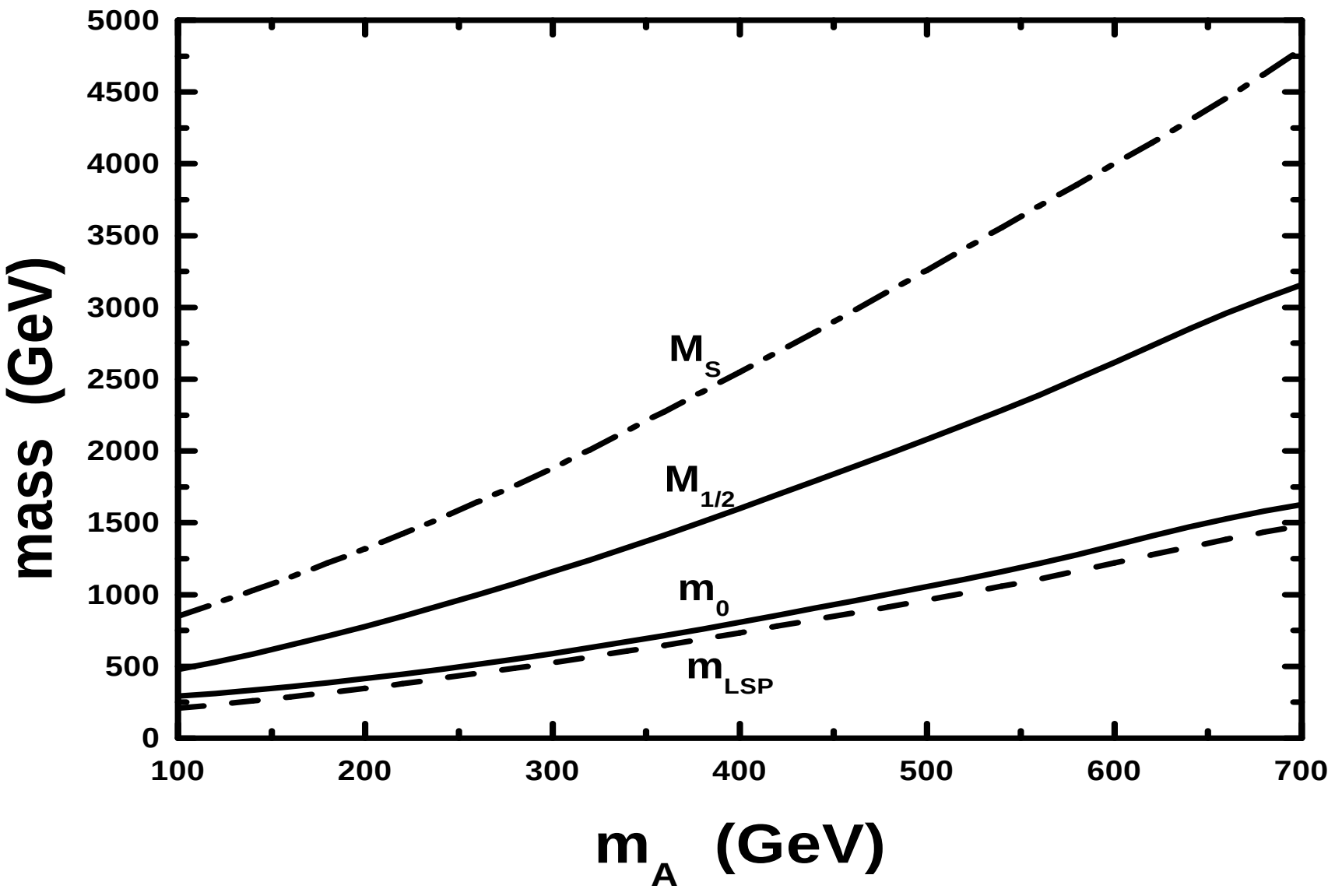

**Σχήμα 6.5:** *Οι τιμές των $m_{LSP}$, $m_0$, $M_{1/2}$ και $M_S$ ως συναρτήσεις της $m_A$ για $\Delta_{NLSP} = 0$ και $\mu > 0$.*

### 6.4.2 Διαγραμματικά Αποτελέσματα

Η ορθότητα του υπολογισμού μας ελέγχεται από το γεγονός ότι αναπαράγει με αξιοπρόσεκτη ακρίβεια το διάγραμμα του Σχ. 1 της Αν. [70], το οποίο έχει παράχθει χωρίς τη χρήση της Εξ. (6.13), όπως προέκυψε από ιδιωτική επικοινωνία με έναν από τους συγγραφείς. Στο διάγραμμα αυτό φαίνονται οι τιμές που πρέπει να έχουν οι ασυμπτωτικές τιμές των μαζών ασθενούς παραβίασης της SUSY για διάφορες τιμές του $m_t(m_t)$ με $\Delta_{NLSP} = 0$, $m_A = M_Z$ και $M_S = 1\,\mathrm{TeV}$. Αν και οι τιμές αυτές είναι ξεπέρασμένες από τα νεώτερα πειραμάτικα ευρήματα ($m_t(m_t) \simeq 166\,\mathrm{GeV}$ και $m_A > M_Z$), η μορφή που επιτεύχθηκε από το δικό μας πρόγραμμα παρέχεται για συγκριση στο Σχ. 6.3. Η μεταβολή στη $m_t(m_t)$ μπορεί να επιτευχθεί προφανώς με μεταβολή της τιμής της ενοποιημένης ζεύξης Yukawa στην Εξ. (6.7).

Επιστρέφοντας στην οικεία διαδικασία (με σταθεροποιημένη την τιμή της $m_t(m_t)$) τα αποτελέσματα του αριθμητικού προγράμματος σε αυτό το σκέλος της μελέτης καταγράφονται παρακάτω:

**α.** Περίπτωση $\mu < 0$. Στην περίπτωση αυτή, ο φαινομενολογικός περιορισμός που αναδύεται από το $\mathrm{BR}(b \to s\gamma)$, επιτρέπει τη χρήση σχετικά ελαφρού φάσματος. Για το λόγο αυτό είναι δυνατόν να σταθεροποιηθεί η τιμή του $M_S = 1\,\mathrm{TeV}$. Η επιλογή αυτή δεν έχει ουσιατική επίπτωση στα αποτελέσματα γιατί αυτά είναι ευσταθή με μικρή μετακίνηση του $M_S$ . Στο Σχ. 6.4 εικονίζονται οι αρχικές συνθήκες που τίθενται στο αριθμητικό πρόγραμμα ($M_{1/2}$ και $m_0$), ώστε να επιτυγχάνεται για $95\,\mathrm{GeV} \leq m_A \leq 210\,\mathrm{GeV}$, $\Delta_{NLSP} = 0.2$ ή $0.8$. Οι επιλογές αυτές των τιμών του $\Delta_{NLSP}$ είναι ενδεικτικές.

**β.** Περίπτωση $\mu > 0$. Στην περίπτωση αυτή, ο φαινομενολογικός περιορισμός που αναδύεται από το $\mathrm{BR}(b \to s\gamma)$, επιβάλλει τη χρήση σχετικά βαρύτερου φάσματος από το προηγούμενο σημείο. Για το λόγο αυτό χρησιμοποιείται μεταβλητή τιμή για το $M_S$. Στο Σχ. 6.5 εικονίζονται οι τιμές των επίμαχων ποσοτήτων του προγράμματος $m_{LSP}$, $m_0$, $M_{1/2}$ και $M_S$ ως συναρτήσεις της $m_A$ για $\Delta_{NLSP} = 0$

Συμπερασματικά, και στις δύο περιπτώσεις παρατηρείται ότι $M_{1/2} > m_0$ και με σταθερό $m_A$ αύξηση του $\Delta_{NLSP}$ συνεπάγεται αύξηση των απαιτούμενων $M_{1/2}$ και $m_0$ ενώ με σταθερό $\Delta_{NLSP}$ αύξηση του $m_A$ συνεπάγεται αύξηση των απαιτούμενων $M_{1/2}$ και $m_0$

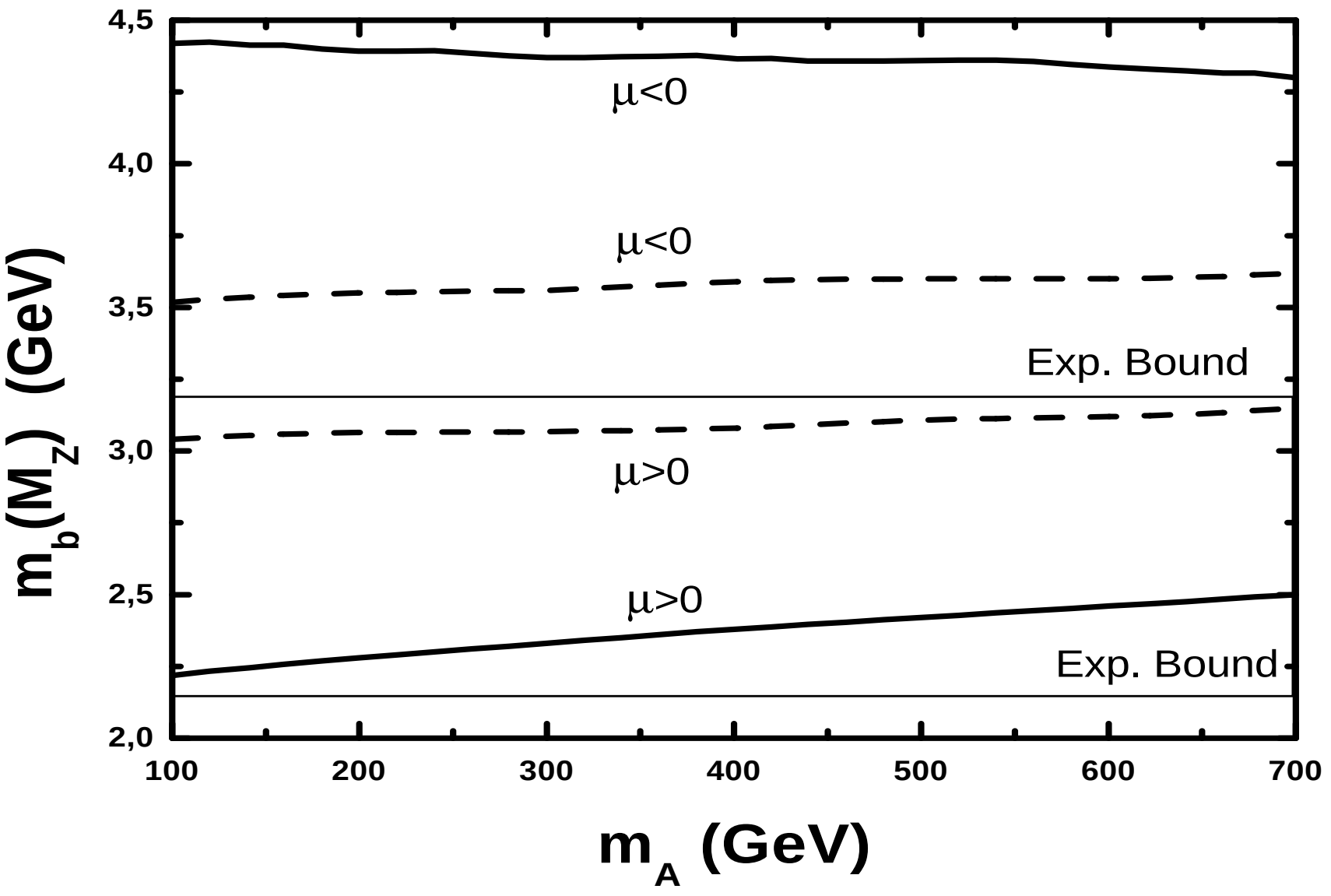

**Σχήμα 6.6:** *Η δενδρικού επιπέδου (διακεκομένη γραμμή) και η διορθωμένη (συνεχής γραμμή) μάζα του b-quark ως συνάρτηση της $m_A$, για $\Delta_{NLSP} \simeq 0$. Τα πειραματικά όρια έχουν επίσης τοποθετηθεί.*

## 6.5 Φαινομενολογική μελέτη

Μια πολύ ενδιαφέρουσα προβληματική ανακύπτει με την υιοθέτηση της ενοποίησης Yukawa. Το ζήτημα που πρέπει να διερευνηθεί είναι η συμβατότητα της ενοποίησης Yukawa με τα πειραματικά δεδομένα για τη μάζα του $b$-quark και το $\mathrm{BR}(b \to s\gamma)$. Αυτός ο προβληματισμός αποκτά βαρύνουσα αξία στην περίπτωση της ενοποίησης Yukawa γιατί, λόγω της υψηλής τιμής που αποκτά η παράμετρος $\tan\beta$, οι διορθώσεις στη μάζα του $b$-quark είναι ιδιαιτέρως ισχυρές και η επιλογή σημείου για την παράμετρο $\mu$ καθίσταται κρίσιμη. Συνακόλουθα, επειδή αποφασιστική είναι αυτή η επιλογή σημείου για το $\mu$, στη συμπεριφορά του $\mathrm{BR}(b \to s\gamma)$ η ταυτόχρονη ικανοποίηση και των δύο περιορισμών προβάλλει ως ένα πολύ ελκυστικό εγχείρημα. Οι περιορισμοί αυτοί μελετώνται στα Εδ. 6.5.1 και 6.5.2. Τέλος, η φαινομενολογία που προκύπτει από τις διορθώσεις στο $h$ μελετάται στο Εδ. 6.5.3.

### 6.5.1 Διορθώσεις στη μάζα του $b$-quark

Εφαρμόζοντας το τυπολόγιο του Εδ. 3.3.1 , υπολογίζεται η μάζα του $b$-quark έχοντας συμπεριλάβει τις SUSY διορθώσεις για $\Delta_{NLSP} \simeq 0$. Τα αποτελέσματα καταγράφονται στο Σχ. 6.6. Παρατηρείται ότι για $100\,\mathrm{GeV} \leq m_A \leq 700\,\mathrm{GeV}$:

- Όταν $\mu < 0$, λαμβάνεται για την δενδρικού επιπέδου και τη διορθωμένη μάζα του $b$-quark αντίστοιχα:

$$3.5\,\mathrm{GeV} \lesssim m_b(M_Z) \lesssim 3.6\,\mathrm{GeV} \tag{6.17}$$

$$4.4\,\mathrm{GeV} \gtrsim m_b^c(M_Z) \gtrsim 4.3\,\mathrm{GeV}. \tag{6.18}$$

  Συνεπώς, σε αυτή την περίπτωση, η διόρθωση που λαμβάνεται είναι θετική και παραβιάζονται τα πειραματικά όρια που ισχύουν για τη μάζα του $b$-quark, όπως προκύπτει με σύγκριση των πιο πάνω τιμών με την Εξ. (3.18). Το γεγονός αυτό δε θεωρείται καταστροφικό για το εξεταζόμενο πρότυπο, γιατί σε μια πλήρη $SO(10)$ θεωρία υπάρχουν και άλλες διορθώσεις (Αν. [71]) που μπορούν να ταπεινώσουν ή και να αναιρέσουν τις συμπεριλαμβανόμενες στον εδώ υπολογισμό. Σίγουρα, όμως αυτό είναι ένα σημείο που χρήζει βαθύτερης και διεξοδικότερης έρευνας.

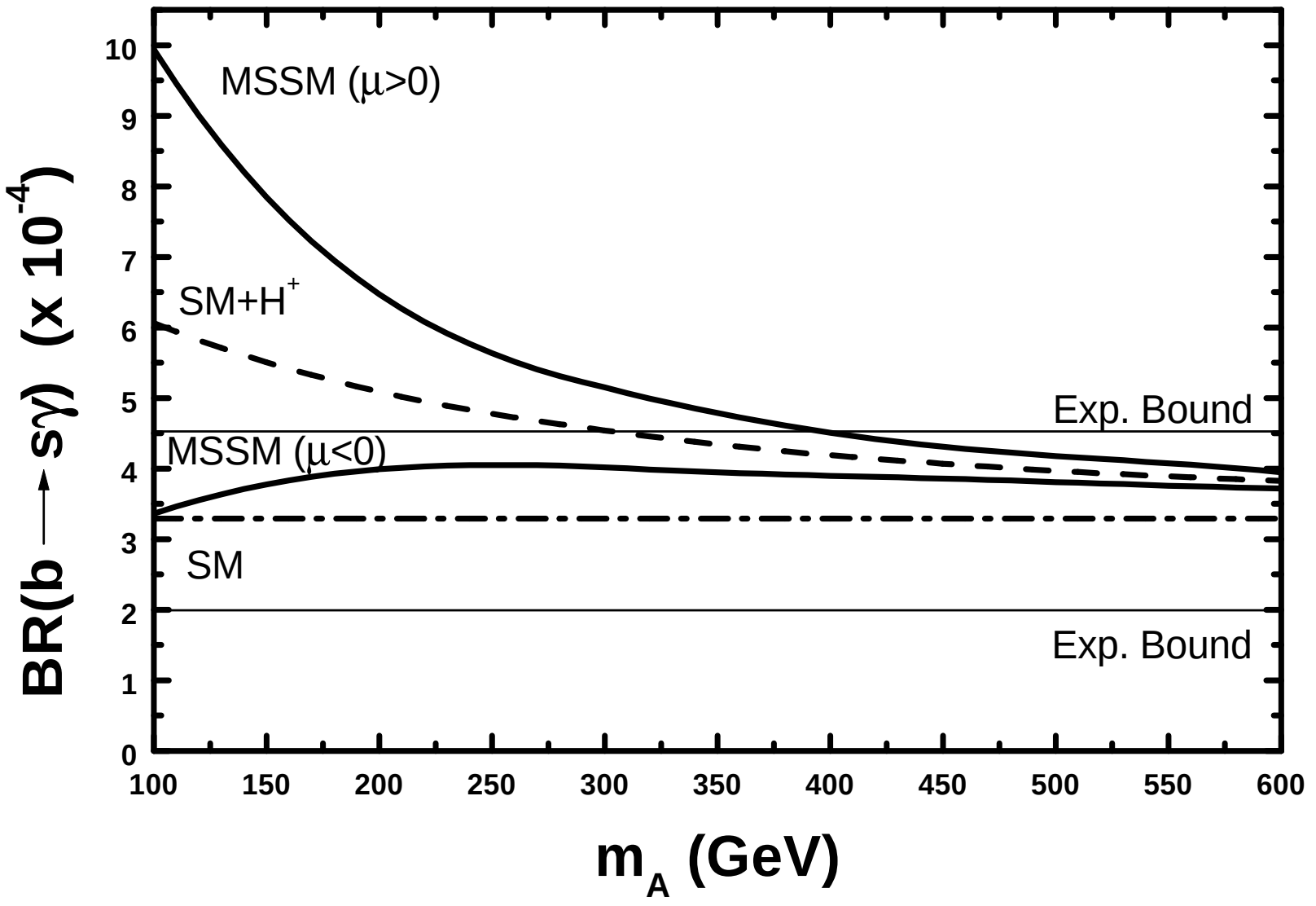

**Σχήμα 6.7:** *Το* $\mathrm{BR}(b \to s\gamma)$ *ως συνάρτηση της* $m_A$*, για* $\Delta_{NLSP} \simeq 0$ *έχοντας συμπεριλάβει τις συνεισφορές του MSSM για* $\mu > 0$ *και* $\mu < 0$*, του SM και του SM συν των φορτισμένων Higgs (SM+*$H^+$*). Τα πειραματικά όρια έχουν επίσης τοποθετηθεί.*

- Όταν $\mu > 0$, λαμβάνεται για την δενδρικού επιπέδου και τη διορθωμένη μάζα του $b$-quark αντίστοιχα:

$$3.04\,\mathrm{GeV} \lesssim m_b(M_Z) \lesssim 3.15\,\mathrm{GeV} \tag{6.19}$$

$$2.2\,\mathrm{GeV} \lesssim m_b^c(M_Z) \lesssim 2.5\,\mathrm{GeV} \tag{6.20}$$

  Συνεπώς, σε αυτή την περίπτωση, η διόρθωση που λαμβάνεται είναι αρνητική και οι τιμές που επιτυγχάνονται για τη μάζα του $b$-quark, βρίσκονται οριακά, μέσα στις πειραματικές προβλεψεις της Εξ. (3.18)

Η παρατηρούμενη διαφορά μάζας του $b$-quark σε δενδρικό επίπεδο στις δύο προηγούμενες περιπτώσεις οφείλεται στη διαφορετική τιμή που επιτυγχάνεται για την $\tan\beta$, σε κάθε μια απο αυτές τις περιπτώσεις, όπως έχει αναφερθεί στο Εδ. 6.3.2. Επίσης, και στις δύο περιπτώσεις, η διόρθωση στη μάζα του $b$-quark έχει πρόσημο αντίθετο του χρησιμοποιούμενου για το $\mu$.

### 6.5.2 Μελέτη του $\mathrm{BR}(b \to s\gamma)$

Εφαρμόζοντας το τυπολόγιο του Εδ. 3.4.2, υπολογίζεται και απεικονίζεται στο Σχ. 6.7 το $\mathrm{BR}(b \to s\gamma)$ ως συνάρτηση της $m_A$ για $\Delta_{NLSP} \simeq 0$. Από το αριθμητικό πρόγραμμα συμπεραίνεται οτι η επιλογή του $\Delta_{NLSP}$ επηρεάζει πολύ λίγο τα αποτελέσματα στο $\mathrm{BR}(b \to s\gamma)$. Ακόμα παρατηρείται ότι όταν $\mu <$ [>]0 οι συνεισφορές των βρόχων που μετέχουν chargino είναι απ[επ]οικοδομητικές στην συνεισφορά από το SM συν τα φορτισμένα Higgs. Οπότε,

- Όταν $\mu < 0$, το συνολικό αποτέλεσμα κυμαίνεται σε πειραματικώς ανεκτά επίπεδα. Επομένως, η περίπτωση αυτή είναι επιτρεπτή για κάθε τιμή των $m_A$, $\Delta_{NLSP}$.

- Όταν $\mu > 0$, το συνολικό αποτέλεσμα κυμαίνεται σε επίπεδα ανώτερα των πειραματικώς προβλεπομένων για μικρές τιμές της $m_A$. Επειδή οι συνεισφορές, εκτός αυτής του SM, φθίνουν με αύξηση της μάζας των μετεχώντων σωματίων, βρίσκεται τιμή του $m_A$ μετά την οποία το $\mathrm{BR}(b \to s\gamma)$ μπαίνει και

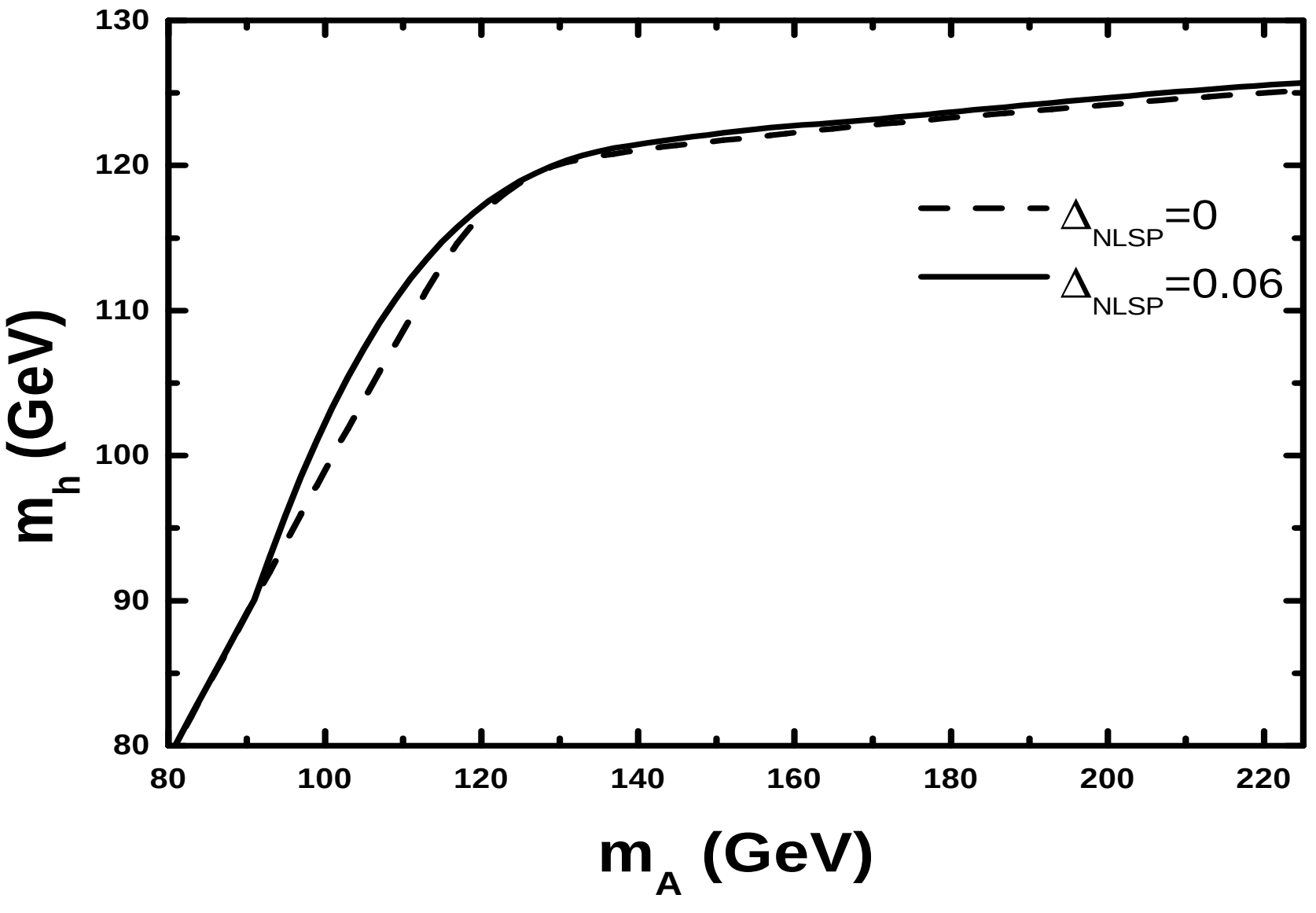

**Σχήμα 6.8:** *Η $m_h$ ως συνάρτηση της $m_A$, για $\mu < 0$ και $\Delta_{NLSP} \simeq 0,\ 0.06$.*

παραμένει στην πειραματικά επιτρεπτή περιοχή. Αυτή η τιμή, παρέχει ένα κάτω όριο στις επιτρεπτές τιμές που μπορεί να πάρει το $m_A$. (ή το $m_{LSP}$). Δηλαδή:

$$m_A \gtrsim 402\,\text{GeV} \;\; \text{ή} \;\; m_{LSP} \gtrsim 738\,\text{GeV}. \tag{6.21}$$

Είναι αξιοσημείωτο το γεγονός ότι οι περιορισμοί που αντλούνται από τη διόρθωση στη μάζα του $b$-quark και του $\text{BR}(b \to s\gamma)$ είναι αλληλοσυγκρουόμενοι με την έννοια ότι το σημείο του $\mu$ που ευνοεί την ικανοποίηση του ενός περιορισμού αντιβαίνει στην ικανοποίηση του άλλου. Συγκεκριμένα, η $\mu <$ [$>$]0 επιλογή είναι δυσμενής [ευμενής] για τη μάζα του $b$-quark αλλά ευμενής [δυσμενής] για το $\text{BR}(b \to s\gamma)$.

### 6.5.3 Διορθώσεις στη μάζα του $h$ Higgs

Αξιοποίηση του τυπολόγιου του Εδ. 3.2.2 γίνεται με σκοπό τον προσδιορισμό της μάζας του $h$ Higgs από το μελετούμενο πρότυπο. Από το τυπολόγιο συμπεραίνεται ότι το σημείο του $\mu$ δεν επηρεάζει αισθητά τα αποτελέσματα. Ενδεικτικά εκτίθενται στο Σχ. 6.8, η $m_h$ ως συνάρτηση του $m_A$, με $\mu < 0$ για $\Delta_{NLSP} \simeq 0$ και 0.06. Παρατηρείται ότι υπάρχει σαφώς ένα μέγιστο όριο που αντιστοιχεί στα 125 GeV, πράγμα απόλυτα σύννομο με τα πειραματικά όρια της Εξ. (3.10) και ότι η διαφοροποίηση της καμπύλης για διάφορες τιμές του $\Delta_{NLSP}$ είναι σχεδόν αδιόρατη, ειδικά σε ό,τι αφορά την τιμή του μεγίστου. Λαμβάνοντας υπόψη μας τα πρόσφατα φαινομενολογικά όρια της Εξ. (3.10), μπορεί να τεθεί ένα κάτω όριο για τη μάζα $m_A$ (ή τη $m_{LSP}$) που καταγράφεται παρακάτω:

$$m_A \gtrsim 95\,\text{GeV} \;\; \text{ή} \;\; m_{LSP} \gtrsim 209\,\text{GeV}. \tag{6.22}$$

## 6.6 Κοσμολογικά αποτελέσματα

Η Κοσμολογική μελέτη του σωματιδιακού προτύπου που περιγράφηκε, περιλαμβάνει τον υπολογισμό της CRD του LSP, $\Omega_{LSP}\,h^2$. Αυτός γίνεται εφαρμόζοντας το τυπολόγιο των Εδ. 5.5.1 και 5.5.2 συμπεριλαμβάνοντας ANE και CAE, αντίστοιχα. Ειδικότερα, στις περιπτώσεις $\mu > 0$ και $\mu < 0$ είναι αφιερωμένα

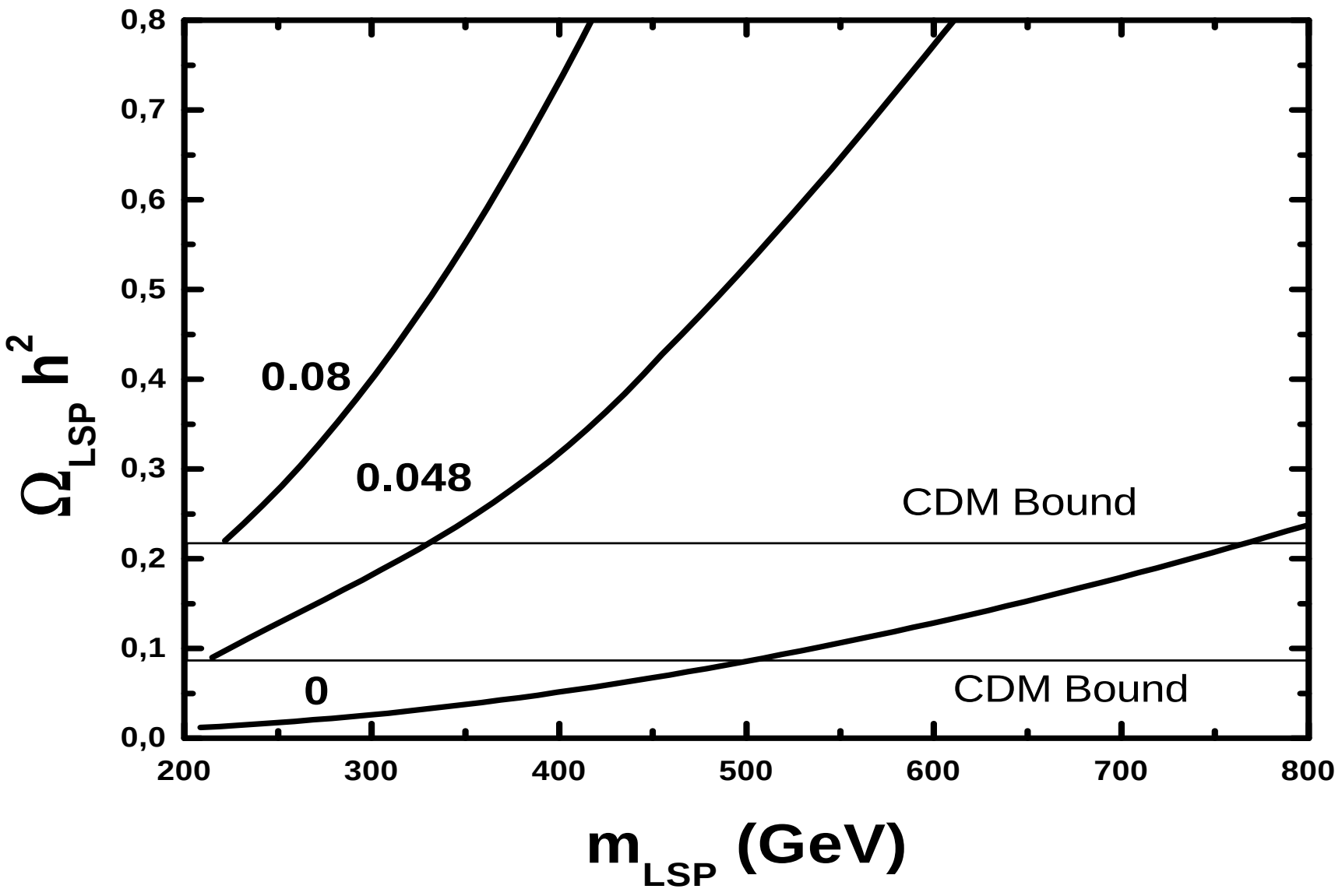

**Σχήμα 6.9:** *Το $\Omega_{LSP}\,h^2$ ως συνάρτηση του $m_{LSP}$ για $\mu < 0$ και $\Delta_{NLSP}$=0, 0.048 και 0.08. Τα όρια από τα CDM σενάρια έχουν επίσης τοποθετηθεί.*

τα Εδ. 6.6.1 και 6.6.2. Η συμβολή κάθε διαδικασίας στο τελικό αποτέλεσμα αξιολογείται γενικά στο Εδ. 6.6.3.

### 6.6.1 Περίπτωση $\mu < 0$

Στην περίπτωση αυτή, ο φαινομενολογικός περιορισμός που πρέπει να ληφθεί υπόψη είναι αυτός της Εξ. (6.22). Δύο είναι τα βασικά διαγράμματα που με σαφήνεια καταγράφουν τη φυσική του προβλήματος:

**α.** Το διάγραμμα του $\Omega_{LSP}\,h^2$ ως συνάρτηση του $m_{LSP}$ για καθορισμένα $\Delta_{NLSP}$, που απεικονίζεται στο Σχ. 6.9. Από αυτά συμπεραίνεται ότι με σταθερό $m_{LSP}$ [$\Delta_{NLSP}$] αύξηση του $\Delta_{NLSP}$ [$m_{LSP}$] συνεπάγεται αύξηση του $\Omega_{LSP}\,h^2$. Επομένως, η $\Omega_{LSP}\,h^2$ είναι αύξουσα συνάρτηση των $m_{LSP}$ και $\Delta_{NLSP}$. Οι τιμές του $\Delta_{NLSP}$, που έχουν επιλεγεί είναι:

- 0. Πάνω σε αυτή την καμπύλη λαμβάνεται η μέγιστη συνεισφορά των CAE. Το σημείο τομής της καμπύλης αυτής με το άνω CDM όριο, παρέχει τη μέγιστη επιτρεπτή μάζα του LSP της θεωρίας που είναι $m_{LSP}^{max} \simeq 770\,\mathrm{GeV}$.
- 0.048. Με αυτή την τιμή, επιτυγχάνεται το κατώτερο CDM όριο με την ελάχιστη δυνατή τιμή στη μάζα του LSP για αυτό το $\Delta_{NLSP}$ που είναι 215 GeV.
- 0.08. Με αυτή την τιμή, επιτυγχάνεται το ανώτερο CDM όριο με την ελάχιστη δυνατή τιμή στη μάζα του LSP για αυτό το $\Delta_{NLSP}$ που είναι 222 GeV. Επομένως, αυτό μπορεί να χαρακτηριστεί ως το μέγιστο επιτρεπτό $\Delta_{NLSP}$, $\Delta_{NLSP}^{max}$.

**β.** Το διάγραμμα της επιτρεπτής περιοχής στο επίπεδο $m_{LSP} - \Delta_{NLSP}$, που απεικονίζεται στο Σχ. 6.10. Αυτό το διάγραμμα αποτελεί ουσιαστικά την προβολή του Σχ. 6.9 στο επίπεδο $m_{LSP} - \Delta_{NLSP}$. Συγκεκριμένα:

- Η κατώτερη [ανώτερη] συνοριακή καμπύλη αντιστοιχεί στο κατώτερο [ανώτερο] CDM όριο ($\Omega_{LSP}\,h^2 = 0.09\,[0.22]$).

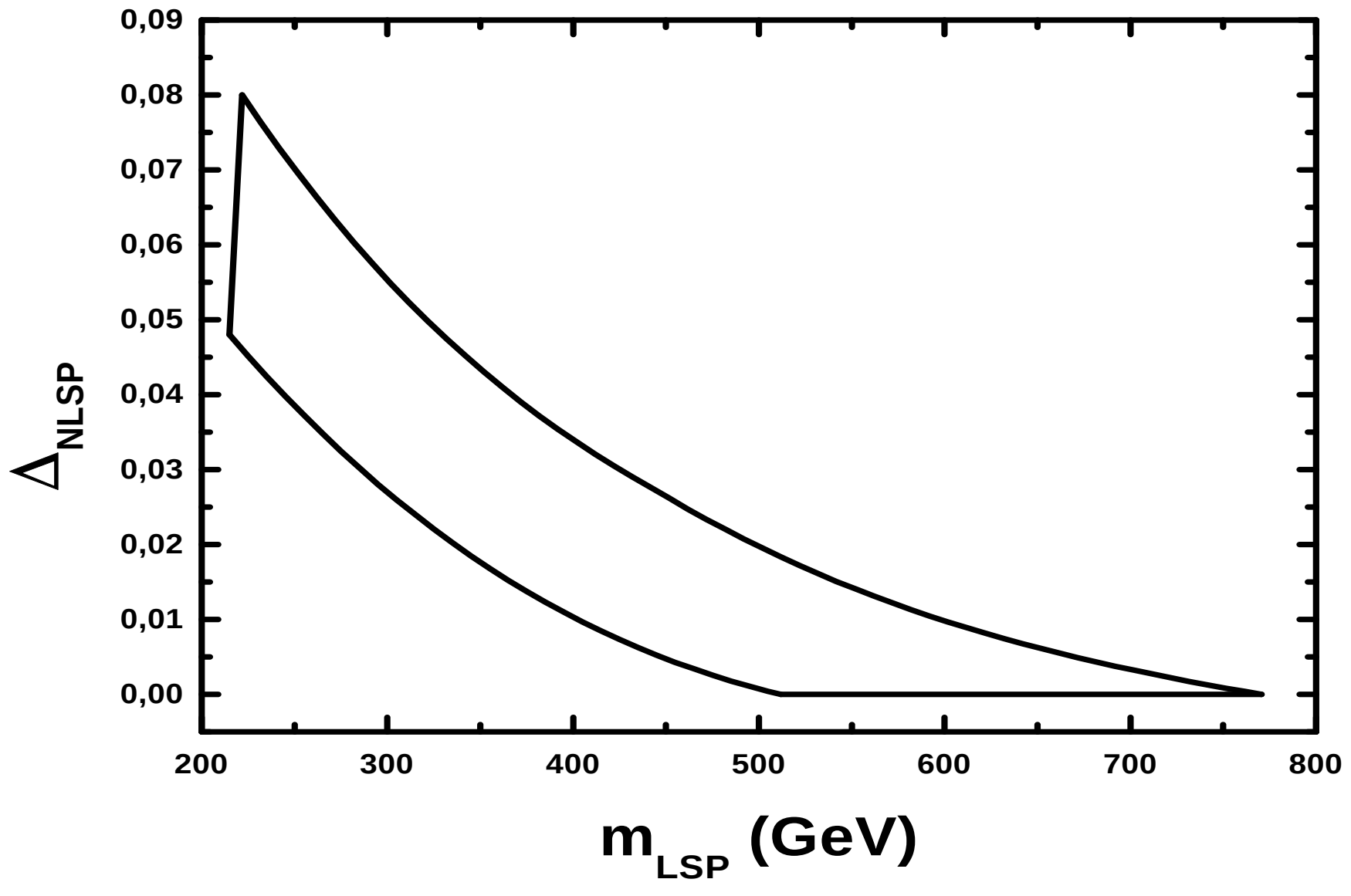

**Σχήμα 6.10:** *Η επιτρεπτή περιοχή στο επίπεδο $m_{LSP} - \Delta_{NLSP}$, για $\mu < 0$.*

- Η κάτω αριστερή γωνία, που σχηματίζει η κατώτερη συνοριακή καμπύλη με την ευθεία την σχεδόν κατακόρυφη, αντιστοιχεί στην τομή της καμπύλης $\Delta_{NLSP} = 0.048$ με το κατώτερο CDM όριο στο Σχ. 6.9. Από το σημείο αυτό, προκύπτει το $m_{LSP}^{min}$
- Η πάνω αριστερή γωνία, που σχηματίζει η ανώτερη συνοριακή καμπύλη με την ευθεία την σχεδόν κατακόρυφη, αντιστοιχεί στην τομή της καμπύλης $\Delta_{NLSP} = 0.08$ με το ανώτερο CDM όριο στο Σχ. 6.9. Από το σημείο αυτό, προκύπτει το $\Delta_{NLSP}^{max}$.
- Η κάτω αριστερή γωνία, που σχηματίζει η κατώτερη συνοριακή καμπύλη με την οριζόντια ευθεία, αντιστοιχεί στην τομή της καμπύλης $\Delta_{NLSP} = 0$ με το κατώτερο CDM όριο στο Σχ. 6.9.
- Η κάτω δεξιά γωνία, που σχηματίζει η ανώτερη συνοριακή καμπύλη με την οριζόντια ευθεία, αντιστοιχεί στην τομή της καμπύλης $\Delta_{NLSP} = 0$ με το ανώτερο CDM όριο στο Σχ. 6.9. Από το σημείο αυτό προκύπτει το $m_{LSP}^{max}$.

Προσεκτικός αναγνώστης της διατριβής θα παρατηρήσει ότι τα Σχ. 6.9 και 6.10 αποτελούν βελτιωμένη έκδοση τών Σχ. 3, 4 της Αν. [44]. Μεταβλητό $M_S$ έχει χρησιμοποιηθεί στα νέα, που επιτρέπει τη μετάβαση σε υψηλότερες τιμές της παραμέτρου $m_A$ ή $m_{LSP}$. Επομένως, μπορούν να γίνουν σαφέστερες προβλέψεις για τις ανώτερες κοσμολογικά επιτρεπτες τιμές των παραμέτρων αυτών.

### 6.6.2 Περίπτωση $\mu > 0$

Στην περίπτωση αυτή, ο φαινομενολογικός περιορισμός που πρέπει να ληφθεί υπόψη είναι αυτός της Εξ. (6.21), αφού είναι ισχυρότερος από αυτό της Εξ. (6.22). Επειδή μάλιστα το κάτω όριο που επιβάλλεται είναι υψηλό, θα χρησιμοποιηθεί η μέγιστη δυνατή συμβολή των CAE που επιτυγχάνεται με $\Delta_{NLSP} = 0$. Τα CDM σενάρια θα δώσουν ένα άνω όριο για την τιμή του $m_A$. Το προς εξέταση πρόβλημα αν υπάρχει χώρος συναλήθευσης των δύο ανισοτήτων που θα προκύψουν.

Το κρίσιμο διάγραμμα είναι αυτό του Σχ. 6.11 όπου απεικονίζεται το $\Omega_{LSP}\, h^2$ ως συνάρτηση του $m_A$. Από αυτό προκύπτει ότι για να είναι η υπολογιζόμενη $\Omega_{LSP}\, h^2$ συμβατή με τα CDM σενάρια, πρέπει:

$$m_A \lesssim 416\,\mathrm{GeV} \ \text{ ή } \ m_{LSP} \lesssim 770\,\mathrm{GeV}. \tag{6.23}$$

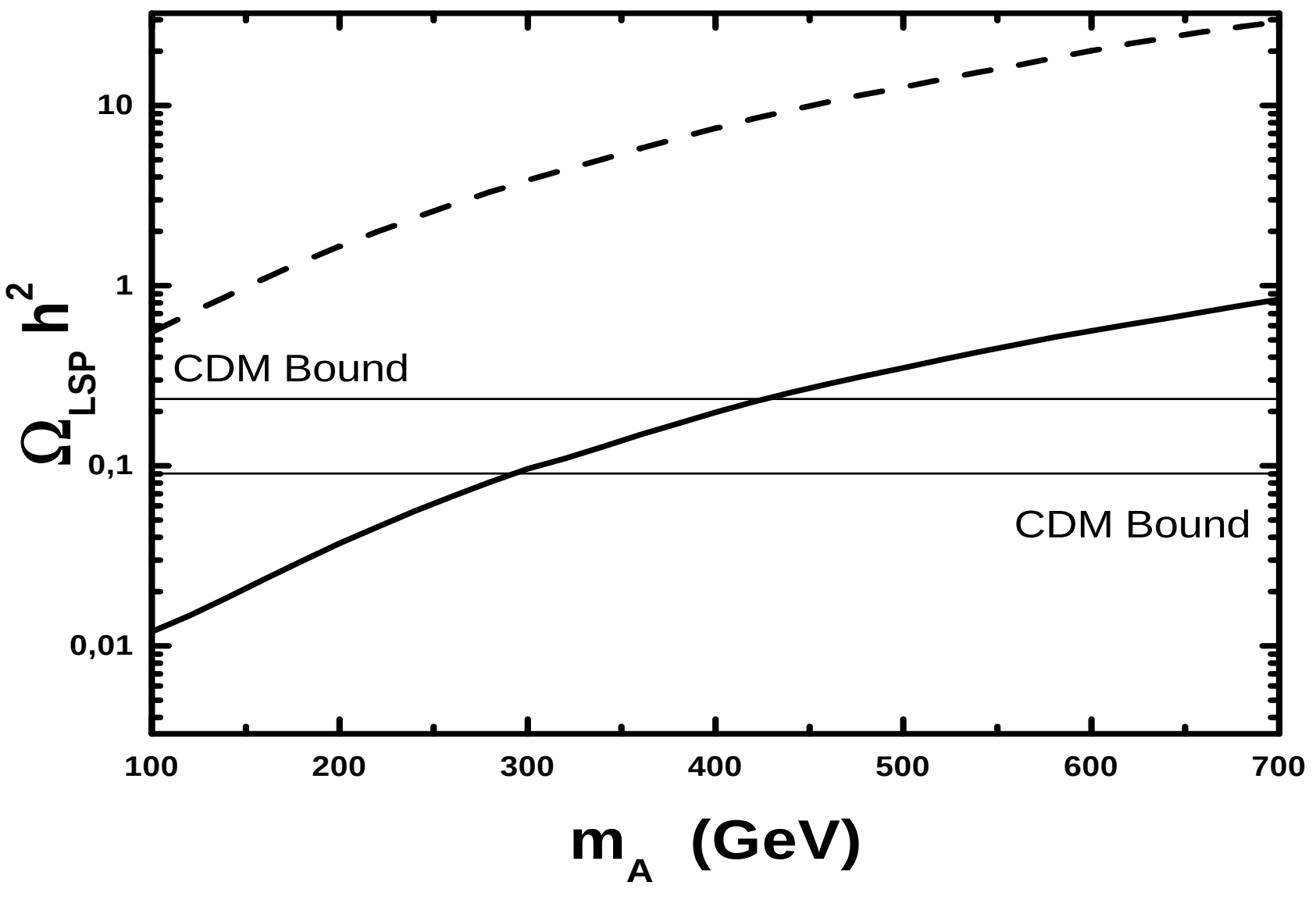

**Σχήμα 6.11:** *Το $\Omega_{LSP}\, h^2$, έχοντας συμπεριλάβει CAE (συνεχής γραμμή) και χωρίς αυτά (διακκεκομένη γραμμή) ως συνάρτηση του $m_A$ για $\mu > 0$ και $\Delta_{NLSP} = 0$. Τα όρια από τα CDM σενάρια έχουν επίσης τοποθετηθεί.*

Από τις Εξ. (6.21) και (6.23) προκύπτει ότι υπάρχουν τιμές της παραμέτρου $m_A$ που επιτρέπουν τη συνύπαρξη των περιορισμών από το BR($b \to s\gamma$) και τα CDM σενάρια και αυτές είναι:

$$402\,\mathrm{GeV} \lesssim m_A \lesssim 416\,\mathrm{GeV} \quad \text{ή} \quad 738\,\mathrm{GeV} \lesssim m_{LSP} \lesssim 770\,\mathrm{GeV}. \tag{6.24}$$

Ειδικότερα, για την κατώτερη επιτρεπτή τιμή του $m_A$ από τον περιορισμό του BR($b \to s\gamma$), $m_A \simeq 402\,\mathrm{GeV}$ λαμβάνεται $\Omega_{LSP}\, h^2 \simeq 0.2$ αν συμπεριληφθούν CAE στον υπολογισμό, ή $\Omega_{LSP}\, h^2 \simeq 7.55$ αν δεν συμπεριληφθούν. Παράλληλα, για την ανώτερη επιτρεπτή τιμή του $m_A$ από τον περιορισμό των CDM σεναριων, $m_A \simeq 416\,\mathrm{GeV}$ λαμβάνεται BR($b \to s\gamma$) $\simeq 4.44 \times 10^{-4}$, που είναι μέσα στα όρια της Εξ. (3.50).

Αν και η περιοχή της Εξ. (6.24) είναι πολύ περιορισμένη, ο συνυπολογισμός των σφαλμάτων στον προσδιορισμό του BR($b \to s\gamma$) αναμένεται να τη διευρύνει. Και αυτό, διοτι η τομή του άνω κλάδου του BR($b \to s\gamma$) με το άνω πειραματικό όριο γίνεται υπό οξεία γωνία οπότε μια μικρή μετακίνηση λόγω σφαλμάτων συνεπάγεται ισχυρή ελάττωση του κάτω ορίου στη μεταβλητή $m_A$. Μόνο η θεωρητική αβεβαιότητα στον καθορισμό του BR$^{\mathrm{SM}}$($b \to s\gamma$), που δίνεται από την Εξ. (3.33), επιφέρει μια μετακίνηση του κάτω επιτρεπτού ορίου του $m_A$ περίπου στα 320 GeV. Η επιτρεπτή περιοχή στο επίπεδο $m_{LSP} - \Delta_{NLSP}$ θα είναι της μορφής του Σχ. 6.10 μόνο που το σχεδόν κατακόρυφο τμήμα στα αριστερά της περιοχής θα έχει μετακινηθεί προς τα δεξιά αφού $m_{LSP}^{min} \simeq 525\,\mathrm{GeV}$ οπότε και $\Delta_{NLSP}^{max} \simeq .02$. Εντούτοις μια πιο ακριβής διαπραγμάτευση θα απαιτούσε τον συνυπολογισμό και των υπολοίπων σφαλμάτων που υπεισέρχονται στον υπολογισμό του BR($b \to s\gamma$), η προέλευση των οποίων έχει ήδη αναφερθεί στο Εδ. 3.4.3. Τελικά αποφεύχθηκε η εμπλοκή σε αυτή τη διαδικασία, γιατί από τη μία θα απαιτούσε μεγάλη υπολογιστική ακρίβεια και από την άλλη το συμπέρασμα βίωσιμότητας της $\mu > 0$ περίπτωσης αποτελεί μια σημαντική κατάληξη. Και αυτό, διότι για την εξεταζόμενη περίπτωση υπήρχε ήδη πριν την δημοσίευση της Αν. [72] μια απορριπτική καταγραφή στην Αν. [30]. Σε αυτή ο υπολογισμός της CRD γινόταν με θεώρηση μόνο των ΑΝΕ. Τα CDM όρια της εποχής δεν ήταν συμβατά με τα αντίστοιχα από το BR($b \to s\gamma$). Ο νεωτερισμός της Αν. [72] ήταν ο συνυπολογισμός των CAE στην εξαγωγή της CRD και των NLO διορθώσεων της QCD που επέφερε τελικά την αναβίωση του $\mu > 0$ σεναρίου.

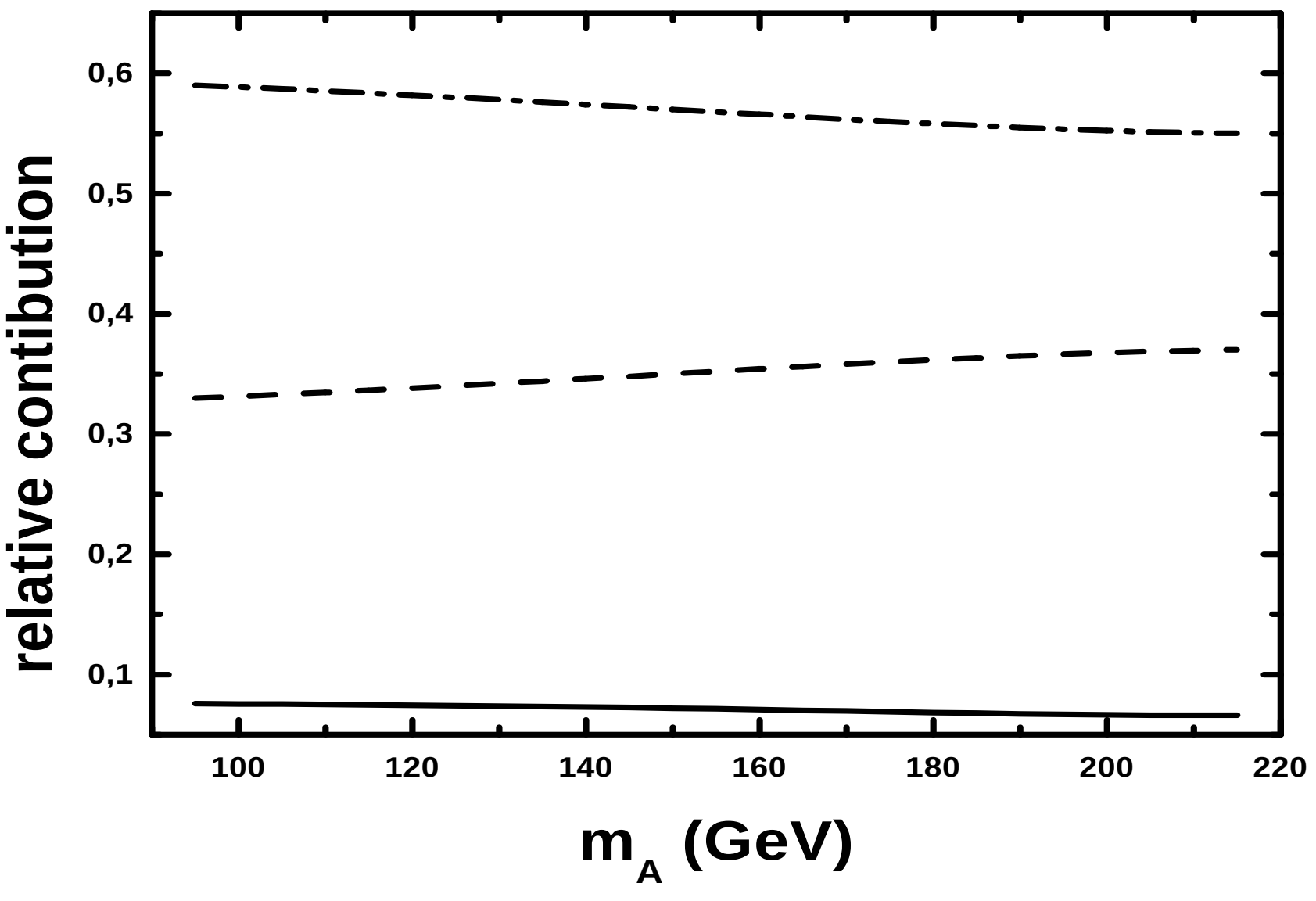

**Σχήμα 6.12:** *Οι σχετικές συνεισφορές* $J_{(\tilde\chi\tilde\chi)}/J_{\rm eff}$ *(συνεχής γραμμή),* $J_{(\tilde\tau_2\tilde\tau_2^{(*)})}$ *(διακεκομένη γραμμή) και* $J_{(\tilde\chi\tilde\tau_2)}$ *(εστιγμένα διακεκομένη γραμμή) για* $\Delta_{NLSP} = 0.047$, $\mu < 0$ *και* $M_S = 1\,{\rm TeV}$.

### 6.6.3 Σχετικές συνεισφορές στην $J_{\rm eff}$

Μια πολύ διαφωτιστική εικόνα για τη συμμετοχή κάθε μίας από τις διαδικασίες αλληλοκαταστροφής και συγγενικής καταστοφής στο τελικό αποτέλεσμα για την ανηγμένη δρώσα ενεργό διατομή $J_{\rm eff}$ της Εξ. (5.83) δίνεται στο εδάφιο αυτό. Αν και η ανάλυση έγινε για την περίπτωση $\mu < 0$, τα συμπεράσματα έχουν αυξημένη γενικότητα, γιατί, όπως προκύπτει με σύγκριση των καμπύλων για $\Delta_{NLSP} = 0$ από τα Σχ. 6.11 και 6.9, η αλλαγή προσήμου του $\mu$ αφήνει σχεδόν ανεπηρέαστη την υπολογιζόμενη CRD.

Με συγκριση της συνεισφοράς κάθε CAE ή ANE στο $J_{\rm eff}$, μπορούν να γίνουν τα παρακάτω σχόλια:

- Όταν $\Delta_{NLSP} = 0$, η συνεισφορά των ANE $\tilde\chi\tilde\chi$ στο $J_{\rm eff}$ είναι πολύ μικρή (0.4%). Οι αντίστοιχες συνεισφορά των CAE στις $J_{(\tilde\chi\tilde\tau_2)}$ και $J_{(\tilde\tau_2\tilde\tau_2^{(*)})}$ κυμαίνεται στο διάστημα $27-24\%$και $73-76\%$ αντίστοιχα καθώς το $m_A$ κυμαίνεται από 95 σε 224GeV. Όταν $\Delta_{NLSP} = 0.1$, εντούτοις, η συνεισφορά των ANE $\tilde\chi\tilde\chi$ γίνεται πολύ σημαντική παρέχοντας περίπου το $33-31\%$ της $J_{\rm eff}$. Η πιο σπουδαία συνεισφορά (το 58% της $J_{\rm eff}$), σε αυτή την περίπτωση, προέρχεται από τις CAE $\tilde\chi\tilde\tau_2$, ενώ οι CAE $\tilde\tau_2\tilde\tau_2^{(*)}$ παρέχουν περίπου το $9-11\%$ της $J_{\rm eff}$. Συμπερασματικά, οι ANE είναι αμεληταίες για μικρές τιμές του $\Delta_{NLSP}$ αλλά είναι πολύ σημαντικές για μεγαλύτερες τιμές του $\Delta_{NLSP}$. Αυτό οφείλεται στο ότι οι αριθμητικές πυκνότητες των $\tilde\tau_2$ ελαττώνονται σχετικά με αυτή του $\tilde\chi$ καθώς το $\Delta_{NLSP}$ αυξάνει.

- Και οι πέντε CAE του $\tilde\chi$ με $\tilde\tau_2$ που αναγράφονται στον Πίνακα 5.3 δίνουν παραπλήσιας αξίας συνεισφορές στο $a_{\tilde\chi\tilde\tau_2}$ (η προεξάρχουσα συνεισφορά προέρχεται, γενικά από την $\tilde\chi\tilde\tau_2 \to \tau h$). Η σχετική συνεισφορά του $b_{\tilde\chi\tilde\tau_2}$ στο $J_{(\tilde\chi\tilde\tau_2)}$ προκύπτει ότι είναι ουσιαστικά ανεξάρτητη της τιμή του $m_A$ ($95{\rm GeV} \leq m_A \leq 215\,{\rm GeV}$). Αυτή η συνεισφορά κυμαίνεται από 5% σε περίπου 8% καθώς το $\Delta_{NLSP}$ αυξάνει από 0 σε 0.1.

- Η μέγιστη συνεισφορά στο $a_{\tilde\tau_2\tilde\tau_2^{(*)}}$ προέρχεται από τις CAE $\tilde\tau_2\tilde\tau_2^* \to hh$, $t\bar t$ και $\tilde\tau_2\tilde\tau_2 \to \tau\tau$. Εντουτοις, πολλές από τις άλλες CAE (όπως οι $\tilde\tau_2\tilde\tau_2^* \to ZZ$, $\gamma\gamma$, $HH$, $AA$, $H^+H^-$, $\gamma Z$) έχουν, γενικά σπουδαία συνεισφορά που δεν μπορεί να αμεληθεί (η $\tilde\tau_2\tilde\tau_2^* \to ZZ$, για μεγάλες τιμές των $\Delta_{NLSP}$ και $m_A$ δίνει μέγιστη συνεισφορά). Επίσης, η CAE $\tilde\tau_2\tilde\tau_2^* \to hH$ $(W^+W^-)$ είναι αυξημένη για μικρές τιμές των

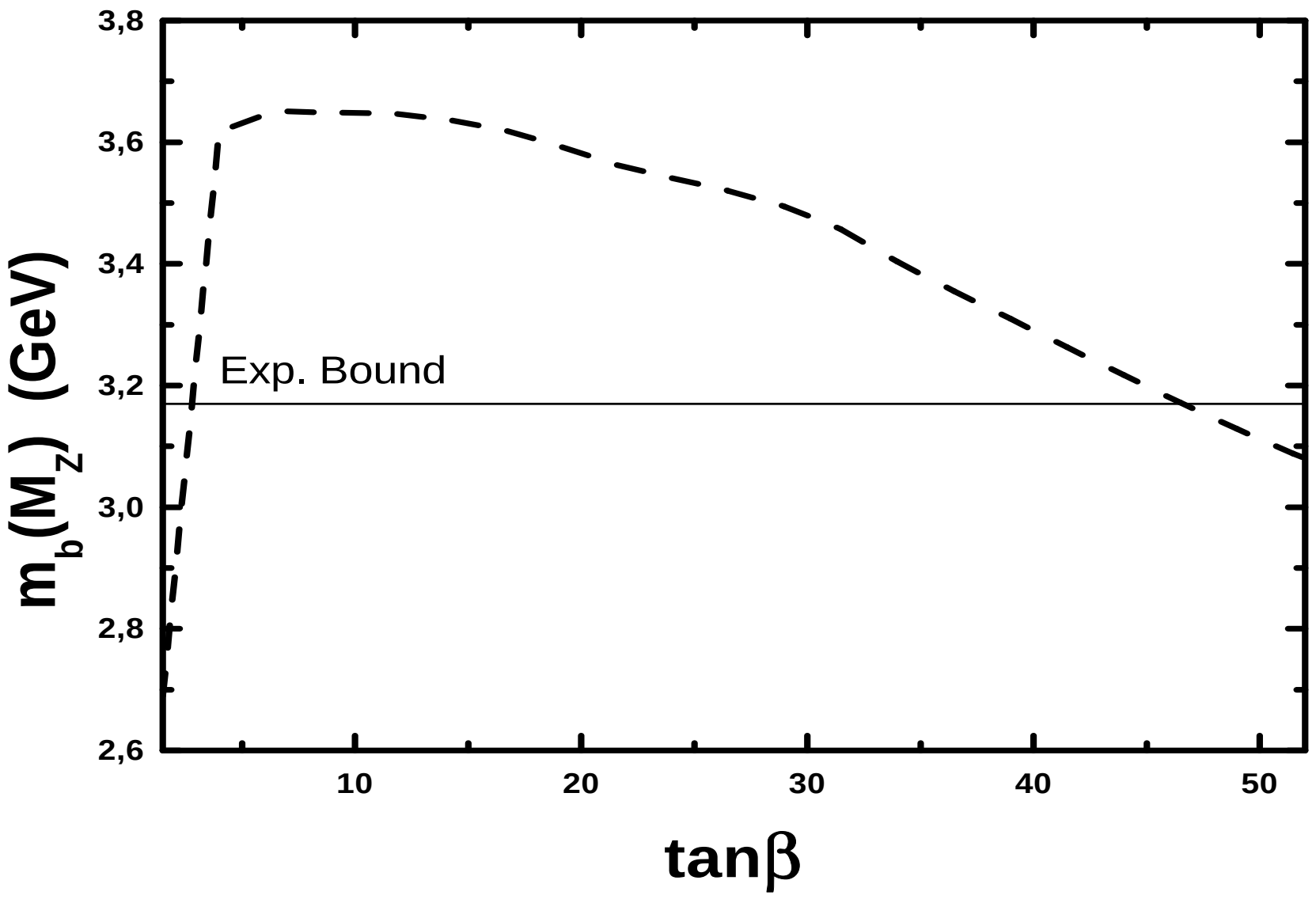

**Σχήμα 6.13:** *Η δενδρικού επιπέδου μάζα του b-quark ως συνάρτηση της* $\tan\beta$.

$m_A$ (και $\Delta_{NLSP}$). Η σχετική συνεισφορά του $b_{\tilde{\tau}_2\tilde{\tau}_2^{(*)}}$ στη $J_{(\tilde{\tau}_2\tilde{\tau}_2^{(*)})}$, είναι μικρότερη από 1% για όλες τις σχετικές τιμές των άλλων παραμέτρων.

- Όταν $\Delta_{NLSP} = 0$, οι συνεισφορές των $b_{\tilde{\chi}\tilde{\tau}_2}$ και $b_{\tilde{\tau}_2\tilde{\tau}_2^{(*)}}$ στη $J_{\rm eff}$ αλληλοδιαγράφονται μερικώς και επομένως, ένα ακριβές αποτέλεσμα λαμβάνεται (με λάθος μέχρι 0.5%) αγνοώντας τη συμβολή των $b$ στο τελικό απότέλεσμα. Όταν $\Delta_{NLSP} = 0.1$, εντούτοις, η συνεισφορά του $b_{\tilde{\chi}\tilde{\tau}_2}$ κυριαρχεί στην $b_{\tilde{\tau}_2\tilde{\tau}_2^{(*)}}$ που παρέχει το $4-5\%$ του $J_{\rm eff}$. Επομένως, τα αριθμητικά αποτελέσματα μπορούν να αναπαραχθούν με μια ακρίβεια καλύτερη από $\approx 5\%$ χρησιμοποιώντας για τις CAE μόνο τους $a_{ij}$. Οι αναλυτικές τους εκφράσεις δίνονται στο Εδ. 5.5.2. Αντιθέτως, το $b_{\tilde{\chi}\tilde{\chi}}$ δεν μπορεί να αγνοηθεί, αφού η συνεισφορά του στη $J_{(\tilde{\chi}\tilde{\chi})}$ ανέρχεται στο 80% και οι ΑΝΕ $\tilde{\chi}\tilde{\chi}$ είναι πολύ σημαντικές για μεγαλύτερες τιμές του $\Delta_{NLSP}$.

Οι σχετικές συνεισφορές $J_{(ij)}/J_{\rm eff}$ $((ij) = (\tilde{\chi}\tilde{\chi}), (\tilde{\chi}\tilde{\tau}_2), (\tilde{\tau}_2\tilde{\tau}_2^{(*)}))$ των τριών ειδών συμπεριλαμβανομένων διαδικασιών στο $J_{\rm eff}$ δίνονται στο Σχ. 6.12 σαν συνάρτηση του $m_A$ για την κεντρική τιμή του $\Delta_{NLSP} \simeq 0.047$ και για $M_S = 1\,{\rm TeV}$ όπου το επιτρεπτό πεδίο τιμών του $m_A$ είναι $95\,{\rm GeV} \lesssim m_A \lesssim 215\,{\rm GeV}$.

## 6.7 Συμπεράσματα

Τα συμπεράσματα από τη διερεύνηση του μελετούμενου προτύπου καταγράφονται παρακάτω:

**α.** Περίπτωση $\mu < 0$. Η μάζα του $b$-quark με την προσθήκη των SUSY διορθώσεων παραβιάζει τα πειραματικά όρια. Αντιθέτως, εφαρμογή του ${\rm BR}(b \to s\gamma)$ δεν θέτει περιορισμούς στο πεδίο ορισμού των παραμέτρων του προτύπου. Τα όρια των CDM σεναρίων ικανοποιούνται όταν το ελαφρότερο stau είναι 0-8% βαρύτερο από το LSP που μπορεί να κυμαίνεται από $215-770\,{\rm GeV}$. Καταληκτικά, αν δεχθεί κανείς ότι η απόκλιση στη μάζα του $b$-quark δεν είναι καταστροφική για τη θεωρία, η περίπτωση αυτή παρέχει έναν ευρύ χώρο παραμέτρων ‘φιλόξενο¨.

**β.** Περίπτωση $\mu > 0$. Η μάζα του $b$-quark με την προσθήκη των SUSY διορθώσεων βρίσκεται οριακά μέσα στα πειραματικά όρια. Αντιθέτως, εφαρμογή του ${\rm BR}(b \to s\gamma)$ επιβάλλει ένα ισχυρό κάτω όριο

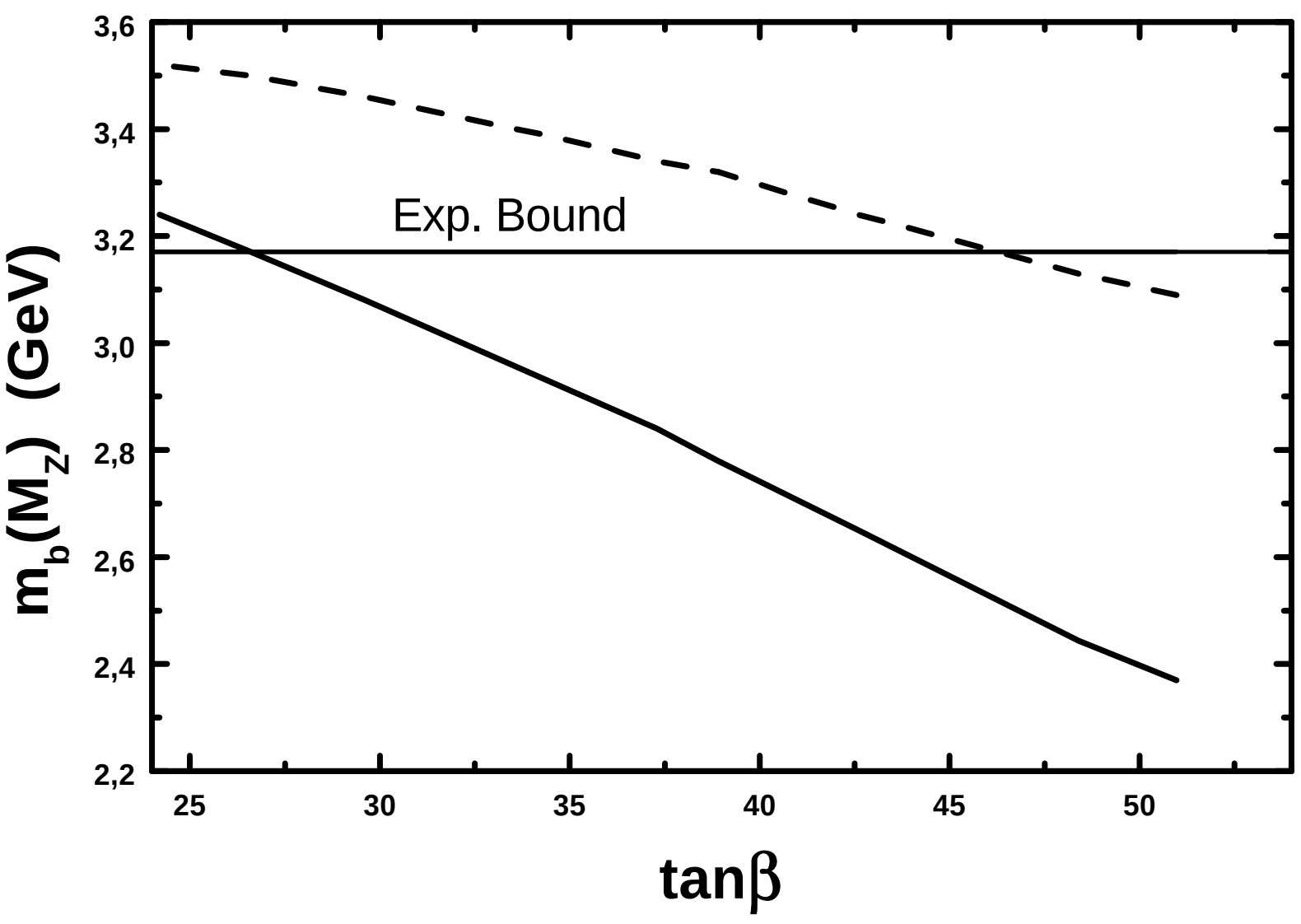

**Σχήμα 6.14:** *Η δενδρικού επιπέδου (διακεκομένη γραμμή) και η διορθωμένη (συνεχής γραμμή) μάζα του b-quark ως συνάρτηση της* $\tan\beta$, *για* $\Delta_{NLSP} \simeq 0$, $\mu > 0$ *και* $\mathrm{BR}(b \to s\gamma) \simeq 4.5 \times 10^{-4}$.

στη παράμετρο $m_A$. Τα όρια των CDM σεναρίων ικανοποιούνται με τον συνυπολογισμό των CAE, για αυτές τις επιτρεπτές τιμές του $m_A$. Καταληκτικά, η απαίτηση ταυτόχρονης ικανοποίησης των περιορισμών από το $\mathrm{BR}(b \to s\gamma)$ και τα CDM σενάρια παρέχει χώρο παραμέτρων πιο περιορισμένο από την προηγούμενη περίπτωση.

## 6.8 Επέκεινα της ενοποίησης Yukawa

Επέκεινα της Ενοποίησης Yukawa υπαρχούν δύο βασικές πιθανότητες. Η Ενοποίηση $b-\tau$, στην οποία είναι αφιερωμένο το Εδ. 6.8.1 και η μη Ενοποίηση των ζεύξεων Yukawa για την οποία γίνεται αναφορά στο Εδ. 6.8.2. Και στις δύο περιπτώσεις, υπάρχει χώρος παραμέτρων που ικανοποιεί ταυτόχρονα τους φαινομενολογικούς και τους κοσμολογικούς περιορισμούς. Το κόστος που πληρώνει κανεις εγκαταλείποντας τη Ενοποίηση Yukawa είναι η εισαγωγή μιας νέας παραμέτρου στο πρόβλημα. Η $\tan\beta$ δεν προσδιορίζεται πλέον μονότιμα αλλά γίνεται ελεύθερη παράμετρος. Ώστε, οι ελεύθερες παράμετροι του προγράματος θα είναι:

$$m_A,\ \Delta_{NLSP},\ \mathrm{sign}\mu\ , \tan\beta.$$

Προφανώς, η Εξ. (6.13) ισχύει με την ίδια μορφή σε κάθε περίπτωση. Αλλάζουν μόνο οι τιμές των συντελεστών. Επιπλέον η χρήση μεταβλήτου $\tan\beta$ κάνει το αριθμητικό πρόγραμμα πιο δυσκίνητο, γιατί για κάθε διαφορετική τιμή του $\tan\beta$ πρέπει να χρησιμοποιείται διαφορέτική ασυμπτωτική συνθήκη στην Εξ. (6.7). Σε όλα τα αριθμητικά αποτελεσματα που θα εκτεθούν σε αυτό και στο επόμενο κεφάλαιο της διατριβής οι διορθώσεις στη μάζα του λεπτονίου $\tau$ δεν θα ληφθούν υπόψη. Άλλωστε, καθώς η ενοποίηση Yukawa εγκαταλείπεται σταδιακά, η τιμή της $\tan\beta$ θα ελλατώνεται, όποτε και η διόρθωση της Εξ. (3.22) θα γίνεται όλο και πιο επουσιώδης.

### 6.8.1 Ενοποίηση $b-\tau$

Η δυνατότητα της $b-\tau$ Ενοποίησης αναδύεται στις περιπτώσεις υιοθέτησης μιας Ομάδας Ενοποίησης, όπως η $SU(5)$, στην οποία τα φερμιόνια κάτω τύπου μπορούν να διευθετηθούν στην ίδια αναπαράσταση. Οι

**Πίνακας 6.2: Ενοποίηση ζεύξεων Βαθμίδας και $b-\tau$ $(\mu>0)$**

| $M_S$ (GeV) | 1800 | 1500 | 1300 |
|---|---|---|---|
| **Μεταβλητές Εισόδου** ($m_\tau(M_S) = 1.78\,$GeV) | | | |
| $M_G$ ($10^{16}$ GeV) | 1.35 | 1.39 | 1.39 |
| $\alpha_G^{-1}$ | 25.57 | 25.43 | 25.3 |
| $h_t(M_G)$ | .628 | .593 | .564 |
| $h_b(M_G) = h_\tau(M_G)$ | .419 | .296 | .188 |
| **Μεταβλητές Εξόδου** | | | |
| $\alpha_3(M_Z)$ | .1201 | .1202 | .1204 |
| $s_W^2(M_Z)$ | .2306 | .2308 | .2310 |
| $\alpha_{em}^{-1}(M_Z)$ | 127.98 | 128.00 | 127.91 |
| $\tan\beta$ | 42.6 | 34.4 | 24.2 |
| $m_t(m_t)$ (GeV) | 166.025 | 166.046 | 166.002 |
| $m_b^c(M_Z)$ (GeV) | 2.65 | 2.94 | 3.24 |

ασυμπτωτικές τιμές των ζεύξεων Yukawa ικανοποιούν σε αυτή την περίπτωση, μια σχέση της μορφής:

$$h_t(M_G) > h_b(M_G) = h_\tau(M_G) := h_0\ . \tag{6.25}$$

Η διαδικασία προσδιορισμού του $h_0$ και της $\tan\beta$ εξελίσσεται όπως ακριβώς περιγράφεται στο Εδ. 6.3.2 με τη διαφορά ότι η δυνατότητα μεταβολής του λόγου $h_t(M_G)/h_0$ παρέχει τη δυνατότητα λήψης πολλών τιμών για την $\tan\beta$. Στον Πίνακα 6.2 εκτίθενται ενδεικτικά κάποιες τιμές που χρησιμοποιήθηκαν ως input στο αριθμητικό πρόγραμμα με τα αντίστοιχα output. Μερικά από τα φάσματα που χρησιμοποιήθηκαν για τον υπολογισμό του $m_b^c(M_Z)$ παρέχονται στον Πίνακα 6.4 με σημείο συσχέτισης την τιμή της $\tan\beta$.

Αρχικά το θέμα που εξετάζεται είναι το αποτέλεσμα για τη δενδρικού επιπέδου μάζα του $b$-quark με τη μεταβολή της $\tan\beta$. Σε αυτό το ερώτημα είναι αφιερωμένο το Σχ. 6.13, όπου εικονίζεται η δενδρικού επιπέδου μάζα του $b$-quark ως συνάρτηση της $\tan\beta$. Οι άλλες παράμετρες του προγράμματος δεν χρειάζεται να καθοριστούν γιατί η κατασκευή του διαγράμματος αυτού δεν προυποθέτει τον υπολογισμό του SUSY φάσματος αλλά μόνο την λύση του πρώτου συνόλου των RGE των ζεύξεων βαθμίδας και Yukawa. Σε αρμονία με τα ευρήματα της Αν. [73], από το Σχ. 6.13, παρατηρείται ότι η μάζα του $b$-quark αυξάνει με μείωση της $\tan\beta$, πράγμα που θα ευνοούσε την $\mu>0$ περίπτωση, κατά την οποία μια αρνητική διόρθωση θα συνέβαλλε να επαναφέρει τη μάζα του $b$-quark σε πειραματικά αποδεκτά επίπεδα.

Το παραπάνω σενάριο υλοποιείται στο Σχ. 6.14, όπου εικονίζεται η δενδρικού επιπέδου και η διορθωμένη μάζα του $b$-quark ως συνάρτηση της $\tan\beta$ για $\Delta_{NLSP}\simeq 0$ και BR$(b\to s\gamma)\simeq 4.5\times 10^{-4}$.Η τελευταια επιλογή προσδιορίζει ουσιαστικά την τιμή της παραμέτρου $m_A$, η οποία όμως δεν είναι καθοριστική για το διερευνούμενο θέμα, γιατί όπως φαίνεται και στο Σχ. 6.6, η $m_b$ παραμένει σχεδόν αμετάβλητη με μεταβολή του $m_A$. Εξαρτάται ισχυρά μόνο από το εκάστοτε $\tan\beta$. Από το Σχ. 6.14 συνάγεται επίσης ότι η διορθωμένη και η δενδρικού επιπέδου μάζα $b$-quark αυξάνει καθώς η $\tan\beta$ ελαττώνεται. Επιπλέον, η SUSY διόρθωση ελαττώνεται καθώς το $\tan\beta$ ελατώνεται αφού είναι σχεδόν ανάλογη του $\tan\beta$ όπως φαίνεται από την Εξ. (3.11) και (3.13). Συνεπώς, οι δεσμοί που πρέπει να πληρούνται για να βρίσκεται η (διορθωμένη) μάζα του $b$-quark μέσα στην επιτρεπόμενη πειραματικά περιοχή της Εξ. (3.18), είναι:

$$26.5 \lesssim \tan\beta \ \text{ και } \ \mu > 0 \tag{6.26}$$

Επομένως, το σύνολο των ανεξάρτητων παραμέτρων της θεωρίας (έχοντας επιλέξει $A_0=0$) είναι:

$$m_A,\ \Delta_{NLSP}, \tan\beta.$$

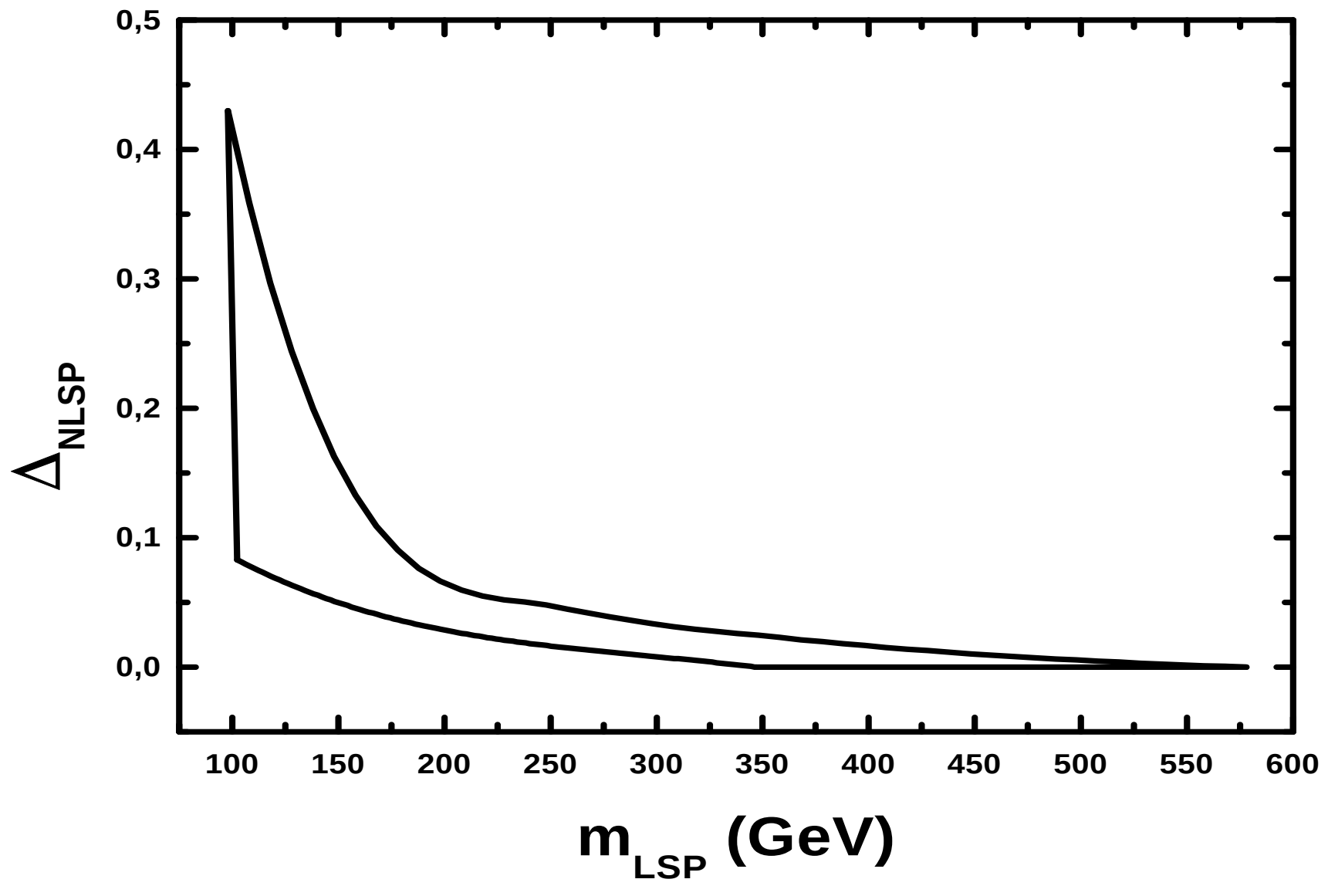

**Σχήμα 6.15:** *Η επιτρεπτή περιοχή στο επίπεδο $m_{LSP} - \Delta_{NLSP}$, για $\mu < 0$ και $\tan\beta \simeq 15.3$.*

Για τη συνέχεια της μελέτης του προτύπου, ακολουθούνται τα βήματα που περιγράφηκαν στα Εδ. 6.5.2 και 6.6.2. Το κάτω επιτρεπτό όριο του $m_A$ καθορίζεται από την συνθήκη $\mathrm{BR}(b \to s\gamma) \simeq 4.5 \times 10^{-4}$. Η κατάσταση, δηλαδή, είναι παρόμοια με αυτή του Σχ. 6.7, με $\mu > 0$. Αξιοποιώντας το τυπολόγιο του Εδ. 5.5.1 για τις ΑΝΕ και του Εδ. 5.5.2 για τις CAE υπολογίζεται το $\Omega_{LSP}h^2$ Παρατηρείται ότι το μέγιστο επιτρεπτό $\Delta_{NLSP}$ από τα CDM σενάρια, κυμαίνεται από 0.023 μέχρι 0, καθώς το $\tan\beta$ αυξάνεται από το κάτω όριο της Εξ. (6.26). Επιπλέον τα CDM όρια της Εξ. (5.4) θέτουν ένα άνω όριο στη $\tan\beta$ που λαμβάνεται στο σημείο που με $\Delta_{NLSP} = 0$ επιτυγχάνεται $\Omega_{LSP}h^2 = .22$ και είναι:

$$\tan\beta \lesssim 49.5 \tag{6.27}$$

Όπως παρατηρείται και από τις αντίστοιχες στήλες του Πίνακα 6.4 με ελάττωση της τιμής του $\tan\beta$ οι φαινομενολογικά επιτρεπτές τιμές του $m_{LSP}$ μειώνονται αλλά οι αντίστοιχες του $m_A$ αυξάνονται.

## 6.8.2 Μη Ενοποίηση Yukawa

Η πιο απελευθερωμένη από περιορισμούς άνωθεν περίπτωση είναι αυτή που εξετάζεται στο εδάφιο αυτό. Η εξέλιξη των ζεύξεων Yukawa γίνεται και πάλι από την κλίμακα ενέργειας GUT. Στη διαδικασία του Εδ. 6.3.2 επιβάλλεται ένας ακόμα περιορισμός. Επιδιώκεται με κατάλληλη επιλογή της $h_b(M_G)$ να επιτυγχάνεται η κεντρική πειραματική τιμή για τη μάζα του $b$-quark από την Εξ. (3.18). Έχοντας υπόψη μας το γεγονός ότι οι διορθώσεις στη μάζα του $b$-quark είναι προσήμου αντιθέτου από αυτό του $\mu$, οι ασυμπτωτικές τιμές των ζεύξεων Yukawa, πρέπει να ικανοποιούν σχέσεις της μορφής:

$$h_t(M_G) > h_\tau(M_G) \;\;\text{και}\;\; h_b(M_G) > h_\tau(M_G) \;\;\text{όταν}\;\; \mu > 0\,, \tag{6.28}$$

$$h_t(M_G) > h_\tau(M_G) \;\;\text{και}\;\; h_b(M_G) < h_\tau(M_G) \;\;\text{όταν}\;\; \mu < 0\,. \tag{6.29}$$

Πιο συγκεκριμένα αποτελέσματα των προσπαθειών μας καταγράφονται στον Πίνακα 6.3. Μερικά από τα φάσματα που χρησιμοποιήθηκαν για τον υπολογισμό του $m_b^c(M_Z)$ παρέχονται στον Πίνακα 6.4 με σημείο συσχέτισης την τιμή της $\tan\beta$. Συνεπώς, και στο πρόγραμμα αυτό η $\tan\beta$ είναι μια μεταβλητή εξόδου. Παρόλαταύτα, για την αποφυγή συγχύσεως, στην παρουσίαση των αποτελεσμάτων θα παρουσιάζεται ως

**Πίνακας 6.3: Ενοποίηση ζεύξεων Βαθμίδας**

| sign$\mu$ | $-1$ | $+1$ | $-1$ | $+1$ |
|---|---|---|---|---|
| $M_S$ (GeV) | 400 | 900 | 500 | 1000 |
| **Μεταβλητές Εισόδου** ($m_\tau(M_S) = 1.78$ GeV) | | | | |
| $M_G$ ($10^{16}$GeV) | 1.7 | 1.5 | 1.48 | 1.48 |
| $\alpha_G^{-1}$ | 24.54 | 25.06 | 24.7 | 25.1 |
| $h_t(M_G)$ | .547 | .547 | .547 | .544 |
| $h_b(M_G)$ | .048 | .055 | .074 | .083 |
| $h_\tau(M_G)$ | .075 | .073 | .1094 | .1088 |
| **Μεταβλητές Εξόδου** | | | | |
| $\alpha_3(M_Z)$ | .1206 | .1205 | .1202 | .1209 |
| $s_W^2(M_Z)$ | .2319 | .2313 | .2318 | .2311 |
| $\alpha_{em}^{-1}(M_Z)$ | 128.00 | 127.9 | 127.99 | 127.99 |
| $\tan\beta$ | 10.32 | 10.16 | 15.24 | 14.95 |
| $m_t(m_t)$ (GeV) | 166.083 | 166.025 | 166.054 | 166.069 |
| $m_b(M_Z)$ (GeV) | 2.67 | 2.73 | 2.71 | 2.70 |

μεταβλητή εισόδου. Γιαυτό και η τιμή της θα είναι προσεγγιστικά ίση με τις αναγραφόμενες τιμές. Οι ελεύθερες παράμετροι του προβλήματος, λοιπόν, είναι ($A_0 = 0$):

$$m_A,\ \Delta_{NLSP},\ \text{sign}\mu\ , \tan\beta.$$

Έχοντας απαλλαγεί από το πρόβλημα της μάζας του $b$-quark, με τη δια χειρός διευθέτηση, οι φαινομενολογικοί περιορισμοί που πρέπει να ληφθούν υπόψη είναι ασφαλώς πιο ελαστικοί. Για το μεγαλύτερο μέρος του παραμετρικού χώρου το κάτω όριο του $m_A$ λαμβάνεται από τη θεώρηση του BR($b \to s\gamma$). Επειδή μάλιστα για $\mu < 0$, μικρότερες τιμές του $m_A$ είναι επιτρεπτές, το σενάριο $\mu < 0$ είναι πιο προνομιακό από το $\mu > 0$ όπως αναλυτικά θα εξηγηθεί στο Εδ. 7.5 του επόμενου κεφαλαίου.

Ενδεικτικά, για $\tan\beta \simeq 15.2$ γίνεται το διάγραμμα της επιτρεπτής περιοχής στο επίπεδο $m_{LSP} - \Delta_{NLSP}$, για $\mu < 0$ που εικονίζεται στο Σχ. 6.15. Για το σκοπό αυτό, φυσικά αξιοποιείται το τυπολόγιο του Εδ. 5.5.1 για τις ΑΝΕ και του Εδ. 5.5.3 για τις CAE. Παρατηρείται ότι το επιτρεπτό φάσμα τιμών του LSP είναι: 100 GeV $\lesssim m_{LSP} \lesssim$ 578 GeV με $\Delta_{NLSP}^{max} \simeq 0.4$. Επίσης, υπάρχει μικρή περιοχή για 100 GeV $\lesssim m_{LSP} \lesssim$ 120 GeV στην οποία είναι δυνατόν να ληφθούν κοσμολογικά αποδεκτά αποτελέσματα χωρίς την ενεργοποίηση των CAE. Όλες οι CAE με κάποιο higgs στις τελικές καταστάσεις είναι αποκλεισμένες κινηματικά αφού ο τομέας higgs του προτύπου είναι αρκετά βαρύς (360 GeV $\lesssim m_A \lesssim$ 1650 GeV), γιαυτό και το μέγιστο επιτρεπτό $m_{LSP}$ (578 GeV) είναι αρκετά μικρότερο από το αντίστοιχο του Εδ. 6.6.1, (770 GeV) αφού η ταπείνωση της CRD λόγω των CAE είναι ασθενέσθερη.

Συμπερασματικά, η περίπτωση μη ενοποίησης Yukawa διαθέτει μεγαλύτερη ευκαίρια διαχείρησης των ελεύθερων παραμέτρων του προβλήματος. Στον Πίνακα 6.4 δίνονται κάποια φάσματα μαζών των sparticles και higgs που αντιστοιχούν σε τιμές των μεταβλητών των Πινάκων 6.1, 2, 3. Παρατηρείται ότι με ελλάτωση της $\tan\beta$ το ελάχιστο δυνατό $m_{LSP}$ ελλατώνεται αλλά το αντίστοιχο $m_A$ αυξάνεται και ακόμα ότι με $\tan\beta \leq 15$ οι μάζες των $\tilde{e}_R$ πλησιάζουν αυτή του $\tilde{\tau}_2$ οπότε λαμβάνουν μέρος στις CAE. Επίσης για σταθερό $\tan\beta$ το ελάχιστο δυνατό $m_{LSP}$ είναι μικρότερο στην περίπτωση $\mu < 0$.

**Πίνακας 6.4: Το φάσμα των sparticles και Higgs (Μάζες σε GeV)**

| Αρχικές Συνθήκες ($A_0 = 0$) | | | | | | | | | |
|---|---|---|---|---|---|---|---|---|---|
| sign$\mu$ | $-1$ | $-1$ | $-1$ | $+1$ | $+1$ | $+1$ | $-1$ | $+1$ | $+1$ |
| $\tan\beta$ | 46.28 | 45.96 | 45.89 | 52.8 | 34.4 | 24.2 | 15.24 | 14.95 | 10.1 |
| $M_{1/2}$ | 477.5 | 652 | 813 | 1600 | 896.4 | 731.4 | 244.8 | 578 | 493.5 |
| $m_0$ | 292.4 | 395 | 457 | 805 | 334.7 | 227.9 | 77.1 | 152.9 | 139.1 |
| $\tilde{\chi}$ | 208.7 | 288.6 | 363.5 | 733.5 | 401.8 | 325.9 | 102.5 | 254.8 | 217.0 |
| $\tilde{\chi}_2^0$ | 387.4 | 537.0 | 674.9 | 1350.3 | 748.4 | 607.7 | 185.9 | 477.4 | 406.4 |
| $\tilde{\chi}_3^0$ | 586.4 | 769.1 | 926.6 | 1668.3 | 1037.9 | 867.2 | 334.1 | 717.2 | 625.3 |
| $\tilde{\chi}_4^0$ | 601.5 | 782.3 | 938.9 | 1677.3 | 1047.1 | 876.9 | 355.9 | 726.4 | 633.9 |
| $\tilde{\chi}_1^\pm$ | 387.3 | 536.9 | 674.8 | 1350.3 | 1047.4 | 877.5 | 356.4 | 727.4 | 635.5 |
| $\tilde{\chi}_2^\pm$ | 601.8 | 782.4 | 938.8 | 1677.3 | 748.4 | 607.7 | 185.2 | 477.4 | 406.4 |
| $\tilde{g}^A$ | 1125.4 | 1502.2 | 1836.5 | 3439.7 | 2018.3 | 1666.5 | 604.5 | 1348.6 | 1159.8 |
| $\tilde{t}_1$ | 933 | 1217 | 1466.1 | 2676.6 | 1637.7 | 1384.5 | 570.7 | 1150.0 | 1000.5 |
| $\tilde{t}_2$ | 777.9 | 1052.6 | 1290.8 | 2430.7 | 1405.8 | 1155.3 | 396.9 | 931.7 | 798.8 |
| $\tilde{b}_1$ | 908.6 | 1196.1 | 1447.3 | 2664.4 | 1630.5 | 1381.3 | 529.5 | 1150.6 | 997.3 |
| $\tilde{b}_2$ | 793.6 | 1073.9 | 1315.1 | 2465.8 | 1563.8 | 1337.4 | 500.0 | 1114.4 | 964.1 |
| $\tilde{u}_L$ | 1052 | 1403 | 1704.3 | 3157.4 | 1835.3 | 1509.8 | 552.2 | 1221.2 | 1052.2 |
| $\tilde{u}_R$ | 1011 | 1345.6 | 1631.1 | 3008.7 | 1752.3 | 1442.1 | 530.5 | 1168.2 | 1007.2 |
| $\tilde{d}_L$ | 1055 | 1405.5 | 1706.1 | 3158.4 | 1835.3 | 1509.8 | 552.2 | 1221.2 | 1052.2 |
| $\tilde{d}_R$ | 1006 | 1338.7 | 1621.9 | 2988.7 | 1741.2 | 1433.1 | 527.7 | 1168.2 | 1001.3 |
| $\tilde{\tau}_1$ | 438.8 | 577.3 | 689.3 | 1268.04 | 688.7 | 550.8 | 201.0 | 430.5 | 372.2 |
| $\tilde{\tau}_2$ | 208.8 | 311.4 | 380.5 | 733.2 | 401.3 | 325.9 | 111.4 | 258.4 | 232.4 |
| $\tilde{\nu}_\tau$ | 404.3 | 550.9 | 667.1 | 1256.5 | 674.5 | 538.6 | 175.3 | 419.7 | 361.9 |
| $\tilde{e}_L$ | 443.8 | 599.9 | 721.4 | 1346.9 | 698.1 | 550.9 | 188.0 | 427.3 | 368.6 |
| $\tilde{e}_R$ | 347.8 | 469.5 | 553.9 | 1009.4 | 478.1 | 360.3 | 121.6 | 268.8 | 234.5 |
| $\tilde{\nu}_e$ | 436.6 | 594.6 | 717.0 | 1344.6 | 695.1 | 547.2 | 176.7 | 422.5 | 363.0 |
| $A$ | 100 | 150 | 200 | 400 | 865 | 865 | 360 | 800 | 710 |
| $h$ | 99.6 | 121.7 | 123.2 | 126.9 | 124.3 | 122.5 | 114.2 | 120.5 | 118.3 |
| $H$ | 119.4 | 150.1 | 200.2 | 399.9 | 864.9 | 865.0 | 360.0 | 800.1 | 710.3 |
| $H^\pm$ | 127.9 | 169.9 | 215.5 | 407.7 | 868.6 | 868.6 | 368.6 | 803.9 | 714.4 |

# Κεφάλαιο 7

# MSSM προερχόμενο από τη Θεωρία Hořava-Witten

## 7.1 Εισαγωγή

Μια έκδοση του MSSM προερχόμενη από μια θεωρία Υπερχορδών με το όνομα Hořava-Witten είναι το αντικείμενο μελέτης του τελευταίου αυτού κεφαλαίου της διατριβής. Στο Εδ. 7.2 γίνεται μια περιληπτική παρουσίαση της θεωρίας Hořava-Witten. Στο Εδ. 7.3 εκτίθενται τα βήματα του αριθμητικού υπολογισμού. Έπονται η παραμετρική μελέτη του προτύπου στο Εδ. 7.4 και η φαινομενολογική στο Εδ. 7.5. Τα κοσμολογικά αποτελέσματα που παρουσιάζονται στο Εδ. 7.6 και το κεφάλαιο ολοκληρώνεται με την καταγραφή των συμπερασμάτων στο Εδ. 7.7. Ο συγγραφέας της διατριβής αυτής δεν έχει μελετήσει Θεωρία Υπερχορδών. Η συνεισφορά του στην εργασία της Αν. [45], στην οποία στηρίζεται το κεφάλαιο αυτό, ήταν ο έλεγχος του προτύπου σε χαμηλές ενέργειες, επιδιώκοντας ταυτόχρονη ικανοποίηση κριτιρίων φαινομενολογικών και κοσμολογικών. Γιαυτό και το κύριο βάρος της παρουσίασης θα επικεντρωθεί στην ανάλυση των αποτελεσμάτων, περιορίζοντας τις θεωρητικές αναφορές στο ελάχιστο δυνατό, αφού οι έννοιες που υπεισέρχονται δεν είναι οικείες στον γράφοντα.

## 7.2 Το πρότυπο Hořava-Witten

Προσφάτως οι Hořava-Witten στην Αν. [74] έδειξαν ότι οι πέντε διαταρακτικές θεωρίες χορδών (τύπου I ανοικτές χορδές, τύπου IIA και IIB κλειστές χορδές και οι δύο κλειστές ετεροτικές χορδές με συμμετρίες βαθμίδας $E_8 \times E_8'$ και $SO(32)$ ) και η 11 διαστάσεων υπερβαρύτητα αντιστοιχούν σε διαφορετικά κενά μιας υπερκείμενης θεμελιώδους θεωρίας που λέγεται M-Θεωρία. Σε αυτή την θεωρία τα παρατηρήσιμα πεδία βαθμίδας περιέχονται στην 10 διαστάσεων ομάδα $E_8$ που βρίσκεται στο ένα άκρο μιας γραμμής μήκους $\rho$, ενώ ο σπασμένος τομέας των πεδίων περιέχεται στο υπόλοιπο τμήμα της ομάδας το $E_8'$ που τοποθετείται στο άλλο άκρο της γραμμής. Συμπαγοποίηση (compactification) του χώρου, επίσης, υποτίθεται σε ένα $S^1/Z_2$ orbifold. Τα πεδία βαρύτητας διαδίδονται στον όγκο αυτού του 11 διαστάσεων κατασκευάσματος που μοιάζει με επίπεδο πυκνοτή και απεικονίζεται παραστατικά στο Σχ. 7.1. Όταν το μήκος $\rho$ της ενδιάμεσης γραμμής τείνει στο μηδέν, τότε οι δύο πλάκες συμπίπτουν και προκύπτει η ασθενώς συζευγμένη θεωρία χορδών.

Η μεγαλύτερη επιτυχία της θεωρίας των Hořava-Witten είναι η λύση με ευφυή τρόπο του προβλήματος της ενοποίησης των ζεύξεων βαθμίδας, δηλαδή ο μηδενισμός της ενεργειακής απόστασης που προέκυπτε ανάμεσα στη κλίμακα της SUSY-GUT, $M_G \simeq 2 \times 10^{16}$ GeV και στην κλίμακα ενοποίησης της χορδής $M_{String} \simeq 5 \times 10^{17}$ GeV που υπολογιζόταν με την ασθενώς συζευγμένη θεωρία χορδών. Στην ισχυρώς συζευγμένη θεωρία χορδών οι επιπλέον Kaluza-Klein καταστάσεις δεν επηρεάζουν την εξέλιξη των ζεύξεων βαθμίδας αλλά μόνο επιταχύνουν την εξέλιξη της βαρυτικής ζεύξης με αποτέλεσμα την μείωση του $M_{String}$ στο $M_G$. Επιπλέον, το SSB της SUSY μέσα στα πλαίσια της M-Θεωρίας οδηγεί σε μάζες των gaugino της τάξης της μάζας των gravitino, σε αντίθεση με την περίπτωση της ασθενώς συζευγμένης θεωρίας χορδών που προέβλεπε πολύ ελαφρές μάζες για τα gaugino.

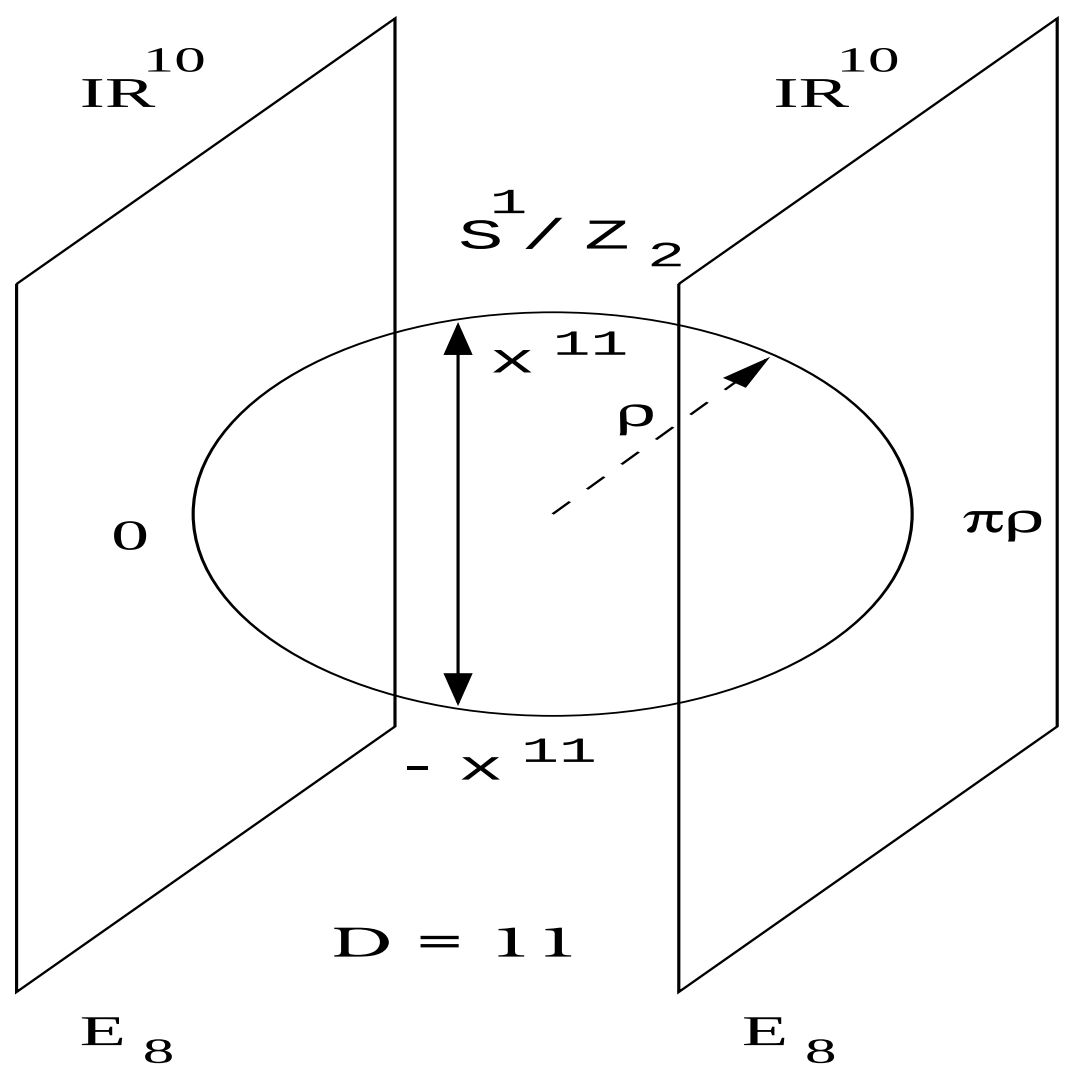

**Σχήμα 7.1:** *Μια παραστατική απεικόνιση του ενδεκαδιάστατου χώρου της Θεωρίας Hořava-Witten παρμένη από την Αν. [75]*

Παρομοίως με την περίπτωση της ασθενώς συζευγμένης θεωρίας χορδών, η συμπαγοποίηση της θεωρίας των Hořava-Witten σε χαμηλότερες διαστάσεις μπορεί να οδηγήσει στο SSB της $E_8$ σε φαινομενολογικά πιο ενδιαφέρουσες ομάδες. Διακρίνονται οι περιπτώσεις:

- Καθιερωμένη Εμφύτευση (SE). Σε αυτή την περίπτωση, το SSB της $E_8$ οδηγεί στην $E_6$. Το SSB των απλών ομάδων ($E_6$, $SO(10)$) επιτυγχάνεται με τη χρήση πεδίων σε συναφή αναπαράσταση, τα οποία όμως δεν περιέχονται στα απομεινάρια των θεωριών χορδών. Οπότε το SSB της $E_6$ μπορεί να πραγματοποιηθεί μόνο μέσω των βρόχων του Wilson και η πρόβλεψη της ενοποίησης των ζεύξεων Yukawa χάνεται. Εντούτοις, αν το SSB της $E_6$ οδηγήσει στην ομάδα τρι-ενοποίησης, $SU(3)_C \times SU(3)_L \times SU(3)_R$, ή στην $SU(6) \times U(1)$ μπορεί να επιτευχθεί αντίστοιχα η πρόβλεψη της πλήρους ή $b-\tau$ ενοποίησης Yukawa σύμφωνα με την Αν. [79].
- Μη Καθιερωμένη Εμφύτευση (NSE). Σε αυτή την περίπτωση, το SSB της $E_8$ οδηγεί σε ομάδες όπως η $SO(10)$ και $SU(5)$, οι οποίες αρχικά ευνοούν την πλήρη ή μερική ενοποίηση Yukawa. Η μη ύπαρξη συναφών υπερπεδίων Higgs, όμως, οδηγεί σε SSB των ομάδων αυτών μέσω των βρόχων Wilson, οπότε η πρόβλεψη για κάποιου είδους ενοποίηση Yukawa χάνεται. Εντούτοις, υπάρχουν κατασκευές με μεγαλύτερο Kac-Moody επίπεδο κατά την Αν. [78], οι οποίες ξεπερνούν την προιούσα αναίρεση. Επίσης, αν κατά το SSB της $E_8$ προκύψει η ομάδα Pati-Salam $SU(4)_c \times SU(2)_L \times SU(2)_R$, τότε δεν απαιτούνται οι βρόχοι Wilson γιατί αυτή η ομάδα περιέχει και τη $SU(4)_c$ και την $SU(2)_R$ οι οποίες διαθέτουν συναφή υπερπεδία Higgs.

Συμπερασματικά, όλες οι ασυμπτωτικές σχέσεις μεταξύ των ζεύξεων Yukawa είναι επιτρεπτές μέσα στα πλαίσια της θεωρίας Hořava-Witten.

Οι ασυμπτωτικές τιμές των SBT της θεωρίας αυτής δίνονται στις Αν [76] και [77] είναι οι ίδιες και στην περίπτωση του SE και του NSE. Η θεωρία προτείνει παγκόσμιες αρχικές συνθήκες για τους SBT, δηλαδή:

- Μια κοινή μάζα για τα gaugino, $M_{1/2}$,

$$M_{1/2} = \frac{\sqrt{3}m_{3/2}}{1+\epsilon}(\sin\theta + \frac{\epsilon}{\sqrt{3}}\cos\theta). \tag{7.1}$$

- Μια κοινή μάζα για τα βαθμωτά πεδία, $m_0$,

$$m_0^2 = m_{3/2}^2 - \frac{3m_{3/2}^2}{(3+\epsilon)^2}\left(\epsilon(6+\epsilon)\sin^2\theta + (3+2\epsilon)\cos^2\theta - 2\sqrt{3}\epsilon\cos\theta\sin\theta\right). \tag{7.2}$$

- Μια κοινή τριγραμμική ζεύξη, $A_0$,

$$A_0 = -\frac{\sqrt{3}m_{3/2}}{3+\epsilon}\left((3-2\epsilon)\sin\theta + \sqrt{3}\epsilon\cos\theta\right). \tag{7.3}$$

όπου $m_{3/2}$ η μάζα του gravitino και $\theta$ η γωνία του goldstino με $0<\theta<\pi/2$ και $0<\epsilon<1$ στο SE, $-1<\epsilon<1$ για το NSE. Η αξιοποίηση των αρχικών αυτών συνθηκών στα πλαίσια του MSSM είναι ο στόχος της επόμενης παραφράφου.

## 7.3 Αριθμητική επεξεργασία

Η μελέτη του MSSM με τη χρησιμοποίηση των αρχικών συνθηκών των Εξ. (7.1), (7.2), (7.3) είναι ο στόχος αυτού του κεφαλαίου της διατριβής. Η αριθμητική επεξεργασία γίνεται μέσα στα πλαίσια που έχουν διεξοδικά περιγραφεί στα Εδ. 6.3 6.4 του προηγούμενου κεφαλαίου. Περιληπτικά, τα βήματα του υπολογισμού καταγράφονται παρακάτω:

- Εύρεση της κλίμακας ενοποίησης $M_G$ και της ενοποιημένης ζεύξης, $g_G$. Λύνονται αριθμητικά οι RGE των ζεύξεων βαθμίδας του Εδ. Α'.2 με δοκιμαστικές αρχικές συνθήκες μέχρι την επίτευξη των πειραματικών τιμών των Εξ. (6.5) και (6.6).

- Υπολογισμός του $\tan\beta$ στο $M_S$. Παράλληλα με τη λύση των RGE για τις ζεύξεις βαθμίδας, λύνονται οι RGE για τις ζεύξεις Yukawa, πάλι με δοκιμαστικές αρχικές συνθήκες. Χρησιμοποιώντας τις τιμές $m_\tau(M_S) = 1.78$ GeV και της Εξ. (3.21), ως εισαγόμενα στο πρόγραμμα, προσδιορίζεται η $\tan\beta$ στο $M_S$. Επομένως η $\tan\beta$ είναι αποτέλεσμα των ασυπτωτικών σχέσεων που ικανοποιούν οι ζεύξεις Yukawa και όχι μια ελεύθερη παράμετρος. Αυτές οι σχέσεις καθορίζονται είτε από ασυμπτωτικές υποθέσεις (πλήρους ή μερικής Yukawa ενοποίσης) είτε από την απαίτηση η προκύπτουσα σε κλίμακα ενέργειας $M_Z$ μάζα του $b$-quark να είναι στην κεντρική πειραματική τιμή της Εξ. (3.18).

- Υπολογισμός του $\mu$. Λύνονται αριθμητικά οι RGE των SBT με αρχικές συνθήκες των Εξ. (7.1), (7.2) και (7.3). Σε κλίμακα ενέργειας $M_S$ υποτίθεται επιτυχές SSB και προσδιορίζεται η παράμετρος $\mu$ μέσω της Εξ. (2.45), χρησιμοποιώντας την προσέγγιση του tree-level renormalization group improved effective potential σύμφωνα με τα αναφερόμενα στο Εδ. 3.2.1.

- Υπολογισμός του SUSY φάσματος της θεωρίας. Με εφαρμογή του τυπολόγιου του Εδ. 2.5 υπολογίζεται το SUSY φάσμα του προτύπου. Το LSP της θεωρίας προκύπτει ότι είναι μορφής bino με καθαρότητα 98% ενώ το NLSP είναι το ελαφρότερο stau μορφής δεξιόστροφης. Επίσης, συμπεριλαμβάνονται οι διορθώσεις στο $h$ με χρησιμοποίηση του τυπολόγιου του Εδ. 3.2.2.

- Επανακαθορισμός του $M_S$. Έχοντας υπολογίσει το SUSY φάσμα, μπορεί να καθοριστεί η κλίμακα $M_S$ μέσω της Εξ. (3.1) οπότε με αυτό ως δεδομένο η διαδικασία επαναλαμβάνεται μέχρι την επίτευξη ευσταθούς λύσεως.

## 7.4 Παραμετρική μελέτη

Αρχικά, οι ελεύθερες παράμετρες του προβλήματος, είναι:

$$\tan\beta,\ \mathrm{sign}\mu,\ \theta,\ \epsilon,\ m_{3/2}.$$

Έχοντας υπόψη μας την εξάρτηση του $m_A$ από τις ασυμπτωτικές μάζες των scalar και gauginos , μπορεί να θεμελιωθεί αριθμητικά μια παρόμοια σχέση βασισμένη σε τεχνικές που έχουν ήδη περιγραφεί στο Εδ. 6.4.1 και οι οποίες ισχύουν με μια ακρίβεια της τάξεως .02%:

$$m_A^2 \simeq c_{3/2}m_{3/2}^2 + c_s m_{3/2}^2\sin^2\theta + c_{2s}m_{3/2}^2\sin 2\theta - M_Z^2, \tag{7.4}$$

με συντελέστες $c_{3/2}\sim 0.1$, $c_s$, $c_{2s}\sim 1$, καθοριζόμενους από το αριθμητικό πρόγραμμα για κάθε τιμή των $\epsilon$, $M_S$, $\tan\beta$. Από τη σχέση αυτή υπολογίζεται το $m_{3/2}$ χρησιμοποιώντας ως ελεύθερες παραμετρους

**Πίνακας 7.1: Συσχετισμός Παραμέτρων** ($\mu < 0$)

| **Μεταβλητές Εισόδου** | | | | | | |
|---|---|---|---|---|---|---|
| | $\tan\beta = 15$ | | | $\tan\beta = 31$ | | |
| | $m_A = 450$ GeV, $\Delta_{NLSP} = 0.2$ | | | $m_A = 660$ GeV, $\Delta_{NLSP} = 0.06$ | | |
| **Τιμές Παραμέτρων** | | | | | | |
| $\epsilon$ | 0.5 | 0.8 | 0.9 | 0.3 | 0.6 | 0.9 |
| $\theta$ | $\pi/2.25$ | $\pi/3.33$ | $\pi/3.56$ | $\pi/2.75$ | $\pi/4.4$ | $\pi/4.95$ |
| $m_{3/2}$ (GeV) | 231.93 | 267.18 | 276.68 | 349.37 | 460.9 | 496.34 |
| **Τιμές Ασθενών Όρων** | | | | | | |
| $M_{1/2}$ (GeV) | 277.17 | 277.8 | 277.98 | 456.92 | 457.36 | 457.6 |
| $m_0$ (GeV) | 120.75 | 120.63 | 120.46 | 248.58 | 248.08 | 247.71 |
| $A_0$ (GeV) | −243.32 | −237.12 | −235.6 | −439.91 | −435.55 | −433.54 |

τις $\tan\beta$, $\epsilon$, $\theta$, $m_A$. Η επόμενη παράμετρος που πρέπει να μεταβάλλεται κατά βούληση είναι η σχετική διαφορά μάζας ανάμεσα στο LSP και στο NLSP, $\Delta_{NLSP}$ που ορίζεται στην Εξ. (5.11). Και αυτό, διότι η παράμετρος αυτή είναι κρίσιμη για την οικοδόμηση ενός επιτυχούς CDM σεναρίου αφού ρυθμίζει την ένταση των CAE.

Στο σημείο αυτό γίνεται μια παρατήρηση που αναδύεται από το αριθμητικό πρόγραμμα και βοηθάει στην περαιτέρω συστολή του παραμετρικού χώρου. Υπάρχουν ισοδύναμα σύνολα τιμών των παραμέτρων $\theta$, $\epsilon$, $m_{3/2}$, δηλαδή τιμών που παρέχουν πολύ παρόμοια σύνολα τιμών για τους όρους $m_0$, $M_{1/2}$, $A_0$, μέσω των Εξ. (7.1)-(7.3). Η πιο μεγάλη απόκλιση στα ισοδύναμα αυτά σύνολα τιμών παρατηρειται στο $A_0$ και είναι μικρότερη του 1%. Του λόγου το ασφαλές επιβεβαιώνεται με αναφορά στον Πίνακα 7.1, όπου δίνονται δύο τυχαία παραδείγματα για τις αναγραφόμενες τιμές εισαγόμενων στο πρόγραμμα παραμέτρων $\tan\beta$, $m_A$, $\Delta_{NLSP}$. Παρατηρείται ότι υπάρχουν τρεις διαφορετικές επιλογές $\theta$, $\epsilon$, $m_{3/2}$ σε κάθε περίπτωση που παρέχουν παραπλήσιες τιμές για τους όρους ασθενούς παραβίασης της SUSY, $m_0$, $M_{1/2}$, $A_0$ και επομένως, σχεδόν ταυτόσημα φάσματα των sparticles. Τα παραδείγματα αναφέρονται σε $\mu < 0$, επιλογή που δεν περιορίζει καθόλου τη γενικότητα της παρατήρησής μας. Όπως φαίνεται, το $m_{3/2}$ μεταβάλλεται εντόνως στις περιπτώσεις ισοδύναμου φάσματος, πράγμα που σημαίνει ότι η πρατήρηση αυτή ταξινόμησης των ισοδύναμων αρχικών συνθηκών δε θα γινόταν, αν δεν είχε γίνει η μετάθεση των παραμέτρων από το $m_{3/2}$ στο $m_A$, μέσω της αριθμητικής Εξ. (7.4).

Διαγραμματικά η παρατήρηση αυτή μπορεί να αναδειχθεί με την παρουσίαση καμπύλων σταθερού $\Delta_{NLSP}, m_A$ στο επίπεδο $\epsilon - \theta$ για επιλεγμένες τις υπόλοιπες παραμέτρους του προβλήματος. Οι καμπύλες αυτές μπορούν να ονομασθούν ισοφασματικές, γιατί πάνω σε αυτές, το φάσμα της θεωρίας παραμένει σταθερό, εκτός από το $m_{3/2}$. Δίνονται δύο χαρακτηριστικά παραδείγματα στα Σχ. 7.2, 7.3 διαλέγοντας $\mu < 0$ και $\tan\beta \simeq 10, m_A = 450\,\mathrm{GeV}$ για το πρώτο και $\tan\beta \simeq 36, m_A = 775\,\mathrm{GeV}$ για το δεύτερο. Παρατηρείται ότι:

- Υπάρχει περιοχή στο επίπεδο $\epsilon$, $\theta$ που αποκλείεται λόγω της απαίτησης θετικού ορισμού του $m_0$ μέσω της Εξ. (7.2). Αυτή εικονίζεται στο πάνω δεξί τμήμα των Σχ. 7.2, 7.3 και είναι ανεξάρτητη από τις επιμέρους επιλογές παραμέτρων.

- Με σταθερό $\epsilon$, αύξηση του $\theta$ συνεπάγεται μείωση του $\Delta_{NLSP}$

- Με σταθερό $\theta$, όταν $\theta >$ [<]$\pi/6$, αύξηση του $\epsilon$, συνεπάγεται μείωση [αύξηση]του $\Delta_{NLSP}$.

- Όταν $\theta < \pi/6$, με σταθερό $\epsilon$, το μέγιστο $\Delta_{NLSP}$ επιτυγχάνεται όταν $\theta \simeq \pi/9$. Επομένως, και λαμβάνοντας υπόψη το τρίτο σημείο των παρατηρήσεων που έγιναν παραπάνω, το μέγιστο δυνατό

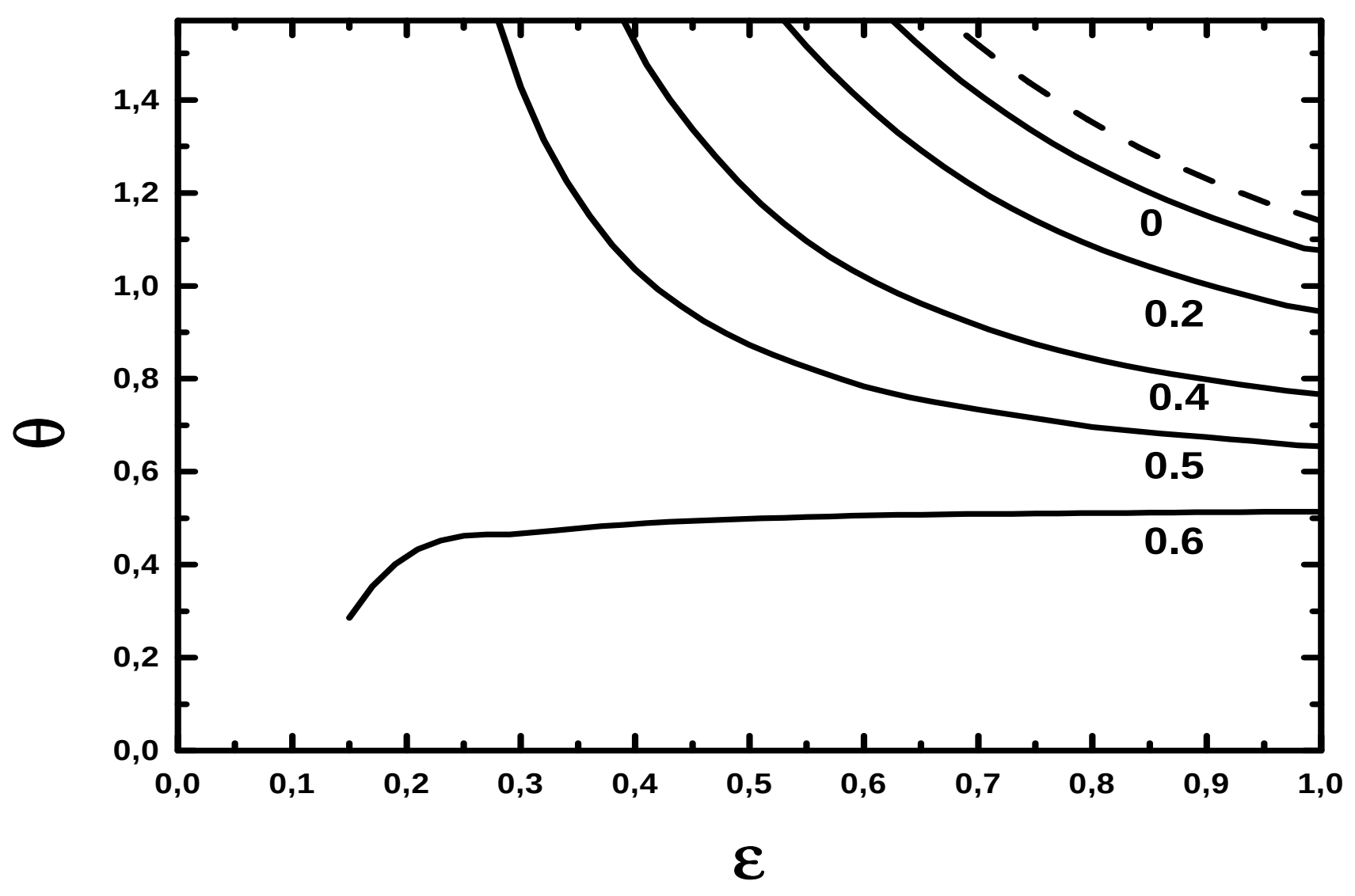

**Σχήμα 7.2:** *Καμπύλες σταθερού $\Delta_{NLSP}$ (0, 0.2, 0.4, 0.5, 0.6) στο επίπεδο $\epsilon - \theta$, για $m_A = 450\,\mathrm{GeV}, \tan\beta = 10$ και $\mu < 0$ ($M_S = 500\,\mathrm{GeV}$). Η περιοχή πάνω από τη διακεκομένη γραμμή αποκλείεται.*

$\Delta_{NLSP}$ επιτυγχάνεται για $\epsilon = .99$ και $\theta \simeq \pi/9$. Με τον όρο μέγιστο δυνατό $\Delta_{NLSP}$ νοείται το μέγιστο $\Delta_{NLSP}$ που μπορεί να επιτευχθεί με όλους τους δυνατούς συνδυασμούς $\epsilon$, $\theta$ χωρίς τον συνυπολογισμό κάποιων κοσμολογικών περιορισμών. Ο συνυπολογισμός αυτών των περιορισμών θα παπάσχει το μέγιστο επιτρεπτό $\Delta_{NLSP}$, $\Delta_{NLSP}^{max}$.

- Όταν $\theta < \pi/6$, υπάρχει ελάχιστο $\epsilon$ για κάθε επιθυμητό $\theta$. Αυτό βρίσκεται όταν $\theta \simeq \pi/9$. Γιαυτό το λόγο, οι καμπύλες αυτές έχουν το ένα άκρο τους στο $\theta \simeq \pi/9$ και δεν εξελίσσονται πιο κάτω.

- Συγκρίνοντας τα δύο διαγράμματα συμπεραίνεται ότι με αύξηση του $\tan\beta$ το μέγιστο δυνατό $\Delta_{NLSP}$ μειώνεται και συνεπώς υπάρχει $\tan\beta$ στο οποίο το μέγιστο δυνατό $\Delta_{NLSP}$ θα είναι μηδέν. Από το σημείο αυτό, $\tan\beta \simeq 46$, η θεωρία γίνεται αδιάφορη κοσμολογικά γιατί διαθέτει φορτισμένο LSP

- Ο συσχετισμός του $m_A$ με το μέγιστο δυνατό $\Delta_{NLSP}$ δεν έχει μελετηθεί σε βάθος. Για $\tan\beta \gtrsim 15$ και για τις φαινομενολογικά επιτρεπτές περιοχές παραμέτρων από τον περιοσμό του $\mathrm{BR}(b \to s\gamma)$, έχει παρατηρηθεί ότι με αύξηση του $m_A$ αυξάνεται το μέγιστο δυνατό $\Delta_{NLSP}$, πράγμα που αξιοποιείται στην κατασκευή του Σχ. 7.7 όπως εξηγείται στο Εδ. 7.6.1. Αντιθέτως, για $\tan\beta \lesssim 15$ παρατηρήθηκε αντίθετη συμπεριφορά, δηλαδή, αύξηση του $\Delta_{NLSP}$ με μείωση του $m_A$. Το γεγονός αυτό, αναδυκνύεται στην κατασκευή του Σχ. 7.4, όπως εξηγείται στο Εδ. 7.5.

- Με σταθερά $m_A$, $\Delta_{NLSP}$ και $\tan\beta$, το ελάχιστο $m_{3/2}$ λαμβάνεται με το μέγιστο $\theta$, όπως παρατήρειται και στο σχετικό Πίνακα 7.3

Επομένως, το τελικό σύνολο ανεξάρτητων παραμέτρων που καθορίζει πλήρως το πρόβλημα, είναι:

$$\tan\beta,\ \mathrm{sign}\mu,\ \Delta_{NLSP},\ m_A.$$

Η παρατήρηση της συστολής του παραμετρικού χώρου του εξεταζόμενου προτύπου είναι πολύ σημαντική και αδιαμφισβήτητη σύμφωνα με τα στοιχεία που παρατέθηκαν, αν και δεν έχει κατανοηθεί πλήρως. Δεν έχει βρεθεί, δηλαδή, η προέλευση της ιδιότητας αυτής που παρουσιάζουν οι αρχικές συνθήκες των SBT των Εξ. (7.1)-(7.3).

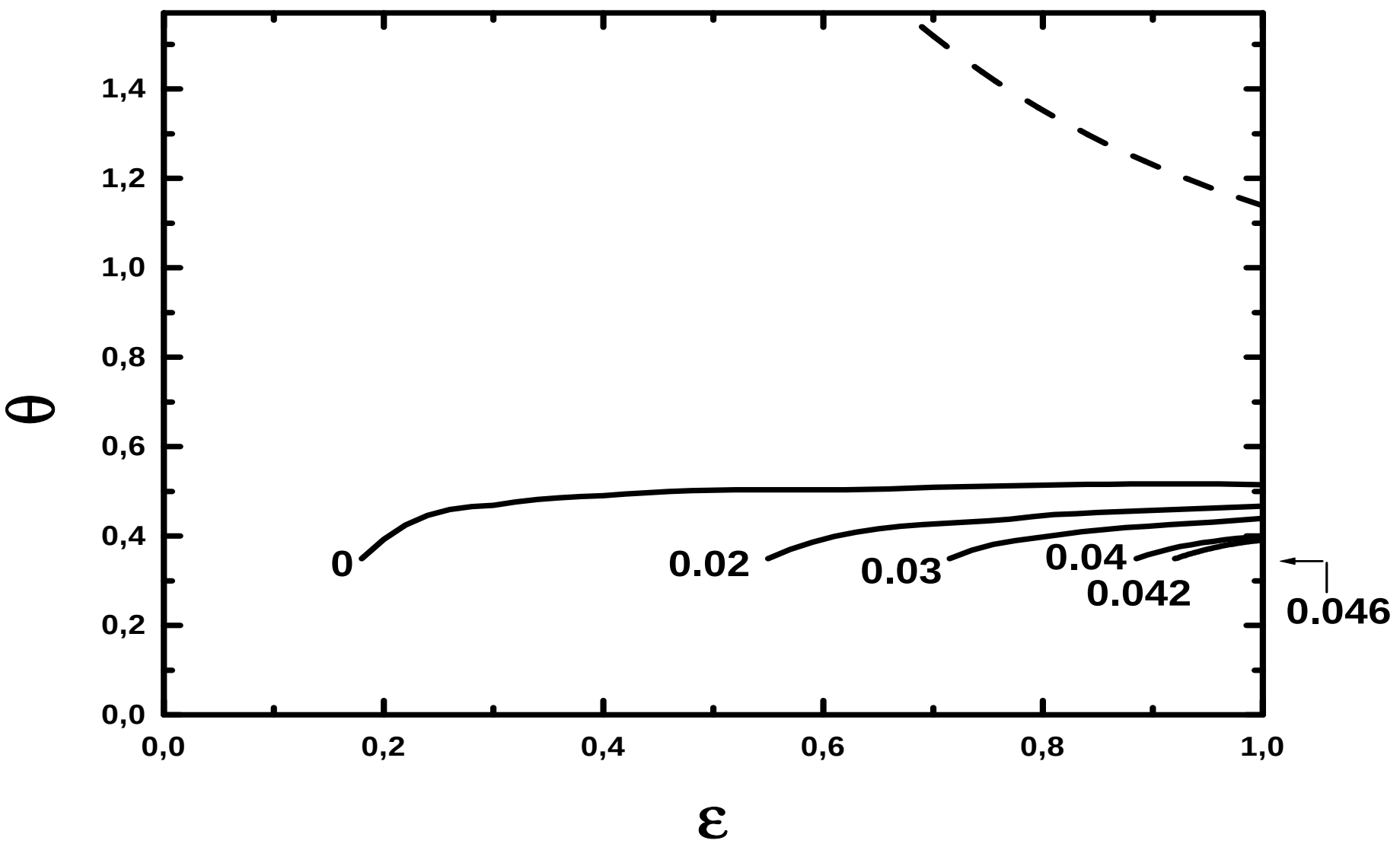

**Σχήμα 7.3:** *Καμπύλες σταθερού $\Delta_{NLSP}$ (0, .02, .03, .04, .042, .046) στο επίπεδο $\epsilon-\theta$, για $m_A =$ 775GeV, $\tan\beta = 37$ και $\mu < 0$ ($M_S = 900$GeV). Η περιοχή πάνω από τη διακεκομένη γραμμή αποκλείεται.*

## 7.5 Φαινομενολογική μελέτη

Η φαινομενολογική μελέτη του προτύπου ξεκινά με τον υπολογισμό του BR($b \to s\gamma$) ακολουθώντας το τυπολόγιο του Εδ. 3.4.1. Επιβεβαιώνεται και στην περίπτωση αυτή ότι για $\mu > 0 (< 0)$ το BR($b \to s\gamma$) είναι αυξημένο (ελλατωμένο) σε σχέση με αυτό που προκύπτει από τις συνεισφορές του SM και των φορτισμένων μποζονίων Higgs . Επειδή η συνεισφορά των chargino σβήνει καθώς το $m_A$ αυξάνει, συμπεραίνεται ότι μπορεί να βρεθεί τιμή του $m_A$ μετά από την οποία το BR($b \to s\gamma$) μπαίνει και παραμένει στην πειραματικά επιτρεπτή περιοχή της Εξ. (3.50). Ώστε, στο μεγαλύτερο μέρος του χρησιμοποιούμενου παραμετρικού χώρου $3 \lesssim \tan\beta$, το κατώτερο όριο στο $m_A$ μπορεί να προσδιοριστεί από την απαίτηση BR($b \to s\gamma$) $\simeq 4.5 \times 10^{-4}$ (ή $\simeq 2 \times 10^{-4}$) για $\mu > 0$ (ή $\mu < 0$).

Η λογική αυτή, εξηγείται καλύτερα με αναφορά στο Σχ. 7.4. Σε αυτό εκτίθενται ως συνάρτηση του $m_{LSP}$ οι συνεισφορές στο BR($b \to s\gamma$) από το MSSM για για $\tan\beta \simeq 10$ και για $\Delta_{NLSP} = 0.065$, όταν $\mu > 0$ ενώ για $\Delta_{NLSP} = 0.6$ όταν $\mu < 0$. Οι επιλογές για το $\Delta_{NLSP}$ προέρχονται από τους κοσμολόγικούς περιορισμούς όπως θα φανεί στο Εδ. 7.6.1. Βεβαίως η τιμή του BR($b \to s\gamma$) ισχυρά εξαρτάται από την τιμή του $m_A$ και πολύ ασθενέστερα από τη τιμή του $\Delta_{NLSP}$. Περιλαμβάνονται οι συνεισφορές από το SM, και από το SM συν το φορτισμένο higgs (SM+$H^+$). Η γραμμή η αναφερόμενη στη συνεισφορά από το MSSM για $\Delta_{NLSP} = 0.6$, $\mu < 0$ σταματά την εξέλιξη της γύρω στο $m_{LSP} \simeq 212$ GeV γιατί δεν είναι δυνατή η επίτευξη του (κοσμολογικά) απαιτούμενου $\Delta_{NLSP} = 0.6$ με μεγαλύτερες τιμές του $m_{LSP}$. Αυτό είναι σύμφωνο με τις παρατηρήσεις του Εδ. 7.4.

Σε περιοχές του παραμετρικού χώρου, όπου η $\tan\beta$ παίρνει τιμές μεγάλες ή ενδιάμεσες, ο δεύτερος φαινομενολογικός περιορισμός που πρέπει να ληφθεί υπόψη, είναι οι SUSY διορθώσεις στη μάζα του $b$-quark, που λαμβάνονται από το τυπολόγιο του Εδ. 3.3.1. Το αποτέλεσμα εξαρτάται ισχυρά από την επιλεγμένη τιμή του $\tan\beta$, ελαφρότατα μειώνεται με την αύξηση του $m_A$ και έχει σημείο αντίθετο του χρησιμοποιούμενου για το $\mu$.

Τέλος, για μικρές τιμές του $\tan\beta$ το κατώτερο όριο στο $m_A$ βρίσκεται προνοώντας να ισχύουν τα τρέχοντα πειραματικά όρια για τις μάζες των neutralino (50 GeV), chargino (100 GeV) και ελαφρότατου higgs της Εξ. (3.10). Συνήθως ο τελευταίος περιορισμός καλύπτει τους δύο προηγούμενους.

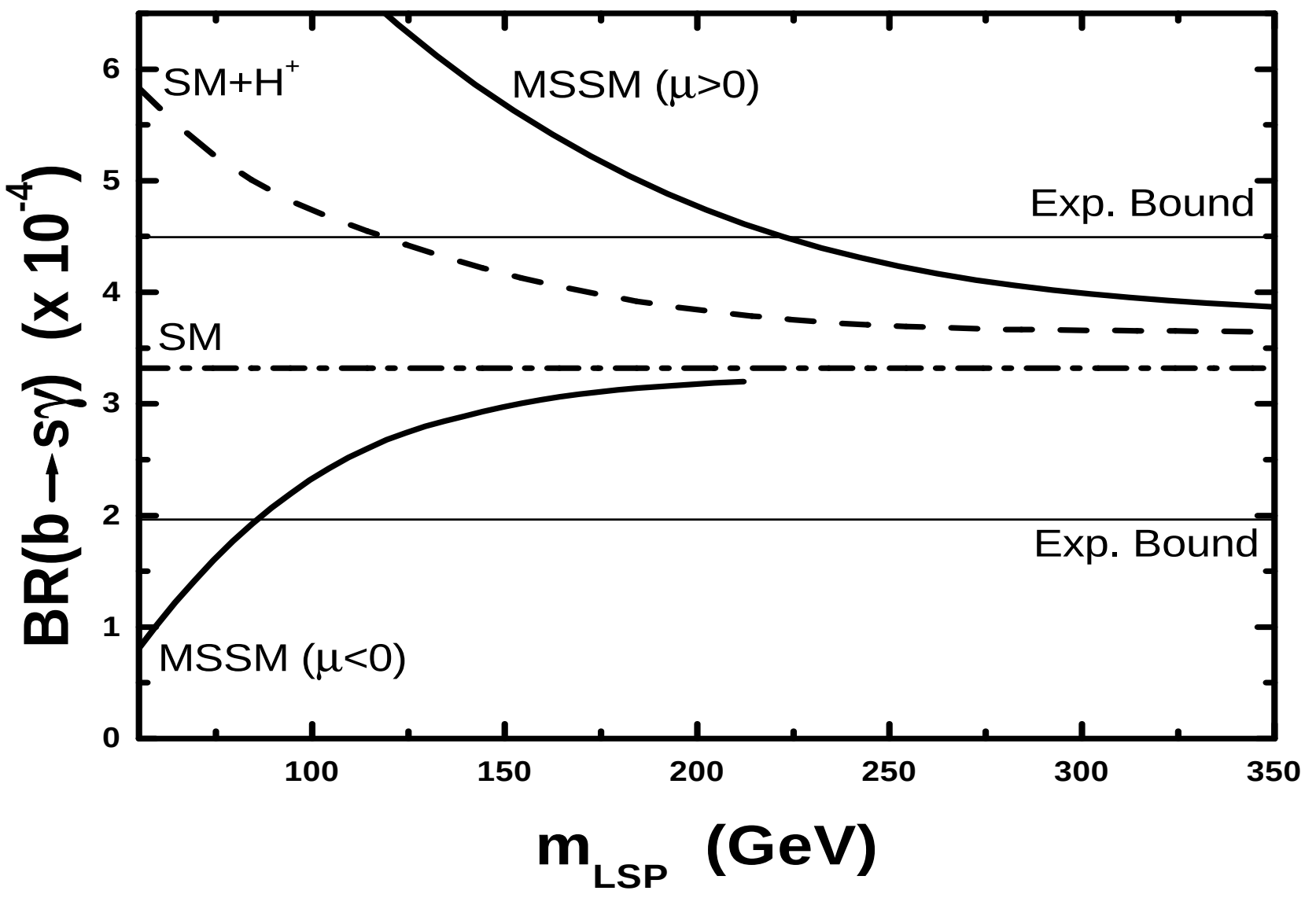

**Σχήμα 7.4:** *Το* BR$(b \to s\gamma)$ *ως συνάρτηση του* $m_{LSP}$ *για* $\tan\beta \simeq 10, \epsilon = 0.65$ *και* $\Delta_{NLSP} = 0.065, \mu > 0$ *και* $\Delta_{NLSP} = 0.6$, $\mu < 0$. *Περιλαμβάνονται οι συνεισφορές από το SM, και από το SM συν το φορτισμένο higgs (SM+*$H^+$*) καθώς και τα πειραματικά όρια.*

## 7.6 Κοσμολογικά αποτελέσματα

Μέτα την φαινομενολογική έπεται η κοσμολογική θεώρηση του προτύπου που φυσικά, περιλαμβάνει τον υπολογισμό της $\Omega_{LSP}h^2$. Ο φορμαλισμός του κεφαλαίου 5 και ειδικότερα της της Εδ. 5.5 εφαρμόζεται για το σκοπό αυτό. Είναι εφαρμόσιμος γιατί, λόγω των παγκόσμιων αρχικών συνθηκών, το LSP προκύπτει να είναι ισχυρά μορφής Bino στο μεγαλύτερο μέρος του παραμετρικού χώρου. Σε περιοχές με μικρό $\tan\beta$, η Bino καθαρότητα του LSP ελαττώνεται από 98% σε 94%. Σύνεπως, σφάλμα τινά υπεισέρχεται στον υπολογισμό της ενεργού διατομής της αλληλοκαταστροφής, όπου το LSP έχει θεωρηθεί καθαρό Bino . Το NLSP προκύπτει να είναι stau μορφής δεξιόστροφης σχεδόν για όλες τις τιμές του $\tan\beta$.

Τα χαρακτηριστικά της $\Omega_{LSP}h^2$ που παρατηρήθηκαν στο Εδ. 6.6.1 ισχύουν και στην προκείμενη περίπτωση. Επιπρόσθετα, το $\Omega_{LSP}h^2$ παραμένει σταθερό στις ισοφασματικές καμπύλες. Απαιτώντας το $\Omega_{LSP}h^2$ να βρίσκεται μέσα στην κοσμολογικά επιτρεπτή περιοχή της Εξ. (5.4), μπορεί να καθοριστεί το δυνατό εύρος του $\Delta_{NLSP}$. Αυτός θα είναι ο τελικός προορισμός του υπολογισμού αυτής της ενότητας.

Σύμφωνα με τα αναφερόμενα στην εισαγωγή, το εξεταζόμενο πρότυπο είναι επιδεκτικό όλων των δυνατών σεναρίων σχετικά με τη ασυμπτωτική συμπεριφορά των ζεύξεων Yukawa. Συνεπώς, ευκρινέστερη παρουσίαση των αποτελέσματων μπορεί να γίνει, διακρίνοντας τις παρακάτω περιπτώσεις:

### 7.6.1 Χωρίς ενοποίηση Yukawa

Σε αυτή την περίπτωση, ο καθορισμός των ασυμπτωτικών τιμών των ζεύξεων Yukawa γίνεται όπως περιγράφεται στο Εδ. 6.8.2. Επειδή διατρέχεται όλος ο δυνατός χώρος τιμών της $\tan\beta$ οι CAE Bino-stau-selectron επιβάλλεται να συμπεριληφθουν στον υπολογισμό μας, οπότε το τυπολόγιο του Εδ. 5.5.3 αξιοποιείται.

Καθώς δεν υπάρχουν περιορισμοί σχετικά με το sign$\mu$, εξετάζονται και οι δύο δυνατές επιλογές. Η ουσιαστική διαφορά ανάμεσα στις δύο, είναι ότι όταν $\mu > 0$, μεγάλες τιμές για το $m_A$ (και κατά συνέπεια για το $m_{LSP}$) απαιτούνται για την είσοδο στην πειραματικά επιτρεπτή περιοχή του BR$(b \to s\gamma)$, από το πάνω ¨κατώφλι¨, οπότε υψηλός εκφυλισμός είναι αναγκαίος ($\Delta_{NLSP} \simeq 0$) για να ικανοποιηθούν τα

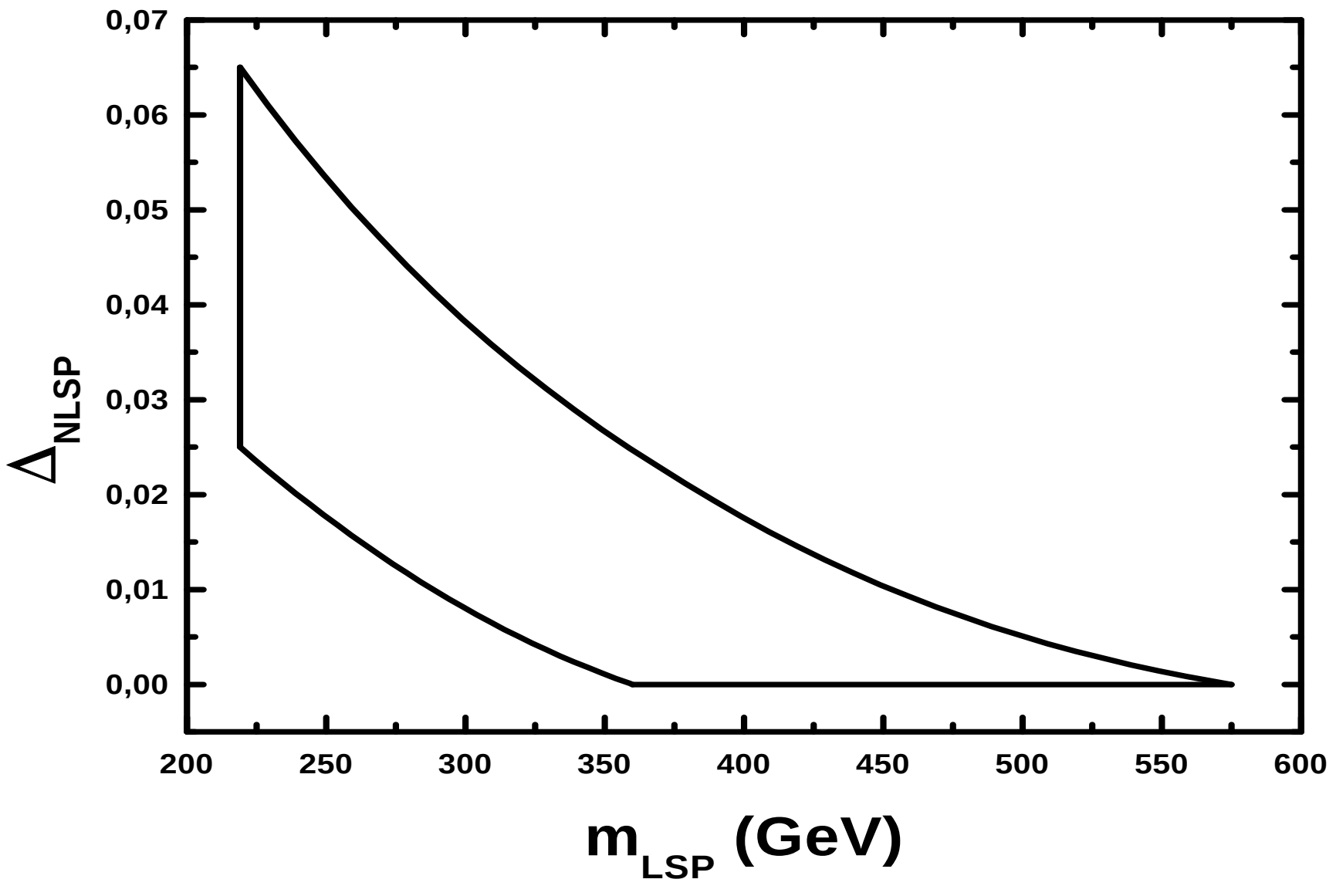

**Σχήμα 7.5:** *Η επιτρεπτή περιοχή στο επίπεδο* $m_{LSP} - \Delta_{NLSP}$, *για* $\tan\beta = 10,\ \epsilon = 0.65$ *και* $\mu > 0$.

κοσμολογικά όρια για την $\Omega_{LSP}h^2$. Αντιθέτως, όταν $\mu < 0$ απαιτούνται μικρότερες για το $m_A$ (και κατά συνέπεια για το $m_{LSP}$) για είσοδο στην πειραματικά επιτρεπτεί περιοχή από το κατώ ¨κατώφλι¨, οπότε όχι μόνο ο αναγκαίος εκφυλισμός είναι ασθενέστερός ($\Delta_{NLSP} \gg 0$) αλλά ακόμα, υπάρχουν περιοχές, όπου επιτυγχάνονται τιμές για την $\Omega_{LSP}h^2$ εντός των αποδεκτών κοσμολογικών ορίων, χώρις καθόλου χρήση του μηχανισμού συγγενικής καταστροφής. Οπωσδήποτε τέτοιες περιοχές είναι αρκέτα περιορισμένες, για $\tan\beta \lesssim 15$ και φυσικά ακολουθούνται από περιοχές στις οποίες οι CAE είναι προεξάρχουσες.

Και στις δύο περιπτώσεις, ο τομέας higgs του προτύπου διαθέτει μάζες βαρύτερες των LSPκαι NLSP ($m_A \geq 660$ GeV για $\mu > 0$ και $m_A \geq 340$ GeV για $\mu < 0$). Επομένως διαδικάσιες με τελικές καταστάσεις $\tau H, \tau A, hH, HH, H^+H^-, AA$ είναι κινηματικά αποκλεισμένες.

Η κατάσταση αυτή περιγράφεται στα Σχ. 7.5 και 7.6. Στα σχήματα αυτά απεικονίζεται η επιτρεπτή περιοχή στο επίπεδο $m_{LSP} - \Delta_{NLSP}$ για $\tan\beta \simeq 10$ και $\mu > 0$ ή $\mu < 0$ αντίστοιχα. Χάριν καλύτερου καθορισμού του προτύπου, και στα δύο διαγράμματα έχει επιλεγεί $\epsilon = .65$ οπότε το $\Delta_{NLSP}$ ρυθμίζεται μόνο μέσω της παραμέτρου $\theta$. Στά διαγράμματα αυτού του τύπου, το $m_{LSP}$ μπορεί να μεταβάλλεται ανάμεσα σε δύο όρια. Το κατώτερο όριο στο $m_{LSP}$ αντίστοιχεί σε $\mathrm{BR}(b \to s\gamma) \simeq 4.5 \times 10^{-4}$ γία $\mu > 0$ ή σε $\mathrm{BR}(b \to s\gamma) \simeq 2 \times 10^{-4}$ για $\mu < 0$. Γιαυτή την τιμή του $m_{LSP}$ επιτυγχάνεται η μέγιστη επιτρεπτή διαφορά μάζας, $\Delta^{max}_{NLSP}$ η οποία αντιστοιχεί στην ανώτερη αριστερή γωνία του διαγράμματος της επιτρεπτής περιοχής. Σε αυτό το σημείο λαμβάνεται $\Omega_{LSP}h^2 \simeq .22$ για το ελάχιστο $m_{LSP}$. Το κατώτερο (ανώτερο) σύνορο της επιτρεπτής περιοχής αντιστοιχεί σε $\Omega_{LSP}h^2 \simeq .09(.22)$. Το ανώτερο όριο στη $m_{LSP}$ λαμβάνεται στην κατώτερη δεξιά γωνία της επιτρεπτής περιοχής. Βρίσκεται ότι το ανώτερο όριο στη μάζα του LSP είναι περίπου $m_{LSP} \lesssim 600$GeV, για $\mu < 0$ και $\mu > 0$.

Ειδικότερα, στο Σχ. 7.5, για $\mu > 0$, παρατηρείται ότι 220 GeV $\lesssim m_{LSP} \lesssim$ 630 GeV. Το κάτω όριο αντιστοιχεί στην τομή του κλάδου $\mu > 0$ του $\mathrm{BR}(b \to s\gamma)$ με το άνω πειραματικό όριο στο Σχ. 7.4. Επομένως, το LSP σε αυτή την περίπτωση είναι σχετικά βαρύ και $\Delta^{max}_{NLSP} \simeq .065$ σχετικά μικρό. Φυσικά, $\Delta^{min}_{NLSP} = 0$. Οι CAE είναι σημάντικές σε όλη την περιοχή του διαγραμμάτος αφού $\Delta_{NLSP} \lesssim .25$. Αντίθετα, στο Σχ. 7.6, για $\mu < 0$, ελαφρότερες μάζες για το LSP είναι δυνατό να επιτευχθούν καθώς 85 GeV $\lesssim m_{LSP} \lesssim$ 625 GeV. Το κάτω όριο αντιστοιχεί στην τομή του κλάδου $\mu < 0$ του $\mathrm{BR}(b \to s\gamma)$ με το κάτω πειραματικό όριο στο Σχ. 7.4. Στο σημείο αυτό βρίσκεται ότι $\Delta^{max}_{NLSP} \simeq .6$ τιμή πολύ υψηλότερη από την προηγούμενη και συνεπώς, υπάρχει περιοχή 85GeV $\lesssim m_{LSP} \lesssim$ 120GeV, όπου οι CAE επιτρέπεται

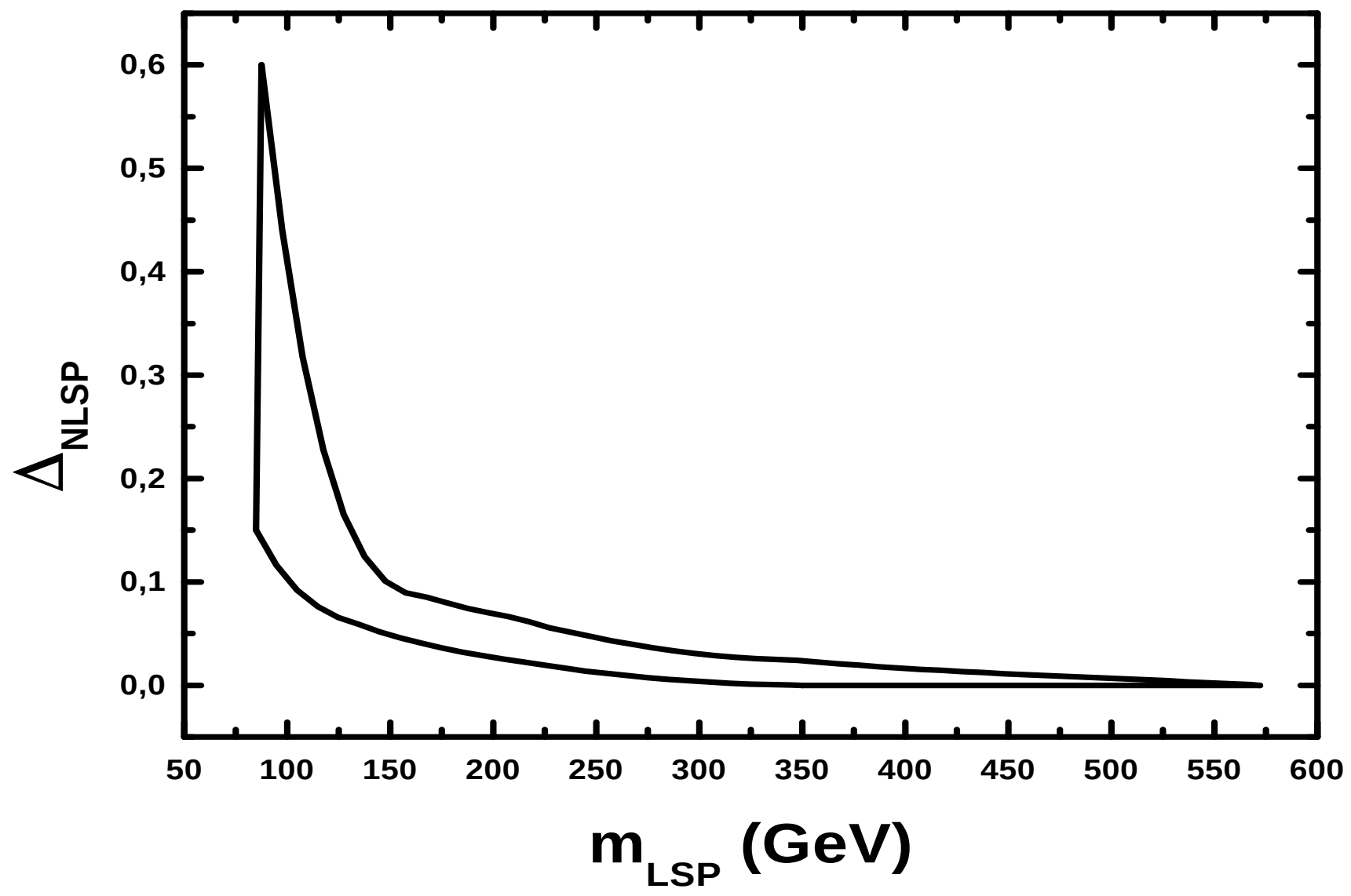

**Σχήμα 7.6:** *Η επιτρεπτή περιοχή στο επίπεδο* $m_{LSP} - \Delta_{NLSP}$, *για* $\tan\beta = 10$, $\epsilon = 0.65$ *και* $\mu < 0$.

να είναι αμεληταίες, αφού εκεί είναι δυνατό να επιζεί $\Delta_{NLSP} \geq .25$. Παρόλα αυτά, ο μηχανίσμος των CAE είναι παρών σε ευρύτερη περιοχή μαζών του LSP , $120\,\mathrm{GeV} \lesssim m_{LSP} \lesssim 625\,\mathrm{GeV}$. Πάλι $\Delta^{min}_{NLSP} = 0$.

Υπάρχουν περιοχές του παραμετρικού χώρου, όπου το $\Delta^{max}_{NLSP}$ δεν αντιστοιχεί στην ελάχιστη τιμή του $m_{LSP}$. Αυτή η κατάσταση απεικονίζεται στο Σχ. 7.7. Στο σχήμα αυτό δίνεται η επιτρεπτή περιοχή στο επίπεδο $m_{LSP} - \Delta_{NLSP}$ για $\tan\beta \simeq 36$ και $\mu < 0$. Χάριν καλύτερου καθόρισμού, επιλέγεται $\epsilon = .99$ και το $\Delta_{NLSP}$ ρυθμίζεται μέσω της παραμέτρου $\theta$. Το $m_{LSP}$ μπορεί να μεταβάλλεται από 220 GeV μέχρι 614 GeV. Το κατώτερο όριο στό $m_{LSP}$, το οποίο αντιστοιχεί σε $\mathrm{BR}(b \to s\gamma) \simeq 2 \times 10^{-4}$, είναι περίπου 220 GeV. Παρατηρείται ότι για αυτή την τιμή του $m_{LSP}$ η μέγιστη διαφορά μάζας (.038), η οποία μπορεί να επιτευχθεί με όλες τις δυνατές τιμές της $\theta$ δεν επιτρέπει να ληφθεί $\Omega_{LSP}h^2 \simeq .22$ αλλά $\Omega_{LSP}h^2 \simeq .108$. Αυτή η μέγιστη διαφορά μάζας αντιστοιχεί στο σημείο τομής της καθέτου και της κεκλιμένης γραμμής στην αριστερή πλευρά του διαγράμματος στο Σχ. 7.7. Αυξάνοντας το $m_{LSP}$ το μέγιστο επιτυγχόμενο $\Delta_{NLSP}$ αυξάνει (καθώς φαίνεται στην κεκλιμένη αριστερή πλευρά του διαγράμματος στο Σχ. 7.7 στην περιοχή $.038 \lesssim \Delta_{NLSP} \lesssim .056$ ). Μπορεί, λοιπόν να βρεθεί τιμή $m_{LSP} = 245.1\,\mathrm{GeV}$ με $\Delta_{NLSP} \simeq .056$ τέτοια, ώστε $\Omega_{LSP}h^2 \simeq 0.22$. Αυτή λογίζεται πλέον ως η μέγιστη επιτρεπτή σχετική διαφορά μάζας. Στο Σχ. 7.3 εικονίζεται ένα παράδειγμα επίτευξης του μέγιστου δυνατού $\Delta_{NLSP}$ για καθορισμένα $m_A$ ή $m_{LSP}$.

Τα παραπάνω αποτελέματα μπορούν να γενικευτούν βρίσκοντας το $\Delta^{max}_{NLSP}$ και το $\Delta^{min}_{NLSP}$, για όλες τις δυνατές τιμές της $\tan\beta$. Η γενίκευση αυτή εκτίθεται στο Σχ. 7.8. Σε αυτό το διάγραμμα φαίνεται η επιτρεπτή περιοχή στο επίπεδο $\Delta_{NLSP} - \tan\beta$ για $\mu > 0$ (από την συνεχή μέχρι την διακεκομένη γραμμή) και $\mu < 0$ (από την συνεχή μέχρι την εστιγμένη-διακεκομένη γραμμή). Ουσιαστικά, καταγράφεται η μετακίνηση της πάνω αριστερής κορυφής των διαγραμμάτων στα Σχ. 7.5, 7.6 και 7.7 μεταβάλλοντας την $\tan\beta$ και απελευθερώνοντας την παράμετρο $\epsilon$. Είναι δυνατό να γίνει αυτό, γιατί όπως ήδη έχει αναφερθεί στο Εδ. 7.4, οι επιπτώσεις της θεωρίας σε χαμηλές ενέργειες ελέγχονται μόνο από τις παραμέτρους $\Delta_{NLSP}, m_A$ και όχι τις $\theta$, $\epsilon$, $m_{3/2}$ για σταθερό $\tan\beta$ και $\mathrm{sign}\mu$.

Επαληθεύεται αριθμητικά ότι πάντα μπορεί να βρεθεί συνδυασμός παραμέτρων $\theta$, $\epsilon$ ο οποίος εξασφαλίζει ότι $\Delta^{min}_{NLSP} = 0$ για κάθε τιμή της $\tan\beta$ στο διάστημα $1.5 \lesssim \tan\beta \lesssim 40$ και για βαρύ LSP (περίπου 600 GeV) που επιβεβαιώνει ότι $\Omega_{LSP}h^2 \simeq 0.22$. Επομένως, επιτρέπεται $\Delta^{min}_{NLSP} \simeq 0$ για όλες τις χρησιμοποιούμενες τιμές της $\tan\beta$. Αυτό, βεβαίως, δεν αποκλείει την περίπτωση ύπαρξης συνδυασμού τιμών $\theta$, $\epsilon$ με τον οποίο δεν μπορεί να επιτευχθεί $\Delta^{min}_{NLSP} \simeq 0$ ούτε για μεγάλες τιμές του $m_{LSP}$.

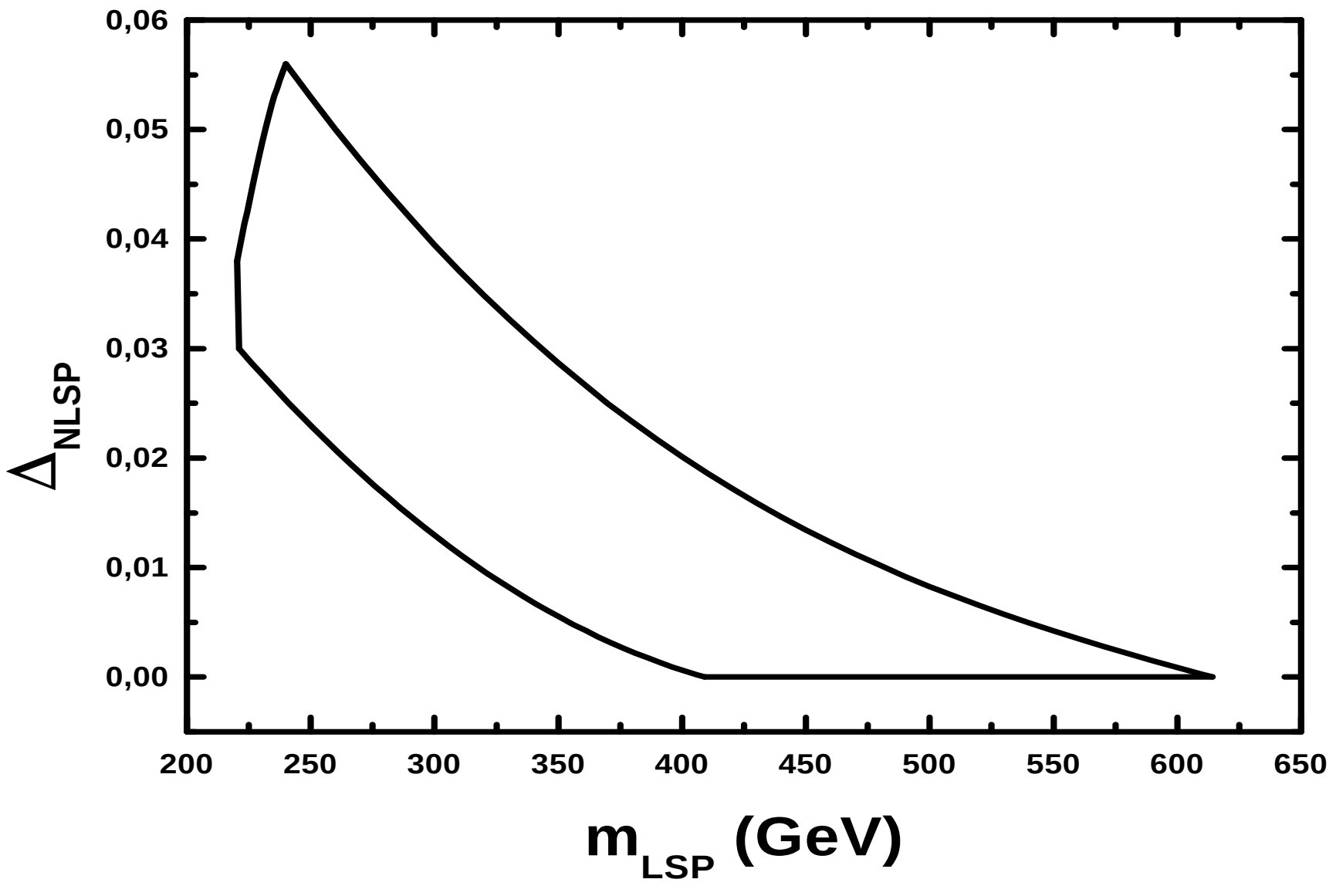

**Σχήμα 7.7:** *Η επιτρεπτή περιοχή στο επίπεδο $m_{LSP} - \Delta_{NLSP}$, για $\tan\beta = 36$, $\epsilon = 0.99$ και $\mu < 0$.*

Σχετικά με το $\Delta_{NLSP}^{max}$ διακρίνονται οι περιπτώσεις:

**α.** Όταν $\mu > 0$ και $3.5 \lesssim \tan\beta \lesssim 42$ ή $\mu < 0$ και $9 \lesssim \tan\beta \lesssim 35.5$, το κατώτερο όριο στο $m_{LSP}$ βρίσκεται από τα όρια της Εξ. (3.50) για το $\mathrm{BR}(b \to s\gamma)$ και το $\Delta_{NLSP}^{max}$ αντιστοιχεί σε αυτή την τιμή του $m_{LSP}$. Διαγράμματα του τύπου των Σχ. 7.5, 7.6 είναι κυρίαρχα. Για $\mu > 0$, καθώς οι τιμές της $\tan\beta$ αυξάνονται, η κάτω συνοριακή καμπύλη της επιτρεπτής περιοχής του Σχ. 7.5 εξαφανίζεται και η σχεδόν κατακόρυφη ευθεία μετακινείται προς την δεξιά γωνία και εκφυλίζεται σε σημείο όταν $\tan\beta \simeq 42$, οπότε λαμβάνεται $\Omega_{LSP}h^2 \simeq .22$ με $m_{LSP} \simeq 670\,\mathrm{GeV}$ και $\Delta_{NLSP} \simeq 0$.

**β.** Όταν $\mu > 0$ και $1.5 \lesssim \tan\beta \lesssim 3.5$ ή $\mu < 0$ και $1.5 \lesssim \tan\beta \lesssim 3$, το κατώτερο όριο στο $m_{LSP}$ βρίσκεται από τα όρια της Εξ. (3.10) για τη μάζα του $h$. Σε αυτές τις περιοχές η $m_h$ γίνεται πολύ μικρή ($< 90\,\mathrm{GeV}$) για τιμές του $m_A$ όχι πάρα πολύ μεγάλες. Επομένως, μεγαλύτερα $m_A$ ή $m_{LSP}$ απαιτούνται για να ανυψωθεί η $m_h$ στα όρια της Εξ. (3.10). Οι επιτρεπτές περιοχές είναι του τύπου των Σχ. 7.5, 7.6 και οι επιτρεπτές τιμές του $\Delta_{NLSP}$ ελαττώνονται ταχέως. Περισσότερα στοιχεία δεν δίνονται γιατί τέτοιες περιοχές θεωρούνται μη ενδιαφέρουσες φαινομενολογικά.

**γ.** Όταν $\mu < 0$ και $3 \lesssim \tan\beta \lesssim 9$ ή $35.5 \lesssim \tan\beta \lesssim 41$, το κατώτερο όριο στο $m_{LSP}$ βρίσκεται από το πειραματικό όριο στο $\mathrm{BR}(b \to s\gamma)$ της Εξ. (3.50), αλλά το $\Delta_{NLSP}^{max}$ δεν αντιστοιχεί στην ελάχιστη τιμή του $m_{LSP}$. Σε αυτές τις περιοχές διαγράμματα του τύπου του Σχ. 7.7 κυριαρχούν. Καθώς η τιμή της $\tan\beta$ αυξάνει έτι περισσότερο, το κατακόρυφο τμήμα στο αριστερό τμήμα του Σχ. 7.7 εκλείπει και το κεκλιμένο τμήμα κινείται από αριστερά πρός τα δεξιά με, και στη συνέχεια, χωρίς την παρουσία της κάτω συνοριακής καμπύλης. Όταν $\tan\beta \simeq 41$, η επιτρεπτή περιοχή του Σχ. 7.7 έχει συρρικνωθεί στην δεξιά γωνία, όπου λαμβάνεται $\Omega_{LSP}h^2 \simeq .22$ με $m_{LSP} \simeq 640\,\mathrm{GeV}$ και $\Delta_{NLSP} \simeq 0$.

Συμπερασματικά, η επιτρεπόμενη περιοχή κίνησης του LSP είναι $144\ \mathrm{GeV} \lesssim m_{LSP} \lesssim 670\ \mathrm{GeV}$ για $\mu > 0$ με $\Delta_{NLSP}^{max} \simeq 0.166$ που επιτυγχάνεται σε $\tan\beta \simeq 3.4$ και $72\,\mathrm{GeV} \lesssim m_{LSP} \lesssim 640\,\mathrm{GeV}$ για $\mu < 0$ με $\Delta_{NLSP}^{max} \simeq 0.93$ που επιτυγχάνεται σε $\tan\beta \simeq 4.5$. Προφανώς, το σενάριο με $\mu < 0$ είναι πιο προνομιακό από αυτό με $\mu > 0$ σε ό,τι αφορά τις χαμηλές μάζες του LSP.

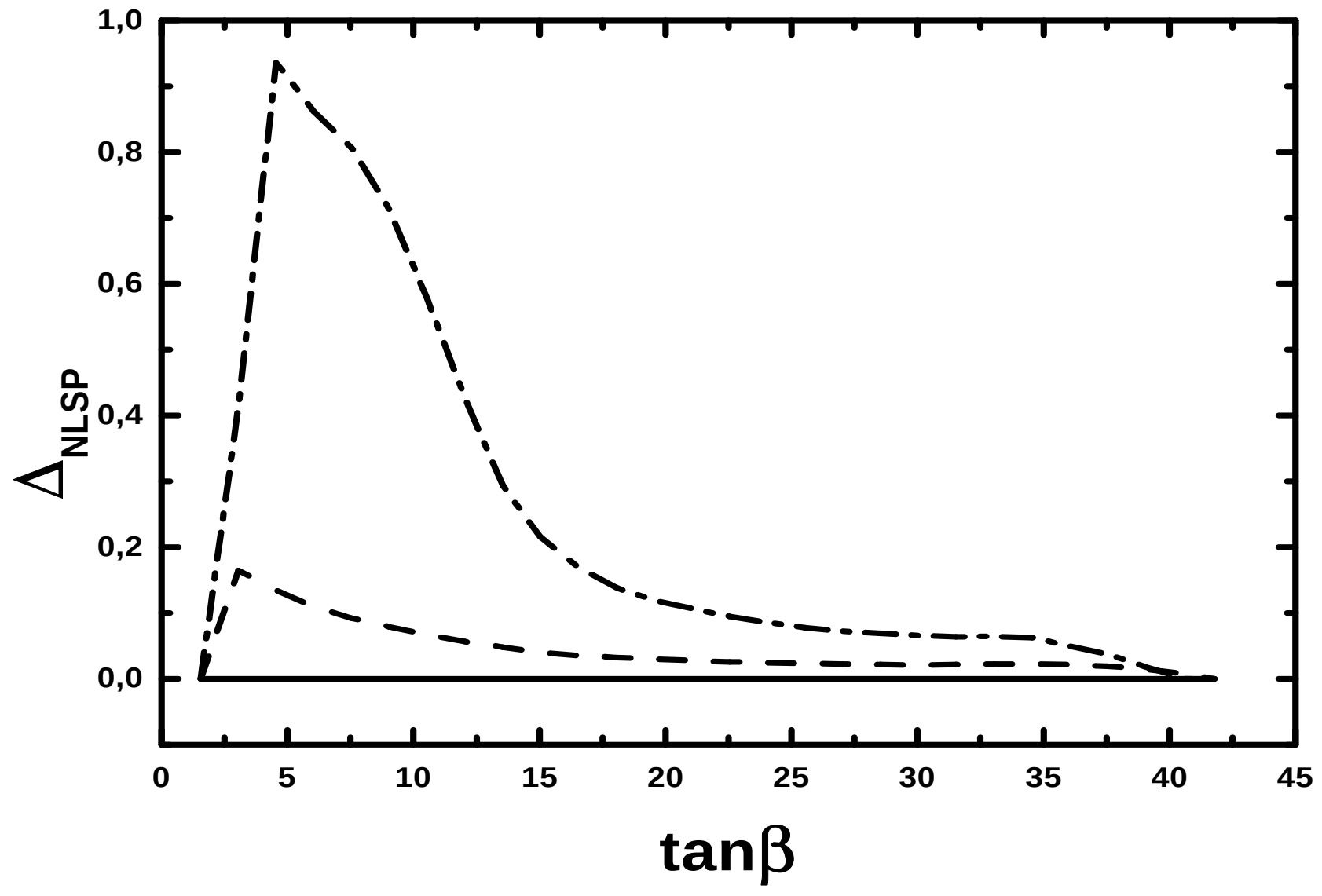

**Σχήμα 7.8:** *Η επιτρεπτή περιοχή στο επίπεδο* $\tan\beta - \Delta_{NLSP}$ *για* $\mu > 0$ *(από τη συνεχή μέχρι τη διακεκομένη γραμμή) και* $\mu < 0$ *(από τη συνεχή μέχρι την εστιγμένα διακεκομένη γραμμή).*

### 7.6.2 Με ενοποίηση $b - \tau$

Σε αυτή την περίπτωση, ο καθορισμός των ασυμπτωτικών τιμών των ζεύξεων Yukawa γίνεται όπως περιγράφεται στο Εδ. 6.8.1. Σύμφωνα με το Σχ. 6.13, η δενδρικού επιπέδου μάζα του $b$-quark προκύπτει μεγαλύτερη από την πειραματικά προβλεπόμενη τιμή της Εξ. (3.18). Επομένως, το σημείο της SUSY διόρθωσης θα πρέπει να είναι αρνητικό, ώστε να μειώνεται η προκύπτουσα μάζα του $b$-quark. Και αφού το σημείο της SUSY διόρθωσης είναι αντίθετο από το sign$\mu$, μόνο η επιλογή $\mu > 0$ είναι βιώσιμη σε αυτό το σενάριο. Διάγραμμα ανάλογο του Σχ. 6.14 μπορεί να κατασκευαστεί που δείχνεται στην Αν. [45], το οποίο οριοθετεί την επιτρεπτή περιοχή των τιμών της $\tan\beta$ ως εξής:

$$22 \lesssim \tan\beta \lesssim 45 \;\; \text{και} \;\; \mu > 0 \tag{7.5}$$

Ο περιορισμός που εδράζεται στη μάζα του $b$-quark μας δίνει επιχείρημα για το κατώτερο όριο της $\tan\beta$. Το ανώτερο όριο τιθέται λόγω του γεγονότος ότι με όλους τους δυνατούς συνδυασμούς παραμέτρων $\theta$, $\epsilon$ δεν μπορεί να επιτευχθεί $\Delta_{NLSP} \simeq 0$ και $\mathrm{BR}(b \to s\gamma) \simeq 4.5 \times 10^{-4}$ για $\tan\beta > 45$. Η περιοχή $1.6 \lesssim \tan\beta \lesssim 22$ αποκλείεται, επειδή ακόμα και η διορθωμένη μάζα του $b$-quark είναι μεγαλύτερη από το ανώτερο πειραματικό όριο. Η περιοχή με $\tan\beta \simeq 1.5$ είναι φαινομενολογικά μη ενδιαφέρουσα λόγω των ελαφρών μαζών για το $m_h$ και συνεπώς, δεν μελετάται περαιτερω.

Όπως εξηγήθηκε στο Εδ. 7.6.1, η επιλογή $\mu > 0$, αναγκαστικά συνοδεύεται με σχετικά βαρύ LSP, πράγμα που σημαίνει ότι η συνδρομή των CAE θα είναι κρίσιμη για την επίτευξη CRD σε αποδεκτά επίπεδα. Λόγω των υψηλών τιμών της $\tan\beta (> 15)$ που χρησιμοποιούνται, ο φορμαλισμός των Εδ. 5.5.2 είναι κατάλληλος για τον υπολογισμό των CAE. Τα αποτελέσματα της διερεύνησης ανάλογα με τις τιμές που παίρνει η $\tan\beta$ ταξινομούνται παρακάτω:

**α.** Όταν $38 \lesssim \tan\beta \lesssim 45$, η μέγιστη επιτρεπόμενη μάζα του LSP είναι ανυψωμένη $m_{LSP} \lesssim 790\,\mathrm{GeV}$. Αυτό γίνεται κατανοητό από την ακόλουθη παρατήρηση. Στην θεωρούμενη περιοχή τιμών της $\tan\beta$, οι διαδικασίες με τελικές καταστάσεις $\tau H$, $\tau A$ είναι κινηματικά επιτρεπτές, γιατί $m_{LSP} > m_A/2, m_H/2$. Επομένως, η σμίκρυνση της CRD, η οφειλόμενη στο μηχανισμό των CAE, είναι ενισχυμένη και συνεπώς, βαρύτερα LSP επιτρέπονται. Το φωτογραφικό υλικό που υπάρχει στην περίπτωση αυτή

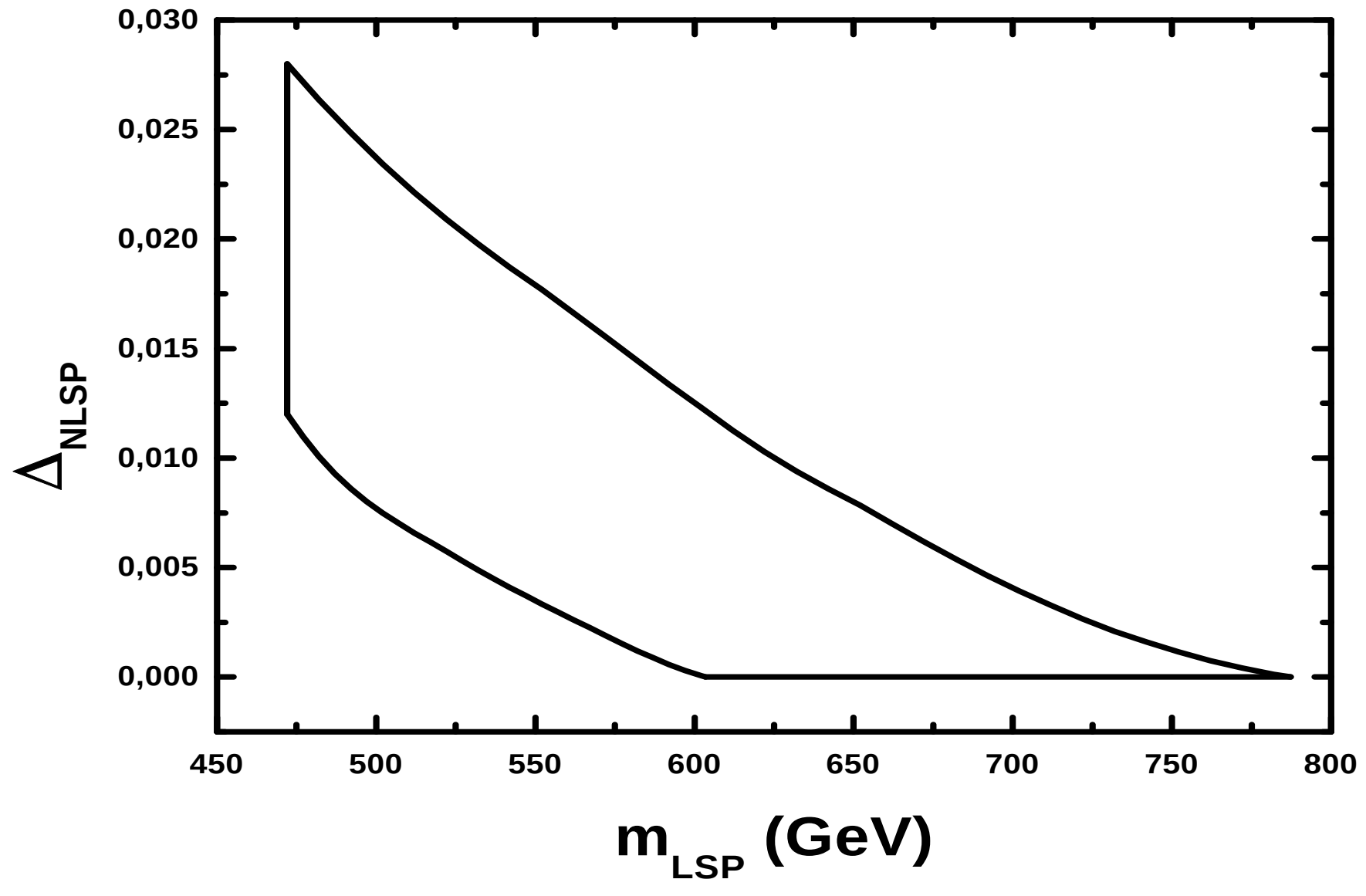

**Σχήμα 7.9:** *Η επιτρεπτή περιοχή στο επίπεδο* $m_{LSP} - \Delta_{NLSP}$, *για* $\tan\beta = 40.7$, $\epsilon = 0.3$ *και* $\mu > 0$.

επιδεικνύεται στο Σχ. 7.9, όπου παρουσιάζεται η επιτρεπτή περιοχή στο επίπεδο $m_{LSP} - \Delta_{NLSP}$, για $\epsilon = 0.3$ και $\tan\beta \simeq 40.7$ με αντιστοιχούσα $m_b^c(M_Z) = 2.63\,\mathrm{GeV}$. Από το Σχ. 7.9 παρατηρείται ότι $470\,\mathrm{GeV} \lesssim m_{LSP} \lesssim 790\,\mathrm{GeV}$ και $\Delta_{NLSP}^{max} \simeq .028$.

**β.** Όταν $22 \lesssim \tan\beta \lesssim 34$, οι διαδικασίες με τελικές καταστάσεις $\tau H$, $\tau A$ γίνονται κινηματικά ανεπίτρεπτες, αφού προκύπτει $m_{LSP} < m_A/2, m_H/2$ και επομένως, το ανώτερο όριο στη μάζα του LSP ελαττώνεται $m_{LSP} \lesssim 580\,\mathrm{GeV}$. Ένα αντιπροσωπευτικό παράδειγμα της διαμορφούμενης κατάστασης σε αυτή την περιοχή δίνεται στο Σχ. 7.10. Επιλέγονται $\epsilon = 0.5$ και $\tan\beta \simeq 28.5$ με αντιστοιχούσα $m_b^c(M_Z) = 2.98\,\mathrm{GeV}$. Παρατηρείται ότι με $\Delta_{NLSP} = 0$ και για την ελάχιστη φαινομενολογικά επιτρεπόμενη τιμή $m_{LSP} = 370\,\mathrm{GeV}$ επιτυγχάνεται το κατώτερο κοσμολογικό όριο .09. Επόμενως, τα σχήματα τεσσάρων πλευρών που συναντήθηκαν μέχρι τώρα στο επίπεδο $m_{LSP} - \Delta_{NLSP}$ μετατρέπονται σε τριών πλευρών. Το πεδίο διακύμανσης του $m_{LSP}$ είναι $370\mathrm{GeV} \lesssim m_{LSP} \lesssim 577\,\mathrm{GeV}$ και $\Delta_{NLSP}^{max} \simeq .021$. Ώστε, ισχυρός εκφυλισμός απαιτείται για να ικανοποιηθούν τα κοσμολογικά όρια.

**γ.** Όταν $34 \lesssim \tan\beta \lesssim 38$ η μάζα του LSP για ορισμένα $m_A$ είναι κοντά στις μάζες $m_A/2$, $m_H/2$ και επομένως, η CRD γίνεται πολύ μικρή, και η επιτρεπτή σχετική διαφορά μάζας τεράστια. Η περιοχή αυτή δεν εξετάζεται περεταίρω, γιατί από τη μια ο φορμαλισμός του Εδ. 5.5 είναι ανεφάρμοστος και από την άλλη γιατί περιοχές παραμέτρων κοντά σε πόλους είναι απομονωμένες και επόμενως, όχι ιδιαιτέρως ενδιαφέρουσες. Η ύπαρξη αυτής της εκκεντρικής περιοχής είναι και η αιτία που δεν γίνεται και διάγραμμα ανάλογο του Σχ. 7.8 στην περίπτωση της ενοποίησης $b - \tau$.

### 7.6.3 Με $b$-$t$ και πλήρη ενοποίηση Yukawa

Η περίπτωση της $b$-$t$ ενοποίησης των ζεύξεων Yukawa μπορεί να απορριφθεί. Και αυτό, διότι η φαινομενολογικά επιτρεπτή περιοχή $42 \lesssim \tan\beta \lesssim 48$ είναι κοσμολογικά απαράδεκτη αφού η μάζα του LSP προκύπτει να παίρνει πολύ μεγάλες τιμές προκαλώντας αύξηση της CRD των LSP σε μη αποδεκτά επίπεδα από τα CDM σενάρια.

Ένα πρόσθετο πρόβλημα ενσκήπτει με την προσπάθεια μελέτης της περίπτωσης της πλήρους ενοποίησης Yukawa που επίσης εξετάστηκε. Η μάζα του ελαφρότερου stau προκύπτει μικρότερη και από αυτή του

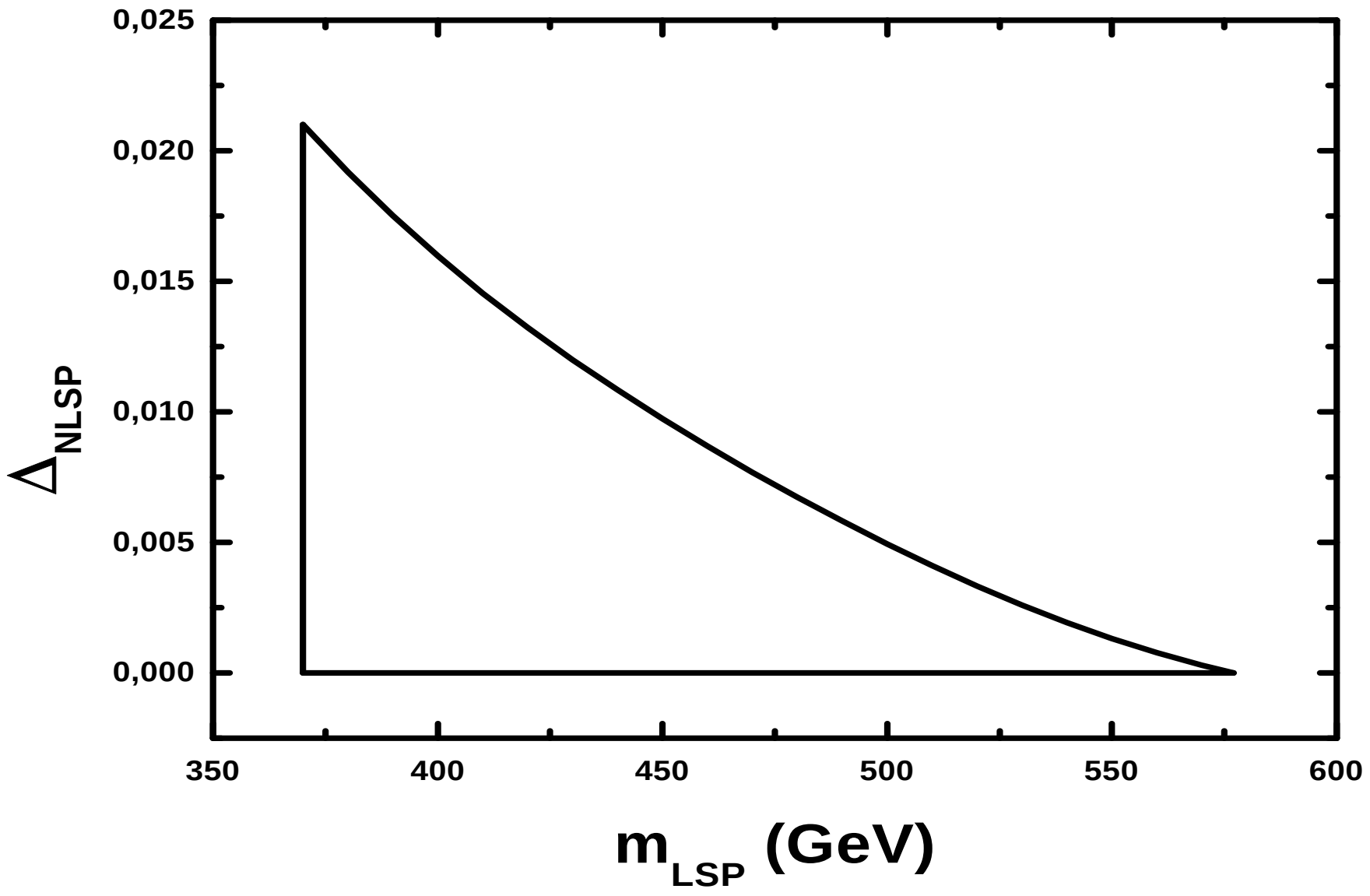

**Σχήμα 7.10:** *Η επιτρεπτή περιοχή στο επίπεδο* $m_{LSP} - \Delta_{NLSP}$, *για* $\tan\beta = 28.5$, $\epsilon = 0.5$ *και* $\mu > 0$.

neutralino. Το πρόβλημα αυτό μπορεί να ατιμετωπιστεί με την εισαγωγή κάποιων $D$-όρων προτεινόμενων στην Αν. [76] οι οποίοι επιτυγχάνουν τη ρύθμιση της σχετικής απόστασης ανάμεσα στις μάζες των δύο ελαφρότατων σωματίων της θεωρίας.

Το επόμενο ερώτημα που αναδύεται είναι η ικανοποίηση του δεσμού για τη μάζα του $b$-quark. Η υψηλή τιμή στη δενδρικού επιπέδου μάζα του $b$-quark μας αναγκάζει στη θεώρηση αποκλειστικά του θετικού σημείου για την παράμετρο $\mu$. Με την επιλογή αυτή η προκύπτουσα μάζα του $b$-quark ελαττώνεται, αλλά η ελάττωση αυτή είναι ισχυρά δραστική με κατάληξη την παραβίαση του κατώτερου πειραμάτικού ορίου (2.17 GeV). Η μη μηδενική ασυμπτωτική τιμή στην τριγραμμική ζεύξη, $A_0$ (σε αντίθεση με το πρότυπο του κεφαλαίου 6) είναι ενάς λόγος για αυτό το αρνητικό αποτέλεσμα. Επομένως, ο περιορισμός για τη μάζα του $b$-quark δεν μπορεί να ικανοποιηθεί στην περίπτωση αυτή.

Επιπρόσθετα, η επίτρεπόμενη μάζα για το LSP από τον περιορισμό του $b \to s\gamma$ αναμένεται να είναι μεγαλύτερη από 800 GeV και συνεπώς, ακόμα και η ενεργοποίηση του μηχανίσμου των CAE στον υπολογισμό της $\Omega_{LSP}h^2$ αποδεικνύεται ατελέσφορη.

## 7.7 Συμπεράσματα

Στο κεφάλαιο αυτό μελετήθηκαν οι χαμηλών ενεργειών σύνεπειες μιας εκδοχής του MSSM με παραβίαση της ηλεκρασθενούς συμμετρίας μέσω κβαντικών διορθώσεων και αρχικές συνθήκες για τους SBT προερχόμενες από τη Θεωρία Hořava-Witten. Ο παραμετρικός χώρος περιορίζεται απαιτώντας ταυτόχρονη ικανοποίηση φαινομενολογικών και κοσμολογικών κριτηρίων. Συγκεκριμένα, απαιτείται η μάζα του $b$-quark με την προσθήκη των SUSY διορθώσεων και το BR($b \to s\gamma$) να είναι συμβατά με τα πειραματικά δεδομένα και επίσης η CRD των LSP να βρίσκεται μέσα στα επιβαλλόμενα από τα CDM σενάρια όρια. Συνάγεται ότι η πλήρης και η $b$-$t$ ενοποίηση των ζεύξεων Yukawa μπορούν να απορριφθούν ενώ η $b$-$\tau$ ενοποίηση δεν είναι ευνοική γιατί απαιτεί μεγάλο εκφυλισμό μαζών ανάμεσα στις μάζες των LSP και NLSP. Αυτό το φαινόμενο μπορεί να αποφευχθεί στην περίπτωση της μη ενοποίηση των ζεύξεων Yukawa όπου η μάζα του LSP μπορεί να είναι μέχρι και 70 GeV.

# Κεφάλαιο 8

# Συμπεράσματα

Σκοπός της διατριβής που παρουσιάστηκε, ήταν η κατάθεση μιας άποψης σχετικά με το πρόβλημα της Σκοτεινής Ύλης αξιοποιώντας το σωματιδιακό περιεχόμενο του MSSM. Για την υλοποίηση αυτού του στόχου επιστρατεύτηκαν γνώσεις από τη φαινομενολογία του MSSM και τη σύγχρονη Κοσμολογία. Αυτές εκτέθηκαν στο πρώτο μέρος της διατριβής, στα κεφάλαια γενικής αναφοράς, 2, 3, 4. Στο κεφάλαιο θεμελίωσης της συνεργασίας Κοσμολογίας και Φυσικής Στοιχειωδών Σωματιδίων, 5 αναπτύχθηκε η μέθοδος υπολογισμού της CRD των LSP. Τέλος, η αριθμητική υλοποίηση των ιδεών αυτών παρουσιάστηκε στα ερευνητικά κεφάλαια της διατριβής, 6, 7. Ο απολογισμός της ερευνητικής αυτής ενασχόλησης είναι το αντικείμενο αυτού του επιλογικού σημειώματος.

Γενικό συμπέρασμα της διατριβής, ανεξαρτήτο της επιλογής σωματιδιακού προτύπου, είναι ότι ένα LSP μορφής Bino αποτελεί επιτυχή υποψήφιο για CDM. Οι CAE ενός τέτοιου LSP με το λίγο βαρύτερο του stau (ή και με τα selectron, αν οι μάζες τους βρίσκονται σε μικρή απόσταση) είναι ένα αποφασιστικής σημασίας φαινόμενο, που συμβάλλει στη μείωση της CRD των LSP σε επιτρεπτά επίπεδα. Λαμβάνοντας υπόψη τα επιβαλλόμενα από τα CDM σενάρια όρια, ο παραμετρικός χώρος του μελετούμενου προτύπου υποβάλλεται σε δραστικούς περιορισμούς που αφορούν τις μάζες των ελαφρότερων σωματίων της θεωρίας. Επιπρόσθετοι περιορισμοί μπορούν να αντληθούν από τη φαινομενολογία των SUSY διορθώσεων στη μάζα του $b$-quark και της διαδικασίας $b \to s\gamma$.

Το πλαίσιο μελέτης εφαρμόστηκε σε δύο εκδοχές του MSSM με παγκόσμιες αρχικές συνθήκες για τους SBT, ενοποίηση των ζεύξεων Βαθμίδας και παραβίαση της ηλεκτρασθενούς συμμετρίας μέσω κβαντικών διορθώσεων. Ειδικότερα, τα πρότυπα που εξετάστηκαν είναι:

- Το MSSM με ενοποίηση Yukawa. Διακρίνονται δύο περιπτώσεις:
  - Αν $\mu < 0$, το πρότυπο είναι συμβατό με τα πειραματικά όρια για το $\mathrm{BR}(b \to s\gamma)$, όμως η μάζα του $b$-quark κείται ανωτέρως των πειραματικών ορίων. Θεωρώντας ότι το γεγονός αυτό μπορεί να ρυθμιστεί με κάποιο τρόπο, η επιβολή των κοσμολογικών δεσμών οδηγεί στο συμπέρασμα ότι η μάζα του LSP μπορεί να κυμαίνεται από 215 GeV ως 770 GeV με τη μάζα του NLSP να μπορεί να είναι από 8-0% μεγαλύτερη.
  - Αν $\mu > 0$, το πρότυπο είναι συμβατό με τα πειραματικά όρια για τη μάζα του $b$-quark και ταυτόχρονη ικανοποίηση των κοσμολογικών περιορισμών και αυτού που πηγάζει από το $\mathrm{BR}(b \to s\gamma)$ μπορεί να επιτευχθεί σε μικρή περιοχή του παραμετρικού χώρου.
- Το MSSM προερχόμενο από τη θεωρία Hořava-Witten. Ταυτόχρονη επιβολή των φαινομενολογικών και κοσμολογικών περιορισμών οδηγεί στο συμπέρασμα ότι το πρότυπο δεν είναι συμβατό με την ενοποίηση Yukawa και $b$-$t$, απαιτεί μεγάλο εκφυλισμό στις μάζες των ελαφρότατων σωματίων της θεωρίας με ενοποίηση $b$-$\tau$, και παρέχει μεγαλύτερη ευκαίρια επιλογών για τις μάζες αυτές στην περίπτωση της μη ενοποίησης Yukawa, κατά τη οποία η μάζα του LSP μπορεί να κυμαίνεται από 70 GeV ως 670 GeV με τη μάζα του NLSP να μπορεί να είναι από 93-0% μεγαλύτερη.

Συμπεράσματα, και μάλιστα ιδιαιτέρως χρήσιμα, γιατί αποτελούν εφαλτήριο για περαιτέρω μελέτη, αποτελούν και οι ανεπάρκειες ενός ερευνητικού πονήματος. Αυτές είναι οι ακόλουθες:

- Σχετικά με την εκδοχή του MSSM με ενοποίηση Yukawa:
  - Ρύθμιση του προβλήματος της μάζας του $b$-quark στην περίπτωση $\mu < 0$. Όπως αναφέρθηκε και στο Εδ. 6.5.1 οι SUSY διορθώσεις δεν είναι οι μόνες που μπορεί να λάβει η μάζα του $b$-quark. Σε ένα πλήρες πρότυπο $SO(10)$ είναι δυνατόν να προκύψουν και άλλες διορθώσεις και επομένως να διευθετηθεί το πρόβλημα της μάζας του $b$-quark, το οποίο, αν ληφθεί σοβαρά υπόψη περιορίζει κατά πολύ τον παραμετρικό χώρο των προτύπων με ενοποίηση Yukawa. Ώστε, μια γενικότερη μελέτη του τομέα των φερμιονίων σε πρότυπα $SO(10)$ είναι επιβεβλημένη.
  - Εξέταση της περίπτωσης όπου $A_0 \neq 0$. Όλη η διαπραγμάτευση του προτύπου με ενοποιήση Yukawa είχε περιοριστεί στην περίπτωση $A_0 = 0$. Η διαφοροποίηση από την τιμή αυτή δεν αναμένεται να έχει σημαντική επίπτωση στη συμπεριφορά των αποτελεσμάτων, και δη, προς το καλύτερο. Ιδιαίτερα τα χαρακτηριστικά του φάσματος της θεωρίας παραμένουν αμετάβλητα. Υπάρχουν, όμως, κάποιες CAE που επηρεάζονται από την επιλογή αυτή, αφού για τον υπολογισμό τους χρησιμοποιούνται σύμβολα $g$ που περιέχουν στις εκφράσεις τους τις τριγραμμικές ζεύξεις. Επομένως, μια μετακίνηση από την επιλογή $A_0 = 0$ θα ήταν μια ενδιαφέρουσα εμπειρία.
  - Εισαγωγή των σφαλμάτων στον υπολογισμό του $\mathrm{BR}(b \to s\gamma)$ και κατασκευή της επιτρεπτής περιοχής στην περίπτωση του $\mu > 0$. Αν και έχει γίνει μια περιγραφή της επιτρεπτής περιοχής που προκύπτει σε αυτή την περίπτωση, στο Εδ. 6.6.2, η εισαγωγή των σφαλμάτων θα συνέβαλλε στη διεύρυνσή της για λόγους που εκεί εξηγήθηκαν.
- Σχετικά με την εκδοχή του MSSM από τη Θεωρία Hořava-Witten θα ήταν χρήσιμη, όχι όμως κρίσιμη, η εισαγωγή των NLO διορθώσεων της QCD στον υπολογισμό του $\mathrm{BR}(b \to s\gamma)$. Αυτή η προσθήκη αναμένεται να ελαττώσει τις δυνατές τιμές των $m_{LSP}^{min}$ και συνεπώς, να αυξήσει τις τιμές των $\Delta_{NLSP}^{max}$. Η διεύρυνση των επιτρεπτών περιοχών που μπορεί να επιτευχθεί με τον τρόπο αυτό, οπωσδήποτε δεν αλλοιώνει αποφασιστικά το χαρακτήρα των συμπερασμάτων του Κεφαλαίου 7.
- Μελέτη της ελαστικής σκέδασης των neutralino από τους πυρήνες. Το LSP μπορεί να ανιχνευτεί από την ελαστική του σύγκρουση με τους πυρήνες. Μάλιστα, πρόσφατα είναι κάποια αποτελέσματα από το πείραμα DAMA, τα οποία βεβαιώνουν τη σύλληψη ενός σήματος που μπορεί να προέρχεται από άμεση ανίχνευση ενός neutralino. Από τέτοια πειραματικά δεδομένα μπορεί κανείς να βγάλει συμπεράσματα για την ταυτότητα του LSP, την καθαρότητα του και τη μάζα του. Επόμενως, ένα επόμενο βήμα της μελέτης για την υποψηφιότητα του LSP για CDM είναι ο υπολογισμός της ενεργού διατομής της σκέδασής του με τους πυρήνες.
- Προσθήκη $D$-όρων στις αρχικές συνθήκες για τους SBT. Σε όλη τη διάρκεια της διατριβής, μελετήθηκε η περίπτωση των παγκόσμιων αρχικών συνθηκών για τους SBT. Εντούτοις, υπάρχει δυνατότητα μετακίνησης από αυτή την επιλογή στα πλαίσια των μελετούμενων προτύπων, σύμφωνα με τις Αν. [67] και [76], με την προσθήκη κάποιων $D$-όρων στις αρχικές συνθήκες για τους SBT. Βεβαίως, με τον τρόπο αυτό, οι αρχικές συνθήκες παύουν να είναι παγκόσμιες και μια επιπλέον αυθαιρεσία εισάγεται. Με αυτή την προσθήκη, όμως, μπορούν να επιτευχθούν οι φαινομενολογικοί και κοσμολογικοί στόχοι ευκολότερα. Μια επιπρόσθετη δυσκολία που ενσκήπτει σε τέτοια κατασκεύασματα είναι ότι το LSP παύει να είναι ισχυρά μορφής Bino, οπότε η διαδικασία του Εδ. 5.5.1 δεν προσφέρει αξιόπιστα αποτελέσματα και πρέπει να επεκταθεί.
- Λεπτομερειακή μελέτη των διαύλων των ΑΝΕ. Η απλοποίηση που επιχειρήθηκε στο Εδ. 5.5 μέσω της Εξ. (5.88), αν και πλήρως αιτιολογημένη, λόγω των αριθμητικών ευρημάτων στις περιπτώσεις που εφαρμόστηκε, δεν παύει να αποτελεί μια προσέγγιση. Επίσης στο Εδ. 7.6 παρατηρήθηκε μια αρκετά ανησυχητική μετατόπιση από αυτή την κατάσταση για κάποιες χαμηλές τιμές της $\tan\beta$. Επίσης, υπάρχει μαρτυρία της Αν. [53] ότι η ζεύξη neutralino-neutralino-A ακόμα και με μια μικρή higgsino συνεισφορά μπορεί να συμβάλλει ενεργά στην ταπείνωση της CRD των LSP. Για τον έλεγχο όλων αυτών των αμφιβολιών θα ήταν χρήσιμη η πιο διεξοδική μελέτη των διαύλων των ΑΝΕ μέσω του παραρτήματος της Αν. [63].

Παρά τις ελλείψεις που επισημάνθηκαν, η διατριβή δεν παύει να αποτελεί μια προσπάθεια έκθεσης των προβλημάτισμών και των αναζητήσεων πάνω σε θέματα Σκοτεινής Ύλης και φαινομενολογίας του MSSM. Και ως τέτοια, τίθεται προς αξιολόγηση.

# Βιβλιογραφία

# Παράρτημα Α΄

# Εξισώσεις Επανακανονικοποίησης

## Α΄.1 Εισαγωγή

Στο παράρτημα αυτό, εκτίθενται οι εξισώσεις επανακανονικοποίησης (RGE) για τις ζεύξεις Βαθμίδας και Yukawa στο σχήμα επανακανονικοποίησης $\overline{\text{MS}}$ και για τους όρους ασθενούς SUSY παραβίασης που χρησιμοποιούνται στο αριθμητικό πρόγραμμα. Βασική πηγή προέλευσης του τυπολόγιου είναι οι Αν. [10] και [11]. Απλοποιήσεις, διευθετήσεις και προσαρμογή του συμβολισμού είναι η δική μας παρέμβαση.

## Α΄.2 Οι RGE για τις ζεύξεις Βαθμίδας και Yukawa

Όπως εχει αναφερθεί στο Εδ. 6.3, δύο συστήματα RGE χρησιμοποιούνται για τις ζεύξεις Βαθμίδας και Yukawa . Αυτό του MSSM για το τρέξιμο των παραμέτρων από την κλίμακα ενοποίησης, $M_G$ μέχρι τη χαμηλή προνομιακή κλίμακα $M_S$ της Εξ. (3.1), και αυτό του SM για το τρέξιμο από την κλίμακα $M_S$ μέχρι την κλίμακα $M_Z$. Επειδή η ενεργειακή απόσταση $M_G$-$M_S$ είναι μεγάλη, ($\sim 10^{15}$ GeV) επιβάλλεται καλή ακρίβεια στην εξέλιξη των παραμέτρων σε αυτή τη διαδρομή. Γι αυτό, και χρησιμοποιούνται οι RGE του MSSM σε επίπεδο δύο βρόχων. Αντιθέτως, λόγω της μικρής ενεργειακής απόστασης $M_S$-$M_Z$, οι RGE του SM σε επίπεδο ενός βρόχου παρέχουν ικανοποιητικά αποτελέσματα. Μεταβλητή ολοκλήρωσης είναι η $t = \ln M$ με $M$ την την τυχαία κλίμακα ενέργειας.

Παρακάτω παρουσιάζονται οι RGE του:

**α.** MSSM σε επίπεδο δύο βρόχων για τις ζεύξεις:

- Βαθμίδας:

$$\frac{dg_i}{dt} = \frac{g_i}{16\pi^2}\left[b_i g_i^2 + \frac{1}{16\pi^2}\left(\sum_{j=1}^{3} b_{ij} g_i^2 g_j^2 - \sum_{j=t,b,\tau} a_{ij} g_i^2 h_j^2\right)\right] . \tag{Α΄.1}$$

- Yukawa :

$$\begin{aligned} \frac{dh_t}{dt} = \frac{1}{16\pi^2} & \left[ \left(-\sum_{j=1}^{3} c_i g_i^2 + 3h_t^2 + h_b^2 + 3h_t^2\right)\right. \\ +\frac{1}{16\pi^2} & \left( \sum_{j=1}^{3} \left(c_i b_i + c_i^2/2\right) g_i^4 + g_1^2 g_2^2 + \frac{136}{45} g_1^2 g_3^2 + 8 g_2^2 g_3^2 \right. \\ & +(\frac{2}{5} g_1^2 + 6 g_2^2) h_t^2 + \frac{2}{5} g_1^2 h_b^2 + (\frac{4}{5} g_1^2 + 16 g_3^2) h_t^2 \\ & \left.\left. -22 h_t^4 - 5 h_b^4 - 5 h_t^2 h_b^2 - h_b^2 h_\tau^2 \right)\right] h_t , \end{aligned} \tag{Α΄.2}$$

$$\frac{dh_b}{dt} = \frac{1}{16\pi^2} \Bigg[ \left( -\sum_{j=1}^{3} c'_i g_i^2 + 6h_b^2 + h_t^2 h_\tau^2 \right)$$

$$+\frac{1}{16\pi^2} \Bigg( \sum_{j=1}^{3} \left(c'_i b_i + c'^2_i/2\right) g_i^4 + g_1^2 g_2^2 + \frac{8}{9} g_1^2 g_3^2 + 8 g_2^2 g_3^2$$

$$+(\frac{4}{5}g_1^2 + 6g_2^2)h_b^2 + \frac{4}{5}g_1^2 h_t^2 + (-\frac{2}{5}g_1^2 + 16g_3^2)h_b^2 + \frac{6}{5}g_1^2 h_\tau^2$$

$$-22h_b^4 - 5h_t^4 - 3h_\tau^4 - 5h_b^2 h_t^2 - 3h_b^2 h_\tau^2 \Bigg) \Bigg] h_b \ , \qquad (Α'.3)$$

$$\frac{dh_\tau}{dt} = \frac{1}{16\pi^2} \Bigg[ \left( -\sum_{j=1}^{3} c''_i g_i^2 + 3h_\tau^2 + 3h_b^2 + h_\tau^2 \right)$$

$$+\frac{1}{16\pi^2} \Bigg( \sum_{j=1}^{3} \left(c''_i b_i + c''^2_i/2\right) g_i^4 + \frac{9}{5} g_1^2 g_2^2$$

$$+6g_2^2 h_\tau^2 + (-\frac{2}{5}g_1^2 + 16g_3^2)h_b^2 + \frac{6}{5}g_1^2 h_\tau^2$$

$$-9h_b^4 - 3h_b^2 h_t^2 - 9h_\tau^2 h_b^2 - 10h_\tau^4 \Bigg) \Bigg] h_\tau \ , \qquad (Α'.4)$$

όπου οι χρησιμοποιούμενες σταθερές είναι στοιχεία των διανυσμάτων:

$$(b_i) = (\frac{33}{5}, 1, -3) \ , \qquad (Α'.5)$$

$$(c_i) = (\frac{13}{15}, 3, \frac{16}{3}) \ , \qquad (Α'.6)$$

$$(c'_i) = (\frac{7}{15}, 3, \frac{16}{3}) \ , \qquad (Α'.7)$$

$$(c''_i) = (\frac{9}{5}, 3, 0) \ , \qquad (Α'.8)$$

$$d_i = c'_i - c''_i \qquad (Α'.9)$$

και των πινάκων:

$$(b_{ij}) = \begin{pmatrix} \frac{199}{25} & \frac{27}{5} & \frac{88}{5} \\ \frac{9}{5} & 25 & 24 \\ \frac{11}{5} & 9 & 14 \end{pmatrix} , \qquad (Α'.10)$$

$$(a_{ij}) = \begin{pmatrix} \frac{26}{5} & \frac{14}{5} & \frac{18}{5} \\ 6 & 6 & 2 \\ 4 & 4 & 0 \end{pmatrix} . \qquad (Α'.11)$$

**β.** SM για τις ζεύξεις:

- Βαθμίδας, σε επίπεδο δύο βρόχων:

$$\frac{dg_i}{dt} = \frac{g_i}{16\pi^2} \left[ b_i^{SM} g_i^2 + \frac{1}{16\pi^2} \left( \sum_{j=1}^{3} b_{ij}^{SM} g_i^2 g_j^2 - \sum_{j=t,b,\tau} a_{ij}^{SM} g_i^2 h_j^2 \right) \right] . \qquad (Α'.12)$$

- Yukawa, σε επίπεδο ενός βρόχου:

$$\frac{dh_t}{dt} = \frac{1}{16\pi^2} \left( -\sum c_i^{SM} g_i^2 + \frac{9}{2} h_t^2 + \frac{3}{2} h_b^2 + h_\tau \right) h_t, \qquad (Α'.13)$$

$$\frac{dh_b}{dt} = \frac{1}{16\pi^2}\left(-\sum c_i'^{SM} g_i^2 + \frac{9}{2}h_b^2 + \frac{9}{2}h_t^2 + h_\tau\right) h_b, \tag{Α΄.14}$$

$$\frac{dh_\tau}{dt} = \frac{1}{16\pi^2}\left(-\sum c_i''^{SM} g_i^2 + 3h_t^2 + 3h_b^2 + \frac{5}{2}h_\tau^2\right) h_\tau\,, \tag{Α΄.15}$$

όπου οι χρησιμοποιούμενες σταθερές είναι στοιχεία των διανυσμάτων:

$$(b_i^{SM}) = (\frac{41}{10}, -\frac{19}{6}, -7)\,, \tag{Α΄.16}$$

$$(c_i^{SM}) = (\frac{17}{20}, \frac{9}{4}, 8)\,, \tag{Α΄.17}$$

$$(c_i'^{SM}) = (\frac{1}{4}, \frac{9}{4}, 8)\,, \tag{Α΄.18}$$

$$(c_i''^{SM}) = (\frac{9}{4}, \frac{9}{4}, 0) \tag{Α΄.19}$$

και των πινάκων:

$$(b_{ij}^{SM}) = \begin{pmatrix} \frac{199}{50} & \frac{27}{10} & \frac{44}{5} \\ \frac{9}{10} & \frac{35}{6} & 12 \\ \frac{11}{10} & \frac{9}{2} & -26 \end{pmatrix}\,, \tag{Α΄.20}$$

$$(a_{ij}^{SM}) = \begin{pmatrix} \frac{17}{10} & \frac{1}{2} & \frac{3}{2} \\ \frac{3}{2} & \frac{3}{2} & \frac{1}{2} \\ 2 & 2 & 0 \end{pmatrix}\,. \tag{Α΄.21}$$

## Α΄.3 Οι RGE για τους όρους ασθενούς παραβίασης της SUSY

Οπως έχει αναφερθεί και στο κυρίως τμήμα της διατριβής, οι όροι ασθενούς παραβίασης της SUSY προέρχονται από την υπερβαρύτητα και είναι μάζες για τα gauginos και τα scalars και οι τριγραμμικές ζεύξεις. Για τις παράμετρες αυτές, ένα σύστημα RGE χρησιμοποιείται, που διέπει την εξέλιξή τους στο ενεργειακό διάστημα $M_G$-$M_S$. Μέτα το $M_S$ οι παράμετρες αυτές είναι παγωμένες. Αξιοπρόσεκτο, επίσης, είναι ότι στην ανάλυση που ακολουθείται, οι RGE των $\mu$ και $B$ δεν χρησιμοποιούνται.

Παρακάτω παρουσιάζονται οι RGE σε επίπεδο ενός βρόχου για τις:

**α.** Ασθενείς μάζες των gauginos :

$$\frac{dM_i}{dt} = \frac{1}{8\pi^2} b_i g_i^2 M_i\,. \tag{Α΄.22}$$

**β.** Τριγραμμικές ζεύξεις των:

- Βαρέων γενεών:

$$\frac{dA_t}{dt} = \frac{1}{8\pi^2}\Big(\sum_{j=1}^{3} c_i g_i^2 M_i + 6h_t^2 A_t + h_b^2 A_b\Big)\,, \tag{Α΄.23}$$

$$\frac{dA_b}{dt} = \frac{2}{16\pi^2}\Big(\sum_{j=1}^{3} c_i' g_i^2 M_i + 6h_b^2 A_b + h_t^2 A_t + h_\tau^2 A_\tau\Big)\,, \tag{Α΄.24}$$

$$\frac{dA_\tau}{dt} = \frac{1}{8\pi^2}\Big(\sum_{j=1}^{3} c_i'' g_i^2 M_i + 3h_b^2 A_b + 4h_\tau^2 A_\tau\Big)\,. \tag{Α΄.25}$$

- Ελαφρών γενεών (εκτίθενται αν και δεν χρησιμοποιούνται στο αριθμητικό πρόγραμμα):

$$\frac{dA_u}{dt} = \frac{1}{8\pi^2}\Big(\sum_{j=1}^{3} c_i g_i^2 M_i + h_t^2 A_t\Big) \ , \qquad (A'.26)$$

$$\frac{dA_d}{dt} = \frac{1}{8\pi^2}\Big(\sum_{j=1}^{3} c_i' g_i^2 M_i + h_b^2 A_b + \frac{1}{3} h_\tau^2 A_\tau\Big) \ , \qquad (A'.27)$$

$$\frac{dA_e}{dt} = \frac{2}{16\pi^2}\Big(\sum_{j=1}^{3} c_i'' g_i^2 M_i + h_b^2 A_b + \frac{1}{3} h_\tau^2 A_\tau\Big) \ . \qquad (A'.28)$$

Οι φάκτορες $c_i$, $c_i'$ και $c_i''$ δίνονται από τις Εξ. (A΄.6), (A΄.7) και (A΄.8).

**β.** Ασθενείς μάζες των scalars των:

- Βαρέων γενεών:

$$\frac{dm_H^2}{dt} = \frac{1}{8\pi^2}\Big(-\frac{3}{5}g_1^2M_1^2 - 3g_2^2M_2^2 + 3h_b^2X_b + h_\tau^2X_\tau\Big) \ , \qquad (A'.29)$$

$$\frac{dm_{\bar{H}}^2}{dt} = \frac{1}{8\pi^2}\Big(-\frac{3}{5}g_1^2M_1^2 - 3g_2^2M_2^2 + 3h_t^2X_t\Big) \ , \qquad (A'.30)$$

$$\frac{dm_{Q_L}^2}{dt} = \frac{1}{8\pi^2}\Big(-\frac{1}{15}g_1^2M_1^2 - 3g_2^2M_2^2 - \frac{16}{3}g_3^2M_3^2 + h_t^2X_t + h_b^2X_b\Big) \ , \qquad (A'.31)$$

$$\frac{dm_{t_R}^2}{dt} = \frac{1}{8\pi^2}\Big(-\frac{16}{15}g_1^2M_1^2 - \frac{16}{3}g_3^2M_3^2 + 2h_t^2X_t\Big) \ , \qquad (A'.32)$$

$$\frac{dm_{b_R}^2}{dt} = \frac{1}{8\pi^2}\Big(-\frac{4}{15}g_1^2M_1^2 - \frac{16}{3}g_3^2M_3^2 + 2h_b^2X_b\Big) \ , \qquad (A'.33)$$

$$\frac{dm_{L_L}^2}{dt} = \frac{1}{8\pi^2}\Big(-\frac{3}{5}g_1^2M_1^2 - 3g_2^2M_2^2 + h_\tau^2X_\tau\Big) \ , \qquad (A'.34)$$

$$\frac{dm_{\tau_R}^2}{dt} = \frac{1}{8\pi^2}\Big(-\frac{12}{5}g_1^2M_1^2 + 2h_\tau^2X_\tau\Big) \ , \qquad (A'.35)$$

όπου

$$X_t = m_{Q_L}^2 + m_{t_R}^2 + m_{\bar{H}}^2 + A_t^2 \ , \qquad (A'.36)$$

$$X_b = m_{Q_L}^2 + m_{b_R}^2 + m_H^2 + A_b^2 \ , \qquad (A'.37)$$

$$X_\tau = m_{L_L}^2 + m_{\tau_R}^2 + m_H^2 + A_\tau^2 \ . \qquad (A'.38)$$

- Ελαφρών γενεών:

$$\frac{dm_{q_L}^2}{dt} = \frac{1}{8\pi^2}\Big(-\frac{1}{15}g_1^2M_1^2 - 3g_2^2M_2^2 - \frac{16}{3}g_3^2M_3^2\Big) \ , \qquad (A'.39)$$

$$\frac{dm_{u_R}^2}{dt} = \frac{1}{8\pi^2}\Big(-\frac{16}{15}g_1^2M_1^2 - \frac{16}{3}g_3^2M_3^2\Big) \ , \qquad (A'.40)$$

$$\frac{dm_{d_R}^2}{dt} = \frac{1}{8\pi^2}\Big(-\frac{4}{15}g_1^2M_1^2 - \frac{16}{3}g_3^2M_3^2\Big) \ , \qquad (A'.41)$$

$$\frac{dm_{l_L}^2}{dt} = \frac{1}{8\pi^2}\Big(-\frac{3}{5}g_1^2M_1^2 - 3g_2^2M_2^2\Big) \ , \qquad (A'.42)$$

$$\frac{dm_{e_R}^2}{dt} = \frac{1}{8\pi^2}\Big(-\frac{12}{5}g_1^2M_1^2\Big) \ . \qquad (A'.43)$$

# Παράρτημα Β′

# Κανόνες Feynman και Εφαρμογές

## Β′.1 Εισαγωγή

Σκοπός του παραρτήματος αυτού είναι η ομαδοποίηση των εργαλείων που χρησιμοποιήθηκαν στην αθέατη πλευρά αυτής της διατριβής, που σε αναλυτικό επίπεδο, ήταν ο υπολογισμός πλατών αλληλεπίδρασης. Καταγραφή των συμβάσεων και των νορμαλισμών που υιοθετήθηκαν γίνεται στο Εδ. Β′.2 και ακολουθεί στο Εδ. Β′.3 η ανάγλυφη παρουσίαση των κανόνων που χρησιμοποιήθηκαν για τον υπολογισμό των ΑΝΕ και CAE. Οι κανόνες αυτοί συνοδεύονται και με τρία απλά παραδείγματα εφαρμογής στο Εδ. Β′.4. Τυπολόγιο Dirac-ολογίας, υπολογισμού ιχνών και σάντουιτς πινάκων $\gamma$, καθώς και μετατροπής των αναλλοίωτων συμπλεγμάτων ορμής με χρήση των μεταβλητών Mandelstam δε χρησιμοποιήθηκε για αυτό και δεν εκτίθεται. Αρωγό στην αναπόφευκτη αυτή διαδικασία είχαμε το πακέτο *FeynCalc* [80] που παρείχε γρήγορα και αξιόπιστα αποτελέσματα.

## Β′.2 Συμβάσεις-Κανονικοποιήσεις

Ο υπολογισμός των ΑΝΕ και CAE βασίστηκε στις Αν. [12] και [13]. Από αυτές επομένως, αντλούνται και οι βασικές συμβάσεις που θα αναφερθούν πιο κάτω. Καταβλήθηκε προσπάθεια περιορισμού του τυπολόγιου μόνο στις τελείως αναγκαίες σχέσεις που θα επιτρέψουν τον εύκολο και ταχύ κατατοπισμό του αναγνώστη.

Η μετρική Minkowski που χρησιμοποιείται είναι:

$$\eta_{\mu\nu} := \text{diag}(+1,-1,-1,-1) \tag{Β′.1}$$

### Β′.2.1 $\gamma$-Πίνακες

Οι πινάκες $\gamma_\mu$, ορίζονται από τις γνωστές αντιμεταθετικές σχέσεις:

$$\gamma^\mu\gamma^\nu + \gamma^\nu\gamma^\mu = 2\eta^{\mu\nu} \tag{Β′.2}$$

και οι προβολικοί εκ-τελεστές ως εξής:

$$P_R = (1+\gamma_5)/2, \quad P_L = (1-\gamma_5)/2, \quad \text{όπου} \quad \gamma^5 = \gamma_5 = i\gamma^0\gamma^1\gamma^2\gamma^3 \tag{Β′.3}$$

με βοηθητική σχέση

$$\gamma^\mu\gamma_5 = -\gamma_5\gamma^\mu \Longrightarrow P_{R[L]}\gamma^\mu = \gamma^\mu P_{L[R]} \tag{Β′.4}$$

Η αναπαράσταση που επιλέγεται είναι:

$$\gamma^\mu = \begin{pmatrix} 0 & \sigma^\mu \\ \bar{\sigma}^\mu & 0 \end{pmatrix} \quad \text{και} \quad \gamma_5 = \begin{pmatrix} -1 & 0 \\ 0 & 1 \end{pmatrix} \tag{Β′.5}$$

όπου

$$\sigma^\mu = (1,\vec{\sigma}\,), \quad \bar{\sigma}^\mu = (1,-\vec{\sigma}\,). \tag{Β′.6}$$

με $\vec{\sigma}(\sigma_i)$ τους γνωστούς $2 \times 2$ πίνακές Pauli.

$$\sigma^1 = \begin{pmatrix} 0 & 1 \\ 1 & 0 \end{pmatrix}, \quad \sigma^2 = \begin{pmatrix} 0 & -i \\ i & 0 \end{pmatrix}, \quad \sigma^3 = \begin{pmatrix} 1 & 0 \\ 0 & 1 \end{pmatrix} \tag{B′.7}$$

Ο πίνακας συζυγίας φορτίου που ορίζεται $C = i\gamma^2\gamma^0$ έχει τις ακόλουθες ιδιότητες:

$$C^\dagger = C^{-1}, \quad C^T = -C, \quad C^{-1}\gamma_\mu[\gamma_5]C = -\gamma_\mu^T[+\gamma_5^T]. \tag{B′.8}$$

### B′.2.2 Spinors

Οι σπίνορες $u(p)$, $v(p)$ για φερμιόνια Majorana ή Dirac σχετίζονται μέσω των σχέσεων

$$u(p) = C\bar{v}^T(p), \qquad v(p) = C\bar{u}^T(p) \tag{B′.9}$$

$$\bar{u}(p) = -v^T(p)C^{-1}, \qquad \bar{v}(p) = -u^T(p)C^{-1} \tag{B′.10}$$

όπου οι σπίνορες φέροντες ¯ ορίζονται κατά τα γνωστά

$$\bar{u}(p) = u(p)^\dagger\gamma_0 \;\; \text{και} \;\; \bar{v}(p) = v(p)^\dagger\gamma_0 \tag{B′.11}$$

Οι δείκτες που αναφέρονται στο σπίν έχουν παραληφθεί από τους σπίνορες $u(p)$, $v(p)$, χάριν απλότητας. Η άθροιση σε αυτούς τους δείκτες για φερμιόνιο συγκεκριμένης ορμής, προσφέρουν τις παρακάτω σχέσεις ορθοκανονοκοποίησης:

$$\sum_{spin} u(p)\bar{u}(p) = \not{p} + m, \qquad \sum_{spin} v(p)\bar{v}(p) = \not{p} - m, \tag{B′.12}$$

$$\sum_{spin} u(p)\bar{v}^T(p) = (\not{p} + m)C^T, \qquad \sum_{spin} v(p)\bar{v}(p) = C^{-1}(\not{p} - m), \tag{B′.13}$$

$$\sum_{spin} \bar{v}^T(p)\bar{u}(p) = C^{-1}(\not{p} + m) \qquad \sum_{spin} v(p)u^T(p) = (\not{p} - m)C^T. \tag{B′.14}$$

Ένας Dirac σπίνορας $\Psi$ μπορεί να γραφεί χρησιμοποιώντας 2 Weyl σπίνορες δύο διαστάσεων ο καθένας. Καθώς η διατριβή δεν εισέρχεται σε λεπτομέριες δομής της SUSY, δε θεωρείται σκόπιμη η εισαγωγή του φορμαλισμού των δύο διαστάσεων spinors. Απλά δίνεται ένας τύπος που θεμελιώνει τη συσχέτιση των δύο αντικειμένων.

$$\Psi = \begin{pmatrix} \psi_L \\ \psi_L^{c\dagger} \end{pmatrix} = \begin{pmatrix} \psi_L \\ \psi_R \end{pmatrix} \tag{B′.15}$$

Επομένως, ένας τυπικός όρος μάζας με σπίνορες Dirac γράφεται:

$$\bar{\Psi}\Psi = \psi_L^c\psi_L + \psi_L^\dagger\psi_L^{c\,\dagger} = \psi_R^\dagger\psi_L + \psi_L^\dagger\psi_R \tag{B′.16}$$

όπου $\psi^c = C\bar{\psi}^T$. Οι μάζες των φερμιονίων του MSSM παρέχονται από τέτοιους όρυς μάζας. Ο δείκτης $L$ συμπιέζεται στο τυπολόγιο του Εδ. 2.2.2.

### B′.2.3 Δίανυσμα πόλωσης

Άλλος χρήσιμος νορμαλισμός αφορά το δίανυσμα πόλωσης ενός μποζωνίου βαθμίδας. Και πάλι χάριν απλότητας, οι δείκτες που αφορούν την κατάσταση πόλωσης του μποζωνίου συμπιέζονται. Η άθροιση πάνω σε αυτούς παρέχει τη σχέση ορθοκανονοκοποίησης, που είναι:

$$\sum_{polarisation} \epsilon_\mu^*(p)\epsilon_\nu(p) = -\eta_{\mu\nu} + \frac{p_\mu p_\nu}{m^2} \tag{B′.17}$$

όπου ο δεύτερος όρος παραλείπεται στην περίπτωση άμαζου σωματίου.

## B′.3 Κανόνες Feynman

Όταν τα εμπλεκόμενα στους υπολογισμούς φερμιόνια αναπαρίστανται με σπίνορες δυο συνιστωσών, οι συνήθεις κανόνες Feynman τροποποιούνται ελαφρώς. Πλήρης περιγραφή γίνεται στο παράρτημα της Αν. [13]. Εδώ αναπαράγονται μόνο οι κανόνες Feynman που αντιστοιχούν στις περιπτώσεις των διαγραμμάτων που υπολογίσθηκαν.

### B′.3.1 Γενικά

**α.** Εξωτερικά τμήματα

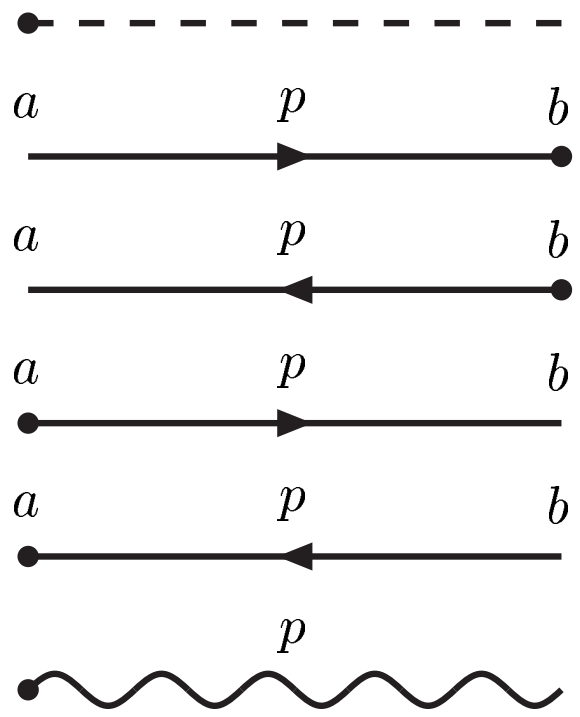

$1$

$u(p)_{ba}$

$\bar{v}(p)_{ab}$

$\bar{u}(p)_{ba}$

$v(p)_{ab}$

$\epsilon_\nu(p)$

**β.** Εσωτερικά τμήματα

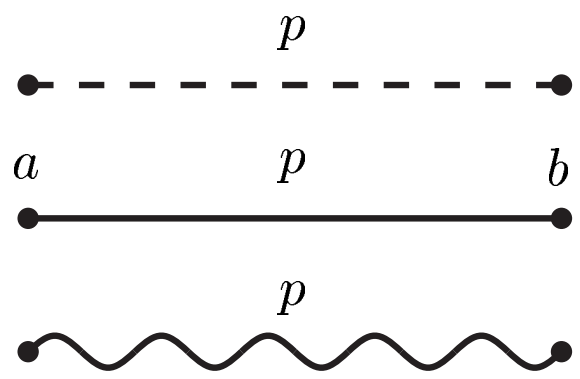

$i/(p^2 - m^2)$

$i\Big((\not{p} + m)/(p^2 - m^2)\Big)_{ab}$

$i(-\eta_{\mu\nu} + p_\mu p_\nu/m^2)/(p^2 - m^2)$

Τα χρησιμοποιούμενα βέλη στα φερμιονικά άκρα αναφέρονται σε ροή φερμιονικού αριθμού και όχι σε φορά ορμής. Η ορμή κατευθύνεται από τα δεξιά προς τα αριστερά. Οι δεικνυόμενοι στο διαδότη και στα άκρα των φερμιονίκων γραμμών δείκτες $a$, $b$ είναι πινακικοί και τρέχουν στις τιμές $1-4$. Προφανώς οι όροι μάζας στο διαδότη των διανυσματικών μποζονίων παραλείπονται όταν αυτά είναι άμαζα.

### B′.3.2 Κόμβοι Feynman

Σε αυτή την ενότητα εκτίθενται οι κόμβοι Feynman που χρησιμοποιήθηκαν για τον υπολογισμό των σημαντικών CAE. Αν και όλοι οι κόμβοι που θα παρουσιαστούν εδώ περιέχονται στις Αν. [12] και [13], η αναδημοσίευσή τους δε θεωρείται πλεονασμός, γιατί αφενός μεν ομαδοποιούνται και εξειδικεύονται για τη μελετούμενη περίπτωση και αφετέρου, με τη δύναμη της είκονας αναδεικνύεται πιο εύγλωττα ο φορμαλισμός που ακολουθήθηκε και υιοθετήθηκε στους πίνακες 5.4, 5.5.

Στις ζεύξεις που υπεισέρχεται slepton, επιλέγεται η παρουσίαση των σχετικών κόμβων με stau, γιατί αυτό παίζει κεντρικότερο ρόλο στον υπολογισμό μας. Η αναγωγή σε κάποιο selectron από τις ελαφρότερες γενεές γίνεται ευκόλως με τις αντικαταστάσεις $c_\tau = 1, s_\tau = 0, m_\tau = 0$. Οι κανόνες για τους κλάδους που φέρουν $\tilde{\tau}_2^*$ λαμβάνονται με αντιστροφή των ορμών στους κλάδους που φέρουν $\tilde{\tau}_2$. Τέλος, το σύμβολο [...] χρησιμοποιείται με διαζευκτική διάθεση.

**α.** stau-tau-bino

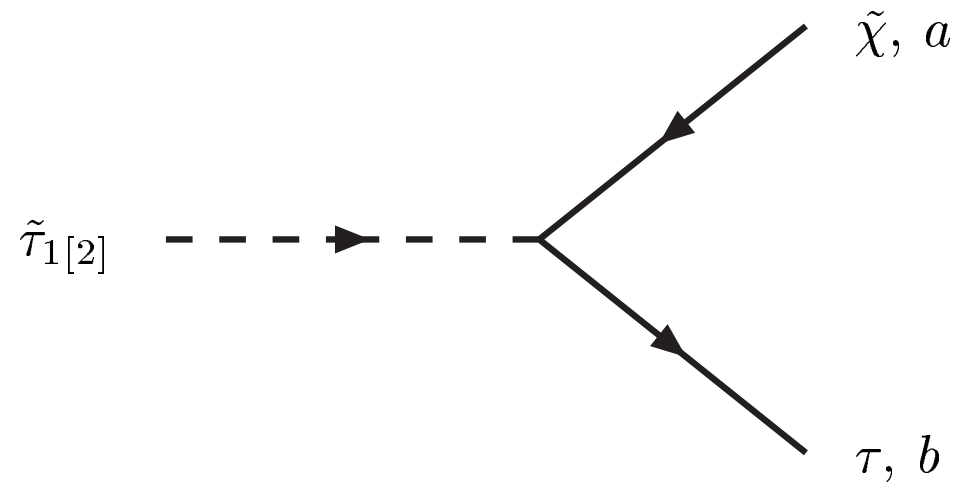

$$i\left(g_{\tau\tilde{\chi}\tilde{\tau}_{1[2]}}\right)_{ba}$$

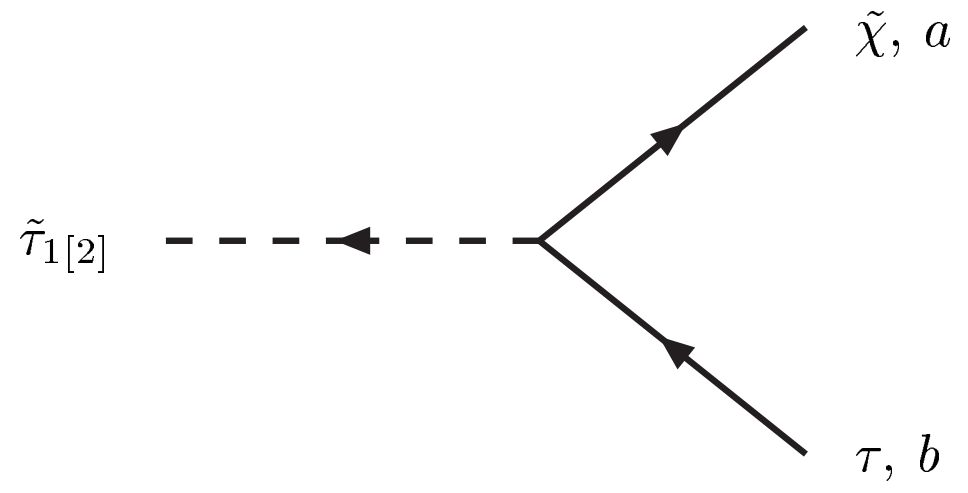

$$i\left(\bar{g}_{\tau\tilde{\chi}\tilde{\tau}_{1[2]}}\right)_{ab}$$

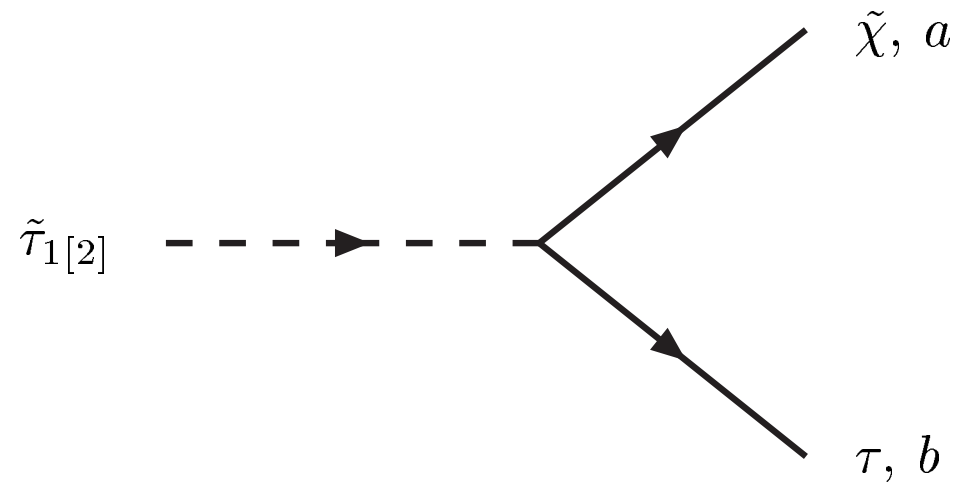

$$i\left(g_{\tau\tilde{\chi}\tilde{\tau}_{1[2]}}C\right)_{ba}$$

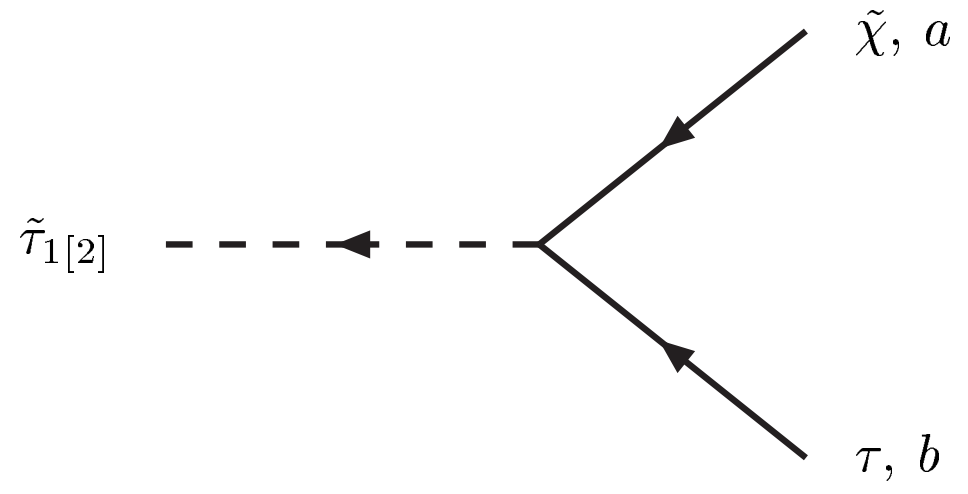

$$-i\left(C^{-1}\bar{g}_{\tau\tilde{\chi}\tilde{\tau}_{1[2]}}\right)_{ab}$$

$$g_{\tau\tilde{\chi}\tilde{\tau}_{1[2]}} = \frac{\sqrt{2}e}{c_W}\Big(Y_R s_\tau[c_\tau]P_L - [+]Y_L c_\tau[s_\tau]P_R\Big)$$
$$\bar{g}_{\tau\tilde{\chi}\tilde{\tau}_{1[2]}} = \frac{\sqrt{2}e}{c_W}\Big(Y_R s_\tau[c_\tau]P_R - [+]Y_L c_\tau[s_\tau]P_L\Big)$$

$$Y_L = -1/2,\ Y_R = 1$$

**β.** gauge boson-tau-tau

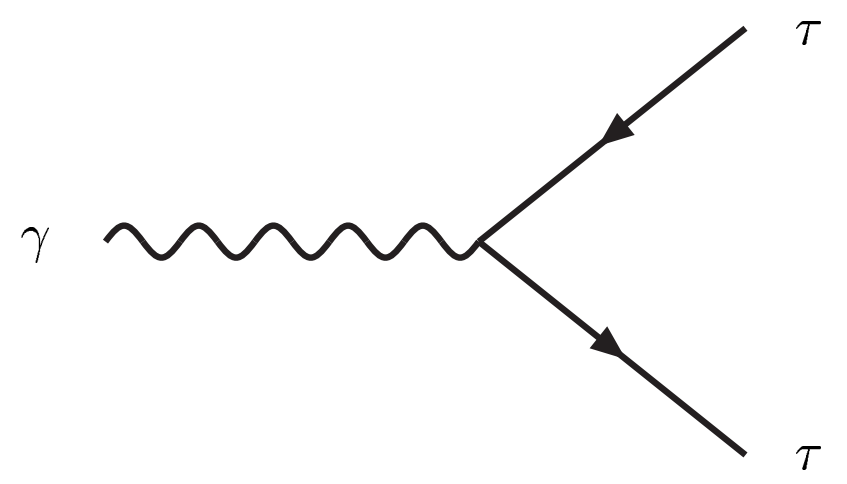

$ie\gamma^{\mu}$

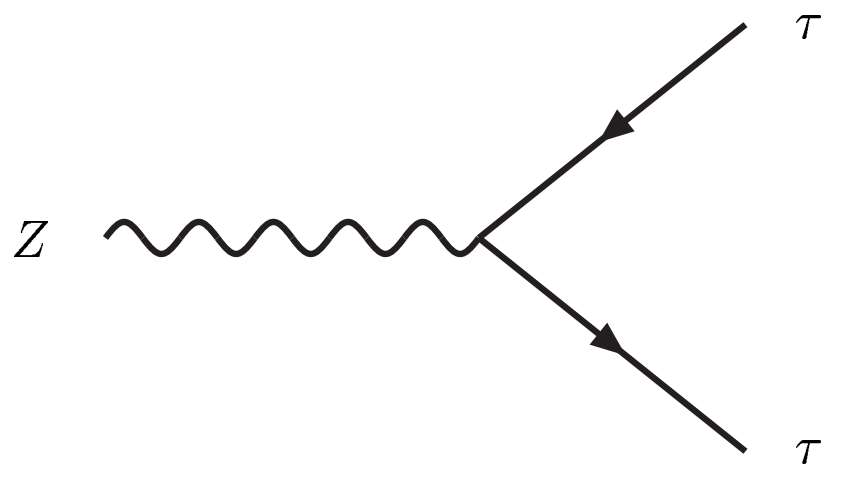

$iZ_{\tau}\gamma^{\mu}$

$$\begin{aligned} Z_{\tau} &= g_Z(L_{\tau}P_L + R_{\tau}P_R), \quad g_Z = g/2c_W \\ L_{\tau} &= 1 - 2s_W^2, \;\; R_{\tau} = -2s_W^2 \end{aligned}$$

**γ.** gauge boson-stau-stau

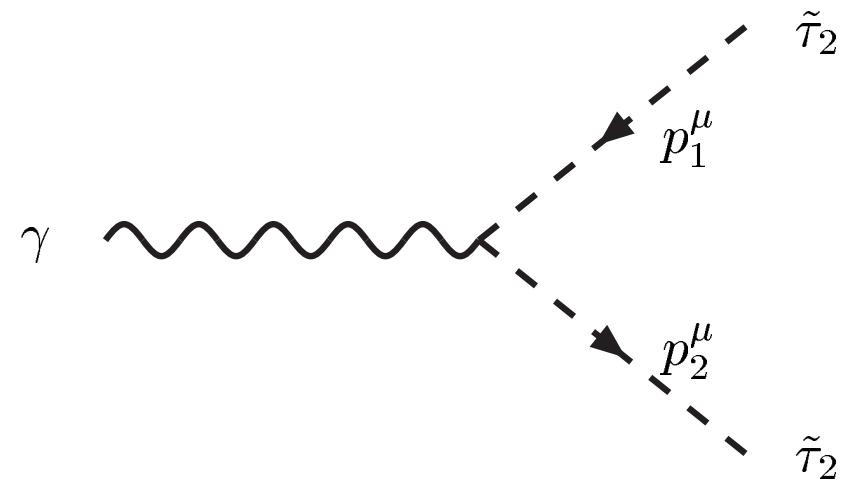

$ie(p_1 + p_2)^{\mu}$

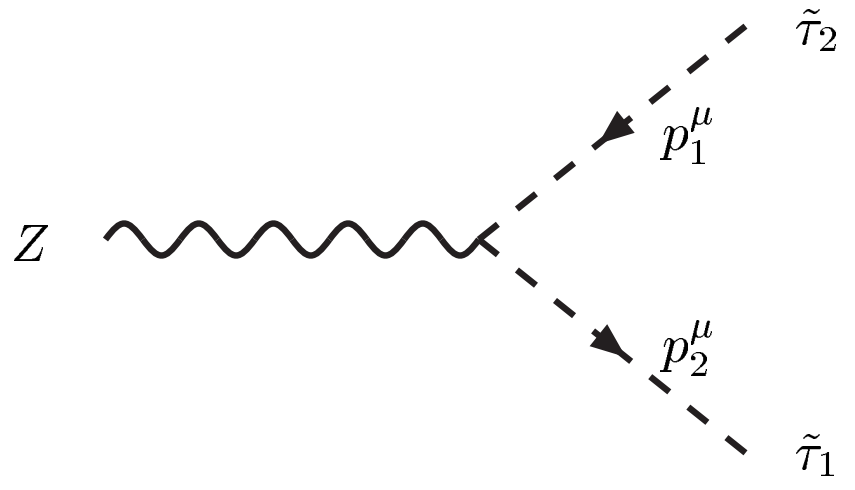

$ig_{\tilde{\tau}_1\tilde{\tau}_2 Z}(p_1 + p_2)^{\mu}$

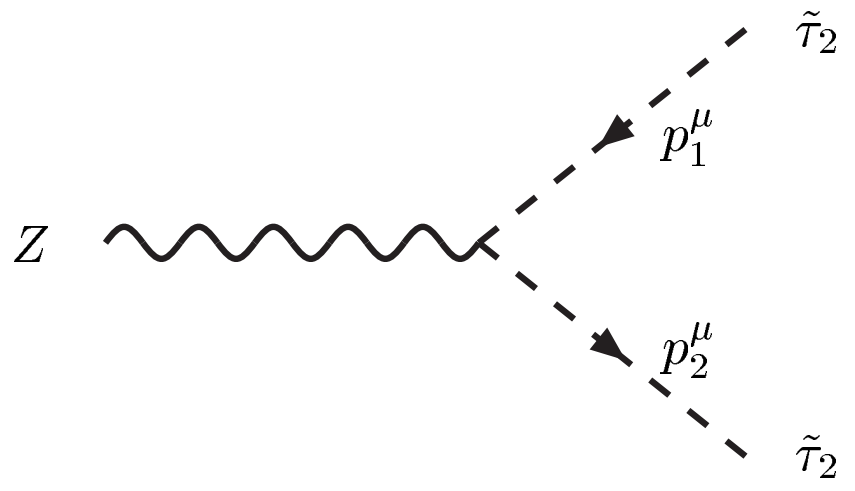

$ig_{\tilde{\tau}_2\tilde{\tau}_2 Z}(p_1 + p_2)^{\mu}$

$$g_{\tilde{\tau}_1\tilde{\tau}_2 Z} = g_Z(-s_{\tau}c_{\tau}), \quad g_{\tilde{\tau}_2\tilde{\tau}_2 Z} = g_Z(s_{\tau}^2 - 2s_W^2)$$

**δ.** gauge boson-gauge boson-stau-stau

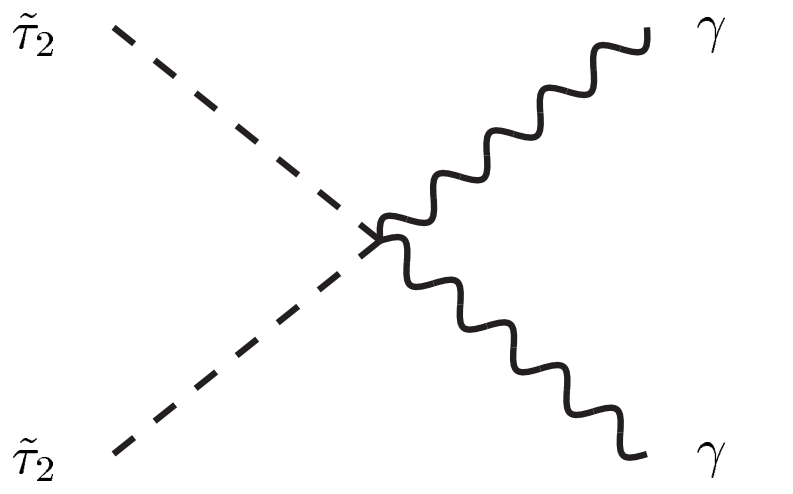

$$2ie^2 g^{\mu\nu}$$

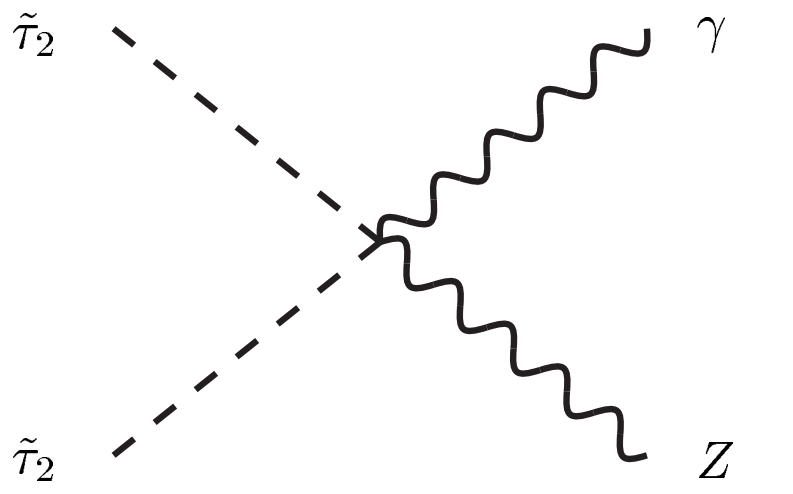

$$2ieg_{\tilde{\tau}_2\tilde{\tau}_2 Z} g^{\mu\nu}$$

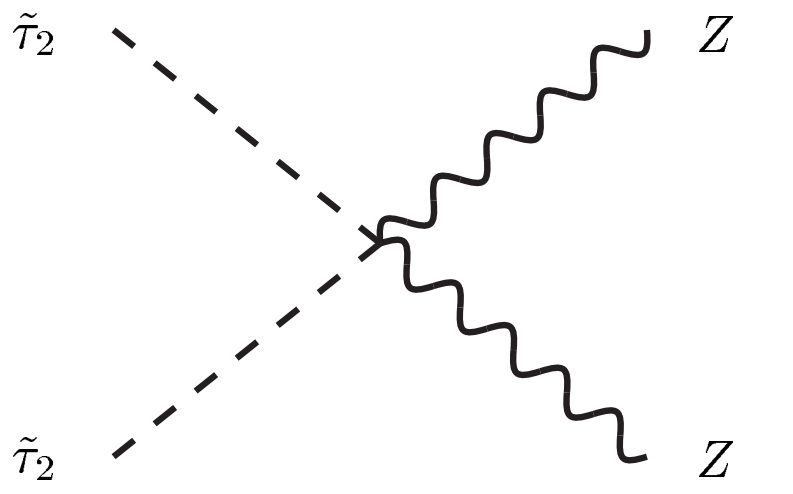

$$2i(g^2_{\tilde{\tau}_2\tilde{\tau}_2 Z} + g^2_{\tilde{\tau}_1\tilde{\tau}_2 Z})g^{\mu\nu}$$

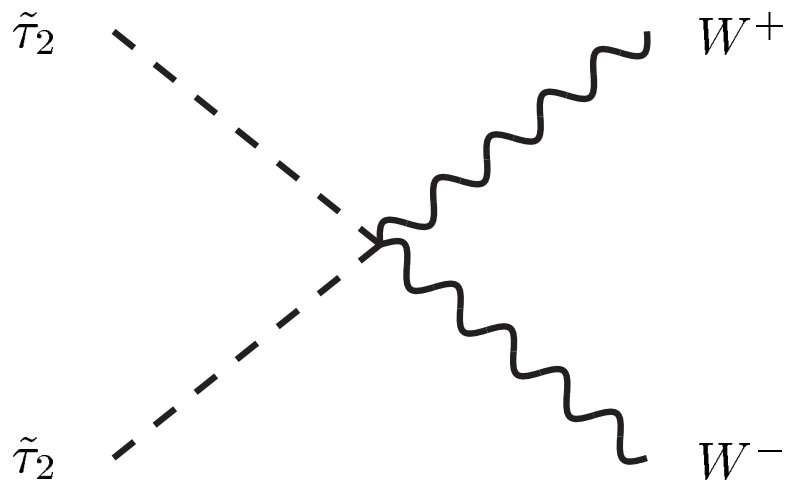

$$ig_{\tilde{\tau}_2\tilde{\tau}_2 W^- W^+} g^{\mu\nu}$$

$$g_{\tilde{\tau}_1\tilde{\tau}_2 Z} = g_Z(-s_\tau c_\tau), \quad g_{\tilde{\tau}_2\tilde{\tau}_2 Z} = g_Z(s_\tau^2 - 2s_W^2), \quad g_{\tilde{\tau}_2\tilde{\tau}_2 W^+ W^-} = g^2 s_\tau^2/2$$

**ε.** gauge boson-higgs-higgs

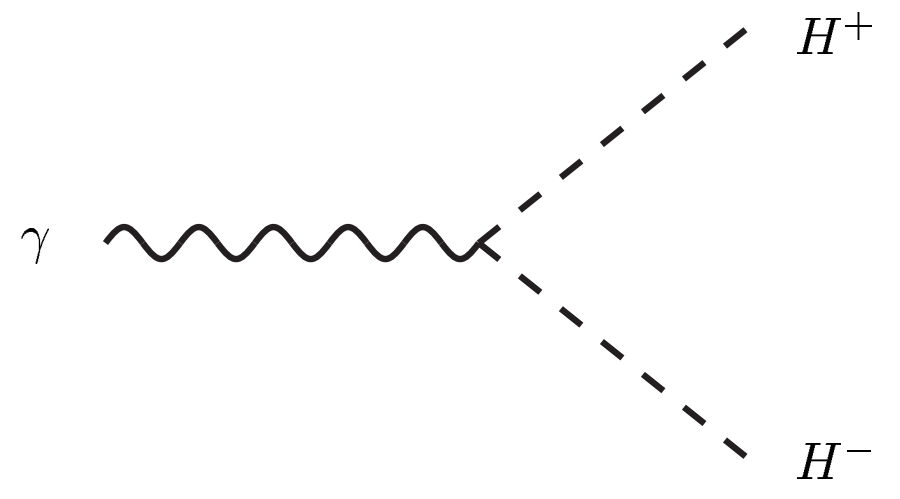

$-ie$

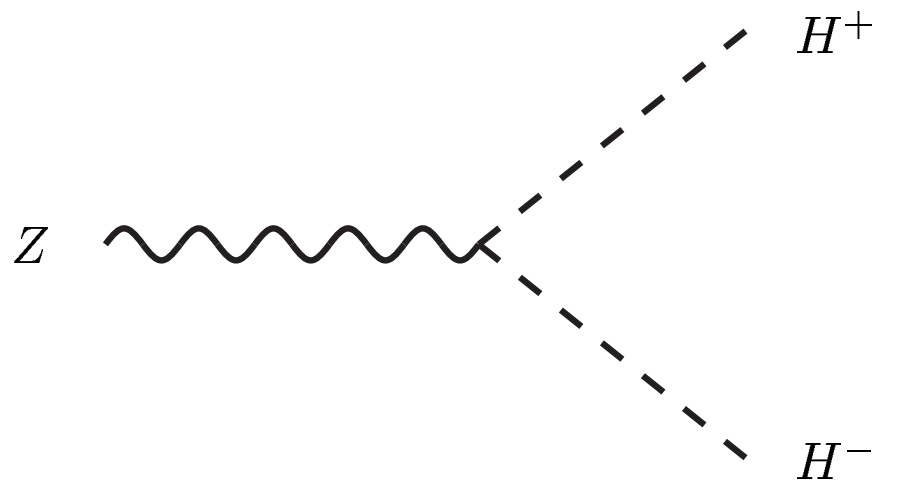

$-ig_Z c_W$

**στ.** gauge boson-gauge boson-gauge boson

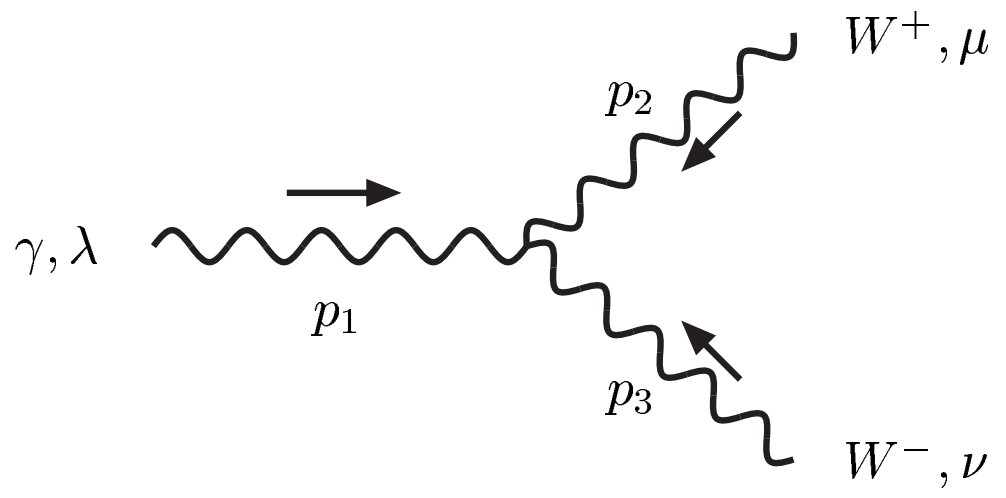

$ieg^{\lambda\mu\nu}_{\gamma W^- W^+}$

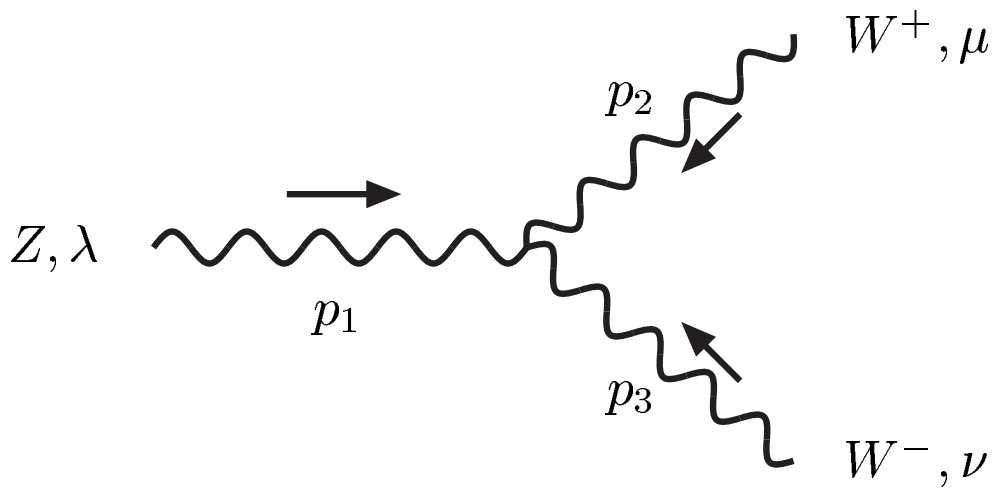

$-igc_W g^{\lambda\mu\nu}_{ZW^- W^+}$

$$g^{\lambda\mu\nu}_{\gamma W^- W^+} = g^{\lambda\mu\nu}_{ZW^- W^+} = \left(\eta^{\lambda\mu}(p_1 - p_2)^\nu + \eta^{\mu\nu}(p_2 - p_3)^\lambda + \eta^{\nu\lambda}(p_3 - p_1)^\mu\right)$$

**ζ.** higgs-stau-stau

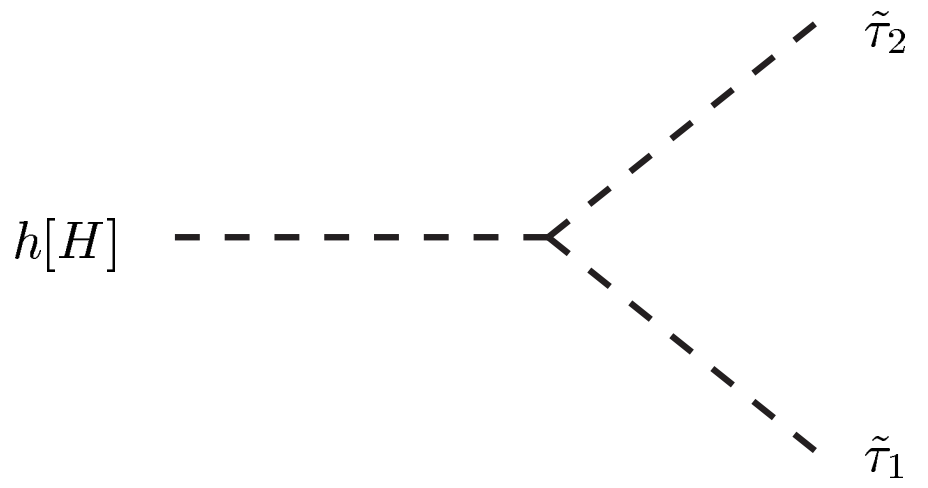

$ig_{h[H]1}$

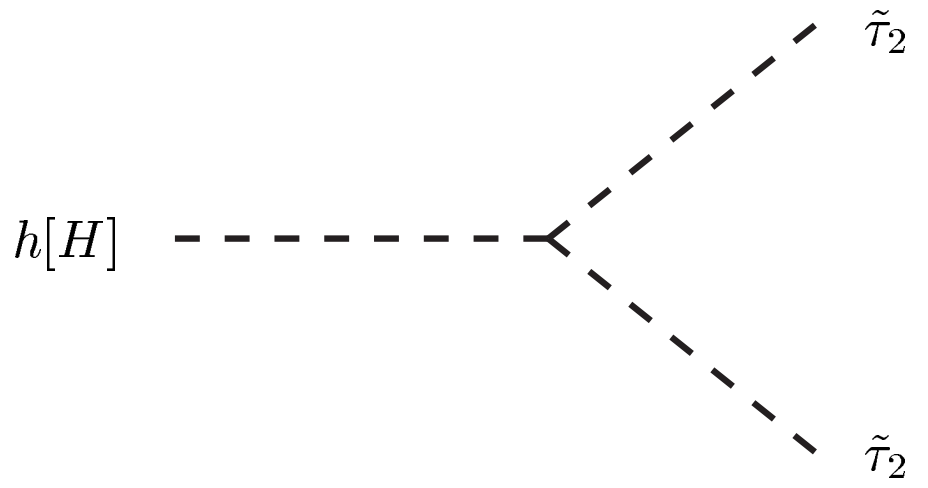

$ig_{h[H]}$

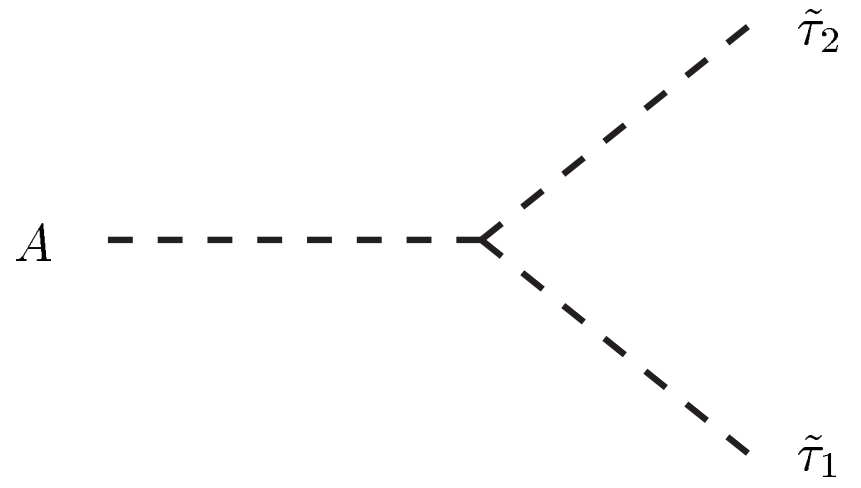

$ig_{A1}$

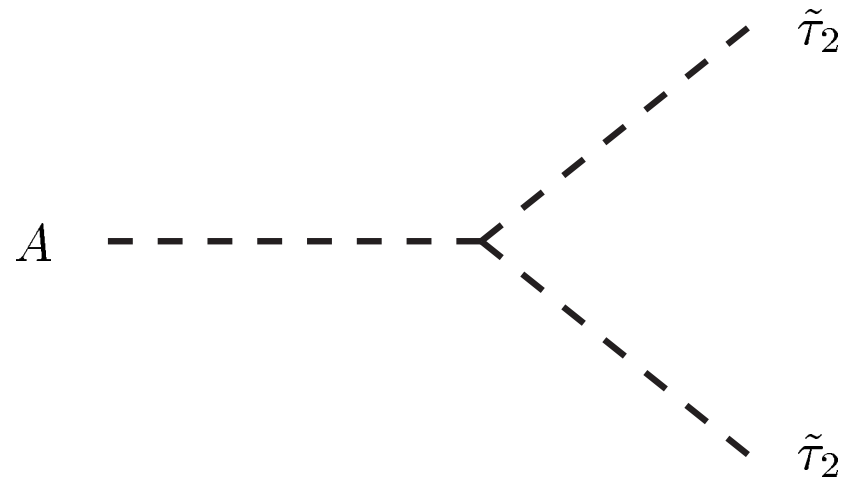

$0$

$$\begin{aligned}
g_{h[H]1}\left(:= g_{\tilde{\tau}_1\tilde{\tau}_2 h[H]}\right) &= g_Z M_Z s_{(\alpha+\beta)}[-c_{(\alpha+\beta)}](L_\tau + R_\tau)s_\tau c_\tau \\
&\quad + \frac{g m_\tau c_{2\tau}}{M_W c_\beta}\Big(A_\tau s_\alpha[-c_\alpha] - \mu c_\alpha[s_\alpha]\Big) \\
g_{h[H]}\left(:= g_{\tilde{\tau}_2\tilde{\tau}_2 h[H]}\right) &= -g_Z M_Z s_{(\alpha+\beta)}[-c_{(\alpha+\beta)}]\left(L_\tau s_\tau^2 - R_\tau c_\tau^2\right) \\
&\quad - \frac{g m_\tau}{M_W c_\beta}\Big(-m_\tau s_\alpha[-c_\alpha] - s_\tau c_\tau\left(A_\tau s_\alpha[-c_\alpha] - \mu c_\alpha[s_\alpha]\right)\Big) \\
g_{A1}\left(:= g_{\tilde{\tau}_1\tilde{\tau}_2 A}\right) &= \frac{g m_\tau}{2M_W}\left(A_\tau \tan\beta - \mu\right)
\end{aligned}$$

**η.** higgs-higgs-stau-stau

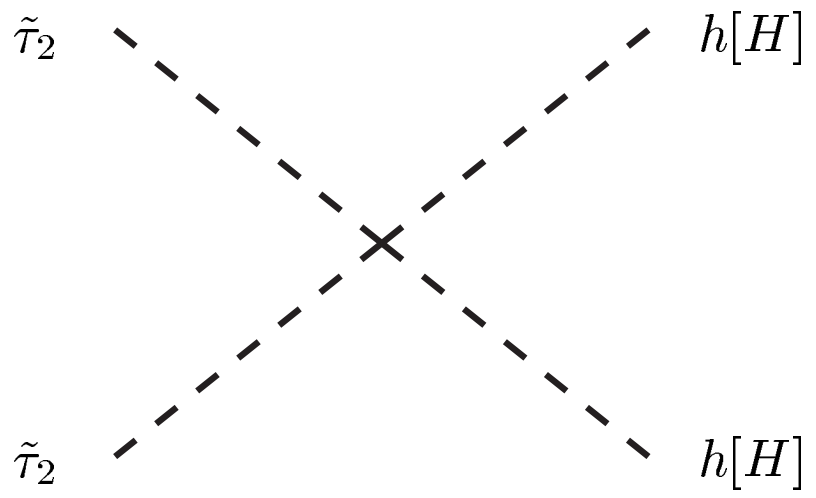

$ig_{\tilde{\tau}_2\tilde{\tau}_2 hh[HH]}$

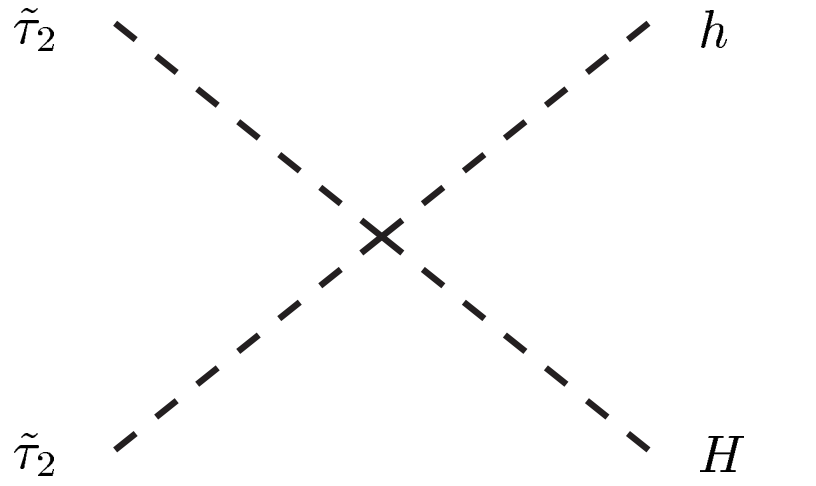

$ig_{\tilde{\tau}_2\tilde{\tau}_2 hH}$

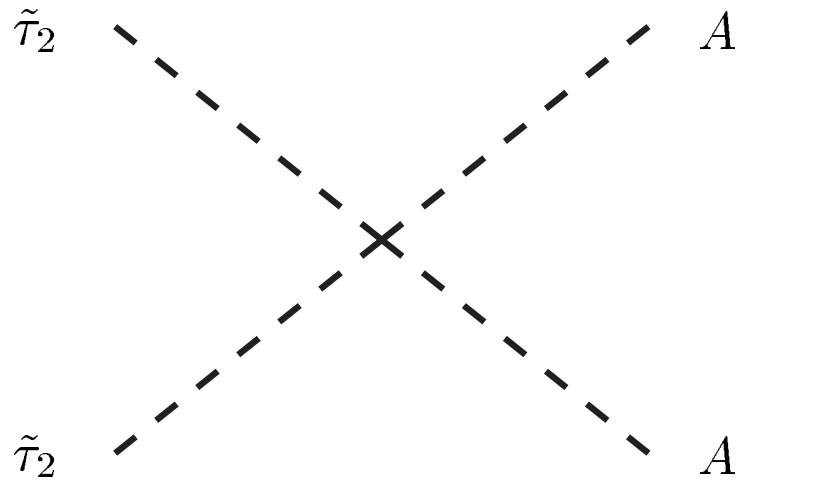

$ig_{\tilde{\tau}_2\tilde{\tau}_2 AA}$

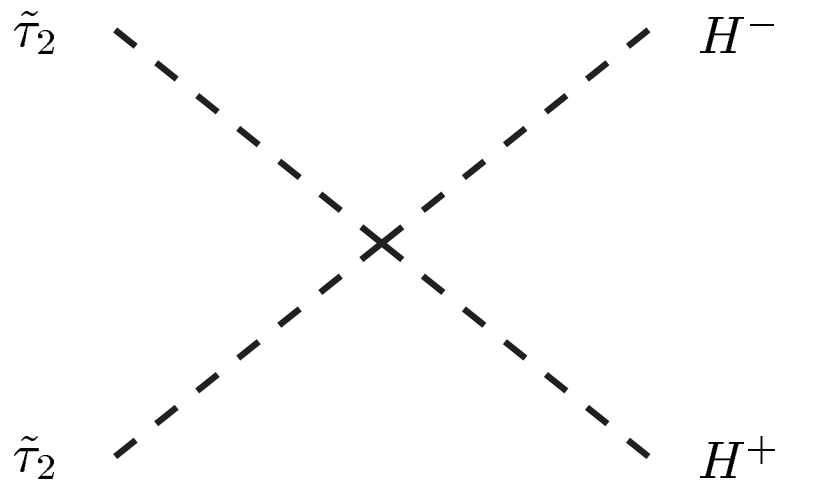

$ig_{\tilde{\tau}_2\tilde{\tau}_2 H^+H^-}$

$$g_{\tilde{\tau}_2\tilde{\tau}_2 hh[HH]} = -[+]g_Z^2 c_{2\alpha}\left(L_\tau s_\tau^2 - R_\tau c_\tau^2\right) - \frac{s_\alpha^2[c_\alpha^2]}{c_\beta^2}\frac{g^2 m_\tau^2}{2M_W^2}$$

$$g_{\tilde{\tau}_2\tilde{\tau}_2 hH} = g^2 s_{2\alpha}\Big( - \frac{L_\tau}{2c_W^2} + \frac{m_\tau^2}{2M_W^2 c_\beta^2}\Big)\frac{s_\tau^2}{2} + g^2 s_{2\alpha}\Big( - \tan^2\theta_W + \frac{m_\tau^2}{2M_W^2 c_\beta^2}\Big)\frac{c_\tau^2}{2}$$

$$g_{\tilde{\tau}_2\tilde{\tau}_2 AA} = -g_Z^2 c_{2\beta}\left(L_\tau s_\tau^2 - R_\tau c_\tau^2\right) - \frac{g^2\tan^2\beta}{2}\frac{m_\tau^2}{M_W^2}$$

$$g_{\tilde{\tau}_2\tilde{\tau}_2 H^+H^-} = g^2 c_{2\beta}\left(\left(1 - \frac{L_\tau}{2c_W^2}\right)\frac{s_\tau^2}{2} - \tan^2\theta_W\frac{c_\tau^2}{2}\right) - g^2\tan^2\beta\,\frac{m_\tau^2}{M_W^2}\frac{c_\tau^2}{2}$$

**ϑ.** higgs-higgs-higgs

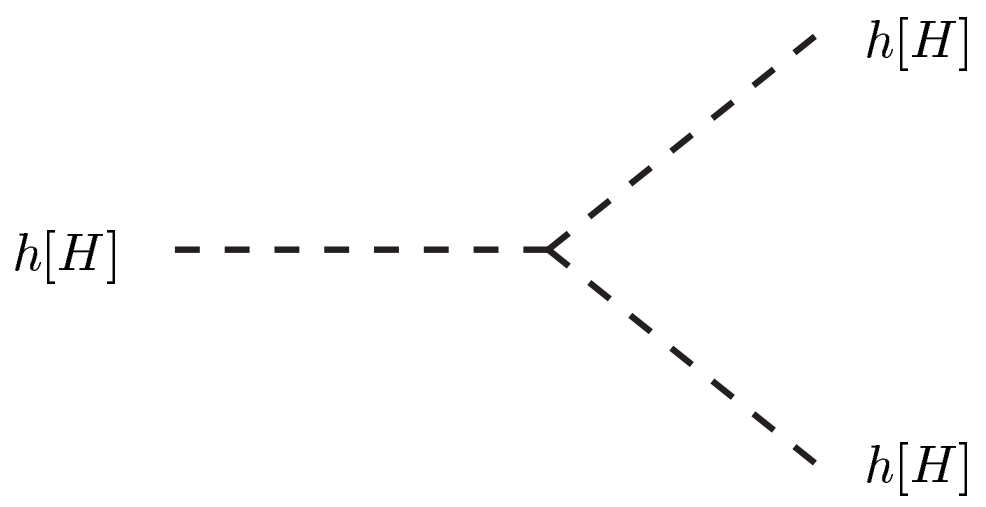

$ig_{hhh[HHH]}$

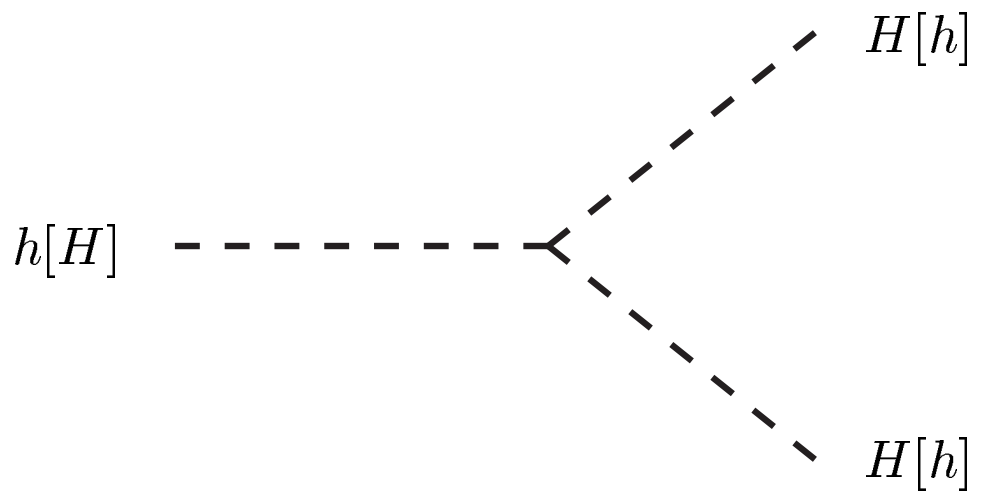

$ig_{hHH[hhH]}$

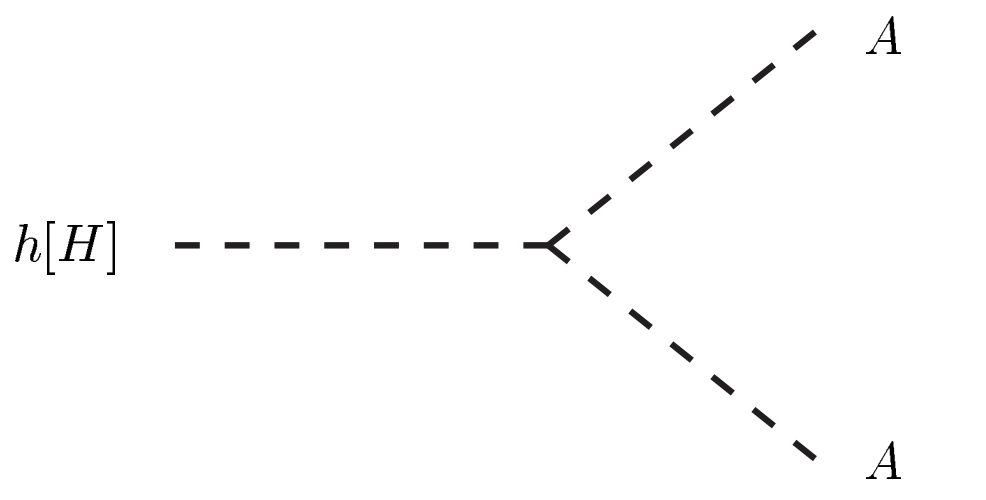

$ig_{h[H]AA}$

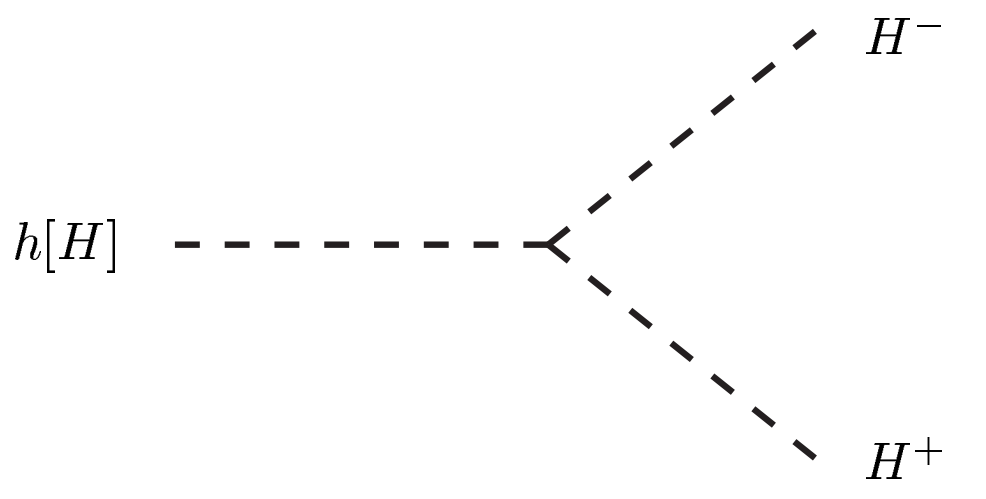

$ig_{h[H]H^+H^-}$

$$
\begin{aligned}
g_{hhh[HHH]} &= -3g_Z M_Z s_{(\alpha+\beta)}[c_{(\alpha+\beta)}]c_{2\alpha} \\
g_{hHH[Hhh]} &= g_Z M_Z\Big(s_{(\alpha+\beta)}[c_{(\alpha+\beta)}]c_{2\alpha} + 2c_{(\alpha+\beta)}[-s_{(\alpha+\beta)}]s_{2\alpha}\Big) \\
g_{h[H]AA} &= -g_Z M_Z s_{(\alpha+\beta)}[-c_{(\alpha+\beta)}]c_{2\beta} \\
g_{h[H]H^+H^-} &= -g\Big(M_W s_{(\beta-\alpha)}[c_{(\beta-\alpha)}] + M_Z s_{(\alpha+\beta)}[-c_{(\alpha+\beta)}]c_{2\beta}/2c_W\Big)
\end{aligned}
$$

**ι.** higgs-gauge boson-gauge boson

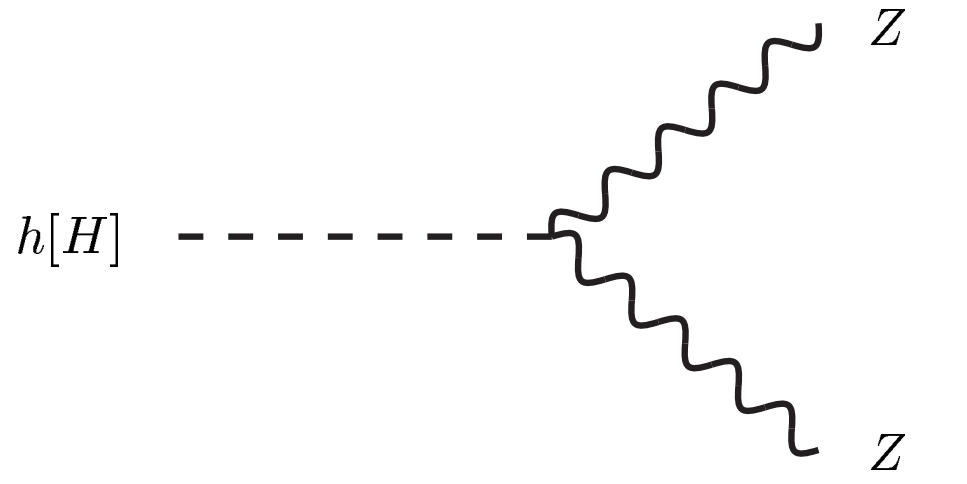

$ig_{h[H]ZZ}g^{\mu\nu}$

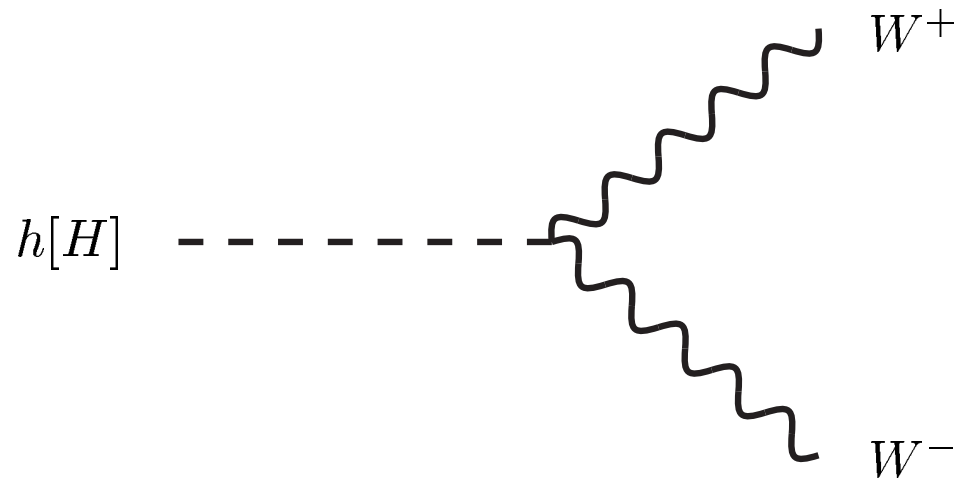

$ig_{h[H]W^+W^-}g^{\mu\nu}$

$$g_{h[H]ZZ} = \frac{g}{c_W M_Z} s_{(\beta-\alpha)}[c_{(\beta-\alpha)}], \quad g_{h[H]W^+W^-} = gM_W s_{(\beta-\alpha)}[c_{(\beta-\alpha)}]$$

**ια.** higgs-fermion-fermion

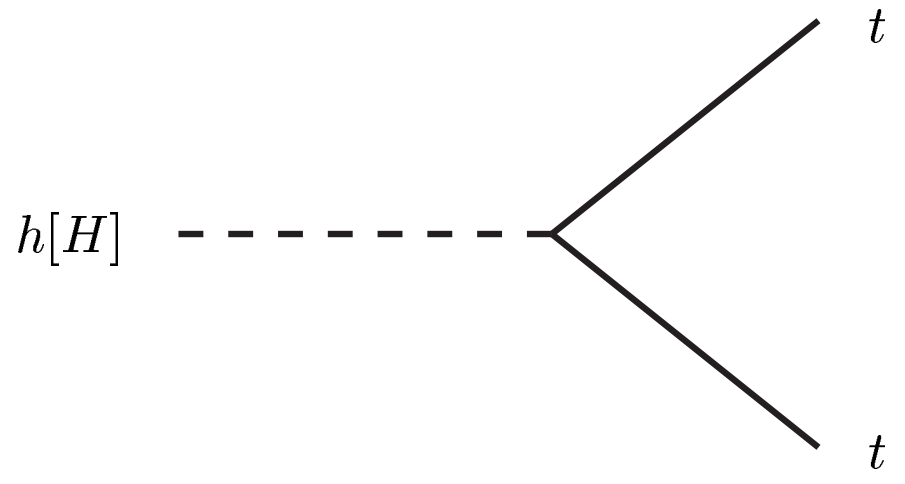

$ig_{h[H]tt}$

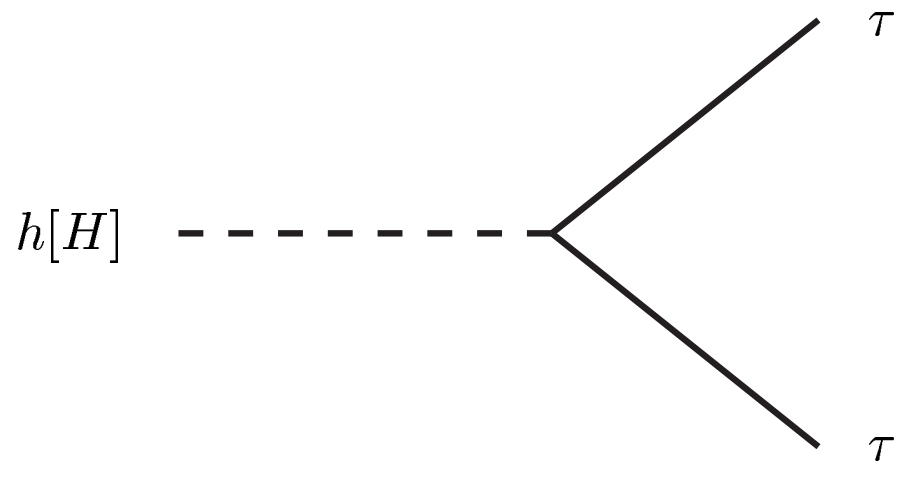

$ig_{h[H]\tau\tau}$

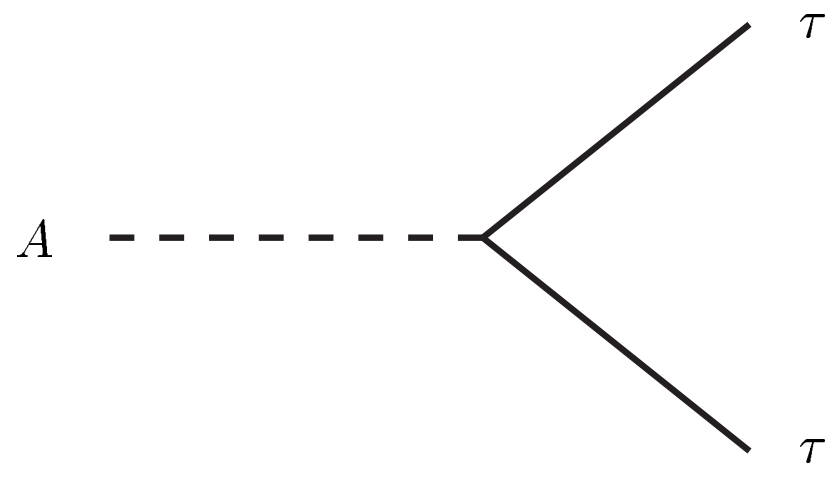

$ig_{A\tau\tau}$

$$g_{h[H]tt} = -\frac{gm_t}{2M_W}\frac{c_\alpha[s_\alpha]}{s_\beta}, \quad g_{h[H]\tau\tau} = \frac{gm_\tau}{2M_W}\frac{s_\alpha[-c_\alpha]}{c_\beta}, \quad g_{A\tau\tau} = -\frac{gm_\tau}{2M_W}\tan\beta$$

## B′.4 Υπολογισμός πλατών αλληλεπίδρασης

Υποδειγματικά και εντελώς ενδεικτικά σκιαγραφείται παρακάτω ο υπολογισμός τριών διαδικασιών από την πληθώρα αυτών που υπολογίσθηκαν κατά τη διαρκεια της εργασίας αυτής. Κριτήρια των επιλογών μας ήταν η αντιπροσωπευτικότητα των παραδειγμάτων, η απλότητα της κατάστρωσης των αρχικών εκφράσεων και η συντομία του τελικού αποτελέσματος.

### B′.4.1 Διαδικασία $\tilde{\chi}\tilde{\tau}_2 \to \tau\gamma$

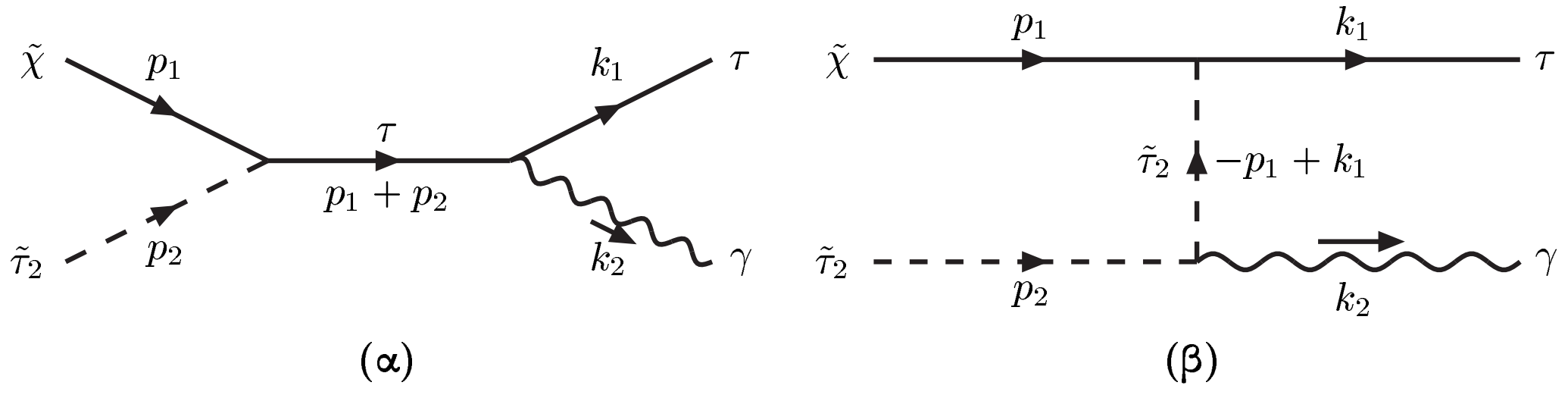

**Σχήμα B′.1:** *Τα διαγράμματα Feynman που συνεισφέρουν στην ενεργό διατομή Συγγενικής Καταστροφής $\tilde{\chi}\tilde{\tau}_2 \to \tau\gamma$ μέσω των διαύλων αλληλεπίδρασης $s$, (α) και $t$, (β)*

Ως εκπρόσωπος των διαδικασιών με $\tilde{\chi} - \tilde{\tau}_2$ στις αρχικές καταστάσεις επιλέγεται ο απλούστερος δύνατος με έξοδο τα σωμάτια $\tau - \gamma$. Βρίσκονται διαδοχικά:

- Το πλάτος αλληλεπίδρασης που αντιστοιχεί στο κανάλι $s$, Σχ. B′.1 (α)

$$\begin{aligned} \mathcal{A}_s &= -i\frac{e}{s}\bar{u}(k_1)\gamma^\mu(\not{p}_1 + \not{p}_2)g_{\tau\tilde{\chi}\tilde{\tau}_2}u(p_1)\epsilon^*_\mu(k_2) \\ \Longrightarrow \mathcal{A}_s^\dagger &= i\frac{e}{s}\bar{u}(p_1)\bar{g}_{\tau\tilde{\chi}\tilde{\tau}_2}(\not{p}_1 + \not{p}_2)\gamma^\nu u(k_1)\epsilon_\nu(k_2) \end{aligned}$$

όπου χρησιμοποιήθηκε η Εξ. (B′.4). Το τετραγωνισμένο πλάτος με χρήση των Εξ. (B′.12) και (B′.17) γράφεται:

$$|\mathcal{A}_s|^2 = -\frac{e^2}{2s^2}\mathrm{Tr}\left[\not{k}_1\gamma^\mu(\not{p}_1 + \not{p}_2)g_{\tau\tilde{\chi}\tilde{\tau}_2}(\not{p}_1 + m_{\tilde{\chi}})\bar{g}_{\tau\tilde{\chi}\tilde{\tau}_2}(\not{p}_1 + \not{p}_2)\gamma_\mu\right] \qquad (B'.18)$$

- Το πλάτος αλληλεπίδρασης που αντιστοιχεί στο κανάλι $t$, Σχ. B′.1 (β)

$$\begin{aligned} \mathcal{A}_t &= -i\frac{e}{(t - m^2_{\tilde{\tau}_2})}\bar{u}(k_1)g_{\tau\tilde{\chi}\tilde{\tau}_2}u(p_1)(-p_1 + p_2 + k_1)^\mu\epsilon^*_\mu(k_2) \\ \Longrightarrow \mathcal{A}_t^\dagger &= i\frac{e}{(t - m^2_{\tilde{\tau}_2})}\bar{u}(p_1)\bar{g}_{\tau\tilde{\chi}\tilde{\tau}_2}u(k_1)(-p_1 + p_2 + k_1)^\nu\epsilon_\nu(k_2) \end{aligned}$$

όποτε παρομοίως με το προηγούμενα, εξάγεται το τετραγωνισμένο πλάτος:

$$|\mathcal{A}_t|^2 = -\frac{e^2}{(t - m^2_{\tilde{\tau}_2})^2}(-p_1 + p_2 + k_1)^2\mathrm{Tr}\left[\not{k}_1 g_{\tau\tilde{\chi}\tilde{\tau}_2}(\not{p}_1 + m_{\tilde{\chi}})\bar{g}_{\tau\tilde{\chi}\tilde{\tau}_2}\right] \qquad (B'.19)$$

- Ο όρος ανάμιξης των καναλιών $s$ και $t$, που με αξιοποίηση των Εξ. (B′.4), (B′.12) και (B′.17) γράφεται:

$$2\mathcal{A}_s\mathcal{A}_t^\dagger = -2\frac{e^2}{2s(t - m^2_{\tilde{\tau}_2})}\mathrm{Tr}\left[\not{k}_1\gamma^\mu(\not{p}_1 + \not{p}_2)g_{\tau\tilde{\chi}\tilde{\tau}_2}(\not{p}_1 + m_{\tilde{\chi}})\bar{g}_{\tau\tilde{\chi}\tilde{\tau}_2}(-p_1 + p_2 + k_1)_\mu\right] \qquad (B'.20)$$

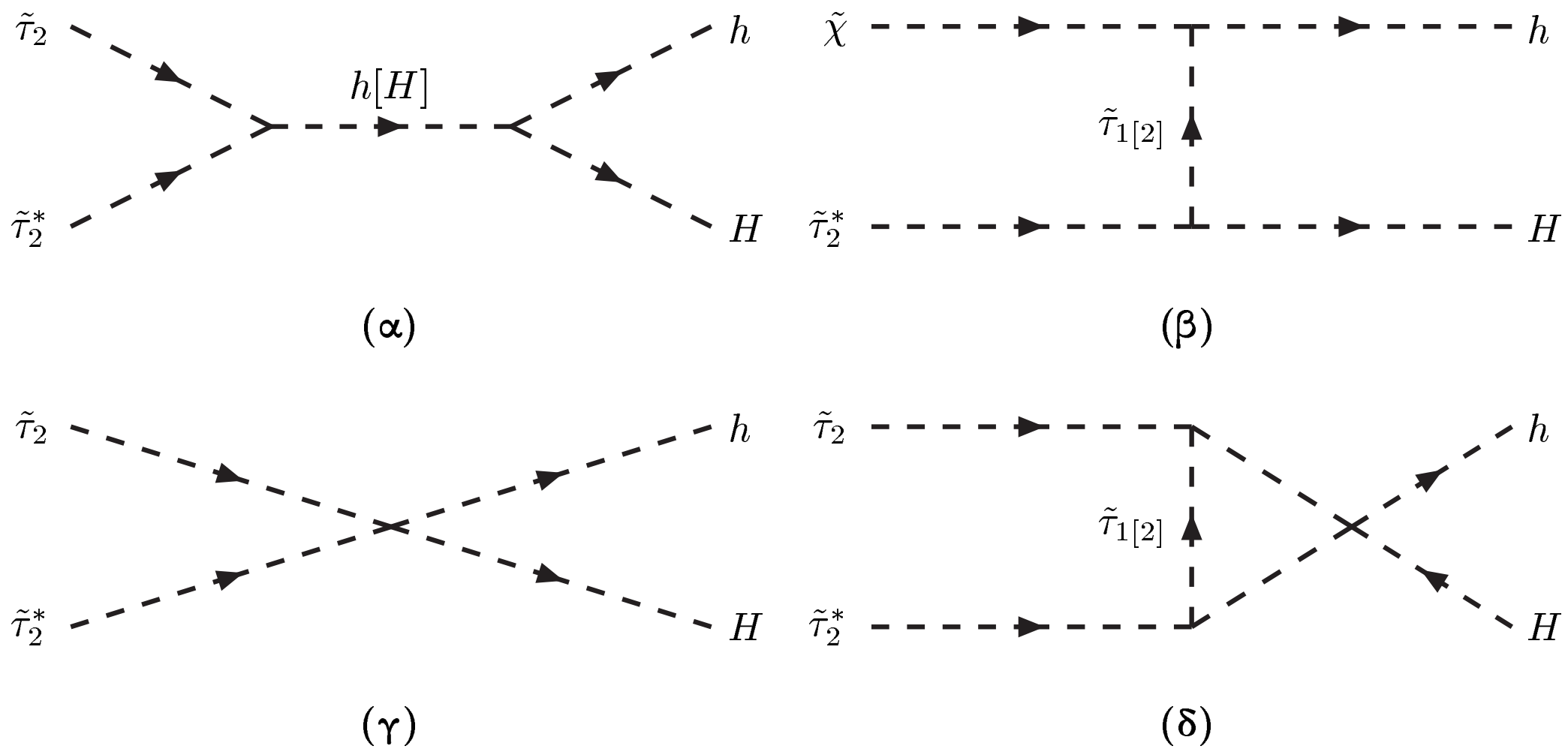

**Σχήμα Β΄.2:** *Τα διαγράμματα Feynman που συνεισφέρουν στην ενεργό διατομή Συγγενικής Καταστροφής* $\tilde{\tau}_2\tilde{\tau}_2^* \to hH$ *μέσω των διαύλων αλληλεπίδρασης s, (α), t, (β), u, (δ), c, (γ)*

Ο παράγοντας 1/2 που φαίνεται στα τετραγωνισμένα πλάτη οφείλεται στη μέση τιμή που λαμβάνεται πάνω στα αρχικά σπίν. Αντιθέτως, τα τελικά σπίν απλώς αθροίζονται. Το *FeynCalc* αναλαμβάνει να υπολογίσει τα ίχνη των Εξ. (Β΄.18), (Β΄.19) και (Β΄.20), να εισάγει τις μεταβλητές Mandelstam και να αθροίσει τα επιμέρους αποτελέσματα, οπότε το τελικό εξαγόμενο, είναι:

$$
\begin{aligned}
|\mathcal{A}|^2 &= \frac{2e^4}{c_W^2}(Y_R^2 c_\tau^2 + Y_L^2 s_\tau^2)\Bigg[ m_{\tilde{\chi}}^4 (m_{\tilde{\tau}_2}^2 - t) \\
&+ m_{\tilde{\chi}}^2 \Big( -3m_{\tilde{\tau}_2}^4 + 2m_{\tilde{\tau}_2}^2 t + t(2s+t) + t^2 u - 2m_{\tilde{\tau}_2}^2 t(s+t+u) + m_{\tilde{\tau}_2}^4 (2t+u)\Big)\Bigg]
\end{aligned}
$$

Η ορθότητα ειδικά αυτού του υπολογισμού έχει επαληθευτεί και σε ιδιωτική επικοινωνία με τον καθηγητή M. Drees μέσα στα πλαίσια σύγκρισης αποτελεσμάτων των Αν. [44] και [57].

## Β΄.4.2 Διαδικασία $\tilde{\tau}_2\tilde{\tau}_2^* \to hH$

Η διαδικασία $\tilde{\tau}_2\tilde{\tau}_2^* \to hH$ αποτελεί τη βάση της γενικευμένης φόρμουλας που δόθηκε για τις διαδικασίες με σωμάτια higgs στις τελικές καταστάσεις. Λόγω των βαθμωτών εξωτερικών ποδιών, στα πλάτη υπεισέρχονται μόνο ζεύξεις και διαδότες. Συνακόλουθα, και οι ορμές μπορούν να μην τεθούν στα διαγράμματα του Σχ. Β΄.2, γιατί είναι προφανείς οι συνεισφορές σε κάθε κανάλι. Μικρή προσοχή χρειάζεται στο πρόσημο του διαγράμματος επαφής που πρέπει να είναι διαφορετικό από τα αλλά, γιατί σε αυτό, εμφανίζεται μόνο ένα $i$ ενώ στά άλλα, εμφανίζονται τρία $i$, ένα από το διαδότη και δύο από τις κορυφές, οπότε το συνολικό πρόσημο είναι $-i$.

Δίνεται απευθείας το τετραγωνισμένο πλάτος, που είναι:

$$
|\mathcal{A}|^2 = \left| \frac{g_h g_{hhH}}{s - m_h^2} + \frac{g_H g_{hHH}}{s - m_H^2} + \frac{g_{h1} g_{H1}}{t - m_{\tilde{\tau}_1}^2} + \frac{g_h g_H}{t - m_{\tilde{\tau}_2}^2} - g_{\tilde{\tau}_2\tilde{\tau}_2 hH} + \frac{g_{h1} g_{H1}}{u - m_{\tilde{\tau}_1}^2} + \frac{g_h g_H}{u - m_{\tilde{\tau}_2}^2} \right|^2
$$

Κάθε όρος του προηγούμενου αθροίσματος αντιστοιχεί στο πλάτος καθενός από τα διαγράμματα του Σχ. Β΄.2 (α)-(δ). Το ίδιο τετραγωνισμένο πλάτος χωρίς τα κανάλια $t(\tilde{\tau}_1), u(\tilde{\tau}_{1[2]})$ δίνεται και στην Αν. [64] καλύπτοντας μισή και πλέον σελίδα, με φορμαλισμό προφανώς πιο δυσνόητο από τον ακολουθούμενο εδώ.

### Β΄.4.3 Διαδικασία $\tilde{\tau}_2\tilde{e}_R \to \tau e$

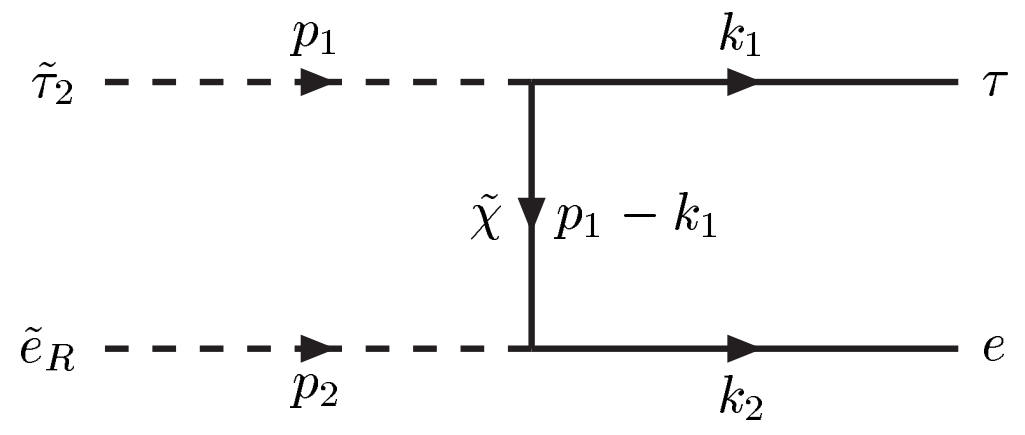

**Σχήμα Β΄.3:** *Το διάγραμματα Feynman που συνεισφέρει στην ενεργό διατομή Συγγενικής Καταστροφής $\tilde{\tau}_2\tilde{e}_R \to \tau e$ μέσω του διαύλου αλληλεπίδρασης $t$*

Το τελευταίο παράδειγμα που επιλέγεται, εμπεριέχει sleptons των ελαφρότερων γενεών. Υπάρχει, επομένως, ζεύξη που προκύπτει από τις εκτεθημένες στο Εδ. Β΄.3.2 με τις αντιστοιχίες που εκεί αναφέρονται. Συγκεκριμένα,

$$g_{e\tilde{\chi}\tilde{e}_R} = Y_R P_L \quad \text{και} \quad \bar{g}_{e\tilde{\chi}\tilde{e}_R} = Y_R P_R \tag{Β΄.21}$$

Εξοπλισμένοι και με αυτό το εφόδιο, μπορούμε να επιτύχουμε την έκφραση για το πλάτος της αλληλεπίδρασης που εικονίζεται και στο Σχ. Β΄.3:

$$\begin{aligned} \mathcal{A} &= \bar{u}(k_2) i g_{e\tilde{\chi}\tilde{e}_R} \frac{i(\not{p}_1 - \not{k}_1 + m_{\tilde{\chi}})}{t - m_{\tilde{\chi}}^2} i g_{\tau\tilde{\chi}\tilde{\tau}_2} C \bar{u}^T(k_2) \\ \Longrightarrow \mathcal{A}^* &= i\bar{v}(k_1) \bar{g}_{\tau\tilde{\chi}\tilde{\tau}_2} \frac{(\not{p}_1 - \not{k}_1 + m_{\tilde{\chi}})}{t - m_{\tilde{\chi}}^2} \bar{g}_{e\tilde{\chi}\tilde{e}_R} u(k_2) \end{aligned}$$

όπου χρησιμοποιήθηκαν οι Εξ. (Β΄.4) και (Β΄.9). Τελικά, για το τετραγωνισμένο πλάτος, λαμβάνεται:

$$|\mathcal{A}|^2 = \frac{1}{(t - m_{\tilde{\chi}}^2)^2} \mathrm{Tr}\left[\bar{g}_{\tau\tilde{\chi}\tilde{\tau}_2}(\not{p}_1 - \not{k}_1 + m_{\tilde{\chi}})\bar{g}_{e\tilde{\chi}\tilde{e}_R}\not{k}_2 g_{e\tilde{\chi}\tilde{e}_R}(\not{p}_1 - \not{k}_1 + m_{\tilde{\chi}}) g_{\tau\tilde{\chi}\tilde{\tau}_2}\not{k}_1\right] \tag{Β΄.22}$$

Το *FeynCalc* αναλαμβάνει να υπολογίσει το ίχνος της προηγούμενης έκφρασης και να εισάγει τις μεταβλητές Mandelstam, οπότε το τελικό αποτέλεσμα είναι:

$$|\mathcal{A}|^2 = \frac{e^4 Y_R^4}{c_W^4} \; \frac{-s_\tau^2 m_{\tilde{e}_R}^2 m_{\tilde{\tau}_2}^2 + c_\tau^2 m_{\tilde{\chi}}^2 s + s_\tau^2 tu}{(t - m_{\tilde{\chi}}^2)^2} \tag{Β΄.23}$$

# Παράρτημα Γ΄

# Ακρονύμια

Με στόχο την αποφυγή παρερμηνειών και τη σαφέστερη διατύπωση, υιοθετήθηκαν κάποια συνθηματικά ακρονύμια σε όλη την πορεία της εξέλιξης του κειμένου. Τα περισσότερα από αυτά είναι εμπνευσμένα από τα αρχικά των αντίστοιχων λατινικών λέξεων. Θεωρείται, λοιπόν επιβεβλημένη η καταγραφή τους μαζί με τις λέξεις καταγωγής τους και τη μετάφραση αυτών, η οποία μερικές φορές ίσως είναι λίγο εξεζητημένη.

- ANE : Annihilation Effects (Διαδικασίες Αλληλοκαταστροφής)
- EWS : Electroweak Scale (Ηλεκτρασθενής κλίμακα ενέργειας)
- BES : Bose Einsten Statistics (Στατιστική Bose Einsten)
- CAE : Coannihilation Effects (Διαδικασίες Συγγενικής Καταστροφής)
- CBR : Cosmic Background Radiation (Κοσμική Ακτινοβολία Υποβάθρου )
- [C, H]DM : [Cold, Hot] Dark Matter ([Ψυχρή, Θερμή] Σκοτεινή Ύλη)
- CRD : Cosmic Relic Density (Κοσμική Εναπομένουσα Πυκνότητα)
- cst. : constant (σταθερά)
- CRF : Cosmic Rest Frame (Κοσμικό Σύστημα Ηρεμίας)
- COBE : Cosmic Background Explorer (Κοσμικός Ανιχνευτής Υποβάθρου)
- FDS : Fermi Dirac Statistics (Στατιστική Fermi Dirac)
- GUT : Grand Unification (Μεγάλη Ενοποίηση)
- h. c : hermitian conjugate (Ερμιτιανό Συζυγές)
- [N]LSP : [Next to] Lightest Supersymmetric Particle
  ([Πρώτο μετά το] Ελαφρότατο Υπερσυμμετρικό Σωμάτιο)
- [N]LO : [Next to] Leading Order ([Πρώτη μετά την] Κύρια Τάξη)
- MBS : Maxwell Boltzamann Statistics (Στατιστική Maxwell Boltzamann)
- MD : Matter Domination (Επικράτηση Ύλης)
- [MS]SM : [Minimal Supersymmetric] Standard Model
  ([Ελάχιστα Υπερσυμμετρικό] Καθιερωμένο Πρότυπο)
- RD : Radiation Domination (Επικράτηση Ακτινοβολίας)
- RGE : Renormalization Group Equations (Εξισώσεις Ομάδας Επανακανονικοποίησης)
- SBB : Standard Big-Bang Model (Καθιερωμένο Κοσμολογικό Πρότυπο)
- SBT : Soft Breaking Terms (Όροι ασθενούς παραβίασης της SUSY)
- [N]SE : [Non] Standard Embending ([Μη] Καθιερωμένη Εμφύτευση)
- SUSY : Supersymmetry (Υπερσυμμετρία)
- SSB : Spontaneus Symmetry Breaking (Αυθόρμητη κατάρρευση της Συμμετρίας)
- QE[C]D : Quantum Electro-[Chromo-]dynamics (Κβαντική Ηλεκτρο-[Χρωμο-]δυναμική )
- VEV : Vacuum Expectation Value (Αναμενόμενη Τιμή Κενού)

Από τους ελληνικούς όρους λιγότερα ακρονύμια προέρχονται. Παρακάτω παρατίθενται αυτά και κάποιες χρήσιμες συντμίσεις:

- Αν. : Αναφορά
- Εδ. : Εδάφιο
- Εξ. : Εξίσωση
- Σχ. : Σχήμα
- ΘΔΙ : Θερμοδυναμική Ισορροπία
- ΙΔΚ : Ιδιοκατάσταση
- ΙΔΤ : Ιδιοτιμή